\documentstyle[preprint,aps,harvard,floats,epsfig]{revtex}
\tightenlines

\newcommand{\beq}{\begin{equation}}
\newcommand{\eeq}{\end{equation}}
\citationstyle{dcu}
\newcommand{\lsim}{\mbox{ \raisebox{-1.0ex}{$\stackrel{\textstyle <}
{\textstyle \sim}$ }}}

\newcommand{\mueg}{$\mu^{+} \rightarrow e^{+} \gamma$~}
\newcommand{\muegg}{$\mu^{+} \rightarrow e^{+} \gamma\gamma$~}
\newcommand{\meee}{$\mu^{+} \rightarrow e^{+}e^{+}e^{-}$~}
\newcommand{\mumu}{$\rm{Mu}-\overline{\rm Mu}$~}
\newcommand{\lagr}{\cal{L}}
\newcommand{\muec}{$\mu^{-}$$-$$e^{-}$~}
\newcommand{\mupc}{$\mu^{-}$$-$$e^{+}$~}

\newcommand{\radmu}{$\mu^{\pm}\rightarrow
e^{\pm}\nu\overline{\nu}\gamma$~}
\newcommand{\muenng}{$\mu^{+}\rightarrow
e^{+}\nu_e\overline{\nu}_{\mu}\gamma$~}
\newcommand{\muennee}{$\mu^{+}\rightarrow
e^{+}e^{+}e^{-}\nu_e\overline{\nu}_{\mu}$~}
\newcommand{\muenn}{$\mu^{+}\rightarrow e^{+}\nu_e\overline{\nu}_{\mu}$~}

\citationstyle{dcu}
\begin{document}

\preprint{\vbox{\baselineskip=3ex
                \hbox{KEK preprint 99-69}
                \hbox{KEK-TH-639}
                \hbox{August 1999}
                \hbox{H}}
                \bigskip}

\title{Muon Decay and Physics Beyond the Standard Model}
\author{Yoshitaka Kuno\thanks{yoshitaka.kuno@kek.jp} 
and Yasuhiro Okada\thanks{yasuhiro.okada@kek.jp}}
\address{Institute of Particle and Nuclear Studies (IPNS),\\
         High Energy Accelerator Research Organization (KEK), \\
         Tsukuba, Ibaraki, Japan 305-0801}
\maketitle

\begin{abstract}
This article reviews the current theoretical and experimental status
of the field of muon decay and its potential to search for new physics
beyond the Standard Model. The importance of rare muon processes with
lepton flavor violation is highly stressed, together with precision
measurements of normal muon decay. Recent up-to-date motivations of
lepton flavor violation based on supersymmetric models, in particular
supersymmetric grand unified theories, are described along with other
theoretical models. Future prospects of experiments and muon sources
of high intensity for further progress in this field are also
discussed.
\end{abstract}

\vspace{2cm}
\centerline{submitted to Review of Modern Physics.}

\newpage
\tableofcontents




\markright{Ekert 8/4/98 Revised}

\section{Introduction}

The muon was discovered in 1937 by Neddermeyer and Anderson
\cite{nedd37} in cosmic rays. The mass of the muon was found to be
about 200 times the mass of the electron. The discovery was made just
after Yukawa postulated the existence of the $\pi$ meson, a force
carrier of the strong interaction, in 1935 \cite{yuka35}. But, it was
demonstrated in 1947 that the muon did not react through the strong
interaction, and thus it could not be the Yukawa $\pi$ meson
\cite{conv47}. The famous comment by Rabi ``Who ordered that ?''
might indicate how puzzling the existence of a new lepton was at that
time. By then, it was known that the muon decays into an electron and
a neutral particle. It was thought that if the muon were a heavy
electron, it would also decay into an electron and a $\gamma$-ray. The
first search for \mueg was made by Hincks and Pontecorvo in 1947 using
cosmic-ray muons~\cite{pont47}. Its negative result set an upper limit
of less than 10\%. This was the beginning of the search for lepton
flavor violation, {\em i.e.} violation of the lepton number
conservation for each generation. In 1948, the continuous spectrum of
electrons was established, suggesting a three-body decay with an
electron accompanied by two neutral particles~\cite{stei48}.  Soon
after, the search for a neutrinoless muon nuclear capture process
($\mu^{-}N\rightarrow e^{-}N$, where $N$ is a nucleus capturing the
muon) was also carried out, but with a negative
result~\cite{laga52}. Such searches were significantly improved when
muons became artificially produced at accelerators. In 1955, the upper
limits of $B(\mu\rightarrow e\gamma)< 2\times 10^{-5}$~\cite{loko55}
and $B(\mu^{-}Cu\rightarrow e^{-}Cu) < 5 \times 10^{-4}$ \cite{stei55}
were set at the Columbia University Nevis cyclotron.

After the discovery of parity violation, it was suggested that the
weak interaction takes place through the exchange of charged
intermediate vector bosons \cite{feyn58}. In 1958, Feinberg pointed
out that the intermediate vector boson, if it exists, would lead to
\mueg at a branching ratio of $10^{-4}$~\cite{fein58}. The absence of
any experimental observation of the \mueg process with
$B(\mu\rightarrow e\gamma)< 2 \times 10^{-5}$ led directly to the
two-neutrino hypothesis \cite{nish57,schw57} in which the neutrino
coupled to the muon is different from that coupled to the electron,
and the \mueg process would be forbidden. The two-neutrino hypothesis
was verified experimentally at Brookhaven National Laboratory (BNL) by
confirming muon production, but no electron production from the
scattering of neutrinos from pion decays \cite{danb62}.  This
introduced the concept of the separate conservation of lepton flavors,
electron number ($L_e$) and muon number ($L_{\mu}$).

Our understanding of modern elementary particle physics is based on
the Standard Model (SM), which is a gauge theory of the strong and
electroweak interactions. The formulation of the SM is based on many
theoretical developments of gauge theory in the 1960s and 1970s.
Since then, the SM has confronted various experimental tests and has
had amazing success in explaining all measurements so far.  In the
minimal version of the SM, where only one Higgs doublet is included
and massless neutrinos are assumed, lepton flavor conservation is an
automatic consequence of gauge invariance and the renormalizability of
the SM Lagrangian. It is the basis of a natural explanation for the
smallness of lepton flavor violation (LFV) in the charged lepton
process.

\begin{figure}[ht!]
\centerline{\epsfig{file=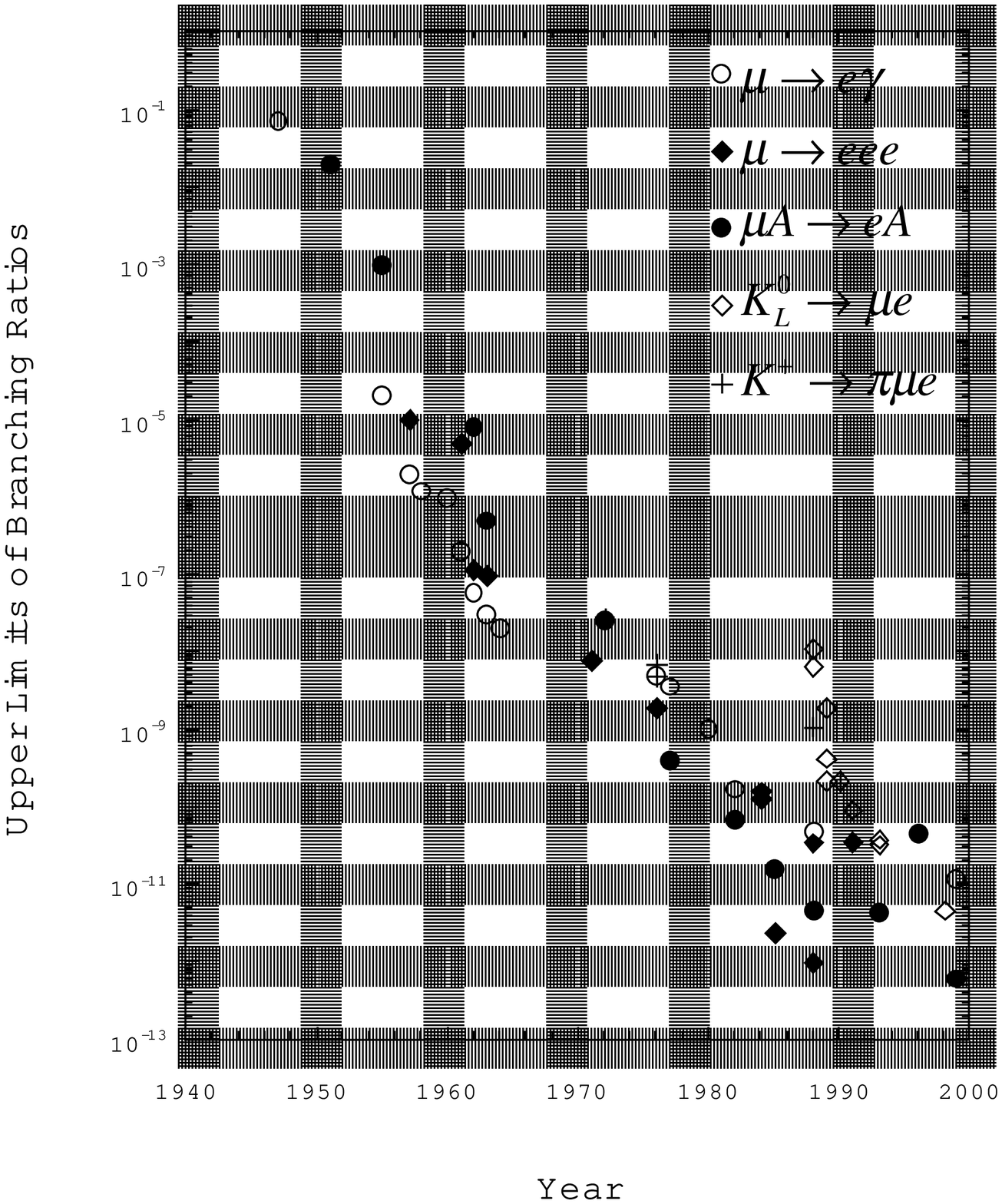,width=14cm}}
\caption{Historical progress of LFV searches for various
processes of muons and kaons.}
\label{fg:lfv}
\end{figure}

In extensions of the minimal SM, however, LFV could occur from various
sources.  In fact, in most cases, new physics or interactions beyond
the SM would predict LFV at some level.  LFV processes with muons of
major interest are such as \mueg, \muec conversion in a muonic atom
($\mu^{-}N\rightarrow e^{-}N$), $\mu^{+} \rightarrow e^{+}e^{+}e^{-}$,
and muonium to anti-muonium conversion (\mumu conversion). Historical
progress in various LFV searches in muon and kaon decays is shown in
Fig.\ref{fg:lfv}, in which the experimental upper limits have been
continuously improved at a rate of about two orders of magnitude per
decade for about 50 years since the first LFV experiment by Hincks and
Pontecorvo.  The current LFV searches with muons have reached a
sensitivity on the order of $10^{-12}-10^{-13}$ in their branching
ratios. In general, searches for rare processes could probe new
interactions at high energy. For example, in the four fermion
interaction, the LFV branching ratios could be scaled by
$(m_W/m_X)^4$, where $m_X$ is the mass of an exotic heavy particle
responsible for the LFV interaction and $m_W$ is the mass of the W
gauge boson. Thus, the present sensitivities for LFV searches in muon
decays could explore a mass scale of several 100 TeV, which is not
directly accessible by the present accelerators.

Recently, considerable interest in LFV processes has arisen based on
supersymmetric (SUSY) extensions to the SM, in particular
supersymmetric grand unified theories (SUSY GUT).  Since the three
gauge coupling constants of the strong, weak, and electromagnetic
interactions, which were measured at LEP and SLC, are shown to be
consistent with the assumption that they are unified to a single
$SU(5)$ gauge coupling constant at the order of $10^{16}$ GeV scale in
SUSY SM, the SUSY GUT model becomes a very attractive candidate for
physics beyond the SM. In the SUSY models, in general, there is a new
source of flavor mixing in the mass matrices of SUSY partners for
leptons and quarks. This will induce LFV processes for charged
leptons.  The predictions of the branching ratios depend on flavor
mixing in the mass matrix of sleptons, which are supersymmetric
partners of leptons. In the SUSY-GUT scenario, the flavor mixing in
the slepton sector is naturally induced at the GUT scale because
leptons and quarks are in the same GUT multiplet \cite{hall86}. It has
been shown \cite{barb94,barb95a} that the surprisingly large top-quark
mass determined recently has an impact on calculations of the
branching ratios of \mueg and \muec conversion in SUSY GUT. The
predictions are as large as just one or two orders of magnitude lower
than the present experimental limits.

There is considerable evidence for the existence of neutrino masses
and their mixing based on the experimental results of the solar
neutrino deficit \cite{clev98,fuku96,hamp99,abdu96,fuku98a} and the
atmospheric neutrino anomaly \cite{fuku98b}.  Since neutrino
oscillations imply that lepton flavor is certainly not conserved, LFV
processes in muon decays are also expected to occur. In non-SUSY
models, however, the neutrino mixing introduces only small effects on
\mueg. For example, the branching ratio of \mueg is on the order of
$10^{-50}$ for a difference of the neutrino mass squared of $\Delta
m_{\nu}^{2}\sim 10^{-3}$ eV$^2$, since it is suppressed by $(\Delta
m_{\nu}^2/m_{W}^2)^2$ \cite{petc77,bile77}. The situation changes
drastically in SUSY models. In SUSY models with a neutrino-mass
generation mechanism of the see-saw type, the Yukawa coupling
constants among the Higgs doublet, lepton doublets and right-handed
neutrinos could induce large flavor mixing effects in the slepton
sector~\cite{borz86,hisa98a,hisa99a}. The resulting LFV rates can be
as large as, or even larger than, the experimental upper bounds,
depending on various parameters, especially on the Majorana mass of
the right-handed neutrino. In such a case, the same Yukawa coupling
constants for the right-handed neutrino are responsible for both the
neutrino oscillation and the LFV processes of charged leptons.

Thus, there are many theoretical scenarios under which the predicted
branching ratios for the muon LFV processes can be close to their
present experimental upper bounds, and therefore they could be
accessible and tested by future experiments. 

Experimentally, there has been much progress. First of all, several
new experimental results have been obtained using the high-intensity
muon beams now available, and on-going and proposed experiments are
aiming at further improvements. Furthermore, in the long term, new
attempts to create high-intensity muon sources have been initiated,
based on the ideas arising from the $\mu^{+}\mu^{-}$ collider project. 
The muon beam intensity aimed at such a new muon source would be about
$10^{12}-10^{13}~\mu^{\pm}$/sec, several orders of magnitude higher
than that presently available. With this increased muon flux,
significant improvements in experimental searches can be anticipated.

In this article, we review the current theoretical and experimental
status of the field of muon decay and its potential for new physics
beyond the Standard Model. We highly stress the importance of rare LFV
processes of muons, especially within the framework of SUSY models.
In addition, we cover precision measurements of normal muon
decay. There have been many excellent review articles on muon decay
and lepton flavor violation
\cite{fran75,sche78,verg86,engf86,depo87,vand93,depo95}.  However,
reflecting current renewed interest, this article has been written to
bring up to date recent topics on muon decay and physics beyond the
SM. The phenomenology and experimental status of some of the important
muon processes are described in detail.

This article is organized as follows. In Chapter 2, we give a short
summary of the SM and the muon properties within the SM. In Chapter 3,
LFV is discussed based on various theoretical models, including the
SUSY models. Chapter 4 deals with the current status of precision
measurements in normal muon decay, such as the muon lifetime, the
Michel decay spectrum and its asymmetry, and $e^{+}$ polarization. In
Chapter 5, the phenomenology and status of the most recent experiments
on various lepton flavor violating muon decay modes, such as
\mueg,~\meee,~\muec and \mupc conversions in a muonic atom, and
\mumu conversion, are described. In Chapter 6, prospects on future
experiments and high-intensity muon sources are briefly discussed.
Some useful formulas are collected in the Appendices.

\section{Basic Properties of the Muon in the Standard Model}

\subsection{Muon in the Standard Model}

\subsubsection{The Standard Model}

The current view of elementary particle physics is based on a gauge
theory of quarks and leptons. In the Standard Model (SM), three
fundamental interactions, the strong, electromagnetic and weak
interactions, are described as a $SU(3)_{C}\times SU(2)_{L}\times
U(1)_{Y}$ gauge theory. Quarks and leptons are classified in three
generations. The three quarks of $\frac{2}{3}e$ electric charge are
the {\em up} quark ($u$), {\em charm} quark ($c$) and {\em top} quark
($t$), and those of $-\frac{1}{3}e$ charge are the {\em down} quark
($d$), {\em strange} quark ($s$) and {\em bottom} quark
($b$). Correspondingly, there are three charged leptons of $-e$
electric charge, electron ($e$), muon ($\mu$) and tau ($\tau$), and
three species of neutrinos of neutral charge are introduced, {\it
i.e.}  electron neutrino ($\nu_{e}$), muon neutrino ($\nu_{\mu}$) and
tau neutrino ($\nu_{\tau}$).  These six quarks and six leptons are
given in Table~\ref{tb:ql}.

\begin{table}[bth!]
\caption{Quarks and leptons in the Standard Model}
\label{tb:ql}
\vspace{5mm}
\begin{tabular}{cccc}
Electric charge& 1st generation& 2nd generation& 3rd generation \cr\hline
${2\over3}e$ & $u$ & $c$ & $t$ \cr 
$-{1\over3}e$ & $d$ & $s$ & $b$ \cr
0 & $\nu_{e}$ & $\nu_{\mu}$ & $\nu_{\tau}$ \cr
$-1$ & $e$ & $\mu$ & $\tau$ \cr 
\end{tabular}
\end{table}      

In the SM, ferminonic fields, gauge fields, and a $SU(2)_L$ doublet
Higgs field are introduced as an elementary field. They are listed
along with their quantum numbers in Table~\ref{tb:efsm}, where the
$SU(3)_{C}$, $SU(2)_{L}$ and $U(1)_{Y}$ gauge fields are denoted as
$G_{\mu}$, $A_{\mu}$ and $B_{\mu}$, respectively. The subscripts of
$L$ and $R$ represent left-handed and right-handed chirality
projections ($P_{L}\equiv (1-\gamma_{5})/2$ and $P_{R}\equiv
(1+\gamma_{5})/2$), respectively. $H$ represents the Higgs doublet
field. The suffix $i~(= 1 - 3)$ for the quark and lepton fields is the
generation index. The $SU(2)_{L}$ doublet fields, such as ${q_i}_{L}$,
${l_{i}}_{L}$, and $H$, have field components given by

\begin{equation}
{q_i}_L=\left( \begin{array}{c}
             {u_i}_L                 \\
             {d_i}_L
                        \end{array}         \right)
,~~
{l_i}_L=\left( \begin{array}{c}
              {\nu_i}_L                  \\
              {e_i}_L
                        \end{array}         \right)
,~~
H=\left( \begin{array}{c}
             \phi^+                   \\
             \phi^0
                        \end{array}         \right)
.
\end{equation}

\begin{table}[tbh!]
\caption{Quantum numbers of elementary fields in the minimal Standard
Model. The $SU(3)_{C}$, $SU(2)_{L}$ representation and $U(1)_{Y}$
charge are given.}
\label{tb:efsm}
\vspace{5mm}
\begin{tabular}{c|ccccccccc}
&~$G_{\mu}$~ &~$A_{\mu}~$ & $~B_{\mu}~$ & ~${q_i}_L$~ & ~${u_i}_R~$ &
~${d_i}_R$~ & ~${l_i}_L$~& ~${e_i}_R$~ &~$H$~ \cr\hline
$SU(3)_C$&\bf{8}& \bf{1}& \bf{1} & \bf{3}&
\bf{3}&\bf{3}& \bf{1} & \bf{1}&\bf{1} \cr 
$SU(2)_L$&\bf{1}& \bf{3}&\bf{1} & \bf{2}&
\bf{1}&\bf{1}& \bf{2} & \bf{1}&\bf{2} \cr 
$U(1)_Y$&0& 0& 0 & $\frac{1}{6}$ &$\frac{2}{3}$ & 
$-\frac{1}{3}$& $-\frac{1}{2}$ & $-1$ & $\frac{1}{2}$\cr
\end{tabular}
\end{table}

The SM Lagrangian, ${\lagr}_{SM}$, consists of three parts, which are
for the gauge interaction, the Higgs potential, and the Yukawa
interaction. It is given as

\begin{equation}
{\lagr}_{SM}= {\lagr}_{gauge}+{\lagr}_{Higgs}+{\lagr}_{Yukawa}.
\end{equation}

The Lagrangian for the gauge interaction, ${\lagr}_{gauge}$, is
presented by

\begin{equation}
{\lagr}_{gauge}= \sum_{SU(3)_C,SU(2)_L,U(1)_Y} F^{(a)}_{\mu \nu}
F^{(a)\mu \nu} +\sum_{quarks,~ leptons}i {\overline
{{\psi}_{i}}_{L(R)}}
\gamma^{\mu}{\cal D_{\mu}} {\psi_{i}}_{L(R)} + |{\cal D}_{\mu} H|^2,
\end{equation}

\noindent where $F_{\mu\nu}$ is the gauge-field strength, and $\cal
D_{\mu}$ is a covariant derivative defined as

\begin{equation}
{\cal D}_{\mu} = \partial_{\mu}+ig_s \frac{\lambda^a}{2} G_{\mu}^a
+ig\frac{\tau^a}{2} A_{\mu}^a +ig' Q_Y B_{\mu}
\end{equation}

\noindent for the representations with $SU(3)_{C}$-triplet and
$SU(2)_{L}$-doublet and $Q_Y$-$U(1)_{Y}$ charge quantum numbers.
$g_s$, $g$, and $g^{'}$ are the gauge coupling constants for
$SU(3)_{C}$, $SU(2)_{L}$, and $U(1)_{Y}$, respectively.  $\lambda^a $
($a=1-8$) is the Gell-Mann matrix for a $SU(3)$ group, and $\tau^a$
($a=1-3$) is the Pauli matrix for a $SU(2)$ group. The terms for a
singlet representation for either the $SU(3)_{C}$ or $SU(2)_{L}$ gauge
groups are absent in the definition of $\cal D_{\mu}$.

The Lagrangian for the Higgs potential, ${\lagr}_{Higgs}$, is given by

\begin{equation}
{\lagr}_{Higgs}= -( -\mu^{2}|H|^2 +\lambda |H|^4).
\end{equation}

\noindent For $ \mu^{2}>0$, the Higgs field develops the following vacuum
expectation:

\begin{equation}
<H>=\left( \begin{array}{c}
             0                  \\
             {v/\sqrt{2}}
                        \end{array}         \right),
\end{equation}

\noindent where $v={\mu}/\sqrt{\lambda}$ ($\cong$ 246 GeV). The physical
Higgs mass is given by $m_{H}=\sqrt{2\lambda}v$. After electroweak
symmetry breaking, the $SU(2)_{L}$ and $U(1)_{Y}$ gauge fields form a
massless photon, and massive $W^{\pm}$ and $Z^0$ bosons. At the tree
level, their masses are given as $m_W = {1\over2}gv$ and $m_Z =
{1\over2}\sqrt{g^2 + {g^{'}}^{2}}v$. The SU(3) gauge boson, gluon,
remains massless.

The Yukawa interaction part of the Lagrangian is given by

\begin{equation}
{\lagr}_{Yukawa} = (y_e)_{ij}H^{\dagger}{\overline{e_i}}_R {l_{j}}_L
+(y_d)_{ij}H^{\dagger}{\overline{{d_i}}_R} {q_{j}}_L
+(y_u)_{ij}\tilde{H}^{\dagger}{\overline{{u_i}}_R} {q_{j}}_L + H.c.,
\label{eq:Yukawa}
\end{equation}

\noindent where $(y_{e})_{ij}$, $(y_{d})_{ij}$, and $(y_{u})_{ij}$
are the Yukawa coupling constants for the charged leptons, the
down-type quarks, and the up-type quarks respectively, and,

\begin{equation}
\tilde{H}=i \tau_2 H^*=\left( \begin{array}{c}
             {\phi^0}^*                   \\
             - \phi^-
                        \end{array}         \right).
\end{equation}

Substituting the vacuum expectation value for the Higgs field, the
Yukawa interaction in Eq.(\ref{eq:Yukawa}) generates the mass terms
for quarks and leptons, as follows:

\begin{equation}
{\lagr}_{mass} = -\Bigl( {\overline{{e_i}}_R}{(m_e)}_{ij}{e_{j}}_L
+{\overline{{d_i}}_R} {(m_d)}_{ij} {d_{j}}_L 
+{\overline{{u_i}}_R} {(m_u)}_{ij} {u_{j}}_L \Bigr) + H.c.,
\label{eq:mass}
\end{equation}

\noindent where $(m_e)_{ij}=-(y_e)_{ij}(v/\sqrt{2})$,
$(m_d)_{ij}=-(y_d)_{ij}(v/\sqrt{2})$ and
$(m_u)_{ij}=-(y_u)_{ij}(v/\sqrt{2})$.  Each mass matrix is
diagonalized by unitary transformations for the left-handed fermions
and the right-handed fermions with the same charge. Since the unitary
matrices for the left-handed up-type quark and the left-handed
down-type quark are generally different, flavor mixing is induced in
the charged weak current interaction for quarks. It is given by

\begin{equation}
{\lagr}_{W {\overline q} q}= -\frac{g}{\sqrt{2}} \Bigl( {\overline
{{u_i}}_L} \gamma^{\mu} (V_{CKM})_{ij} {d_j}_L W^+_{\mu} +{\overline
{{d_i}}_L}\gamma^{\mu} (V_{CKM})_{ji}^* {u_j}_L W^-_{\mu} \Bigr),
\end{equation}

\noindent where the $(V_{CKM})_{ij}$ represents the flavor mixing
matrix for the quark sector, {\it i.e.} Cabibbo-Kobayashi-Maskawa
(CKM) matrix~\cite{koba73}. In this equation and hereafter, the quark
fields are presented in the mass-diagonalized basis.

On the other hand, the charged lepton mass matrix in
Eq.(\ref{eq:mass}), equivalently the lepton Yukawa coupling constant,
is fully diagonalized by unitary transformations on the lepton doublet
fields (${l_i}_L$) and the lepton singlet fields (${e_j}_R$).  In the
mass-diagonalized basis, the charged weak current interaction for
leptons remain diagonal, as follows:

\begin{equation}
{\lagr}_{W {\overline \nu} e}= -\frac{g}{\sqrt{2}} \Bigl( {\overline
{{\nu_i}}_L} \gamma^{\mu}{e_i}_L W^+_{\mu} +{\overline
{{e_i}}_L}\gamma^{\mu}{\nu_i}_L W^-_{\mu} \Bigr).
\end{equation}

\noindent In the above basis, the lepton flavors can be defined for each
generation, and are thus conserved. They are the electron number
($L_{e}$), the muon number ($L_{\mu}$), and the tau number
($L_{\tau}$), as defined in Table~\ref{tb:lfm}.

\begin{table}[htb!]
\caption{Assignment of lepton flavors, the electron ($L_{e}$), the muon
number ($L_{\mu}$), and the tau number ($L_{\tau}$).}
\label{tb:lfm}
\vspace{5mm}
\begin{tabular}{c|rrrrrrrrrrrr}
& $e^{-}$ & $\nu_{e}$ & $\mu^{-}$ & $\nu_{\mu}$ & 
$\tau^{-}$ & $\nu_{\tau}$& 
$e^{+}$ & $\overline{{\nu}_{e}}$ & $\mu^{+}$ & $\overline{{\nu}_{\mu}}$& 
$\tau^{+}$ & $\overline{{\nu}_{\tau}}$ \cr\hline
$L_e$     & +1 & +1 &  0 &  0 & 0 & 0& 
$-1$ & $-1$ & 0 & 0 & 0 & 0 \cr
$L_{\mu}$ &  0 &  0 & +1 & +1 & 0 & 0 &
0 & 0 & $-1$ & $-1$ & 0 & 0 \cr
$L_{\tau}$ &  0 &  0 & 0 & 0 & +1 & +1 &
0 & 0 &0 & 0& $-1$ & $-1$ \cr
\end{tabular}
\end{table}

\subsubsection{Interaction of the muon in the Standard Model}

At the tree level of the SM Lagrangian, the muon has three gauge
interactions, namely those with the photon, the $W^{\pm}$ and $Z^{0}$
bosons, and also the Higgs interaction. They are given by

\begin{eqnarray}
{\lagr}&=& e {\overline{\mu}} \gamma^{\mu} \mu A_{\mu} -\frac{g}{\sqrt{2}} 
(\overline{{\nu_{\mu}}_L} \gamma^{\mu} \mu_L W^+_{\mu} +{\overline {{\mu}_L}}
\gamma^{\mu} {\nu_{\mu}}_L W^-_{\mu}) \nonumber \\ 
&-&\sqrt{g^2+{g'}^2}\Bigl\lbrace{\overline{{\mu}_L}} \gamma^{\mu}
(-\frac{1}{2}+\sin^2{\theta_W}) {\mu}_L +{\overline{{\mu}_R}}
\gamma^{\mu} \sin^2{\theta_W} {\mu}_R\Bigr\rbrace Z^{0}_{\mu}-\frac{m_{\mu}}{v}
{\overline{\mu}}{\mu} H,
\end{eqnarray} 

\noindent where the Weinberg angle $\theta_W$ is defined by
$\sin\theta_W \equiv g^{'}/\sqrt{g^2+{g^{'}}^{2}}$, and also $e =
g\sin\theta_{W}$ at the tree level. $H$ denotes the physical Higgs
boson field.  In addition to the electromagnetic interaction, the
second and third terms describe, respectively, the charged
weak-current interaction mediated by the $W^{\pm}$ boson and the
neutral weak-current interaction mediated by the $Z^{0}$ boson. The
other charged leptons, electron and tau, have the same gauge
interaction as the above, and the coupling constant to the Higgs boson
is proportional to their lepton masses.

The muon decay in the SM is described by a charged weak-current
interaction mediated by the W$^\pm$ gauge boson. The four fermion
interaction is given by

\begin{equation}
{\lagr}_{Fermi}=-\frac{G_F}{\sqrt{2}}[{\overline {\nu_{\mu}}} \gamma^{\mu} 
(1-\gamma_{5})\mu {\overline e} \gamma_{\mu} 
(1-\gamma_{5}){\nu_e}+{\overline {\nu_e}} \gamma^{\mu} 
(1-\gamma_{5})e {\overline \mu} \gamma_{\mu} 
(1-\gamma_{5}){\nu_{\mu}}],
\end{equation} 

\noindent where $G_F$ is the Fermi coupling constant.  At the tree
level of the SM, this is given by

\begin{equation}
G_F=\frac{g^2}{4\sqrt{2}m_W^2},
\end{equation} 

\noindent where $m_W$ is the $W^{\pm}$ boson mass. This interaction
describes the normal muon decays, $\mu^{+}\rightarrow e^{+} \nu_{e}
{\overline \nu_{\mu}}$ and $\mu^{-}\rightarrow e^{-} {\overline
\nu_{e}} \nu_{\mu}$.
 
Lepton universality is a fundamental property of the gauge
interaction. The universality in charged weak currents has been tested
from the combination of leptonic and semi-leptonic decays of $\tau$,
and leptonic decays of $\mu$, $\pi$ and $K$ mesons.  The constraints
on the ratio of the charged weak current coupling constants for
electron ($g_{e}$) and muon ($g_{\mu}$) are obtained from the ratios
of $B(\tau^{-}\rightarrow e^{-} {\overline{\nu_e}}\nu_{\tau}) /
B(\tau^{-}\rightarrow\mu^{-}\overline{\nu}_{\mu}\nu_{\tau} )$ and
$B(\pi^{-}\rightarrow e^{-}\overline{\nu}_e)/B(\pi^{-}\rightarrow
\mu^- {\overline \nu_{\mu}})$.  For the ratio of the charged
weak current coupling of tau ($g_{\tau}$) and $g_{\mu}$,
$\Gamma(\tau^{-}\rightarrow e^{-}\overline{\nu}_{e}\nu_{\tau}) /
\Gamma(\tau^{-}\rightarrow\mu^{-}\overline{\nu}_{\mu}\nu_{\tau})$,
$\Gamma(\tau^-\rightarrow\nu_{\tau} \pi^-)/\Gamma( \pi^{-}\rightarrow
\mu^{-}\overline{\nu}_{\mu})$ and $\Gamma(\tau^{-}\rightarrow \nu_{\tau}
K^-)/\Gamma( K^- \rightarrow \mu^- \overline{\nu}_{\mu})$ were used.
These ratios give a test of the equality of the relevant coupling
constants at the level of 0.1 - 0.5 \%, except for those decays
involving kaons, where the sensitivity is at the 2\% level
\cite{pich97}.  A similar test can be performed for leptonic decays of the
$W$ boson at Tevatron and LEP. A recent measurement of leptonic
branching ratios of $W$ gives $B(W\rightarrow e \nu) = 10.9 \pm 0.4
\%$, $B(W\rightarrow \mu \nu) = 10.2 \pm 0.5 \%$ and $B(W \rightarrow \tau
\nu) = 11.3 \pm 0.8 \%$ \cite{pdg98}. Thus, the sensitivity for the coupling
ratio is a few~\%. The lepton universality for neutral weak currents
has been tested at the $Z^0$ boson pole. At LEP experiments, the
measurements of partial leptonic widths, leptonic forward-backward
asymmetries, tau polarization and its angular distributions are
sensitive to the vector and axial vector coupling constants for
different lepton species.  At SLC, the left-right asymmetry and
forward-backward left-right asymmetry are also measured using a
polarized beam. Lepton universality is now treated at the level of
5-10\% for vector couplings and less than 0.2\% for axial vector
couplings \cite{pich97}. The difference in sensitivity is due to the
fact that the lepton vector coupling is very small compared to the
axial coupling.

\subsubsection{Neutrino mass and mixing}\label{sc:NMM}

Although the minimal SM does not allow massive neutrinos, there is
increasing evidence for the masses of neutrinos from the
solar-neutrino deficit \cite{clev98,fuku96,hamp99,abdu96,fuku98a} and
the atmospheric-neutrino anomaly \cite{fuku98b}. If the solar-neutrino
deficit is explained by neutrino oscillations, the mass-square
difference is in the range of $\Delta m_{\nu}^2 \approx 10^{-6} -
10^{-5}$ eV$^2$ for the MSW solution \cite{wolf78,mikh85} or $\Delta
m_{\nu}^2\approx 10^{-11}$ eV$^2$ for the vacuum oscillation (or
``just-so'' oscillation) \cite{glas87,glas99}.  Also, the
atmospheric-neutrino anomaly suggests that the mass-square difference
between the muon neutrino and the tau (or sterile) neutrino is on the
order of $\Delta m_{\nu}^2 \approx 10^{-3} - 10^{-2}$ eV$^2$
\cite{fuku98b}.  In addition, the LSND (Liquid Scintillator Neutrino
Detector) experiment has reported the
$\overline{\nu}_{\mu}(\nu_{\mu})\rightarrow\overline{\nu}_{e}(\nu_{e})$
oscillation, suggesting $|m_{\nu_{\mu}}^2 - m_{\nu_{e}}^2|\approx
10^{-1} - 10^{0}$ eV$^{2}$ \cite{atha98}, although the KARMEN
experiment has seen no evidence of such an oscillation
\cite{eite99}. If neutrino mixing is the true interpretation of the
anomalies, the SM must be extended.  On the other hand, there are
direct upper bounds on the neutrino mass for each species, which is 15
eV/$c^2$ for the electron neutrino, 170 keV/$c^2$ for the muon
neutrino, and 18.2 MeV/$c^2$ for the tau neutrino
\cite{pdg98}. Recently, improved measurements of the electron neutrino
mass have been reported with a much better sensitivity of a few
eV/$c^2$ \cite{loba98,otte98}.

It is possible to accommodate the Dirac mass terms for the neutrinos
if $SU(2)$ singlet fields of the right-handed neutrinos ${\nu_{i}}_R$
($i = 1 - 3$) are included in the minimal SM field contents.  Then,
the following interaction can be added to Eq.(\ref{eq:Yukawa}):

\begin{equation}
{\lagr}_{\nu_R} =
(y_{\nu})_{ij}\tilde{H}^{\dagger}{\overline{{\nu_i}}_R} {l_{j}}_L +
H.c.,
\end{equation} 

\noindent where $(y_{\nu})_{ij}$ is the Yukawa coupling for
neutrinos. If it is very small, the small masses of neutrinos can be
explained. For example, the Yukawa coupling constant should be
$O(10^{-11}$) for a neutrino mass of 1~eV/$c^2$. Note that the total
lepton number is conserved in this scenario, whereas the lepton flavor
could be generally violated.

A more natural explanation for the small neutrino masses is provided
by the ``see-saw mechanism'' \cite{yana79,gell79}. In this scenario,
the Majorana mass term is also included for the right-handed neutrino,

\begin{equation}
{\lagr}_{\nu_R} = 
(y_{\nu})_{ij}\tilde{H}^{\dagger}{\overline{{\nu_i}}_R} {l_{j}}_L
- {1\over2}{\overline{\nu_{i}}_R} (M_{R})_{ij} {\nu_{j}}_R^{c} +H.c.,
\end{equation}

\noindent where the charge-conjugation field is defined as $ {\overline
\psi^c} = -\psi^T C^{-1}$ and the charge-conjugation matrix ($C$)
satisfies $C^{-1} \gamma^{\mu} C = - {\gamma^{\mu}}^T$.
$(M_{R})_{ij}$ is the right-handed Majorana neutrino
matrix. Substituting the vacuum expectation value for the Higgs field,
the neutrino mass terms become

\begin{equation}
{\lagr}_{\nu~mass} =  -{1\over2}
\left( \,{\overline{(\nu_{iL})^{c}}} \,,  \, {\overline{{\nu_i}}_R} \right)
  \left( \begin{array}{cc}
            0 &  m_{D}^T  \\
            m_{D}  &  M_R           
              \end{array} \right) 
  \left( \begin{array}{c}
            {\nu_j}_L     \\
            ({\nu_j}_R)^c           
              \end{array} \right) 
 + H.c.,
\end{equation}

\noindent where the Dirac mass term is
$({m_{D}})_{ij}=-(y_{\nu})_{ij}(v/\sqrt{2})$.  When the Majorana mass
scale is much larger than the Dirac masses, the lighter neutrino masses
are given by

\begin{equation} 
{\lagr}_{\nu~mass}\approx -{1\over2}{\overline{(\nu_{iL})^{c}}}
(m_{\nu})_{ij}{\nu_{jL}} + H.c.
\end{equation}

\noindent and 

\begin{equation}
m_{\nu}=m_D^T (M_R)^{-1}m_D.
\end{equation}

\noindent For example, if $M_R$ is $10^{15}$ GeV and the Dirac mass
is in the range of 100 GeV, then the neutrino mass becomes naturally
$O(10^{-2})$ eV.

By diagonalization of the charged lepton and neutrino mass matrices,
lepton flavor mixing is induced in the charged weak current
interaction, as follows:

\begin{equation}
{\lagr}_{W \nu e}= -\frac{g}{\sqrt{2}} \Biggl[ {\overline {\nu_i}}_L
\gamma^{\mu} (V_{MNS})^{*}_{ji} {e_j}_L W^+_{\mu} +{\overline
{{e_i}}_L}\gamma^{\mu} (V_{MNS})_{ij}{\nu_j}_L W^-_{\mu} \Biggr],
\label{eq:mns}
\end{equation}

\noindent where the $(V_{MNS})_{ij}$ is the flavor mixing matrix for the
lepton sector, {\it i.e.} Maki-Nakagawa-Sakata (MNS) matrix
\cite{maki62}. Note that as in the case of quarks, the lepton fields in
Eq.(\ref{eq:mns}) are written in the mass-diagonalized basis. The
$(V_{MNS})_{ij}$ matrix element represents neutrino mixing which can
be studied by neutrino oscillations. For a review on the neutrino mass
and mixing, please see other references
\cite{bile87,fuku94,moha98,fish99}.

\subsection{Static properties of the muon}

\subsubsection{Mass and lifetime}

The mass and lifetime of the muon are the fundamental inputs of the
SM. The muon mass is given by \cite{pdg98}

\begin{equation}
m_{\mu}=106.658389(34)~ {\rm MeV}.
\end{equation}

\noindent It is derived from the ratio of the muon mass to the
electron mass, $m_{\mu}/m_{e}$, which is measured in a muonium
($\mu^{+}e^{-}$ atom) with QED corrections \cite{cohe87}.

The experimental value of the muon lifetime is

\begin{equation}
\tau_{\mu}=2.19703(4)\times 10^{-6} ~{\rm sec}.
\end{equation}

\noindent In the framework of the SM, the muon lifetime
($\tau_{\mu}$) is related to the Fermi coupling constant ($G_F$), by
including QED corrections, as follows \cite{kino59,marc88}:

\begin{equation}
\tau_{\mu}^{-1}=\frac{G_F^2 m_{\mu}^5}{192\pi^3}
F(\frac{m_e^2}{m_{\mu}^2})(1+\frac{3}{5}
\frac{m_{\mu}^2}{m_W^2})
\Bigl[ 1+\frac{\alpha(m_{\mu})}{2\pi}( \frac{25}{4}-\pi^2 ) \Bigr],
\label{eq:tmu}
\end{equation}

\noindent where $F(x)=1-8x+8x^3-x^4-12x^2\ln{x}$. $m_{\mu}$ and
$m_{e}$ are masses of the muon and the electron, respectively.  The
$\alpha$ value at the $m_{\mu}$ scale, $\alpha(m_{\mu})$, is given by

\begin{equation}
\alpha(m_{\mu})^{-1} = \alpha^{-1} - {2\over
3\pi}\ln\Bigl({m_{\mu}\over m_{e}}\Bigr) + {1\over 6\pi} \approx 136.
\end{equation}

\noindent From Eq.(\ref{eq:tmu}), the Fermi coupling constant of $G_F =
1.16639(1)\times 10^{-5} $GeV$^{-2}$ is determined \cite{pdg98}. The
higher order two-loop corrections to the muon lifetime have been
calculated \cite{ritb99}.

\subsubsection{Magnetic moment}

Since the muon is a Dirac particle, the $g$ factor of its magnetic
moment is 2, if radiative corrections are ignored. A deviation from 2,
namely $g-2$, is very important to investigate quantum
corrections. The present experimental value of $a_{\mu}=(g_{\mu}-2)/2$
is given by
\cite{pdg98}

\begin{equation}
a_{\mu}^{exp}=11659230(84)\times 10^{-10} \quad(\pm 7{\rm ppm}).
\end{equation}

\noindent A new experiment, E821, is on-going at Brookhaven National
Laboratory (BNL) and the experimental error is aimed to be improved by
a factor of 20.  The result from the first run gives $a_{\mu}^{exp} =
1165925(15) \times 10^{-9}$ ($\pm 13$ ppm)
\cite{care99a}. 

Theoretically, this quantity is calculated very precise
\cite{hugh99}.  The correction is divided into higher-order QED
corrections, hadronic contributions and the electroweak (EW)
contributions.  A recent update of theoretical calculations for them
gives \cite{czar98}

\begin{eqnarray}
a_{\mu}^{QED}&=& 11658470.56(0.29)\times 10^{-10}, \label{eq:aqed}\\
a_{\mu}^{hadron}&=&673.9(6.7)\times 10^{-10},\\
a_{\mu}^{EW}&=&15.1(0.4)\times 10^{-10}. \label{eq:amuew}
\end{eqnarray}

\noindent By adding them, the SM prediction is 

\begin{eqnarray}
a_{\mu}^{SM} &=& a_{\mu}^{QED}+ a_{\mu}^{hadron}+ a_{\mu}^{EW}
\nonumber\\
    &=&11659159.6(6.7)\times 10^{-10}.
\end{eqnarray}

\noindent The theoretical prediction is in good agreement with the
experimental value. 

To calculate the QED correction, the fine structure constant is needed
as an input. Eq.(\ref{eq:aqed}) was estimated by using the fine
structure constant obtained from the Quantum Hall effect, which gives
$\alpha^{-1}(qH)=137.03600370(270)$. It is consistent with that
determined from the electron anomalous magnetic moments ($a_{e}$) by
assuming a theoretical evaluation based on the SM. It is
$\alpha^{-1}(a_e) = 137.03599959(38)$, where the experimental values
of the electron anomalous magnetic moments are $a_{e^-}^{ext}
=1159652188.4(4.3)\times 10^{-12}$ and
$a_{e^+}^{ext}=1159652187.9(4.3)\times 10^{-12}$.

Although the electron $g-2$ factor is better determined than the muon
$g-2$ factor, the latter is much more sensitive to short-distance
physics. For example, the EW correction to $a_{e}$ is $O(10^{-14})$
level compared the $a_{\mu}^{EW}$ quoted in Eq.(\ref{eq:amuew}).
Therefore, the muon $g-2$ experiment is much more important to search
for the effects of new physics. For example, the on-going experiment
E821 is expected to put very strong constraints on supersymmetric
(SUSY) models, since SUSY models could contribute to the muon $g-2$
factor significantly when the slepton, charginos, and neutralinos
exist in the mass range of a few hundred GeV
\cite{lope94,chat96,moro96,carena97,gabr97}.

\subsection{Decay modes of the muon}

\begin{table}[htb!]
\caption{Decay modes of muons.}
\label{tb:mudecays}
\vspace{5mm}
\begin{tabular}{lll}
Decay~mode & Branching~ratio & References \cr\hline 
$\mu^{-}\rightarrow e^-\overline{\nu}_{e} \nu_{\mu}$ & $\sim 100\%$& \cr 
$\mu^-\rightarrow e^-\overline{\nu}_{e} \nu_{\mu}\gamma$& $1.4\pm
0.4$\%~~( for $E_{\gamma}>10$ MeV) & \citeasnoun{crit61} \cr 
$\mu^- \rightarrow e^-\overline{\nu}_{e}\nu_{\mu} e^+ e^-$ & 
$(3.4\pm 0.4) \times 10^{-5}$ & \citeasnoun{bert85} \cr 
$\mu^-\rightarrow e^- \nu_e \overline{\nu}_{\mu}$&
$<1.2\%$ & \citeasnoun{free93} \cr 
$\mu^-\rightarrow e^-\gamma$& $<1.2 \times 10^{-11}$ &
\citeasnoun{broo99} \cr
$\mu^-\rightarrow e^- e^- e^+$ & $<1.0 \times 10^{-12}$ &
\citeasnoun{bell88} \cr
$\mu^-\rightarrow e^-\gamma\gamma$& $<7.2 \times 10^{-11}$ &
\citeasnoun{bolt88} \cr
\end{tabular}
\end{table} 

The measured decay modes of muons are $\mu^- \rightarrow
e^-\overline{\nu}_{e} \nu_{\mu}$ (Michel decay), $\mu^- \rightarrow
e^-\overline{\nu}_{e} \nu_{\mu}\gamma$ (radiative muon decay), and
$\mu^-\rightarrow e^- \overline{\nu}_{e}\nu_{\mu} e^+ e^-$. The
branching ratios for these modes and the upper bounds on the other
exotic decay modes at 90\% confidence level (C.L.)  are summarized in
Table~\ref{tb:mudecays}.  Although these branching fractions have been
measured in experiments with positive muon decays, they are listed for
negative muons by assuming CP invariance.  Since $\mu^-\rightarrow
e^{-}\overline{\nu}_{e}\nu_{\mu}\gamma$ cannot be clearly separated
from $\mu^{-}\rightarrow e^{-}\overline{\nu}_e\nu_{\mu}$ with a soft
photon, the branching ratio for the radiative decay is shown for
$E_{\gamma}>10$ MeV.  There is no evidence of lepton flavor violating
processes, such as $\mu^{-}\rightarrow e^{-}\gamma$ ,
$\mu^{-}\rightarrow e^{-}e^{-}e^{+}$ and $\mu^{-}\rightarrow
e^{-}\gamma\gamma$. Also, an upper bound is set for those with
$|\Delta L_{i}|=2$, such as $\mu^{-}\rightarrow
e^{-}\nu_e\overline{\nu}_{\mu}$ decay, which is allowed if the lepton
flavor is conserved multiplicatively instead of additively.

\subsubsection{Normal muon decay}\label{sc:NMMD}

In the SM, the normal muon decay is described by the V$-$A interaction.
In extensions to the SM, the energy spectrum of a decay electron
(positron), its angular distribution if muons are polarized, and its
spin polarization in $\mu^{-}\rightarrow e^-
\overline{\nu}_{e}\nu_{\mu}$ (or $\mu^+\rightarrow
e^+\nu_e\overline{\nu}_{\mu}$) decay are sensitive to the type of
interaction on muon decays, including new possible interactions
besides the V$-$A interaction. If the general four fermion interactions
with no derivatives are assumed, the muon differential decay rate is
given with a few parameters by \cite{fets98}

\begin{eqnarray}
{{d^2 \Gamma(\mu^{\pm}\rightarrow e^{\pm}\nu\overline{\nu})}\over{dx
d\cos{\theta_{e}}}}&=&
\frac{m_{\mu}}{4\pi^3}W_{e\mu}^4 G_F^2 \sqrt{x^2-x_0^2}~
\Bigl( F_{IS}(x)\pm P_{\mu} \cos{\theta_e}F_{AS}(x) \Bigr)
\Bigl( 1+\vec{P_{e}}(x,\theta_{e})\cdot {\hat{\zeta}} \Bigr),
\label{eq:muenn}
\end{eqnarray}

\noindent where $W_{e\mu}=(m_{\mu}^2+m_e^2)/(2m_{\mu})$,
$x=E_e/W_{e\mu}$ and $x_0=m_e/W_{e\mu} (=9.7 \times 10^{-3}) \leq x
\leq 1$.  $E_{e}$ is the energy of the $e^{\pm}$. $m_{e}$ and $m_{\mu}$
are the masses of the positron and the muon, respectively. The plus
(minus) sign corresponds to $\mu^{+} (\mu^{-})$ decay. $\theta_e$ is
the angle between the muon polarization ($\vec {P_{\mu}}$) and the
electron (or positron) momentum, and ${\hat{\zeta}}$ is the
directional vector of the measurement of the $e^{\pm}$ spin
polarization. $\vec{P}_{e}(x,\theta_e)$ is the polarization vector of
the $e^{\pm}$. The functions $F_{IS}(x)$ and $F_{AS}(x)$ are the
isotropic and anisotropic parts of the $e^{\pm}$ energy spectrum,
respectively. They are given by

\begin{eqnarray}
F_{IS}(x) &=& x(1-x)+{2\over9}\rho(4x^2-3x-x_0^2)+\eta x_0 (1-x), \\
F_{AS}(x) &=& {1\over3}\xi\sqrt{x^2-x_0^2}
\Bigl[ 1-x + {2\over3}\delta (4x-3+(\sqrt{1-x_0^2}-1)) \Bigr],
\end{eqnarray}

\noindent where $\rho$, $\eta$, $\xi$, and $\delta$ are called Michel
parameters \cite{mich50,bouc57}. 

In the SM, these parameters are given as $\rho=\frac{3}{4}$, $\eta=0$,
$\xi=1$ and $\delta=\frac{3}{4}$.  When the electron (positron)
polarization is not measured and $x_0$ is ignored, the differential
branching ratio in the SM in Eq.(\ref{eq:muenn}) leads to a simple
form of

\begin{equation}
{{d^2 \Gamma(\mu^{\pm}\rightarrow e^{\pm}\nu\overline{\nu})}\over{dx
d\cos{\theta_e}}}=
\frac{m_{\mu}^5 G_F^2}{192\pi^3} x^2
[(3-2x) \pm P_{\mu} \cos{\theta_e}(2x-1)].
\label{eq:simple}
\end{equation}

\begin{figure}[ht!]
\centerline{\epsfig{file=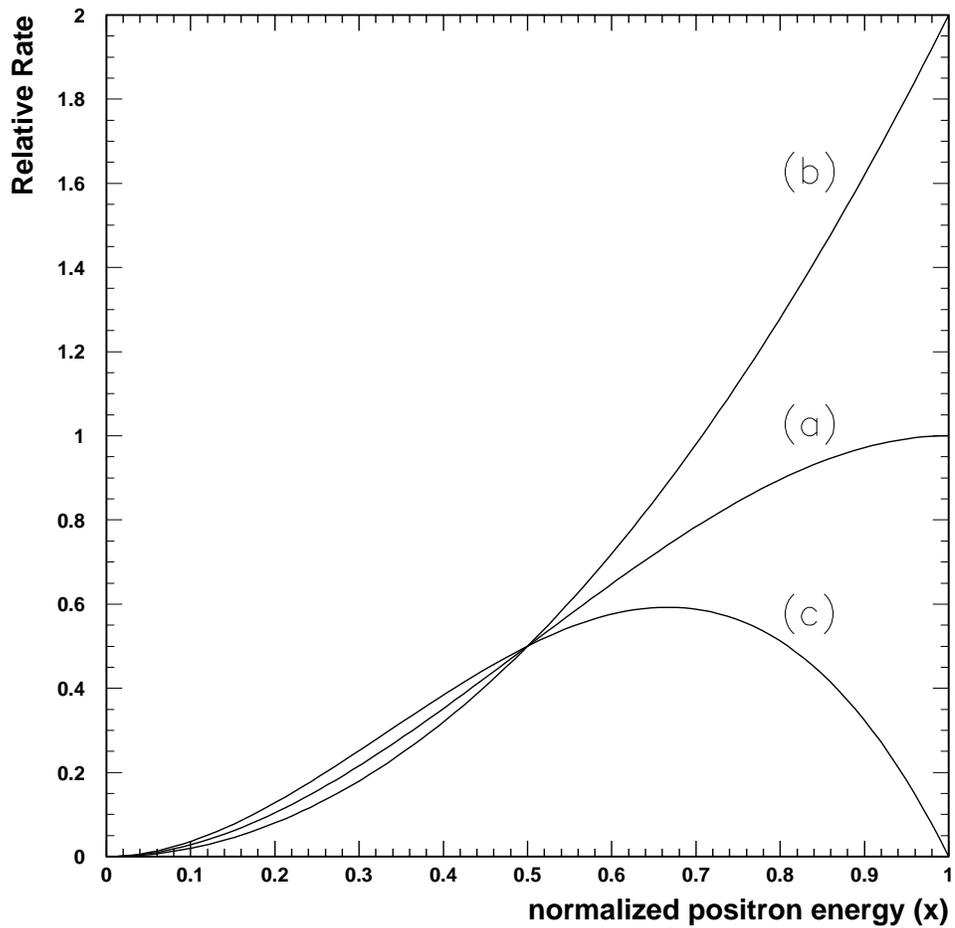,width=14cm}}
\caption{Michel $e^{+}$ energy spectrum of polarized \muenn decay with 
100\% muon polarization ($P_{\mu}=1$). (a) $\cos\theta_{e}=0$, (b)
$\cos\theta_{e}=1$, and (c) $\cos\theta_{e}=-1$.}
\label{fg:michel}
\end{figure}

\noindent Fig.\ref{fg:michel} shows the $e^{+}$ energy spectrum in
\muenn decay in the SM, for the cases of $\cos\theta=0$,
$\cos\theta_{e}=+1$, and $\cos\theta_{e}=-1$ with 100\% polarized
positive muons. As can be seen, the spectrum is high at $x\approx1$,
and the sign of $e^{\pm}$ asymmetry changes at $x=1/2$.

If the SM is not assumed, the muon lifetime in Eq.(\ref{eq:tmu})
should be replaced by \cite{sche78,fets95,pich95}

\begin{eqnarray}
\tau_{\mu}^{-1}&=&\frac{G_F^2 m_{\mu}^5}{192\pi^3}
\Bigl[ F({{m_e^2}\over{m_{\mu}^2}}) + 4\eta {m_e\over m_{\mu}} 
G({{m_e^2}\over{m_{\mu}^2}}) - {32\over3}(\rho -
{3\over4}){m_{e}^2\over m_{\mu}^2}(1-{m_{e}^4\over m_{\mu}^4}) \Bigr]
\times \nonumber \\
&&\quad (1+{3\over5}\frac{m_{\mu}^2}{m_W^2})
[1+\frac{\alpha(m_{\mu})}{2\pi}( \frac{25}{4}-\pi^2 )],
\label{eq:tmunew}
\end{eqnarray}

\noindent where $G(x) = 1 + 9x - 9x^2 - x^3 + 6x(1+x){\rm
ln}x$. Regarding radiative corrections, since it can be assumed that
the SM contribution dominates in the normal muon decay process, those
based on the SM, as in Eq.(\ref{eq:tmu}), are used. From
Eq.(\ref{eq:tmunew}), the correction from the $\eta$ parameter is
proportional to $O(m_e/m_{\mu})$, whereas that from the $\rho$
parameter is very small, in the order of $O(m_e^2/m_{\mu}^2)$. Since
the present experimental accuracy of the $\eta$ parameter is about
1\%, the uncertainty from the $\eta$ correction is about on the order
of $10^{-4}$ to the estimation of the muon lifetime in the non-SM
case.

When the spin polarization of $e^{+}(e^{-})$ in the
$\mu^{+}\rightarrow e^{+}\nu_{e}\overline{\nu}_{\mu}$
($\mu^{-}\rightarrow e^{-}\nu_{\mu}\overline{\nu}_{e}$) decay is
detected, $\vec{P}_{e}(x,\theta_e)$ in Eq.(\ref{eq:muenn}) can be
measured.  It is given by

\begin{equation}
\vec{P}_{e}(x,\theta_e) = P_{T1}\cdot{(\vec{z}\times
\vec{P}_{\mu})\times\vec{z}\over|(\vec{z}\times
\vec{P}_{\mu})\times\vec{z}|} + 
P_{T2}\cdot{\vec{z}\times\vec{P}_{\mu}\over
|\vec{z}\times\vec{P}_{\mu}|} + 
P_L\cdot{\vec{z}\over|\vec{z}|},
\end{equation}

\noindent where $\vec{z}$ is the direction of the $e^{\pm}$ momentum, and
$\vec{P}_{\mu}$ is the muon spin polarization. $P_{L}$, $P_{T1}$, and
$P_{T2}$ are, respectively, the $e^{\pm}$ polarization component
parallel to the $e^{\pm}$ momentum direction, that transverse to the
$e^{\pm}$ momentum within the decay plane, and that transverse to the
$e^{\pm}$ momentum and normal to the decay plane. A non-zero value of
the triple T-odd correction, $P_{T2}$, would imply violation of
time-reversal invariance. They are given by

\begin{eqnarray}
P_{T1}(x,\theta_e) &=& { P_{\mu}\sin\theta_{e} F_{T1}(x) \over 
F_{IS}(x) \pm P_{\mu}\cos\theta_e F_{AS}(x)} \label{eq:t1pol}, \\
P_{T2}(x,\theta_e) &=& { P_{\mu}\sin\theta_{e} F_{T2}(x) \over 
F_{IS}(x) \pm P_{\mu}\cos\theta_e F_{AS}(x)} \label{eq:t2pol}, \\
P_{L}(x,\theta_e) &=& { \pm F_{IP}(x) + P_{\mu}\cos\theta_{e}
F_{AP}(x) \over F_{IS}(x) \pm P_{\mu}\cos\theta_e F_{AS}(x)}, \label{eq:lpol}
\end{eqnarray}

\noindent where the $\pm$ sign corresponds to $\mu^{\pm}$ decays, and

\begin{eqnarray}
F_{T1}(x) &=& {1\over12}\Biggl\lbrace -2\Bigl[\xi^{''} + 12(\rho -
{3\over4})\Bigr](1-x)x_{0} - 3\eta(x^2 - x^2_0) + \eta^{''}(-3x^2 + 4x -
x_0^2) \Biggr\rbrace, \\
F_{T2}(x) &=& {1\over3}\sqrt{x^2 - x_0^2}~\Biggl\lbrace 3{\alpha^{'}\over
A}(1-x) + 2{\beta^{'}\over A}\sqrt{1-x_0^2} \Biggr\rbrace, \\
F_{IP}(x) &=& {1\over54}\sqrt{x^2 - x_0^2}~\Biggl\lbrace
9\xi^{'}(-2x+2+\sqrt{1-x_0^2}) +
4\xi(\delta-{3\over4})(4x-4+\sqrt{1-x_0^2}) \Biggr\rbrace,\\
F_{AP}(x) &=& {1\over6} \Biggl\lbrace \xi^{''}(2x^2 - x - x_0^2) + 4(\rho -
{3\over4})(4x^2 - 3x - x_0^2) + 2\eta^{''}(1-x)x_0 \Biggr\rbrace.
\end{eqnarray}

\noindent where $\xi^{'}$, $\xi^{''}$, $\eta^{''}$, $(\alpha^{'}/A)$, and
$(\beta^{'}/A)$ are newly defined Michel parameters
\cite{kino57b,fets98}. In the SM, $\xi^{'}=\xi^{''}=1$ and
$\eta^{''}=(\alpha^{'}/A)=(\beta^{'}/A)=0$.

The muon-decay Lagrangian for the general four-fermion couplings with
ten complex parameters is expressed as \cite{fets86}

\begin{eqnarray}
{\lagr}_{\mu\rightarrow e\nu\overline{\nu}}
&=&-{4 G_F \over \sqrt{2}}\Bigl[~ 
 g_{RR}^S ({\overline e}_R {\nu_e}_L)({\overline {\nu_{\mu}}}_L {\mu}_R)
+g_{RL}^S ({\overline e}_R {\nu_e}_L)({\overline {\nu_{\mu}}}_R {\mu}_L)
\nonumber \\
 & & 
+g_{LR}^S ({\overline e}_L {\nu_e}_R)({\overline {\nu_{\mu}}}_L {\mu}_R)
+g_{LL}^S ({\overline e}_L {\nu_e}_R)({\overline {\nu_{\mu}}}_R {\mu}_L)
\nonumber \\
 & & 
+g_{RR}^V ({\overline e}_R \gamma^{\mu}{\nu_e}_R)
({\overline {\nu_{\mu}}}_R \gamma_{\mu}{\mu}_R)
+g_{RL}^V ({\overline e}_R \gamma^{\mu}{\nu_e}_R) 
({\overline {\nu_{\mu}}}_L \gamma_{\mu}{\mu}_L)
\nonumber \\
 & & 
+g_{LR}^V ({\overline e}_L \gamma^{\mu}{\nu_e}_L) 
({\overline {\nu_{\mu}}}_R \gamma_{\mu}{\mu}_R)
+g_{LL}^V ({\overline e}_L \gamma^{\mu}{\nu_e}_L)
({\overline {\nu_{\mu}}}_L \gamma_{\mu}{\mu}_L)
\nonumber \\
 & & 
+\frac{g_{RL}^T}{2} ({\overline e}_R \sigma^{\mu \nu}{\nu_e}_L)
({\overline {\nu_{\mu}}}_R \sigma_{\mu \nu}{\mu}_L) 
+\frac{g_{LR}^T}{2} ({\overline e}_L \sigma^{\mu \nu}{\nu_e}_R)
({\overline {\nu_{\mu}}}_L \sigma_{\mu \nu}{\mu}_R) + H.c.~ \Bigr],
\end{eqnarray}  

\noindent where there is a normalization condition of

\begin{eqnarray}
&\frac{1}{4}(|g_{RR}^S|^2+|g_{LL}^S|^2+|g_{RL}^S|^2+|g_{LR}^S|^2)
+(|g_{RR}^V|^2+|g_{LL}^V|^2+|g_{RL}^V|^2+|g_{LR}^V|^2)
\nonumber \\
&+3(|g_{RL}^T|^2+|g_{LR}^T|^2) = 1.
\end{eqnarray} 

\noindent Note that in the V$-$A interaction of the SM, $g_{LL}^V = 1$
and all the others are zero. 

\begin{table}[tb!]
\caption{Experimental values of some of the Michel decay parameters.}
\label{tb:michel}
\vspace{5mm}
\begin{tabular}{ccll}
Michel parameter & SM value & Experimental value &
Sensitive observables  \cr\hline
$\rho$     & $3/4$ & $0.7518 \pm 0.0026$ & $F_{IS}$ \cr
$\eta$     & 0     & $-0.007 \pm 0.013$  & $F_{IS}$ and $P_{T1}$ \cr
$\delta$   & $3/4$ & $0.7486 \pm 0.0038$ & $F_{AS}$ and $P_L$  \cr
$\xi$      & 1     & $1.0027 \pm 0.0084$ & $F_{AS}^{\dagger}$ and $P_L$ \cr
$\xi^{'}$  & 1     & $1.00 \pm 0.04 $    & $P_{L}$ \cr
$\xi^{''}$ & 1     & $0.65 \pm 0.36 $    & $P_{L}$ \cr
\end{tabular}
\vspace{5mm}
$^{\dagger}$ Only the product of $\xi P_{\mu}$ is measured.
\end{table}

The Michel decay parameters of $\rho$, $\eta$, $\xi$ and $\delta$ are
given by

\begin{eqnarray}
\rho&=&\frac{3}{16}(|g_{RR}^S|^2+|g_{LL}^S|^2 +|g_{RL}^S-2g_{RL}^T|^2
+|g_{LR}^S-2g_{LR}^T|^2)+\frac{3}{4}(|g_{RR}^V|^2+|g_{LL}^V|^2)\\
\eta&=& \frac{1}{2}Re[g_{RR}^V g_{LL}^{S*}+g_{LL}^V g_{RR}^{S*}+
g_{RL}^V (g_{LR}^{S*}+6g_{LR}^{T*})+g_{LR}^V (g_{RL}^{S*}+6g_{RL}^{T*})]\\
\xi&=& \frac{1}{4}(|g_{LL}^S|^2-|g_{RR}^S|^2)
-\frac{1}{4}(|g_{LR}^S|^2-|g_{RL}^S|^2) 
+(|g_{LL}^V|^2-|g_{RR}^V|^2)+3(|g_{LR}^V|^2-|g_{RL}^V|^2)
\nonumber \\
&&+5(|g_{LR}^T|^2-|g_{RL}^T|^2)
+4 Re(g_{LR}^S g_{RL}^{T*}-g_{RL}^S g_{LR}^{T*})\\
\xi \delta&=&\frac{3}{16}(|g_{LL}^S|^2-|g_{RR}^S|^2
+|g_{RL}^S-2g_{RL}^T|^2-|g_{LR}^S-2g_{LR}^T|^2))
+\frac{3}{4}(|g_{LL}^V|^2-|g_{RR}^V|^2)
\label{eq:mpara}
\end{eqnarray} 

\noindent Table~\ref{tb:michel} summarizes the present knowledge of
the Michel decay parameters \cite{pdg98}. Precise measurements of the
Michel decay parameters would place constraints on various new
physics, which would induce a small deviation from the V$-$A
couplings.  The current constraints on the general four-fermion
couplings are summarized in
\citeasnoun{fets98}. 

\begin{figure}[ht!]
\centerline{\epsfig{file=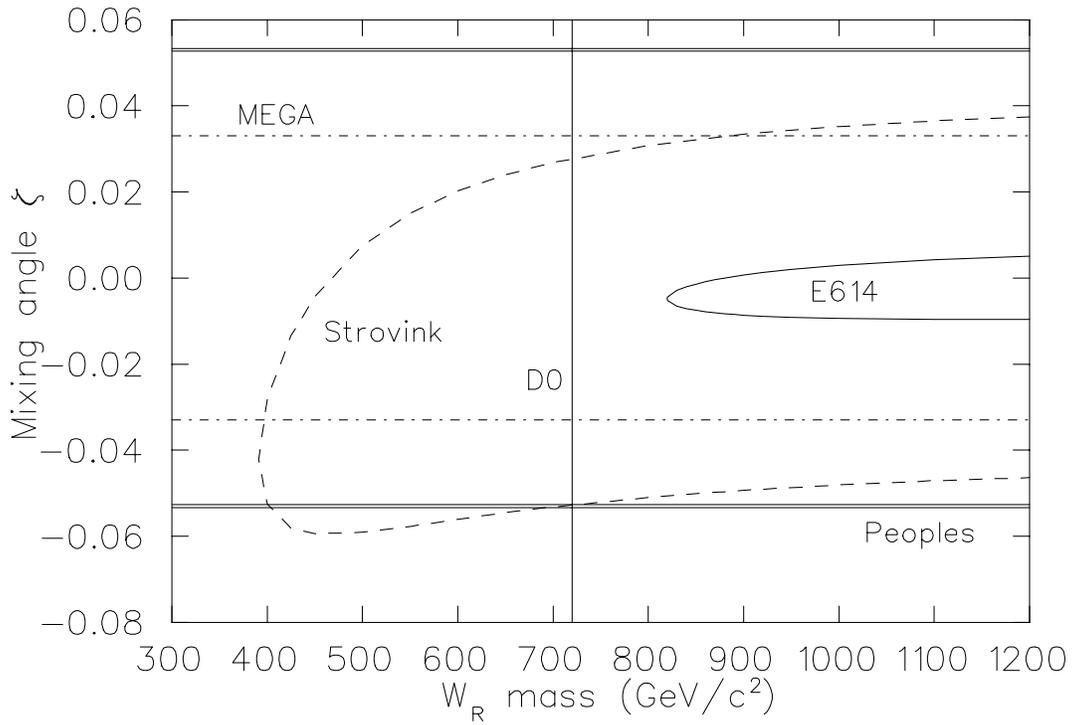,width=14cm}}
\vspace{5mm}
\caption{Constraints on the mass of $W_R$ vs. its mixing angle
($\zeta$) in the manifest left-right symmetric model. The experimental
constraints of ``Strovink'', ``Peoples'', ``MEGA'' and ``D0'' are from
Jodidio, {\em et al.} (1986), Derenzo (1969), the MEGA experiment
(unpublished), and Abachi, {\em et al.} (1996), respectively. The
aimed goal for E614 is also shown (provided by D.R.Gill).}
\label{fg:lrmodel}
\end{figure}

Among many theoretical models which can be studied by precise
measurements of normal muon decay, one example is the left-right
symmetric models of $SU(2)_{L} \times SU(2)_{R} \times U(1)$ symmetry
\cite{herc86,lang89}. In this model, heavy right-handed gauge bosons
(two charged $W^{\pm}_{R}$ and one neutral $Z'$) are added. In
general, they are mixed with each other, and form mass eigenstates
$W^{\pm}_{1,2}$.

\begin{equation}
\left( \matrix{ W^{\pm}_{L} \cr W^{\pm}_{R} } \right) = 
\left( \matrix{\cos\zeta & -\sin\zeta \cr 
e^{i\omega}\sin\zeta & e^{i\omega}\cos\zeta}\right) 
\left( \matrix{ W^{\pm}_{1} \cr W^{\pm}_{2} }\right), 
\end{equation}

\noindent where $\zeta$ is a mixing angle, and $\omega$ is a
CP-violating phase. One of the models is a manifest $SU(2)_{L}
\times SU(2)_{R} \times U(1)$ model, which has the same gauge
couplings and CKM quark mixing matrix elements for both the
$SU(2)_{L}$ and $SU(2)_{R}$ sectors. It, however, can receive a strong
constraint from the $K_{L}$$-$$K_{S}$ mass difference, yielding a
limit of the mass of $W_{R}$, $m_{R}>1.4$ TeV. The constraint can be
much weaker in general cases \cite{lang89}.  In the manifest
$SU(2)_{L}\times SU(2)_{R}\times U(1)$ model, the Michel parameters of
$\rho$ and $\xi$ would be related to $\zeta$ and the masses of
$W^{\pm}_{1,2}$ as follows:

\begin{eqnarray}
{1\over2}(1 - {4\over3}\rho) &=& \zeta^2, \\
{1\over2}(1-\xi) &=& \zeta^2 + \bigl( {m_{1} \over m_{2}} \bigr)^4,
\end{eqnarray}

\noindent where $m_{1,2}$ are the masses of $W^{\pm}_{1,2}$. The
direct lower limit for the $W^{\pm}_{2}$ mass of $m_{2} >$ 720
GeV/$c^2$ is obtained from D0 \cite{abac96}. Together with the
previous measurement of $\rho$ and $\xi$ parameters \cite{jodi86}, the
constraint on the $\zeta-m_{R}$ plane is given in
Fig.\ref{fg:lrmodel}.
  
\subsubsection{Radiative muon decay}

The spectrum of the radiative muon decay, \radmu, has been calculated
by several authors \cite{kino57a,fron59,ecks59}.  Within the framework
of the V$-$A interaction, the differential branching ratio of the
radiative muon decay, where final electron (positron) and photon are
emitted to energy intervals of $dx$ and $dy$ at solid angles of
$d\Omega_e$ and $d\Omega_{\gamma}$ respectively in the muon rest
frame, is expressed by

\begin{equation}
dB(\mu^{\pm}\rightarrow e^{\pm}\nu \overline{\nu} \gamma)
=\frac{\alpha}{64 \pi^3} \beta dx \frac{dy}{y} 
d\Omega_e d\Omega_{\gamma}
[F(x,y,d)\mp \beta \vec{P_{\mu}} \cdot \hat{p_e}G(x,y,d) \mp 
\vec{P_{\mu}} \cdot \hat{p_{\gamma}}H(x,y,d)].
\label{eq:br_radmu}
\end{equation}

\noindent $\vec{P_{\mu}}$ is the muon polarization
vector; $\vec{p_e}$ and $\vec{p_{\gamma}}$ are the momenta of an
electron (positron) and a photon in the muon rest frame respectively;
$\hat{p_e}$ and $\hat{p_{\gamma}}$ are their unit vectors defined by
$\hat{p_e}\equiv\vec{p_e}/|\vec{p_e}|$,
$\hat{p_{\gamma}}\equiv\vec{p_{\gamma}}/|\vec{p_{\gamma}}|$
respectively; $\beta$ is defined as $\beta \equiv |\vec{p_e}|/E_{e}$;
$d$ is given by $d\equiv 1-\beta\hat{p_e}\cdot\hat{p_{\gamma}}$; $x$
and $y$ are normalized electron and photon energies, $x =
2E_{e}/m_{\mu}$ and $y = 2E_{\gamma}/m_{\mu}$ in the muon rest
frame. From the four-body kinematics, the ranges of $x$ and $y$ are
given by

\begin{eqnarray}
2\sqrt{r}< x < 1+r &~~for~~& 0 < y \leq 1-\sqrt{r},\nonumber\\
(1-y) + r/(1-y) \leq x \leq 1+r &for& 1 - \sqrt{r} < y \leq 1 - r,
\label{eq:xyrange}
\end{eqnarray}

\noindent where $r = (m_{e}/m_{\mu})^2$.  $F(x,y,d)$, $G(x,y,d)$ and
$H(x,y,d)$ in the SM are given in Appendix~\ref{sc:ARMD}.

\begin{figure}[ht!]
\centerline{\epsfig{file=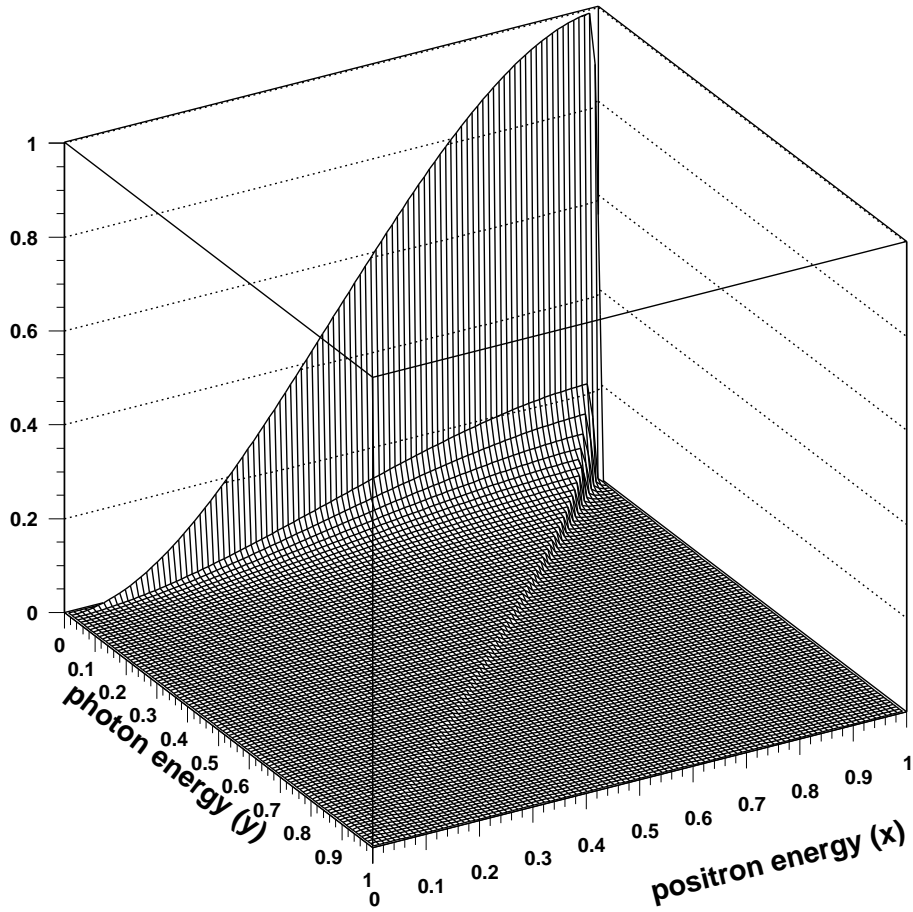,width=14cm}}
\caption{Decay probability distribution of \radmu decay as a function 
of the $e^{\pm}$ energy ($x \equiv {2E_{e}/m_{\mu}}$) and photon
energy $(y \equiv {2E_{\gamma}/m_{\mu}}$) for unpolarized muons. The
decay probability is high at $x\sim1$ and $y\sim0$.}
\label{fg:distribution-rmd}
\end{figure}

The decay probability distribution in the $x$-$y$ plane is shown in
Fig.\ref{fg:distribution-rmd}. The probability is high for an
energetic $e^{\pm}$ with a soft photon, namely $x\approx 1$ and
$y\approx 0$. In the soft-photon limit $(y\rightarrow 0)$, the
distribution has an infrared singularity which is canceled by the
radiative correction of the Michel decay.

\begin{figure}[ht!]
\centerline{\epsfig{file=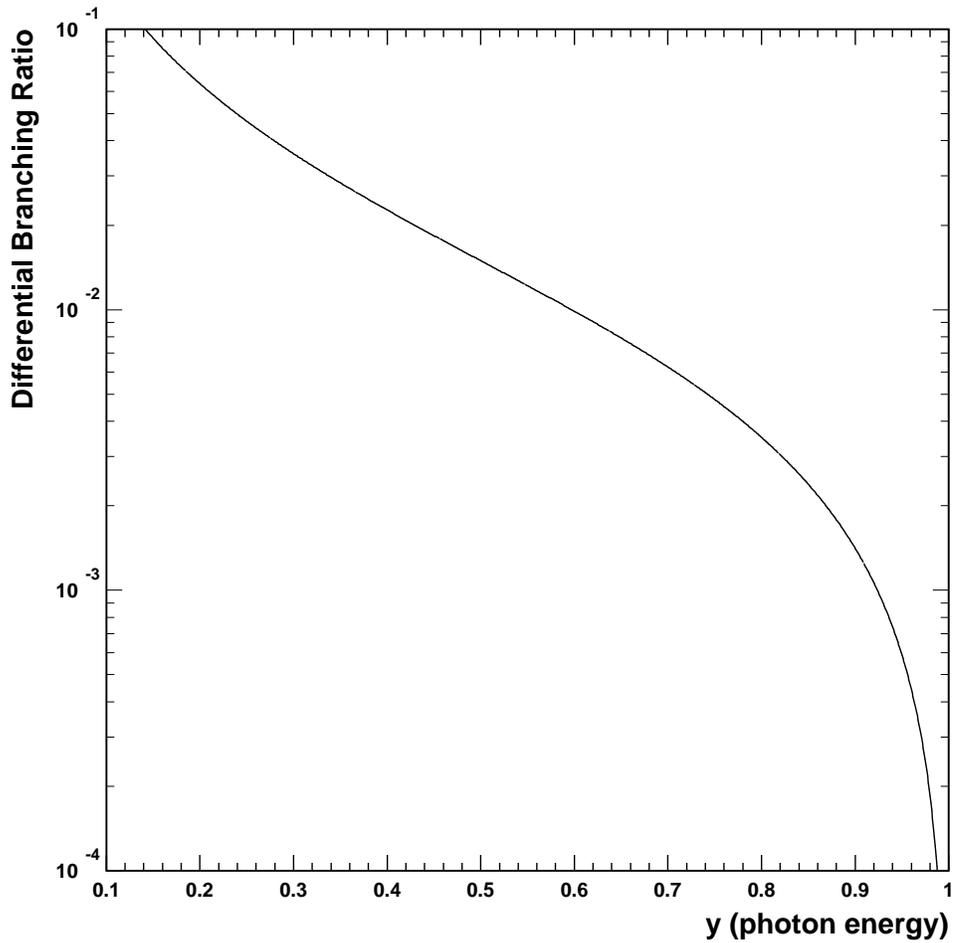,width=14truecm}}
\caption{Differential branching ratio of the \radmu decay as a
function of the photon energy ($y \equiv 2E_{\gamma}/m_{\mu}$). This
branching ratio is obtained by integrating over the $e^{+}$ energy and
the angle between an $e^{+}$ and a photon.}
\label{fg:photon-rmd}
\end{figure}

The photon spectrum is obtained by integrating over the electron
(positron) energy and angle variables. By neglecting the terms
suppressed by $m_e/m_{\mu}$, it is given by \cite{kuno97a}

\begin{equation}
\frac{dB(\mu^{\pm}\rightarrow e^{\pm}\nu \overline{\nu} \gamma)}
{dy d\cos\theta_{\gamma}}
 =\frac{1}{y}
\Bigl[ J_{+}(y)(1\pm P_{\mu}\cos\theta_{\gamma}) + 
       J_{-}(y)(1\mp P_{\mu}\cos\theta_{\gamma}) \Bigr],
\label{eq:radmu}
\end{equation}

\noindent where $J_{+}(y)$ and $J_{-}(y)$ are defined by

\begin{eqnarray}
J_{+}(y) &=& {\alpha \over 6\pi}(1-y)
\Bigl[(  3{\rm ln}{1-y\over r} - {17\over2} ) 
    + ( -3{\rm ln}{1-y\over r} + 7 )(1-y) \nonumber \\
 && + (  2{\rm ln}{1-y\over r} - {13\over3} )(1-y)^2 \Bigr], \\
J_{-}(y) &=& {\alpha \over 6\pi}(1-y)^2
\Bigl[(  3{\rm ln}{1-y\over r} - {93\over12} )
    + ( -4{\rm ln}{1-y\over r} + {29\over3})(1-y)  \nonumber \\
 && + (  2{\rm ln}{1-y\over r} - {55\over12})(1-y)^2 \Bigr],
\label{eq:jj}
\end{eqnarray}

\noindent and $\theta_{\gamma}$ is the angle between the muon spin
polarization and the photon momentum. The photon spectrum for
unpolarized muons is shown in Fig.\ref{fg:photon-rmd}. Note that at
the maximum photon energy $(y\sim 1)$, the photon distribution is
approximately given by $(1 + P_{\mu}\cos\theta_{\gamma})$ for the
$\mu^+ \rightarrow e^+\nu_e\overline{\nu}_{\mu}\gamma$ decay, because
$J_{+}(y)$ has a first-order term in $(1-y)$, but $J_{-}(y)$ only
contains the second and higher order terms.  This fact is important
for the suppression of accidental background in a $\mu^{+}
\rightarrow e^{+}\gamma$ search using polarized muons, as mentioned in
Section~\ref{sc:MP}.

In generalized interactions, the differential branching ratio of
\radmu decay has been calculated \cite{lena53,behr56,fron59}. Here,
the spectra of $e^{\pm}$ and a photon depend not only on the Michel
parameters of $\rho$ and $\delta$ in the normal muon decay, but also on
an additional parameter, $\overline{\eta}$, which should be zero in
the V$-$A interaction in the SM. Also, the asymmetry of $e^{\pm}$ in
\radmu from polarized muons is parameterized by another parameter,
$\xi\cdot\kappa$ \cite{fets95}.  Measurements of these parameters
would give additional constraints on the four fermion coupling
constants \cite{eich84}. Time-reversal violation in radiative muon
decay was also discussed \cite{prat58}, but it was concluded that the
T-odd effects have to include either the $e^{\pm}$ polarization or
those terms suppressed by the electron mass.

\section{Lepton Flavor Violation and Physics beyond the Standard Model}
\label{LFV}

In the minimal SM with vanishing neutrino masses, lepton flavor is
conserved separately for each generation. This is not necessarily true
if new particles or new interactions beyond the SM are introduced.  In
this section, we discuss various theoretical models with LFV in the
charged-lepton processes. In particular, we mention those in which LFV
effects could be large enough to be detected in on-going or future
experiments on \mueg decay, \meee decay, \muec conversion, and other
LFV processes. Among the theoretical models which predict observable
LFV effects, SUSY models have recently received much attention. In
SUSY models, the origin of LFV could be interactions at a very high
energy scale, such as the GUT scale or the mass scale of a heavy
right-handed Majorana neutrino that appears in the sea-saw mechanism.
Searches for rare muon decays, thereby, could provide a hint for
physics at a very high energy scale.  In the following, we first
explain the effective Lagrangians for various muon LFV processes of
$|\Delta L_{i}|=1$, and then discuss LFV in SUSY models and other
theoretical models. Finally, the muon LFV processes with $|\Delta
L_{i}|=2$ are discussed.

\subsection{Effective Lagrangians for lepton flavor violating processes}

The effective Lagrangians for muon LFV processes of $|\Delta
L_{i}|=1$, such as \mueg decay, \meee decay, and \muec conversion in a
muonic atom, are discussed. Possible LFV contributions can be grouped
into two types: photonic interaction and four fermion interaction.

First, the effective Lagrangian for \mueg process is given by

\begin{equation}
{\lagr}_{\mu \rightarrow e \gamma} 
     =  -{4G_F\over\sqrt{2}} \Biggl[ {m_{\mu }}{A_R}\overline{\mu_{R}}
        {{\sigma }^{\mu \nu}{e_L}{F_{\mu \nu}}}
       + {m_{\mu }}{A_L}\overline{\mu_{L}}
        {{\sigma }^{\mu \nu}{e_R}{F_{\mu \nu}}} + H.c. \Biggr],
\label{eq:muegeff}
\end{equation}

\noindent where $A_R$ and $A_L$ are coupling constants that
correspond to $\mu^{+}\rightarrow e^{+}_{R} \gamma$ and
$\mu^+\rightarrow e^{+}_{L} \gamma$ processes, respectively. 

For \meee decay and \muec conversion, off-shell photon emission also
contributes. The general photonic $\mu - e$ transition amplitude is,
then, written as

\begin{eqnarray}
M_{photonic}&=& -\Biggl[ e A^*_{\mu}(q) \overline{u}_{e}(p_{e})
\Bigl[ (f_{E0}(q^2) + \gamma_{5}f_{M0}(q^2))\gamma_{\nu}
(g^{\mu\nu} - {q^{\mu}q^{\nu} \over q^2}) \nonumber\\ && +
(f_{M1}(q^2) + \gamma_{5}f_{E1}(q^2))
{i\sigma_{\mu\nu}q^{\nu}\over m_{\mu}}\Bigr] u_{\mu}(p_{\mu})
\Biggr],
\label{eq:photonic}
\end{eqnarray}

\noindent where $p_{\mu}$ and $p_{e}$ are the $\mu^-$ and $e^-$
four momenta, and $q \equiv p_{\mu}-p_{e}$ is the four-momentum
transfer. The electromagnetic form factors ($f_{E0}$, $f_{E1}$,
$f_{M0}$ and $f_{M1}$) are functions of $q^2$. For \mueg decay, only
$f_{E1}(0)$ and $f_{M1}(0)$ can contribute, whereas all of the four
form factors could contribute to \meee decay and \muec conversion. The
coupling constants $A_R$ and $A_L$ are related to the dipole form
factors as

\begin{eqnarray}
A_{R} &=&-{\sqrt{2}e \over 8G_{F}^2m_{\mu}^2}(f_{E1}^*(0) +
f_{M1}^*(0)), 
\label{eq:ar}\\
A_{L} &=& {\sqrt{2}e \over 8G_{F}^2m_{\mu}^2}(f_{E1}^*(0) - f_{M1}^*(0)). 
\label{eq:al}
\end{eqnarray}

The direct four fermion interactions could introduce \meee decay and
\muec conversion, in addition to the photonic $\mu - e$ transition in
Eq.(\ref{eq:photonic}). For the \meee decay, the general four fermion
couplings are given by

\begin{eqnarray}
{\lagr}_{\mu \rightarrow 3e}^{non-photo}
     &=&-{4G_{F}\over\sqrt{2}} \Biggl[ {g_1}(\overline{{{\mu }_R}}{e_L})
              (\overline{{e_R}}{e_L})
       + {g_2}(\overline{{{\mu }_L}}{e_R})
              (\overline{{e_L}}{e_R}) \nonumber \\
    &&   +{g_3}(\overline{{{\mu }_R}}{{\gamma }^{\mu }}{e_R})
              (\overline{{e_R}}{{\gamma }_{\mu }}{e_R})
       + {g_4}(\overline{{{\mu }_L}}{{\gamma }^{\mu }}{e_L})
              (\overline{{e_L}}{{\gamma }_{\mu }}{e_L})  \nonumber \\
    &&   +{g_5}(\overline{{{\mu }_R}}{{\gamma }^{\mu }}{e_R})
              (\overline{{e_L}}{{\gamma }_{\mu }}{e_L})
       + {g_6}(\overline{{{\mu }_L}}{{\gamma }^{\mu }}{e_L})
              (\overline{{e_R}}{{\gamma }_{\mu }}{e_R})
       +  H.c. \Biggr],
\label{eq:mu3eeff}
\end{eqnarray}

\noindent where the Fierz rearrangement for the four fermion
operators is used.  For the $\mu^{-} - e^{-}$ conversion process, the
relevant interactions are written as

\begin{eqnarray}
{\lagr}_{\mu-e~conv}^{non-photo} &=& -\frac{G_F}{\sqrt{2}}\sum_{q=u,d,s...}
\Biggl[ (g_{LS(q)} \overline{e_L}\mu_{R}+ 
    g_{RS(q)} \overline{e_R}\mu_{L})\overline{q}q
\nonumber\\
&&+  (g_{LP(q)} \overline{e_L}\mu_{R}+ 
    g_{RP(q)} \overline{e_R}\mu_{L})\overline{q}\gamma_{5}q
\nonumber\\
&&+  (g_{LV(q)} \overline{e_L}\gamma^{\mu}\mu_{L}+ 
    g_{RV(q)} \overline{e_R}\gamma^{\mu}\mu_{R})\overline{q}\gamma_{\mu}q
\nonumber\\
&&+  (g_{LA(q)} \overline{e_L}\gamma^{\mu}\mu_{L}+ 
    g_{RA(q)} \overline{e_R}\gamma^{\mu}\mu_{R})
   \overline{q}\gamma_{\mu}\gamma_{5}q
\nonumber\\
&&+\frac{1}{2}(g_{LT(q)} \overline{e_L}\sigma^{\mu \nu}\mu_{R}+ 
    g_{RT(q)} \overline{e_R}\sigma^{\mu \nu}\mu_{L})
    \overline{q}\sigma_{\mu \nu}q + H.c. \Biggr],
\label{eq:conveff}
\end{eqnarray}

\noindent where $g_{LX(q)}$ and $g_{RX(q)}$ are the coupling constants 
for the left-handed and right-handed lepton currents, respectively,
and $X=S,P,V,A,T$ represent scalar, pseudoscalar, vector, axial
vector, and tensor interactions, respectively. Here, the
flavor-changing quark currents are not included.  The four fermion
coupling constants introduced in Eqs.(\ref{eq:mu3eeff}) and
(\ref{eq:conveff}) are determined by specific contributions in some
theoretical models beyond the SM. For examples, they are box diagrams
in supersymmetric models, tree diagrams of $Z^{'}$, supersymmetric
models with $R-$parity breaking, and others.

\begin{figure}[ht!]
\centerline{\epsfig{file=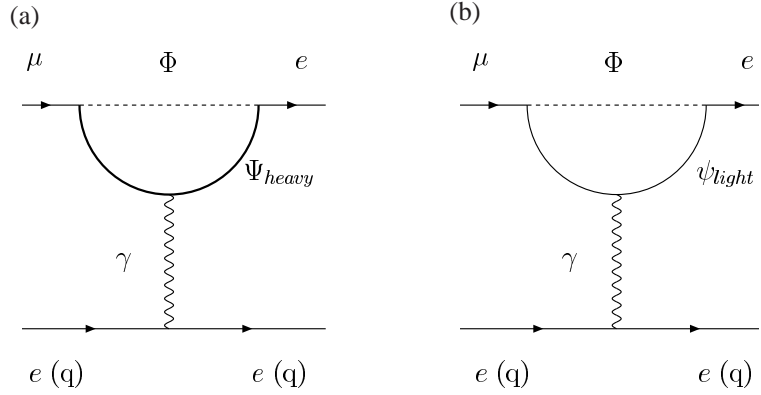,width=10cm}}
\vspace{5mm}
\caption{Photonic penguin diagrams for $\mu-e$ transitions, such as
\meee or \muec conversion. The cases of (a) a heavy particle
($\Psi_{heavy}$) in the loop, and (b) a light fermion
($\psi_{light}$) in the loop are shown. $\Phi$ is a scalar field.}
\label{fg:diagram-penguin}
\end{figure}

The $f_{E0}$ and $f_{M0}$ form factors contribute to off-shell photons
and not to real photon emission. Therefore, they vanish in the
$q^2\rightarrow 0$ limit. They could be rewritten by

\begin{eqnarray}
f_{E0}(q^2) &=& \frac{q^2}{m_{\mu}^2} {\tilde f_{E0}}(q^2), \\
f_{M0}(q^2) &=& \frac{q^2}{m_{\mu}^2} {\tilde f_{M0}}(q^2),
\end{eqnarray}

\noindent where ${\tilde f_{E0}}(q^2)$ and ${\tilde f_{M0}}(q^2)$
are finite at $q^2 \rightarrow 0$. If these transitions are induced by
loop diagrams including heavy particles, ${\tilde f_{E0}}(q^2)$ and
${\tilde f_{M0}}(q^2)$ are regarded as slowly varying functions of
$q^2$.  One example of such diagrams is shown in
Fig.\ref{fg:diagram-penguin}(a). In such a case, these form factors
could be translated into additional contributions which should be
added into the corresponding four-fermion coupling constants in
Eqs.(\ref{eq:mu3eeff}) and (\ref{eq:conveff}). Those additional
contributions are

\begin{eqnarray}
\Delta g_3=\Delta g_5&=&{\sqrt{2}\over4G_{F}}{e^2\over m_{\mu}^2}
( {\tilde f_{E0}^*}(0) +  {\tilde f_{M0}^*}(0)),\\
\Delta g_4=\Delta g_6&=&{\sqrt{2}\over4G_{F}}{e^2\over m_{\mu}^2}
( {\tilde f_{E0}^*}(0) -  {\tilde f_{M0}^*}(0)),
\end{eqnarray}

\noindent for $g_3$, $g_4$, $g_5$, and $g_6$, correspondingly, and

\begin{eqnarray}
\Delta g_{LV(u)}= -2\Delta g_{LV(d)}
&=& -\frac{2}{3}{\sqrt{2}\over G_{F}}{e^2\over m_{\mu}^2} ( {\tilde
f_{E0}}(0) + {\tilde f_{M0}}(0)),\\
\Delta g_{RV(u)}=-2\Delta g_{RV(d)}&=&
-\frac{2}{3}{\sqrt{2}\over G_{F}}{e^2\over m_{\mu}^2}
( {\tilde f_{E0}}(0) -  {\tilde f_{M0}}(0)),
\end{eqnarray}

\noindent for $g_{LV(u)}$, $g_{LV(d)}$, $g_{RV(u)}$ and $g_{RV(d)}$
respectively.  

If these form factors are generated by penguin diagrams with a photon
coupled to an internal line of a light fermion, as seen in
Fig.\ref{fg:diagram-penguin}(b), ${\tilde f_{E0}}(q^2)$ and ${\tilde
f_{M0}}(q^2)$ have a logarithmic dependence of $q^2$ that is cut off
by the light-fermion mass.  Such examples are a model with a doubly
charged Higgs boson and the SUSY models with $R$-parity violation,
which are discussed later in Section \ref{sc:OTM}. The logarithmic
factor could enhance the rates of \meee decay and \muec conversion,
but not that of \mueg decay.

If $f_{E1}(q^2)$ and $f_{M1}(q^2)$ dominate, the following simple
relations among the branching ratios of \mueg, \meee and \muec
conversion can be derived:

\begin{eqnarray} 
{\Gamma(\mu T_i \rightarrow e T_i) \over
 \Gamma(\mu T_i \rightarrow capture)}
&\simeq & {1\over 200} B(\mu^+ \rightarrow e^+ \gamma),
\label{eq:lfv-relation1} \\
B(\mu^+ \rightarrow e^+ e^+ e^-)
&\simeq & {1\over 160} B(\mu^+ \rightarrow e^+ \gamma).
\label{eq:lfv-relation2}
\end {eqnarray}

\noindent These relations hold in some models of SUSY GUT, which
are discussed in the next subsection. Regarding \muec conversion, more
detailed discussions on Eq.(\ref{eq:lfv-relation1}), including the
nuclear dependence, are given in Section \ref{sc:PMUEC}.
 
\subsection{Supersymmetric models}

\subsubsection{Introduction to supersymmetric models}

Phenomenological applications of SUSY theories have been considered
since the late 70's in connection with the naturalness problem (or the
hierarchy problem) in the SM.  The SM model can be regarded as being a
low-energy approximation of a more complete theory, and is replaced by
this at a high energy scale. Supposing that this high energy scale is
close to the Planck scale ($\approx 10^{19}$ GeV), the quadratic
divergence appearing in the radiative corrections to the Higgs scalar
mass becomes problematic, because a precise fine tuning between the
bare mass and the radiative corrections must be necessary to keep the
electroweak scale well below this high energy scale. This problem can
be avoided if SUSY is introduced, since the quadratic divergence is
canceled between the fermionic and bosonic loop contributions. For
general reviews on SUSY models, see other references
\cite{nill84,habe85}.

The minimal SUSY extension of the SM is called the minimal
supersymmetric Standard Model (MSSM). In MSSM, SUSY partners (which
have a different spin by 1/2 from the corresponding ordinary
particles) are introduced for each particle in the SM.  For quarks and
leptons, complex scalar fields, squark ($\tilde{q}$) and slepton
($\tilde{l}$), are introduced.  The superpartner of the gauge boson is
a gauge fermion (or a gaugino), and that of the Higgs field is called
a higgsino($\tilde{H}$).  The superpartners of gluon, $SU(2)$ and
$U(1)$ gauge bosons are a gluino ($\tilde{G}$), a wino ($\tilde{W}$)
and a bino ($\tilde{B}$), respectively. After electroweak symmetry
breaking, the wino, bino and higgsino mix with each other and form two
charged Dirac fermions, called charginos ($\tilde{\chi}^{\pm}_{i}$)
($i=1,2$), and four Majorana fermions, called neutralinos
($\tilde{\chi}^0_{i}$) ($i=1-4$). As for the Higgs sector, the SUSY
models contain at least two Higgs doublet fields.  They are required
separately for the mass terms for up-type quarks, and those for
down-type quarks and charged leptons to eliminate any possible
conflict with SUSY in the Yukawa couplings. The particle contents of
the MSSM are listed in Table~\ref{tb:pc_mssm}.

\begin{table}[htb!]
\caption{Particle Contents in the MSSM.}
\label{tb:pc_mssm}
\vspace{5mm}
\begin{tabular}{ccc|ccc}
\multicolumn{3}{c|}{Ordinary particles} & \multicolumn{3}{c}{SUSY
particles}\cr\hline
Particle & Spin & ~~~& Particle & Spin & ~~~\cr \hline
quark ($q$)  & ${1\over2}$ && squark ($\tilde{q}$) & 0 & \cr\hline
lepton ($l$) & ${1\over2}$ && slepton ($\tilde{l}$) & 0 & \cr\hline
gluon ($G$)  & 1 && gluino $(\tilde{G}$) & ${1\over2}$ & \cr\hline
$W^{\pm}$, $Z^{0}$, $\gamma$ & 1 
&& chargino ($\tilde{\chi}^{\pm}_{i}$)~($i=1-2$) & ${1\over2}$ & \cr
Higgs boson ($h,H,A,H^{\pm}$) & 0 
&& neutralino ($\tilde{\chi}^{0}_{i}$)~($i=1-4$) & ${1\over2}$ &\cr
\end{tabular}
\end{table}      

The MSSM Lagrangian consists of two parts. They are the SUSY invariant
Lagrangian and the soft SUSY-breaking terms, as follows:

\begin{equation}
\cal L \mit  = \cal L \mit _{SUSY~inv} + \cal L \mit _{SUSY~breaking},
\end{equation}

\noindent The MSSM Lagrangian is described in more details in
Appendix~\ref{sc:AMSSM}.  One of the important features of the SUSY
invariant Lagrangian is that various bosonic and fermionic
interactions are related to each other by the requirement of SUSY
invariance. For example, the gauge coupling constants appear not only
in the covariant derivative, but also in the gaugino-scalar-fermion
interactions and the scalar self-couplings.

The ordinary Yukawa coupling constants are included in a scalar
function called the superpotential ($W(\phi_i)$). The Lagrangian
specified by the superpotential (${\lagr}_{superpotential}$) contains
a set of fermionic interactions and scalar potentials, as follows:

\begin{eqnarray}
{\lagr}_{superpotential} &=& - \sum_{i} |\frac {W(\phi)}{\partial
\phi_i}|^2 -\frac{1}{2}\frac{\partial ^2 W({\phi})}{\partial \phi_i
\partial \phi_j}
\overline {(\psi_{i L})^c} \psi_{j L} + H.c.,
\label{eq:superpotential}
\end{eqnarray}

\noindent where the scalar field ($\phi_i$) and the left-handed Weyl
field ($\psi_{i L}$) form a chiral multiplet of SUSY. The
superpotential of the MSSM is given by

\begin{eqnarray}
W_{MSSM} &=& {(y_{e})}_{ij}H_{1} E_{i}^{c} L_{j}
+{(y_{d})}_{ij} H_{1} D_{i}^{c} Q_{j}
+{(y_{u})}_{ij} H_{2} U_{i}^{c} Q_{j}-\mu H_{1}H_{2},
\label{eq:Lagrangian_MSSM}
\end{eqnarray}

\noindent where $E_{i}^{c}$ and $L_{i}$ represent the supermultiplets
of $SU(2)_L$ lepton singlets and doublets, respectively. Also, $Q_{i}$
$U_{i}^{c}$ and $D_{i}^{c}$ are the supermultiplets for quark
doublets, up-type quark singlets, and down-type quark singlets,
respectively. $H_{1}$ and $H_{2}$ are two Higgs doublet fields. From
this superpotential in Eq.(\ref{eq:Lagrangian_MSSM}), the following
Yukawa couplings are induced:

\begin{eqnarray}
\cal{L}_{\mit Yukawa} &=& - \Biggl[ 
 {(y_{e})}_{ij} H_{1} \overline{e_{i}}_{R} l_{j L}
+{(y_{d})}_{ij} H_{1} \overline{d_{i}}_{R} q_{j L}
+{(y_{u})}_{ij} H_{2} \overline{u_{i}}_{R} q_{j L} \Biggr] + H.c.
\label{eq:susy-yukawa}
\end{eqnarray}

\noindent In addition to Eq.(\ref{eq:susy-yukawa}), the
superpotential in Eq.(\ref{eq:Lagrangian_MSSM}) generates the higgsino
mass term, various Yukawa-type couplings, and the three- and
four-point scalar couplings, according to
Eq.(\ref{eq:superpotential}).

\subsubsection{Flavor problems in supersymmetric models}

In the MSSM, the masses of superparticles, {\it i.e.}, squarks,
sleptons, and gaugino, are generated by the soft SUSY-breaking mass
terms, which are defined as SUSY-breaking terms that do not induce
quadratic divergence.  In general, the soft SUSY-breaking mass terms
become a new source of flavor mixing in the MSSM, which is not
necessarily related to the flavor mixing in the Yukawa coupling
constant in Eq.(\ref{eq:Lagrangian_MSSM}). For the slepton sector, the
soft SUSY-breaking mass terms are given by

\begin{eqnarray}
\cal L_{\mit soft} &=&  
-(m^{2}_{E})_{ij}\tilde{e}_{Ri}^{*}\tilde{e}_{Rj}
-(m^{2}_{L})_{ij}\tilde{l_{Li}}^{*} \tilde{l_{Lj}}
-\{m_{0}(A_{e})_{ij}H_{1}\tilde{e}_{Ri}^{*}\tilde{l}_{Lj} + H.c.\},
\end{eqnarray}

\noindent where $(m_{E})_{ij}$ and $(m_{L})_{ij}$ are the mass
matrices for the right-handed sleptons ($\tilde{e}_{R}$) and
left-handed sleptons ($\tilde{l}_{L}$), respectively; $m_{0}$ is a
SUSY-breaking parameter and $A_{e}$ is a dimensionless scalar
trilinear coupling matrix.

\begin{figure}[ht!]
\centerline{\epsfig{file=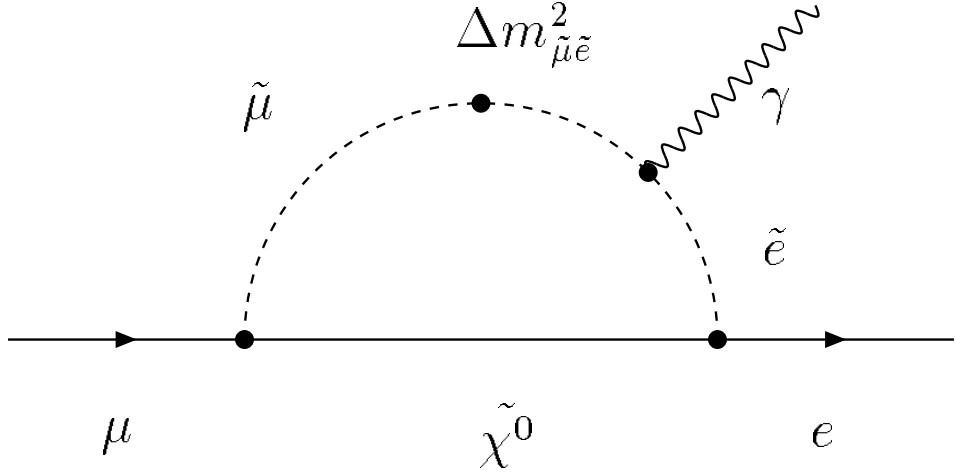,width=10cm}}
\vspace{5mm}
\caption{Feynman diagram for \mueg decay induced by 
slepton flavor mixing ($\Delta m^{2}_{\tilde{\mu}\tilde{e}}$).}
\label{fg:diagram-mueg}
\end{figure}

In the basis where the lepton mass matrix is diagonalized, if there
exist non-zero off-diagonal matrix elements in the slepton mass
matrix, LFV is introduced. From the diagram in
Fig.\ref{fg:diagram-mueg}, the constraint on the off-diagonal elements
on the slepton mass matrix, for instance the $\tilde{\mu}-\tilde{e}$
element ($\Delta m_{\tilde{\mu}\tilde{e}}$), can be placed from the
present upper limit of \mueg decay as follows:

\begin{equation}
\frac{\Delta m_{\tilde{\mu}\tilde{e}}^2}{ m_{\tilde{l}}^2}\lsim 10^{-3}
\left(\frac{ m_{\tilde{l}}}{100 GeV}\right)^2,
\end{equation}
  
\noindent where $m_{\tilde{l}}$ is the mass of a slepton. Similar
constraints on the squark mass matrix elements are obtained from the
flavor-changing neutral current (FCNC) processes in the quark sector.
For example, the observed $K^{0}-\overline{K}^{0}$ mixing places
constraints that the possible SUSY contribution must be small,
resulting in that the squarks in the first and second generations must
be degenerate at the level of a few~\% in the case that the squark
mass is a few 100 GeV and that the squark mixing angle is of similar
magnitude to the Cabibbo angle. These constraints from LFV and FCNC
suggest that there should be a special suppression mechanism on the
flavor mixing of sfermions (squarks and sleptons) from the dynamics
of SUSY breaking. It is called the SUSY flavor problem.

There are several scenarios to solve the SUSY flavor problem:

\begin{itemize}

\item The soft SUSY-breaking mass terms have a universal structure at
a very high energy scale, such as the Planck scale ($\approx 10^{19}$
GeV) ({\em Gravity mediated SUSY breaking scenario}) \cite{nill84}.

\item The SUSY-breaking effects are mediated by the SM gauge
interaction so that squarks and sleptons with the same quantum numbers
receive the same amount of the soft SUSY-breaking mass terms ({\em
Gauge mediated SUSY breaking scenario})
\cite{dine93,dine95,dine96,giud98}.

\item There is some approximate flavor symmetry which produces nearly
degenerate masses for squarks and sleptons, at least for the first two
generations ({\em Flavor Symmetry Scenario}) \cite{barb96}.

\item Squarks and sleptons can be diagonalized in the same basis as
the quarks and lepons ({\em Alignment Scenario}) \cite{nir93}.

\item The squark and slepton masses are heavy enough (10 -100 TeV) to
avoid constraints from FCNC and LFV, at least for the sfermions in the
first two generations ({\em Effective SUSY scenario})
\cite{cohe96}.

\end{itemize}

The minimal supergravity model (SUGRA) is a realization of the first
scenario.  There are many phenomenological analyses based on the
supergravity model. In this model, all squarks and sleptons receive
the same magnitudes of the soft SUSY-breaking mass terms through the
coupling of supergravity at the Planck scale, so that the mass
matrices are diagonal with the same diagonal elements. Therefore,
there is neither FCNC nor LFV at that energy scale.  This does not
necessarily imply that LFV effects should not exist in this
scenario. In fact, if there is some interaction which breaks lepton
flavor conservation between the Planck ($\approx 10^{19}$ GeV) and the
electroweak scales ($\approx 10^{2}$ GeV), the LFV effect could be
induced in the slepton mass matrices through radiative corrections
\cite{hall86}.  In recent years, it has been noticed that such an effect
can induce muon LFV processes at large branching ratios in some models
of SUSY GUT \cite{barb94,barb95a}.

\subsubsection{SUSY GUT and lepton flavor violation}

In SUSY GUT \cite{dimo81,saka81}, the SM gauge groups of $SU(3)_C$,
$SU(2)_L$ and $U(1)_Y$ are assumed to be unified by a larger group at
a high energy scale.  In recent years, SUSY GUT has attracted much
attention because the three gauge coupling constants determined at LEP
and SLC are consistent with the $SU(5)$ GUT prediction if
contributions from SUSY particles are taken into account in the
evolution of the coupling constants. The three coupling constants are
then unified at $2 \times 10^{16}$ GeV \cite{amal91,lang91,ell91}.  It
suggests that SUSY GUTs with the $SU(5)$ group or other gauge groups
which include $SU(5)$ could be a viable candidate of physics beyond
the SM.

Let us first discuss how LFV would be induced in $SU(5)$ SUSY GUT.  In
this model, quarks and leptons are classified in the three generations
of ${\bf \bar{5}}$ and ${\bf 10}$ representations of the $SU(5)$
group, where the ${\bf\bar{5}}$ representation ($\bar{F}_i$) contains
$d_{i L}^c$ and $l_{i L}$ and their superpartners, and the ${\bf 10}$
representation ($T_i$) contains $q_{i L}$, $u_{i L}^c$, and $e_{i
L}^c$ and their superpartners. The Yukawa coupling constants at the
GUT scale are determined by the superpotential,

\begin{equation} 
W_{SU(5)} = \frac{1}{8}(y_u)_{ij}{T_i}\cdot {T_j}\cdot {H(5)}
+(y_d)_{ij}{\bar{F}_i}\cdot{T_j}\cdot {\bar{H}(5)},
\label{su5sp}
\end{equation}

\noindent where ${H(5)}$ and ${\bar{H}(5)}$ are two Higgs
fields associated with the ${\bf 5}$ and ${\bf \bar{5}}$
representations, respectively.  By substituting the fermionic fields
for $\bar{F}_i$ and $T_i$, and also the Higgs boson fields for
${H(5)}$ and ${\bar{H}(5)}$, the Yukawa couplings responsible to the
quark and lepton masses are obtained. Here, the matrix $(y_u)_{ij}$
corresponds to the Yukawa coupling matrix for the up-type quarks, and
$(y_d)_{ij}$ to that for the down-type quarks, and the leptons.

\begin{figure}[ht!]
\centerline{\epsfig{file=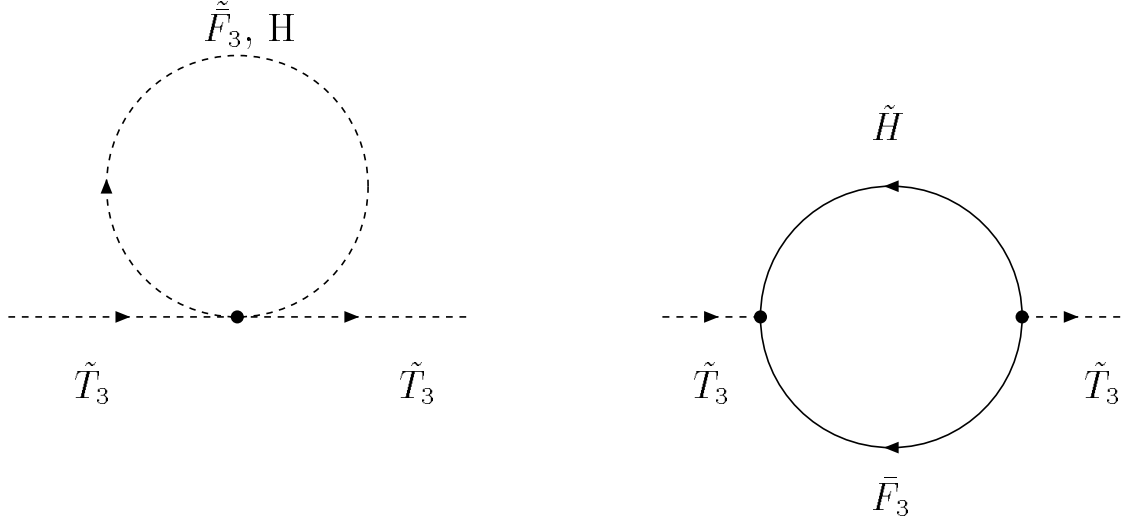,width=14cm}}
\vspace{5mm}
\caption{Feynman diagrams which contribute to the 
renormalization effect on the slepton masses from the Planck to the GUT
energy scales.}
\label{fg:diagram-renorm}
\end{figure}

In the minimal supergravity model, all of the scalar partners of the
quarks and leptons, namely squarks and sleptons, are assumed to have a
common SUSY-breaking mass from the coupling of the gravity
interaction. In addition to the mass terms, the scalar triple
couplings also have a universal structure so that they are
proportional to the corresponding Yukawa coupling constants. At the
Planck scale, the soft SUSY-breaking mass terms are given by

\begin{eqnarray}
\cal L_{\mit soft} &=&  
-m_0^2 \{ \tilde{T}_{i}^{\dag} \tilde{T}_{i}+\tilde{\bar{F}_i}^{\dag} 
\tilde{\bar{F}_{i}}\}
- \Biggl[ m_{0}A_0 \lbrace {1\over8}(y_u)_{ij}{\tilde{T}_i}\cdot
{\tilde{T}_j}\cdot {H(5)}
+(y_d)_{ij}{\tilde{\bar{F}}_i}\cdot{\tilde{T}_j}\cdot
{\bar{H}(5)}\rbrace +H.c.\Biggr],
\end{eqnarray}  

\noindent where $m_{0}$ is the universal scalar mass, and $A_{0}$ is
the universal trilinear coupling. At this stage, there is no LFV in
the slepton sector. When the lepton Yukawa coupling constants (in this
case $y_d$) are diagonalized by unitary transformations on the bosons
and fermions of $\bar{F}_i$ and $T_i$ for each generation indices, the
soft SUSY-breaking mass terms for the sleptons can also become
diagonal in the same basis.  This is no longer true if we take into
account the radiative corrections to the soft SUSY-breaking mass terms
due to the Yukawa coupling constants. In particular, since the Yukawa
coupling constant corresponding to the top-quark mass is surprisingly
large, its effects on the soft SUSY-breaking mass terms are expected
to be sizeable. In the basis where the matrix of up-type Yukawa
coupling constants is diagonal, all members of $T_i$, including the
right-handed slepton masses, are obtained by

\begin{eqnarray}
m_{T}^{2} &\simeq& \left(
\begin{array}{ccc}
m^{2} &       & \\
      & m^{2} & \\
      &       & m^{2}+\Delta m^{2}
\end{array}
\right) \quad {\rm and}\\
\Delta m^{2} &\simeq& -\frac{3}{8\pi^{2}}|(y_{u})_{33}|^{2}m_{0}^{2}(3 +
|A_{0}|^{2}) \ln(\frac{M_{P}}{M_{G}}), 
\label{eq:massdifference}
\end{eqnarray}

\noindent where $M_{P}$ and $M_{G}$ denote the reduced Planck mass
($\sim2\times10^{18}$GeV) and the GUT scale ($\sim2\times10^{16}$GeV).
$\Delta m^{2}$ arises from the evolution of the renormalization group
equation between the Planck and the GUT scales through the diagrams in
Fig.\ref{fg:diagram-renorm}.

Since the physical LFV effect is induced by a mismatch of the lepton
and slepton diagonalization, the off-diagonal terms of the slepton
mass matrix can be examined in the basis where the Yukawa coupling
constant for leptons is diagonalized. If it is diagonalized by 

\begin{eqnarray}
V_{R}y_{e}V_{L}^{\dag} &=& diagonal,
\end{eqnarray}

\noindent the off-diagonal elements of the right handed slepton mass
matrix in this new basis are given by

\begin{eqnarray}
(m_{\tilde{e}_{R}}^{2})_{ij} \simeq -\frac{3}{8\pi^{2}}
(V_{R})_{i3}(V_{R})_{j3}^{*}|y_{u}^{33}|^{2}m_{0}^{2} (3 +
|A_{0}|^{2})\ln(\frac{M_{P}}{M_{G}}).
\label{eq:m_R_su5} 
\end{eqnarray}

\begin{figure}[ht!]
\centerline{\epsfig{file=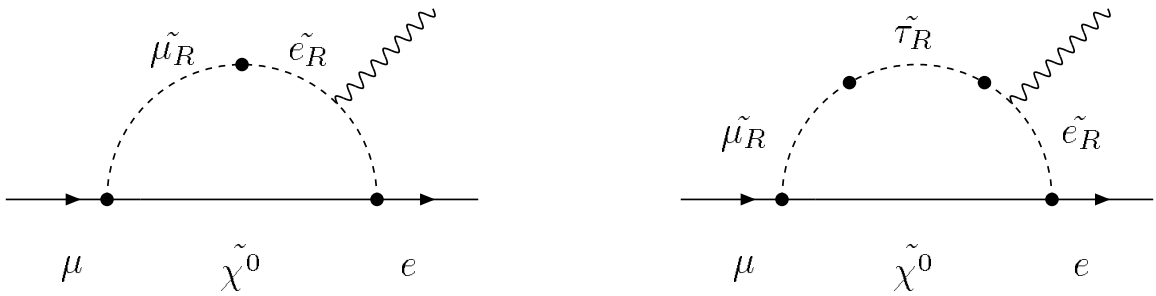,width=14cm}}
\vspace{5mm}
\caption{Feynman diagrams for the \mueg decay in $SU(5)$ SUSY GUT. The 
closed blobs represent the flavor transitions due to the off-diagonal
terms of the slepton mass matrices.}
\label{fg:diagram-su5-mueg}
\end{figure}

\noindent It becomes a source of \mueg decay through the diagrams in
Fig.\ref{fg:diagram-su5-mueg}. If the $SU(5)$ GUT relation for the
down-type quark and lepton Yukawa coupling constants given by

\begin{equation}
y_{e} = y_{d}^{T}, 
\end{equation}

\noindent is used, $V_{R}$ is given by the CKM matrix at the GUT
scale as

\begin{eqnarray}
(V_{R})_{ij} &=& (V_{CKM}^{GUT})_{ji},
\label{eq:ckm_gut}
\end{eqnarray}

\noindent where $V_{CKM}^{GUT}$ can be obtained from the CKM matrix at the
electroweak scale by taking into account the effects of running
coupling constants from the electroweak to the GUT scales.

\begin{figure}[ht!]
\centerline{\epsfig{file=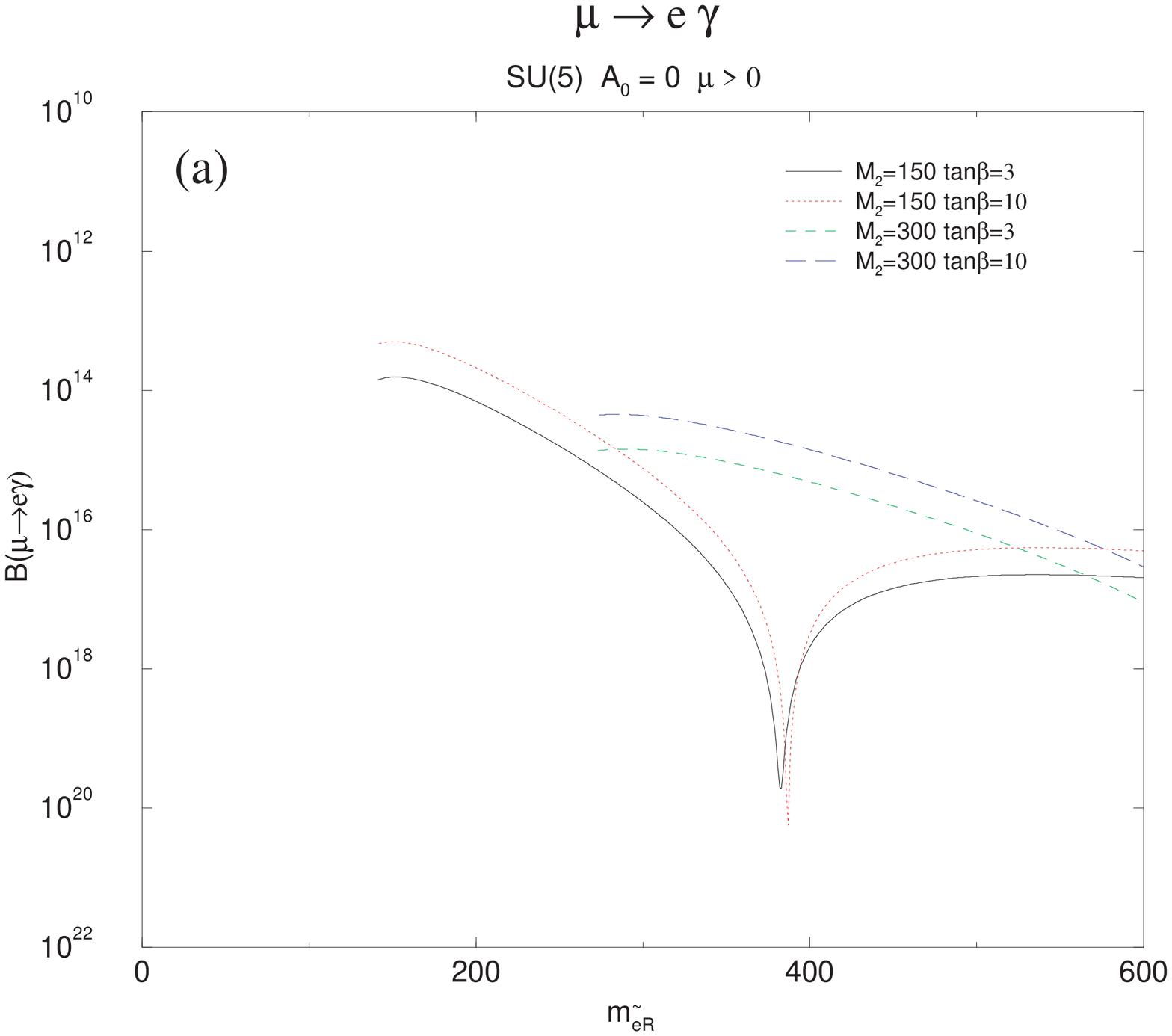,angle=270,width=9.5cm}}
\centerline{\epsfig{file=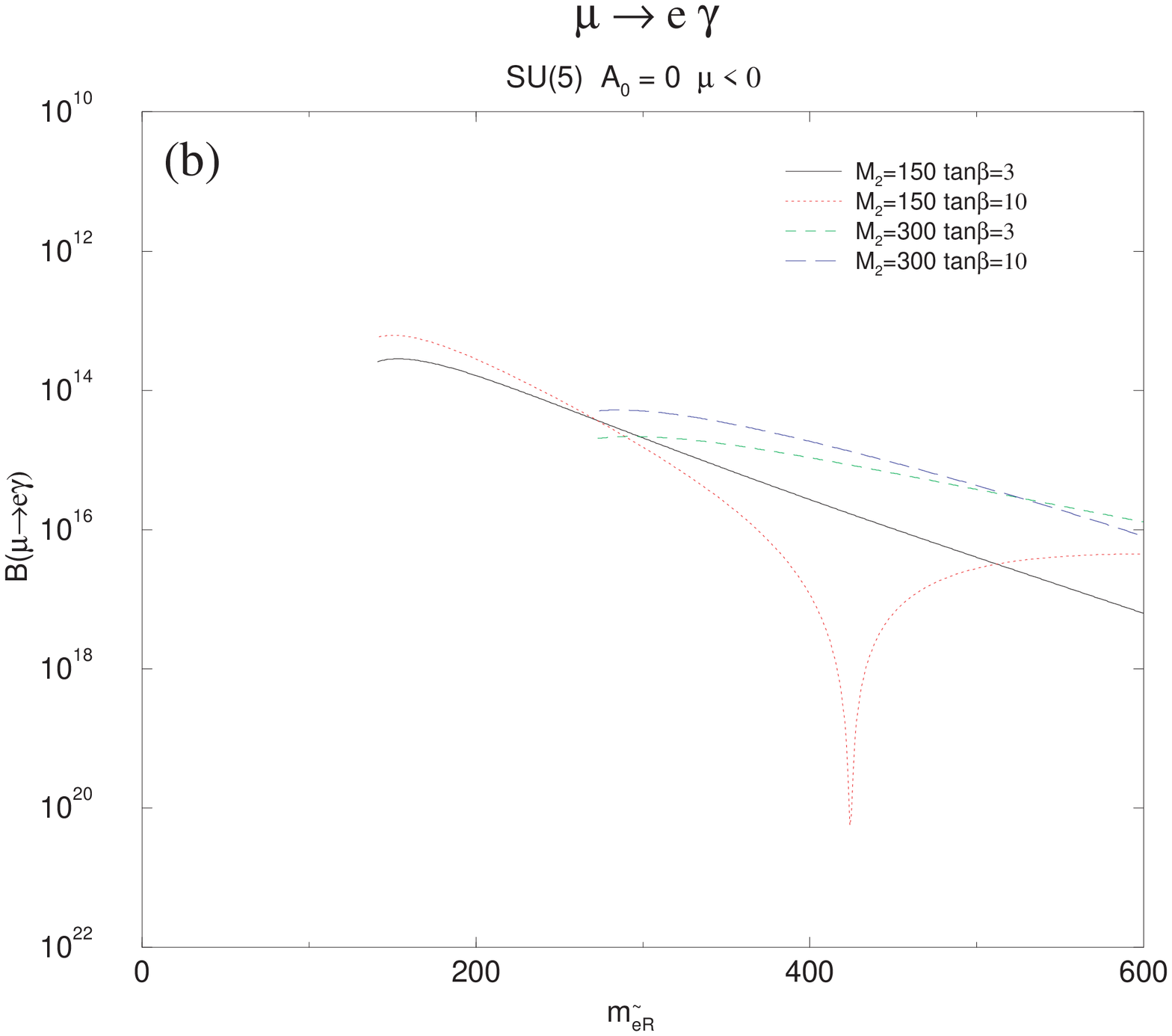,angle=270,width=9.5cm}}
\vspace{5mm}
\caption{Predicted branching ratios for the \mueg decay 
in the $SU(5)$ SUSY GUT based on the minimal supergravity model as a
function of the right-handed slepton mass for four different sets of
the SUSY input parameters of $M_2$ (the $SU(2)$ gaugino mass) and
$\tan{\beta}$ (the ratio of the two Higgs vacuum expectation
values). For the other parameters, the trilinear scalar coupling
constant $A_0=0$ and $m_t =175$ GeV. The following CKM matrix elements
are used: $|(V_{CKM})_{cb}|= 0.04$ and $|(V_{CKM})_{td}|= 0.01$. (a)
and (b) correspond to a positive and negative sign of the higgsino
mass parameter $\mu$, respectively.}
\label{fg:br-mueg-su5}
\end{figure}

The prediction of the branching ratio of \mueg decay is presented in
Fig.\ref{fg:br-mueg-su5} for typical SUSY parameters in $SU(5)$ SUSY
GUT.  The branching ratio reaches the order of $10^{-14}$ for a
slepton mass of a few 100 GeV/$c^2$.

Some remarks on $SU(5)$ SUSY GUT are presented in the following.

\begin{figure}[ht!]
\centerline{\epsfig{file=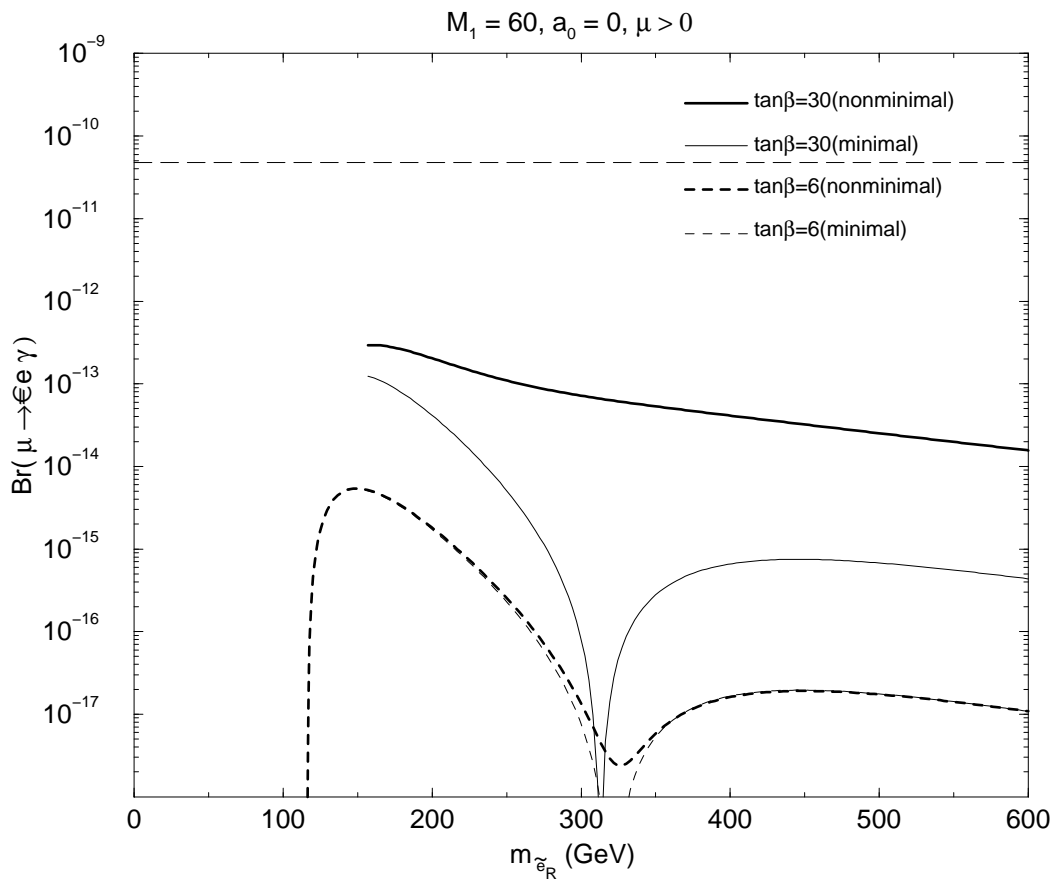,angle=270,width=16cm}}
\vspace{5mm}
\caption{Predicted \mueg branching ratios in the $SU(5)$ SUSY GUT model
with higher dimensional operators in the GUT superpotential. The
branching ratios are shown as a function of the right-handed selectron
mass for $\tan\beta=6$ (dashed lines) and 30 (solid lines). The thick
lines are for the non-minimal case in which $V_{\bar{e}}$ and $V_{l}$
are the same as $V_{CKM}$, and the thin lines are for the minimal case
in which $V_{\bar{e}}=V_{CKM}$ and $V_{l}={\bf 1}$, where
$V_{\bar{e}}$ is the mixing matrix for the right-handed sleptons, and
$V_{l}$ is that for the left-handed sleptons. The bino mass of $M_1 =
60$ GeV/$c^2$, the trilinear scalar coupling constant of $A_0=0$, the
positive higgsino mass ($\mu >0$), and the top quark mass of 175
GeV/$c^2$ are used. The experimental bound shown in the dashed line is
$B(\mu^{+}\rightarrow e^{+}\gamma)\leq 4.9\times 10^{-11}$ (Bolton,
{\em et al.} (1988)), and it is noted that the recent best limit is
$B(\mu^{+}\rightarrow e^{+}\gamma)\leq 1.2\times 10^{-11}$ (Brooks,
{\em et al.} (1999)). For detail on the calculations, see Hisano, {\em
et al.} (1998b) (after Hisano {\em et al.}, (1998b)).}
\label{fg:high-dimension}
\end{figure}

\begin{figure}[ht!]
\centerline{\epsfig{file=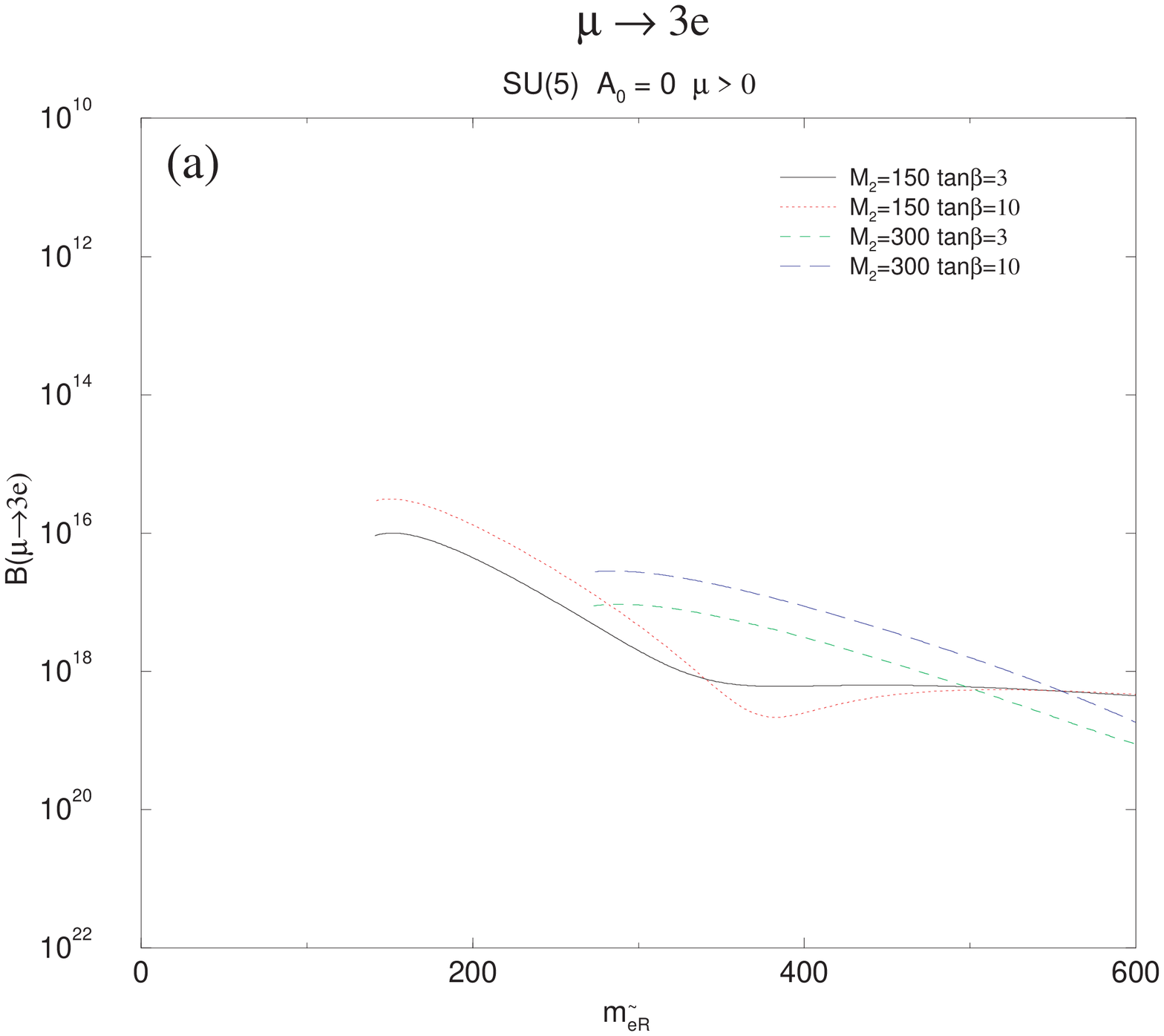,angle=270,width=10cm}}
\centerline{\epsfig{file=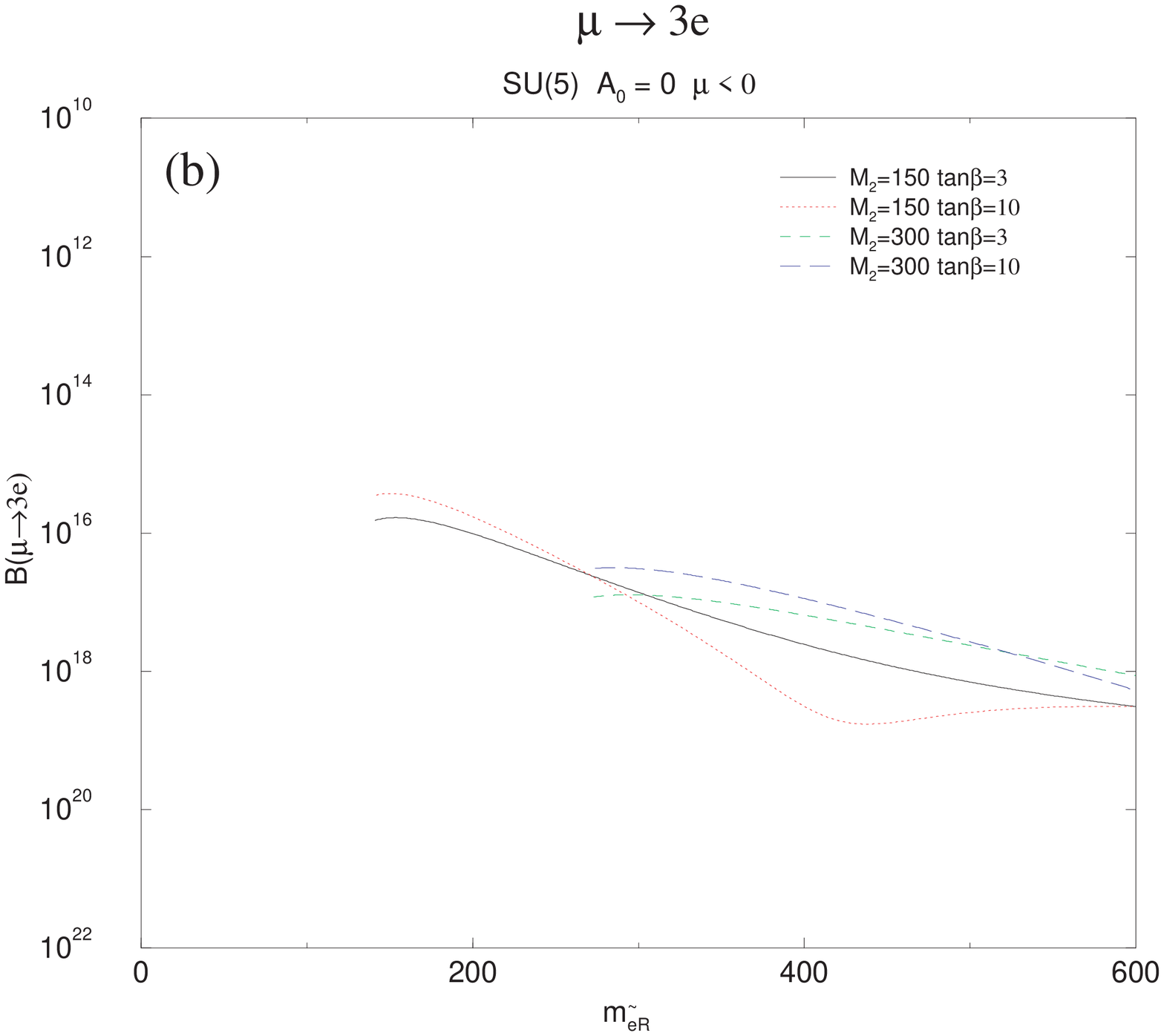,angle=270,width=10cm}}
\vspace{5mm}
\caption{Predicted branching ratios for the \meee decay
in the $SU(5)$ SUSY GUT based on the minimal supergravity model. The
input parameters are the same as in Fig.9.}
\label{fg:br-meee-su5}
\end{figure}

\begin{figure}[ht!]
\centerline{\epsfig{file=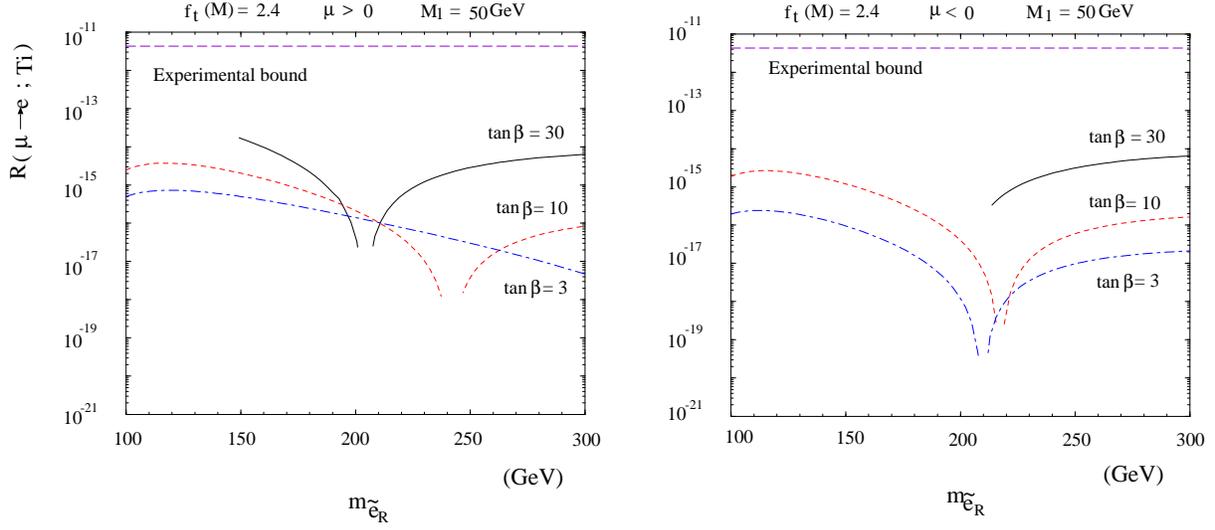,angle=270,width=16cm}}
\caption{Predicted branching ratios for the \muec conversion in $SU(5)$
SUSY GUT. The SUSY gaugino mass of $M_1 = 50$ GeV, and the top Yukawa
coupling of $f_{t}(M)=2.4$ are used. The left and right figures
correspond to a positive and negative sign of the higgsino mass
parameter $\mu$, respectively. The experimental bound shown is
$B(\mu^{-}Ti\rightarrow e^{-}Ti)\leq 4.3\times 10^{-12}$ (Dohmen, {\em
et al.} (1993)), and it is noted that the recent best limit is
$B(\mu^{-}Ti\rightarrow e^{-}Ti)\leq 6.1\times 10^{-13}$ (Wintz
(1998)) (after Hisano {\em et al.} (1997)).}
\label{fg:br-muec-su5}
\end{figure}

\begin{itemize}

\item In the $SU(5)$ SUSY GUT model, LFV appears only in the right-handed
slepton sector for small or moderate values of $\tan{\beta}$, which is
defined by the ratio of two Higgs vacuum expectation values as
$\tan{\beta}\equiv$$<H_2^0>$/$<H_1^0>$.  This is because the
renormalization effects contribute only to $\tilde{e}_R$, and not to
$\tilde{l}_L$. As a result, the helicity of an electron (positron) in
LFV processes becomes only right-handed (left-handed). For instance,
only $\mu^+\rightarrow e^+_L \gamma$ decay occurs, not
$\mu^+\rightarrow e^+_R\gamma$. These two processes can be
distinguished when the angular distribution of the \mueg signal is
measured by using polarized muons.

\item There is partial cancellation among the Feynman diagrams which
contribute to the \mueg amplitudes in the $SU(5)$ SUSY GUT
\cite{hisa97}.  This cancelation can be seen in Fig.\ref{fg:br-mueg-su5}.

\item In Eqs.(\ref{eq:m_R_su5}) and (\ref{eq:ckm_gut}), the
off-diagonal elements of the right-handed slepton mass matrix are
determined by the CKM matrix elements. When the favorable values of
the CKM matrix elements are used, $|\lambda_{\tau}| \equiv
|(V_{R})_{13}(V_{R})_{23}^*|$ of $(3-5)\times 10^{-4}$ is
obtained. This results from the assumption that all of the Yukawa
coupling constants are generated from the superpotential in
Eq.(\ref{su5sp}).  However, it is known that this assumption does not
yield a realistic mass relation for the down-type quarks and charged
leptons in the first and second generations. If higher dimensional
terms or different $SU(5)$ representations of Higgs fields are
included to remedy this problem, the simple relationship between the
slepton mixing matrix and the CKM matrix in Eq.(\ref{eq:ckm_gut})
would be lost, and $\lambda_{\tau}$ would become essentially a free
parameter. As a consequence, the predicted branching ratios could be
different from those in Fig.\ref{fg:br-mueg-su5}. For example, if
$|\lambda_{\tau}|\approx 10^{-2}$, the branching ratio is enhanced by
three orders of magnitude from Fig.\ref{fg:br-mueg-su5}. In addition,
for a large $\tan{\beta}$, a further enhancement can be expected
\cite{arka96a}. One example prediction of the \mueg branching ratio for 
a large $\tan\beta$ is shown in Fig.\ref{fg:high-dimension}
\cite{hisa98b}. When the higher dimensional terms are included, the
mass matrix for left-handed sleptons also has off-diagonal elements
owing to the large bottom Yukawa coupling constant, and therefore the
branching ratio of \mueg decay is enhanced by
$(m_{\tau}/m_{\mu})^{2}\approx 10^{2}$, just as in the case of
$SO(10)$ SUSY GUT discussed below. This enhancement can be seen in the
non-minimal case in Fig.\ref{fg:high-dimension}. Furthermore, the
destructive interference between the different diagrams in the minimal
$SU(5)$ SUSY GUT discussed before may disappear.

\item The \meee decay and the \muec conversion process receive 
contributions from the off-shell photon, $Z$-penguin and box diagrams,
in addition to the photonic penguin diagram contributing to the \mueg
decay. In $SU(5)$ SUSY GUT, the relative magnitudes between different
contributions vary over the SUSY parameter space. This implies that
the relations in Eqs.(\ref{eq:lfv-relation1}) and
(\ref{eq:lfv-relation2}) may change for different SUSY input
values. The predictions of the \meee decay and the \muec conversion in
$SU(5)$ SUSY GUT are shown in Figs.\ref{fg:br-meee-su5} and
\ref{fg:br-muec-su5}, respectively.

\end{itemize}

\begin{figure}[ht!]
\centerline{\epsfig{file=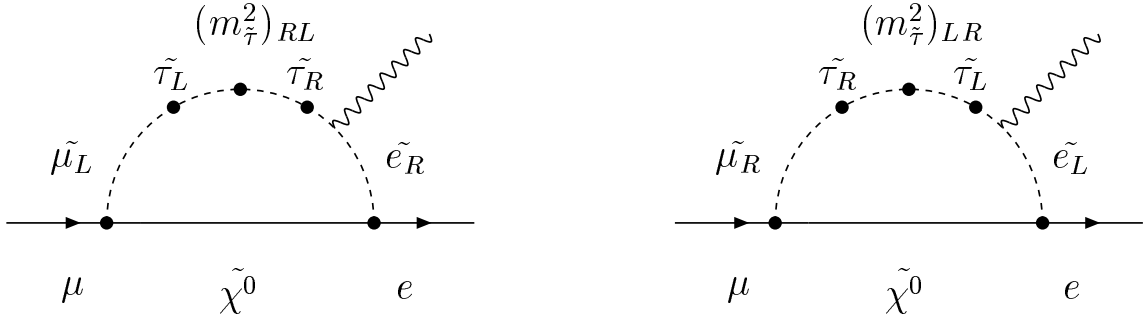,width=14cm}}
\vspace{5mm}
\caption{Feynman diagrams in $SO(10)$ SUSY GUT which give dominant
contributions to the \mueg process. $(m^{2}_{\tilde{\tau}})_{RL}$ and
$(m^{2}_{\tilde{\tau}})_{LR}$ are proportional to $m_{\tau}$.}
\label{fg:diagram-so10}
\end{figure}

\begin{figure}[ht!]
\centerline{\epsfig{file=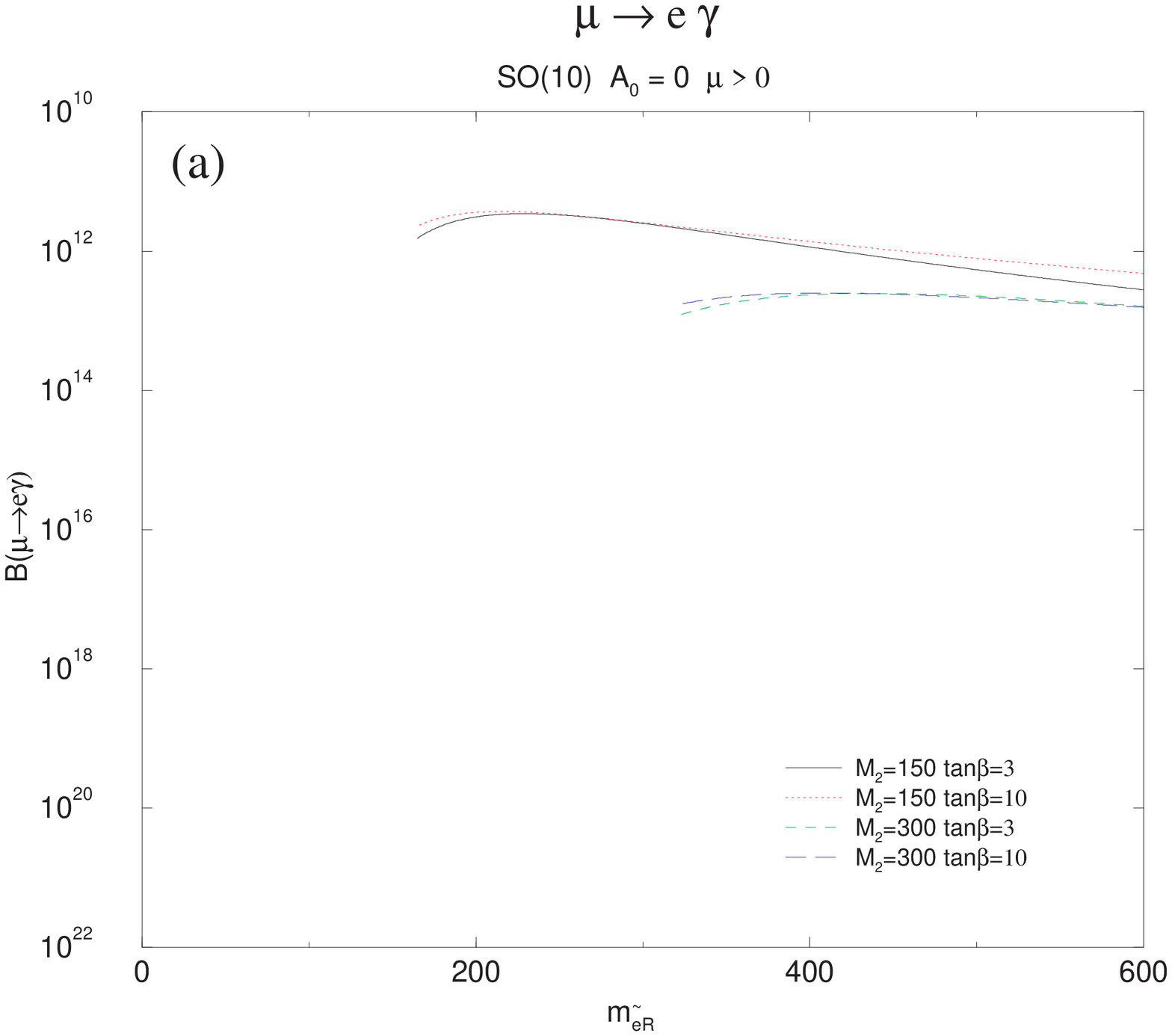,angle=270,width=10cm}}
\centerline{\epsfig{file=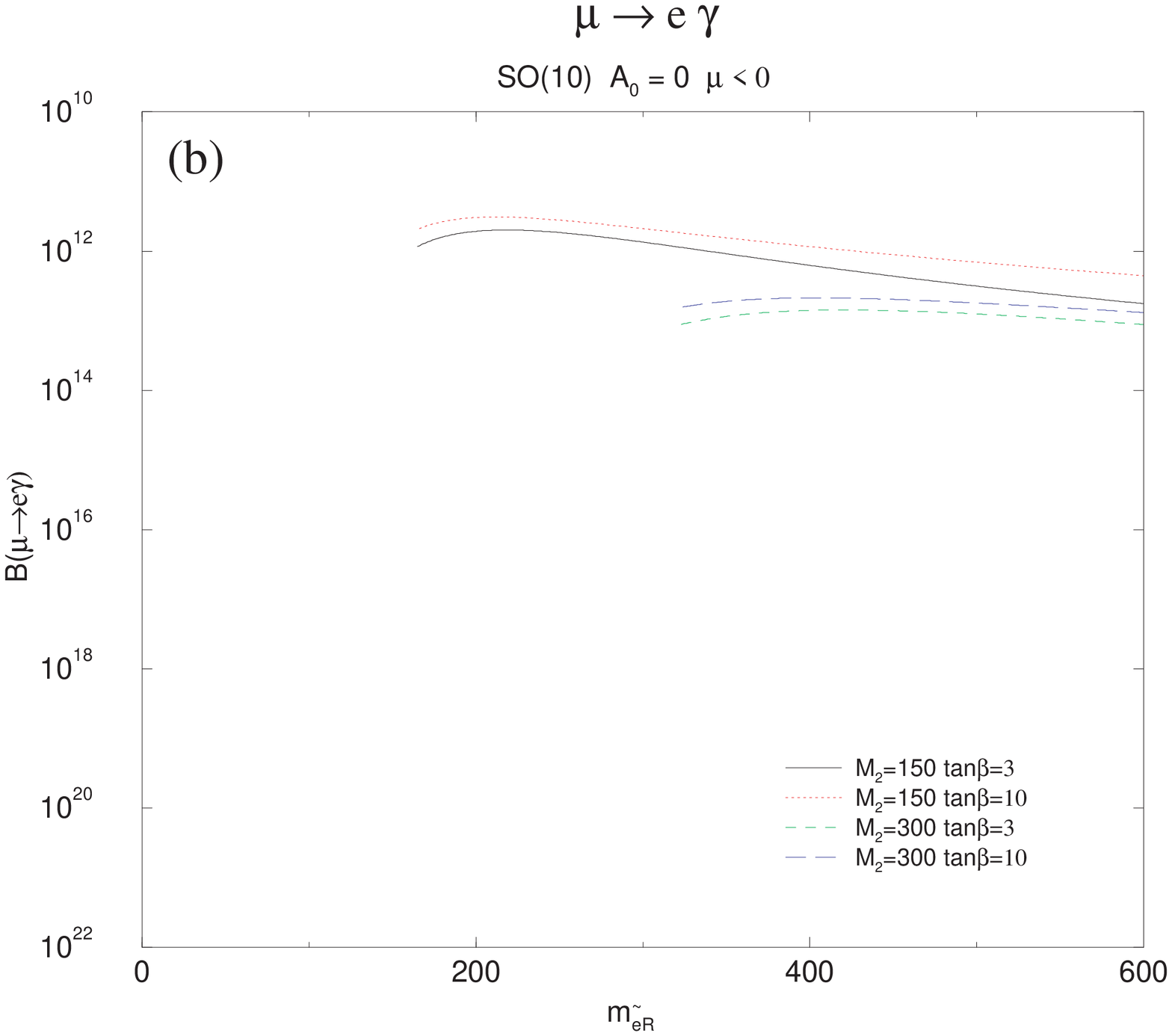,angle=270,width=10cm}}
\vspace{5mm}
\caption{Predicted branching ratios for \mueg decay 
in the $SO(10)$ SUSY GUT based on the minimal supergravity model.
Input parameters are the same as in Fig.9.}
\label{fg:br-mueg-so10}
\end{figure}

Large LFV effects are also expected in $SO(10)$ SUSY GUT. In the
minimal $SO(10)$ SUSY GUT, the superpotential is given by

\begin{equation} 
W_{SO(10)}= \frac{1}{2}(y_u)_{ij}{\Psi_i}\cdot{\Phi_u}\cdot {\Psi_j}
+\frac{1}{2}(y_d)_{ij}{\Psi_i}\cdot {\Phi_d}\cdot{\Psi_j},
\end{equation} 

\noindent where $\Psi_i$ is the 16-dimensional representation of
SO(10) and $\Phi_u$ and $\Phi_d$ are two 10-dimensional Higgs fields.
In this model, both the left-handed and right-handed sleptons receive
LFV effects. In particular, the diagrams shown in
Fig.\ref{fg:diagram-so10} give a large contribution to the amplitude
of \mueg decay because it is proportional to $m_{\tau}$.  Hence, the
branching ratio is enhanced by $(m_{\tau}/m_{\mu})^{2}$ compared to
the minimal $SU(5)$ SUSY GUT. Owing to this enhancement, the branching
ratios for muon LFV processes can become comparable to the present
experimental upper bounds if the slepton mixing matrices are related
to the observed CKM matrix elements. The predictions for the branching
ratio of \mueg decay in $SO(10)$ SUSY GUT are shown in
Fig.\ref{fg:br-mueg-so10}.  In $SO(10)$ SUSY GUT, since the
photon-penguin diagrams dominate in the amplitudes of \meee decay and
\muec conversion, their predicted branching ratios with respect to
\mueg decay would follow Eqs.(\ref{eq:lfv-relation1}) and
(\ref{eq:lfv-relation2}) over a wide SUSY parameter space
\cite{barb95a}. 

The rates of muon LFV processes depend on the structure on the Yukawa
coupling constants at the GUT scale.  The branching ratios for the
muon LFV processes are calculated in various realistic $SO(10)$ SUSY
GUT models \cite{arka96a,ciaf96,duon96,gome96}.  In the $SO(10)$ SUSY
GUT models, the diagrams relevant to the \mueg amplitude would induce
the electric dipole moments (EDM) of the electron and neutron
\cite{dimo95}.  The branching ratios for muon LFV processes are
compared with the prediction of EDMs, FCNC processes in the quark
sector, and CP violations in $B$ and $K$ meson decays in both the
$SU(5)$ and $SO(10)$ SUSY GUT models, and the leptonic signals are
shown to be very sensitive to the interaction at the GUT scale
\cite{barb95b}.  The LFV process is also investigated in the breaking
pattern of $SO(10)\rightarrow SU(3)\times SU_L(2)\times SU_R(2)\times
U_{B-L}(1)\rightarrow SU(3) \times SU_L(2)\times
U_{Y}(1)$\cite{desh96} and in the $SU(4)\times SU_L(2)\times SU_R(2)$
model without GUT unification \cite{king98}.

In some theoretical scenarios in which the mass matrices for squarks
and sleptons at very high energy scale are not universal, but have
some correlations or alignments with corresponding fermion matrices,
it is possible to avoid the SUSY flavor problem and at the same time
still have muon LFV branching ratios large enough to be detected. One
such realization has been investigated concerning the dynamical
alignment mechanism \cite{ratt96}. An interesting possibility is a
class of models based on $U(2)$ flavor symmetry where both the Yukawa
coupling constants and the soft SUSY-breaking mass terms are
controlled by the same approximate symmetry. In this case, the
branching ratio for \mueg decay is expected to receive large SUSY
contributions \cite{barb96}.

\subsubsection{Supersymmetric models with a right-handed neutrino} 

A large LFV effect can be expected if the supermultiplets of
right-handed Majorana neutrino are included in the SUSY Standard Model
\cite{borz86}. As explained in Section~\ref{sc:NMM}, the smallness of
the neutrino masses can be explained by the see-saw mechanism. To
include this see-saw mechanism, part of the lepton sector in the
Lagrangian in Eq.(\ref{eq:Lagrangian_MSSM}) is replaced by

\begin{equation} 
W_N = (y_e)_{ij}{H_1}{E^c_i} { L_j}+(y_{\nu})_{ij}{H_2}{ N_i}{ L_j}
+\frac{1}{2}(M_R)_{ij}{N_i} {N_j},
\end{equation}

\noindent where $N_i$ is the right-handed neutrino supermultiplets and
$(M_{R})_{ij}$ is the Majorana mass matrix, and a new Yukawa coupling
constant matrix, $y_{\nu}$, is introduced. Since there are two Yukawa
coupling matrices ($y_{e}$ and $y_{\nu}$) in the lepton sector, flavor
mixing would arise and lepton flavor would no longer be conserved
separately for each generation, just as in the quark sector. In SUSY
models with universal soft SUSY-breaking at the Planck scale, the
flavor mixing in left-handed sleptons would induce sizeable LFV
effects in muon and tau decays through the renormalization effects
from the Planck to the Majorana mass scales.

The expected magnitudes of the LFV effects depend on the Yukawa
coupling constant and the flavor mixing in the lepton sector. In the
basis where the Yukawa coupling constant matrix for charged leptons is
diagonalized, the mass matrix of light neutrinos is given by

\begin{equation}
(m_{\nu})_{ij} = (y_{\nu})_{ki}(M_R^{-1})_{kl}(y_{\nu})_{lj}
{v ^2\sin^2{\beta}\over 2},
\label{eq:m_L} 
\end{equation}

\noindent where $v$ is the Higgs vacuum expectation value, and $\beta$ 
is the Higgs vacuum angle for the two Higgs doublet. On the other
hand, the off-diagonal terms of the left-handed slepton mass matrix
induced by the renormalization effect is

\begin{eqnarray}
(m_{\tilde{l}_{L}}^{2})_{ij} \simeq -\frac{1}{8\pi^{2}}
(y_{\nu})_{ki}^* (y_{\nu})_{kj} m_{0}^{2}
(3 + |A_{0}|^{2})\ln({M_{P}\over M_{R}}).
\label{eq:mssmrn}
\end{eqnarray}

\begin{figure}[ht!]
{\epsfig{file=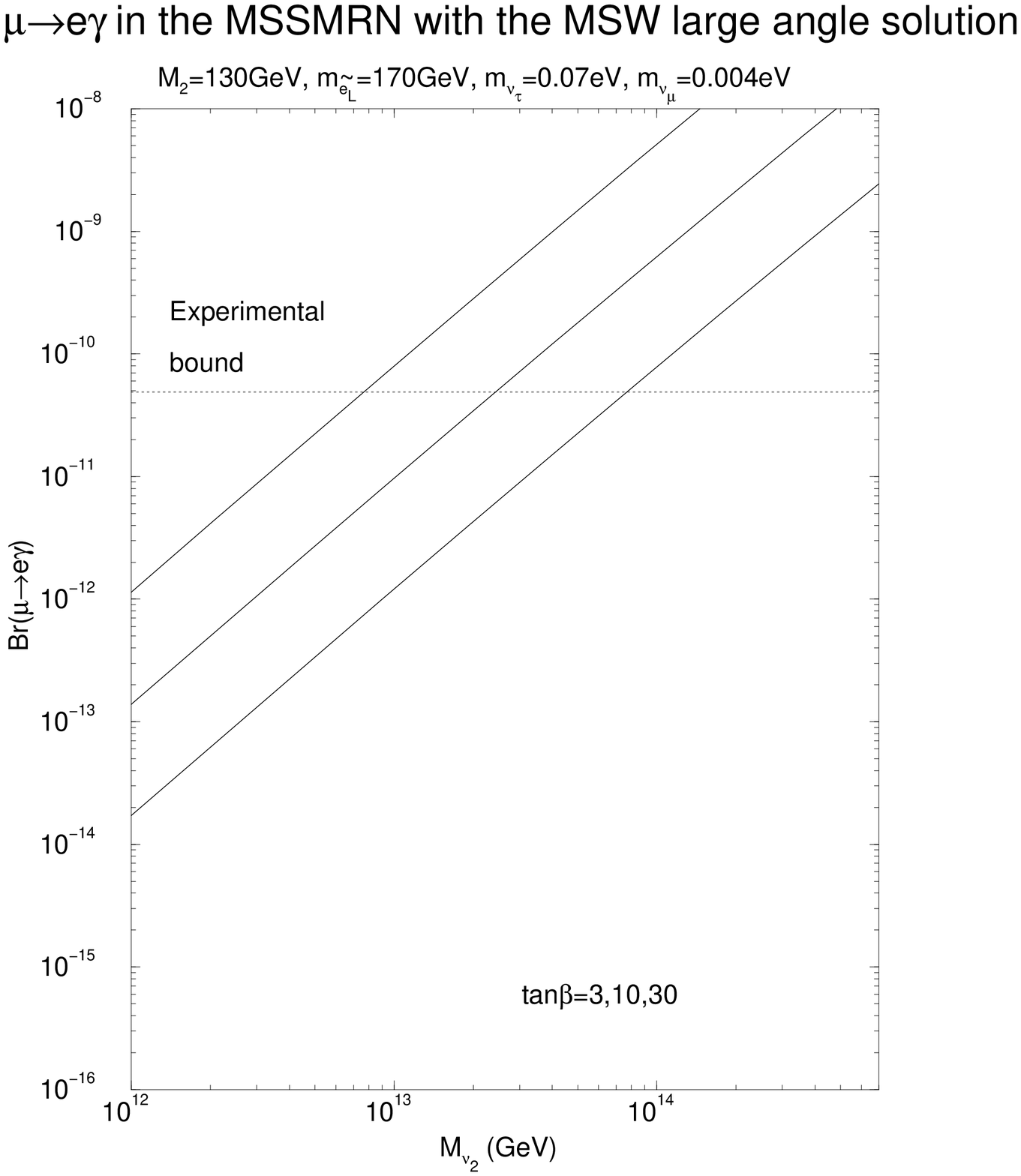,width=8cm}}
{\epsfig{file=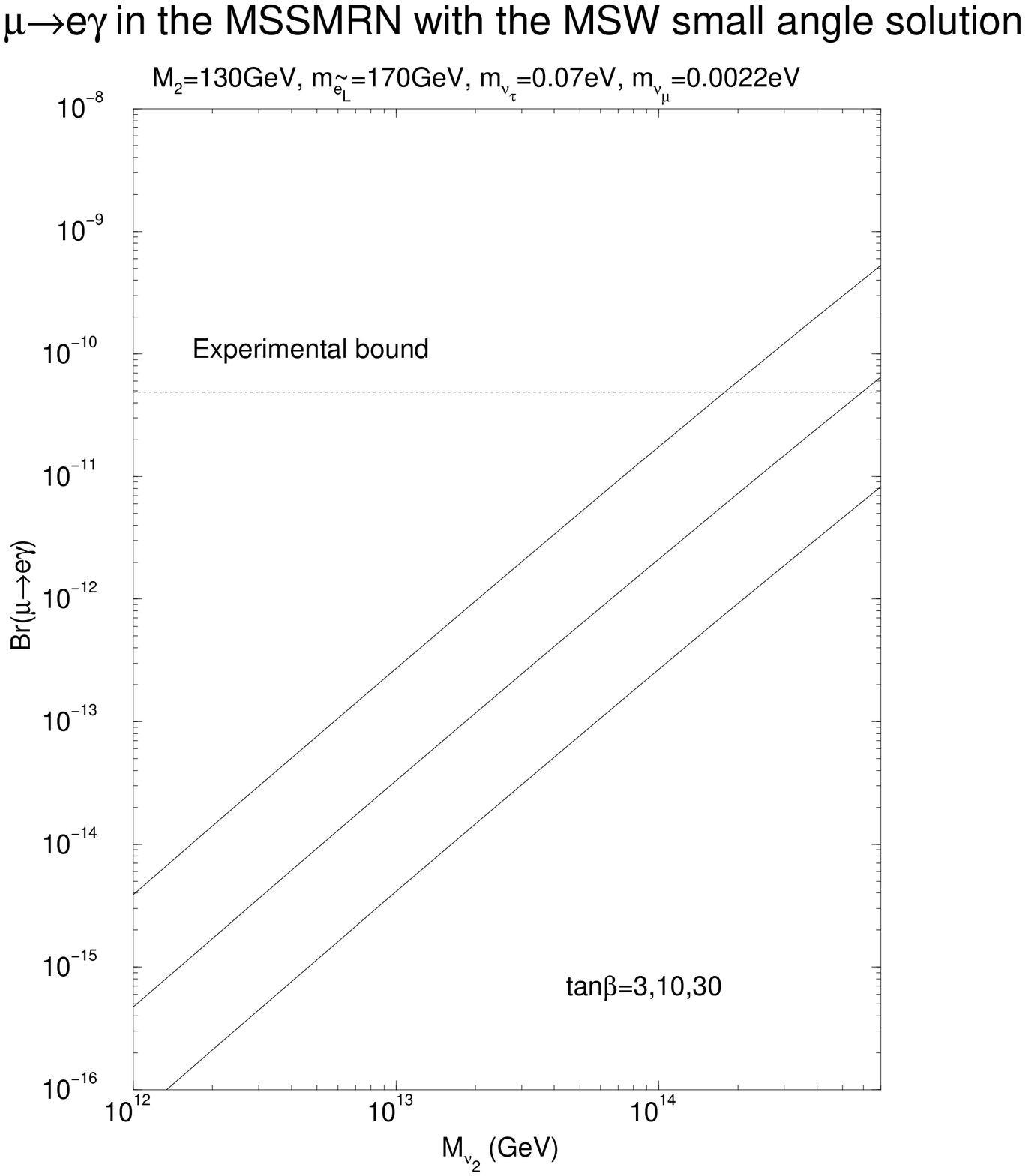,width=8cm}}
\vspace{5mm}
\caption{Predicted branching ratios of \mueg decay as a function of
the Majorana mass of the second-generation right-handed neutrino
($M_{\nu_2}$) in the MSSM model with right-handed neutrino. They are
given for the MSW large angle solution and the MSW small angle
solution. The three curves correspond to $\tan\beta$ = 30, 10, and 3
from top to bottom for both figures. The other parameters are shown in
the top of the figures (after Hisano and Nomura, (1999)).}
\label{fg:br-mssmrn}
\end{figure}

\noindent In general, there is no direct relationship between the
neutrino mixing in Eq.(\ref{eq:m_L}) and the slepton mixing relevant
to $\mu\rightarrow e \gamma$, $\tau\rightarrow \mu\gamma$ and
$\tau\rightarrow e \gamma$ in Eq.(\ref{eq:mssmrn}).  If, however, we
assume that the neutrino mixing mostly originates from the neutrino
Yukawa coupling constants, $(y_{\nu})_{ij}$, the information from the
atmospheric and solar neutrinos can be related to the slepton
mixing. Then, the branching ratios for \mueg and $\tau
\rightarrow \mu\gamma$ decays can be evaluated by using the neutrino mixing
parameters \cite{hisa95,hisa96,hisa98a,hisa99a}. Fig.\ref{fg:br-mssmrn}
shows the predicted branching ratio for the \mueg decay for different
solutions of the solar neutrino mixing. As can be seen in
Fig.\ref{fg:br-mssmrn}, it can reach the present experimental bound
for the Majorana mass larger than $O(10^{14})$ GeV if the large-angle
MSW solution for the solar-neutrino problem is chosen.  It is noted
that from Eq.(\ref{eq:m_L}), given a fixed value of the light neutrino
mass ($m_{\nu}$), the Yukawa coupling constant ($y_{\nu}$) becomes
larger for a larger value of the Majorana mass scale ($M_R$),
resulting in that the LFV rate becomes larger in Eq.(\ref{eq:mssmrn}). 
The LFV rates increases approximately as the second power of $M_R$
and, therefore, it could possibly probe the mass scale of the
right-handed Majorana neutrino ($M_R$) in this scenario. Note that
this prediction is in contrast to the see-saw mechanism without SUSY,
in which the LFV rates for charged leptons are extremely suppressed,
as discussed in Section~\ref{sc:OTM}.

\subsubsection{Other supersymmetric models}

Observable effects of muon LFV processes may arise through the
renormalization effects in the slepton mass matrix.  Because these
effects may come from anywhere between the Planck and the electroweak
scale, it is possible to consider some other interaction at a
high energy scale as a new source of LFV.

Such an example is given in the context of the gauge-mediated SUSY
breaking. If the mixing between the messenger fields of gauge
mediation and the ordinary matter fields is allowed, a sizeable LFV
effect can be generated through the renormalization of the slepton
mass matrix \cite{dine97,dubo98}. A similar effect appears in a
supersymmetric $E_6$ type model, where the mixing between the ordinary
leptons and the exotic leptons in the $E_6$ ${\bf 27}$ representation
could induce a large LFV effect via the slepton mixing \cite{kita99}.

\subsubsection{LFV in slepton production and decay}

If the charged sleptons are discovered in experiments at future
colliders (like LHC and LC), LFV due to slepton mixing could be
directly searched in their production and decay processes
\cite{kras94,kras96,arka96b}.  For example, a process like
$e^{+}e^{-}\rightarrow \tilde{l}^{+} \tilde{l}^{-} \rightarrow e^{\pm} 
\tilde{\chi}_{1}^{0} \mu^{\mp} \tilde{\chi}_{1}^{0}~$ breaks lepton
flavor conservation, where the slepton $\tilde{l}^{\pm}$ is assumed to
decay to a lepton and the neutralino $\tilde{\chi}_{1}^{0}$.  Direct
searches for the $e\mu$ final states in $e^{+}e^{-}$ and $e^{-}e^{-}$
collisions can probe the slepton-mixing angle between the selectron
($\tilde{e}$) and the smuon ($\tilde{\mu}$) of less than 0.1, which
could be comparable to indirect searches such as the muon LFV
processes \cite{arka96b}.

The production cross section for the slepton LFV processes depends on
the masses and mixing of the sleptons. If two sleptons, like
$\tilde{e}$ and $\tilde{\mu}$, are almost degenerate in their masses,
a possible oscillation between them would occur in the decay
process. Another interesting possibility is a CP violating signal in
the slepton oscillation, which may arise due to a new complex phase in
the slepton mixing matrices \cite{arka97,bows97}.

In $SU(5)$ SUSY GUT, LFV in the slepton pair production and decays, in
particular with a tau in the final state, has been investigated, and
it is found to be more sensitive than the searches for
$\tau\rightarrow e(\mu)\gamma$ \cite{hiro98}. LFV signals in the
production of left-handed sleptons at $\mu^{+}\mu^{-}$ and
$e^{+}e^{-}$ colliders are also considered in the framework of MSSM
with right-handed neutrino motivated by the atmospheric-neutrino data. 
It is shown that the decay modes of either $\tau + \mu + 4$ {\em jets
+ missing energy~} or $\tau + \mu + l + 2$ {\em jets + missing
energy~} could be useful to see the signals with suppressing any
potential background \cite{hisa99b}. Other possible LFV searches at a
$e\mu$ collider \cite{choi98} and also a $e\gamma$ collider
\cite{cao99} are also considered.

\subsection{Other theoretical models}\label{sc:OTM}

In the late 70's, especially in 1977, there were many theoretical
papers on models with heavy neutrinos to discuss LFV
\cite{chen77a,chen77b,bjor77,leep77,lee77,alta77}, after responding to
the false rumor on \mueg signals at SIN (Schweizerisches Institut
f\"ur Nuklearforschung). Since then, various theoretical models have
appeared. In this section, we discuss LFV effects in theoretical
models other than SUSY models. Although many possibilities to induce
LFV effects might exist, we discuss only a few examples of different
types of LFV. For other references, see, for instance,
\citeasnoun{verg86}, and \citeasnoun{depo95}.

\subsubsection{Models with a massive neutrino}

The simplest way to violate lepton flavor conservation is to introduce
neutrino masses and mixing.  However, it has been known that the
branching ratio of \mueg decay from ordinary neutrino mixing is very
suppressed if the neutrino mass and mixing suggested by the
atmospheric and solar-neutrino experiments are used. For example, the
predicted branching ratio from the Dirac neutrino masses and mixing is
given by \cite{petc77,bile77}

\begin{equation}
B( \mu \rightarrow e \gamma)=\frac{3\alpha}{32\pi}\left|\sum_i
(V_{MNS})_{\mu i}^* (V_{MNS})_{e
i}\frac{m_{\nu_{i}}^2}{m_{W}^2}\right|^2,
\label{eq:dirac-neutrino-mixing}
\end{equation}

\noindent where $(V_{MNS})_{a i}$ is the lepton flavor mixing matrix
(MNS matrix) defined in Eq.(\ref{eq:mns}). It is represented by

\begin{equation}
\nu_{L a} = \sum_{i} (V_{MNS})_{a i}\nu_{L i},
\end{equation}

\noindent where $\nu_{L a}$ is the neutrino field in the weak flavor
basis, and $\nu_{L i}$ in the mass-diagonalized basis. Even if a 1 eV
neutrino mass with maximal mixing is considered,
Eq.(\ref{eq:dirac-neutrino-mixing}) only gives a branching ratio on
the order of $10^{-47}$.

For the Majorana neutrino model of see-saw type \cite{chen80}, the
suppression factor of $(m_{\nu_{i}}^2/m_{W}^2)$ in
Eq.(\ref{eq:dirac-neutrino-mixing}) is replaced by a factor of
$O(m_{\nu_{i}}/{M_R})$, where $M_R$ is the mass of a heavy Majorana
neutrino. Then, the branching ratio is still $O(10^{-40})$ or less for
$m_{\nu}=1$ eV and $M_R= 10^{10}$ GeV. It is, therefore, difficult to
expect observable LFV effects from the ordinary neutrino masses and
mixing indicated by the atmospheric and solar neutrinos.

\subsubsection{Models with a doubly charged Higgs}

There is a wide class of theoretical models which have an interesting
enhancement for \meee decay and \muec conversion. If LFV is induced by
a penguin diagram with a heavy boson and a light charged fermion, the
photonic form factors $f_{E0}(q^2)$ and $f_{M0}(q^2)$ receive an
enhancement factor of $ \ln{(m_{boson}/m_{fermion})}$ from the loop
diagrams in which a photon is attached to the internal light fermion
line \cite{marc77a,marc77b,wilc77,raid98}. Since this factor is absent
for $f_{E1}(q^2)$ and $f_{M1}(q^2)$, the branching ratio of
\mueg decay has no enhancement. One of such models is 
a double charged Higgs boson which has the following interaction of

\begin{equation}  
\cal L \mit = h_{ij}\overline{e_{iR}^c} e_{jR}\phi^{++}
+ H.c.,
\end{equation}

\noindent where the loop diagrams with internal charged leptons and
doubly charged scalar $ \phi^{++}$ could induce a logarithmic
enhancement \cite{raid98}.

\subsubsection{Supersymmetric models with $R$-parity
violation}\label{sc:smrv}

Another important class of models which might induce logarithmic
enhancement is the SUSY models with $R$ parity violation. In the MSSM,
if we only require gauge invariance to write all possible
superpotentials, the following interactions would also be allowed:

\begin{equation}
W = \lambda_{ijk}L_{i} L_{j} E^{c}_{k} +
\lambda^{'}_{ijk}L_{i} Q_{j} D^{c}_{k} +
\lambda^{''}_{ijk}U_{i} D^{c}_{j} D^{c}_{k} - \mu_{i}L_{i}H_{2}.
\label{eq:rbreak}
\end{equation}

\noindent These interactions violate the baryon- or lepton-number
conservations.  To forbid proton decays that are too fast, a parity
which distinguishes superparticles from ordinary particles is often
required. This parity is called the $R$ parity, and is defined as $R
\equiv (-1)^{3B-L+2S}$, where $B$, $L$, and $S$ are the baryon number,
lepton number, and spin, respectively. We can, however, consider
models with $R$-parity violation when combinations of the coupling
constants which induce proton decays are highly suppressed, but the
other coupling constants can be large.

\begin{figure}[ht!]
\centerline{\epsfig{file=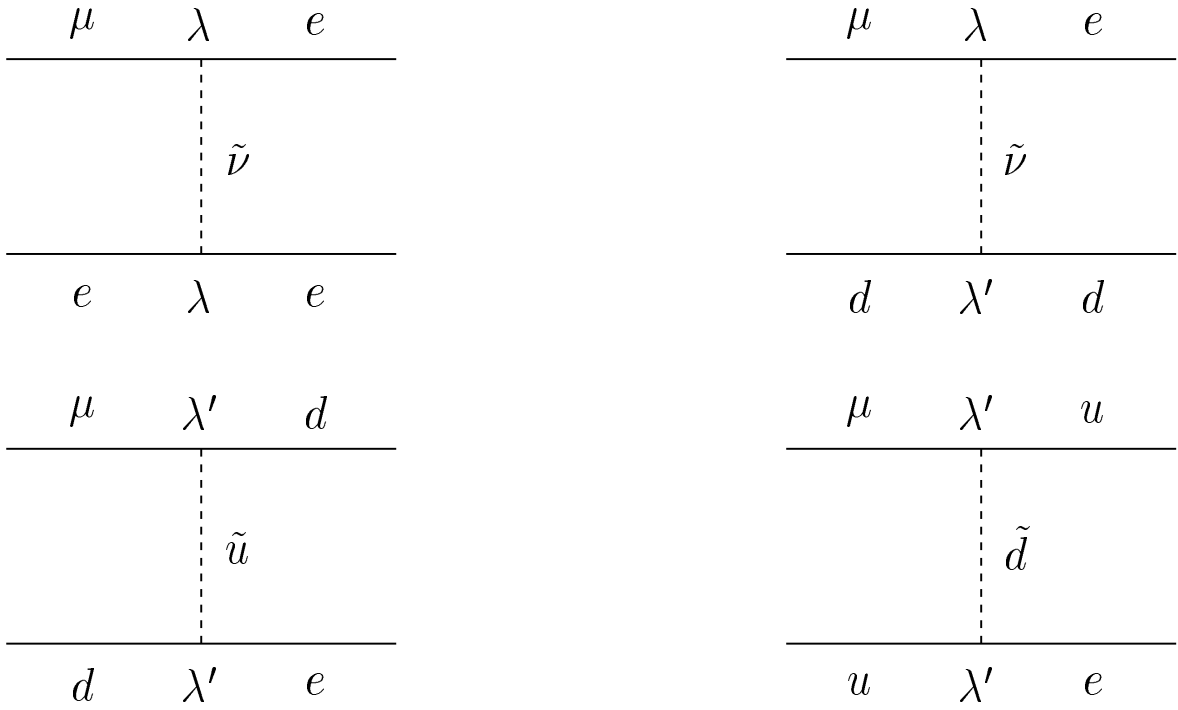,width=14cm}}
\vspace{5mm}
\caption{Tree diagrams for LFV processes in SUSY models 
with $R$-parity violation.}
\label{fg:diagram-rp-tree}
\end{figure}

Some combinations of the coupling constants are severely constrained
by the LFV processes. It is known that the allowed values of
$\lambda\lambda$ and $\lambda\lambda'$ and $\lambda'\lambda'$ still
give large contributions at a tree level to the \meee and \muec
conversion processes \cite{kim97,huit98,faes99}. Typical relevant tree
diagrams are shown in Fig.\ref{fg:diagram-rp-tree}.

\begin{figure}[ht!]
\centerline{\epsfig{file=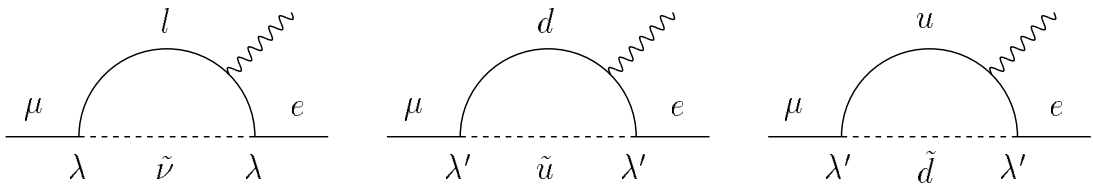,width=14cm}}
\vspace{5mm}
\caption{One-loop diagrams for LFV processes in SUSY models 
with $R$-parity violation.}
\label{fg:diagram-rp-loop}
\end{figure}

In SUSY models with $R$-parity violation, there also exist loop
contributions for muon LFV processes, such as \mueg decay
\cite{chai96}, \meee decay and \muec conversion. The latter two
processes would receive a logarithmic enhancement
\cite{huit98}. Fig.\ref{fg:diagram-rp-loop} shows typical loop
diagrams. From this one-loop diagram, the \muec conversion process can
also be induced from the $\lambda\lambda$ coupling constants. For
example, from the loop diagrams with internal leptons, the following
four form factors are given as

\begin{eqnarray}
f_{E0}(q^2)&=&\pm f_{M0}(q^2)=-\frac{2(\lambda \lambda)}{3(4\pi)^2}
\frac{-q^2}{m_{\tilde{\nu}}^2}
\left( \ln{\frac{-q^2}{m_{\tilde{\nu}}^2}} + F(r)\right),
\noindent \\         
f_{M1}(0)&=&\mp f_{E1}(0)= -\frac{(\lambda \lambda)}{3(4\pi)^2}
\frac{m_{\mu}^2}{m_{\tilde{\nu}}^2},
\end{eqnarray}

\noindent where $r = m_{\tilde{\nu}}^2/(-q^2)$. $F(r)$ takes a value
of the order of $O(1)$, except for $r\gg 1$,
$F(r)=\ln{r}+\frac{4}{3}$. As a result, some of the $\lambda\lambda$
combinations are more severely constrained by the \muec conversion
than the tree-level process of \meee decay.

The present constraints for the combinations of $\lambda\lambda$ and
$\lambda^{'}\lambda^{'}$ couplings are summarized in
\citeasnoun{huit98} and \citeasnoun{faes99}.

\subsubsection{Models with $Z^{'}$}

There are many models where non-photonic LFV transitions occur at tree
levels. Typical examples are a model with a $Z'$ that has flavor
off-diagonal couplings, or models with extra fermion families which
mix with the three families at the tree level so that the $Z$ boson has
a LFV couplings \cite{bern93}. In such cases, the \meee and \muec
conversion processes are expected to be much more important than \mueg
decay.

\subsubsection{Models with Lorentz non-invariance}

Recently, a possible violation of the Lorentz invariance has been
suggested \cite{cole99}. In this context, the Lorentz transformation
is not invariant, but only the translational and rotational symmetries
are assumed to be exact in a preferred system. Thus, the maximum
attainable velocity could be different for each species of particles,
and this would cause many unique phenomena in particle physics and
cosmic-ray physics.

One of the good tests of the violation of Lorentz invariance is the
muon LFV processes \cite{cole99}. If a small Lorentz-non-invariant
interaction exists in the SM Lagrangian, flavor mixing couplings can
be generally allowed in the photon-fermion interaction. The current
limit on the \mueg branching ratio puts a strong constraint on the
relevant coupling constants. Another interesting effect is a change of
the muon lifetime at a high energy. Since the $\mu\rightarrow e\gamma$
decay width due to the Lorentz non-invariant interaction would
increase with $\gamma^{4}$, where $\gamma$ is the Lorentz factor, it
would dominate over the ordinary muon decay that increases with
$\gamma$. Therefore, the muon lifetime might start decreasing as
$\gamma^{-3}$ at a sufficiently-high energy. The current limit on the
energy dependence of the muon lifetime has been given by the
experiment of the muon anomalous magnetic moment.

\subsection{Lepton flavor violation with Polarized Muons}

In this subsection, we discuss the usefulness of polarized muons in
searches for \mueg and \meee decays. As discussed later in Section
\ref{sc:NMD}, highly-polarized $\mu^+$s (surface muons) are available
experimentally. Therefore, it would be useful to examine what kind of
new information can be obtained by measuring the angular distribution
of decay products with respect the muon polarization.

\begin{figure}[ht!]
\centerline{\epsfig{file=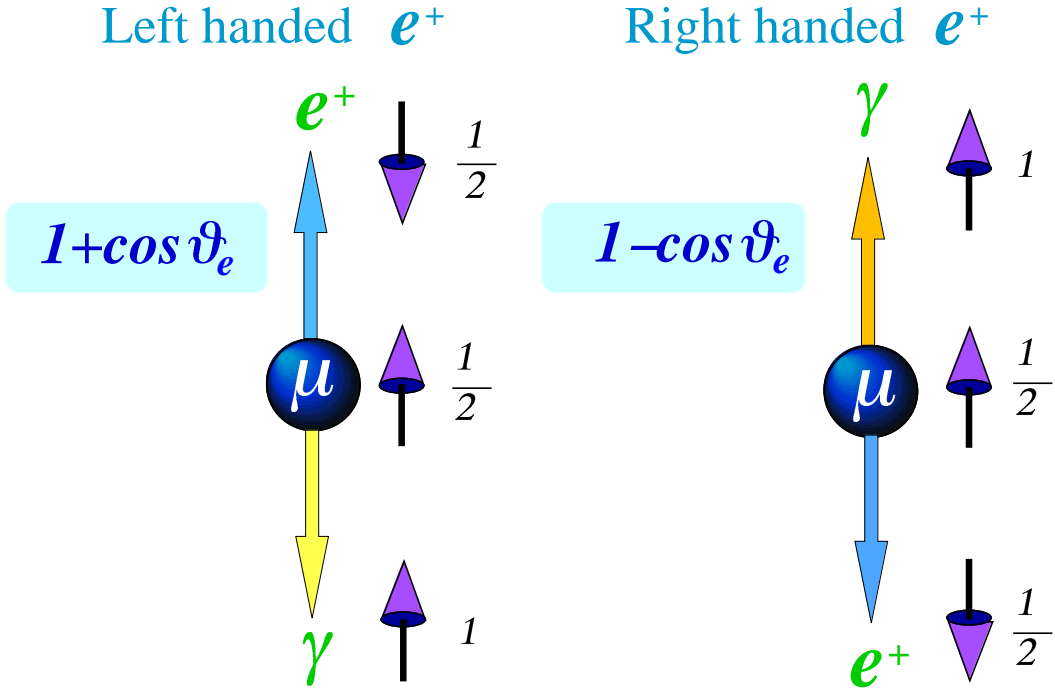,width=14cm}}
\vspace{5mm}
\caption{Angular distribution of $e^{+}$ in polarized \mueg decay. }
\label{fg:ang-mueg}
\end{figure}

When the initial muon is polarized in $\mu^{+} \rightarrow e^{+}
\gamma$ decay, the angular distribution of the positron is given by

\begin{equation}  
{{dB(\mu^{+}\rightarrow e^{+}\gamma)}\over{d \cos{\theta_e}}} =
{192\pi^2}
\left(
\left| A_R \right|^2 (1-P_{\mu}\cos{\theta_e}) + 
\left| A_L \right|^2 (1+P_{\mu}\cos{\theta_e}) \right),
\label{eq:ang-mueg}
\end{equation}

\noindent where $\theta_e$ is the angle between the muon polarization
and the positron momentum in the muon rest frame. $A_{R}$ and $A_{L}$
are given in Eqs.(\ref{eq:ar}) and (\ref{eq:al}). $P_{\mu}$ is the
magnitude of the muon polarization.  A measurement of the $e^{+}$
polarization would give the relative amplitudes of $A_R$ and $A_L$,
which correspond to the emission of right-handed $e^{+}$
($\mu^{+}\rightarrow e^{+}_{R}\gamma$) and left-handed $e^{+}$
($\mu^{+}\rightarrow e^{+}_{L}\gamma$), respectively. This is shown
schematically in Fig.\ref{fg:ang-mueg}.

Since $A_L$ and $A_R$ are model-dependent, it would be useful to
discriminate between different LFV mechanisms. For instance, the
minimal $SU(5)$ SUSY GUT model predicts a vanishing $A_R$ and a
non-zero $A_L$, yielding a $(1+P_{\mu}\cos\theta_e)$ distribution. On
the other hand, the $SO(10)$ SUSY-GUT model predicts the helicity
amplitudes for both right-handed and left-handed $e^{+}$s. For
non-unified supersymmetric models with right-handed neutrino, $A_R$ is
non-zero but $A_L$ is vanishing, giving a $(1-P_{\mu}\cos\theta_{e})$
distribution. Thus, a measurement of the angular distribution of
$e^{+}$ with respect to the direction of muon polarization would
provide a valuable means to clearly discriminate between these models.

The \meee decay with polarized muons would provide us with an
interesting possibility of measuring T violation
\cite{trei77,zee85}. A T-odd triple vector correlation,
$\vec{\sigma}_{\mu}\cdot (\vec{p}_1\times\vec{p}_2)$, can be defined,
where $\vec{\sigma}_{\mu}$ is the muon spin, and $\vec{p}_1$ and
$\vec{p}_2$ are two independent momenta of the $e^{+}$ in the final
state. If CPT invariance holds, the information on CP violation in the
LFV interaction may be obtained from this T-odd correlations of the
decay products.  The T-odd asymmetry would arise as an interference
between the photon-penguin terms and the four-fermion terms.  Details
are discussed in Section \ref{sc:PM3ED}. In particular, the T-odd
asymmetry is evaluated in the $SU(5)$ SUSY GUT model based on
supergravity \cite{okad98}. It is shown that an asymmetry up to 20\%
is possible if CP-violating phases are introduced in the soft
SUSY-breaking mass terms. It would give independent information from
the electric dipole moment (EDM) of the electron and neutron.

In the \meee decay, parity-odd asymmetries can be also defined if the
initial muons are polarized. These asymmetries are sensitive to the
chiralities of the terms in the effective Lagrangian: both the
photon-penguin terms ($A_L$ and $A_R$) and the four fermion coupling
terms ($g_i$). Measurements of the parity-odd asymmetries in \meee
decay, together with the branching ratios of \mueg and \meee decays,
are useful to distinguish different SUSY GUT models \cite{okad99}.
 
\subsection{$|\Delta L_{i}|=2$ processes}\label{sc:DL2P}
 
The LFV processes with $|\Delta L_{i}| = 2$ are like the muonium to
anti-muonium conversion (\mumu conversion), the wrong-flavor muon
decay ($\mu^{+}\rightarrow e^{+}\nu_{\mu}\overline{\nu}_{e}$), and
others.  The phenomenology of the \mumu conversion can be described by
an effective four fermion interaction. As an example, the interaction
of $(V-A)(V-A)$ type was considered by Feinberg and Weinberg
\cite{fein61}.  It is given by

\begin{equation}
H_{\rm Mu\overline{Mu}} = \Bigl({G_{\rm Mu\overline{Mu}} \over
\sqrt{2}} \Bigr)
\overline{\mu}\gamma_{\lambda}(1-\gamma_5)e
\overline{\mu}\gamma^{\lambda}(1-\gamma_5)e + H.c.,
\end{equation}   

\noindent in which $G_{\rm Mu\overline{Mu}}$ is a coupling constant
characterizing the strength of the \mumu conversion.  As described in
Section~\ref{sc:mumuex}, the present experimental limit on $G_{\rm
Mu\overline{Mu}}\leq 3.0 \times 10^{-3} G_{F}$ is given
\cite{will99}.

\begin{figure}[ht!]
\centerline{\epsfig{file=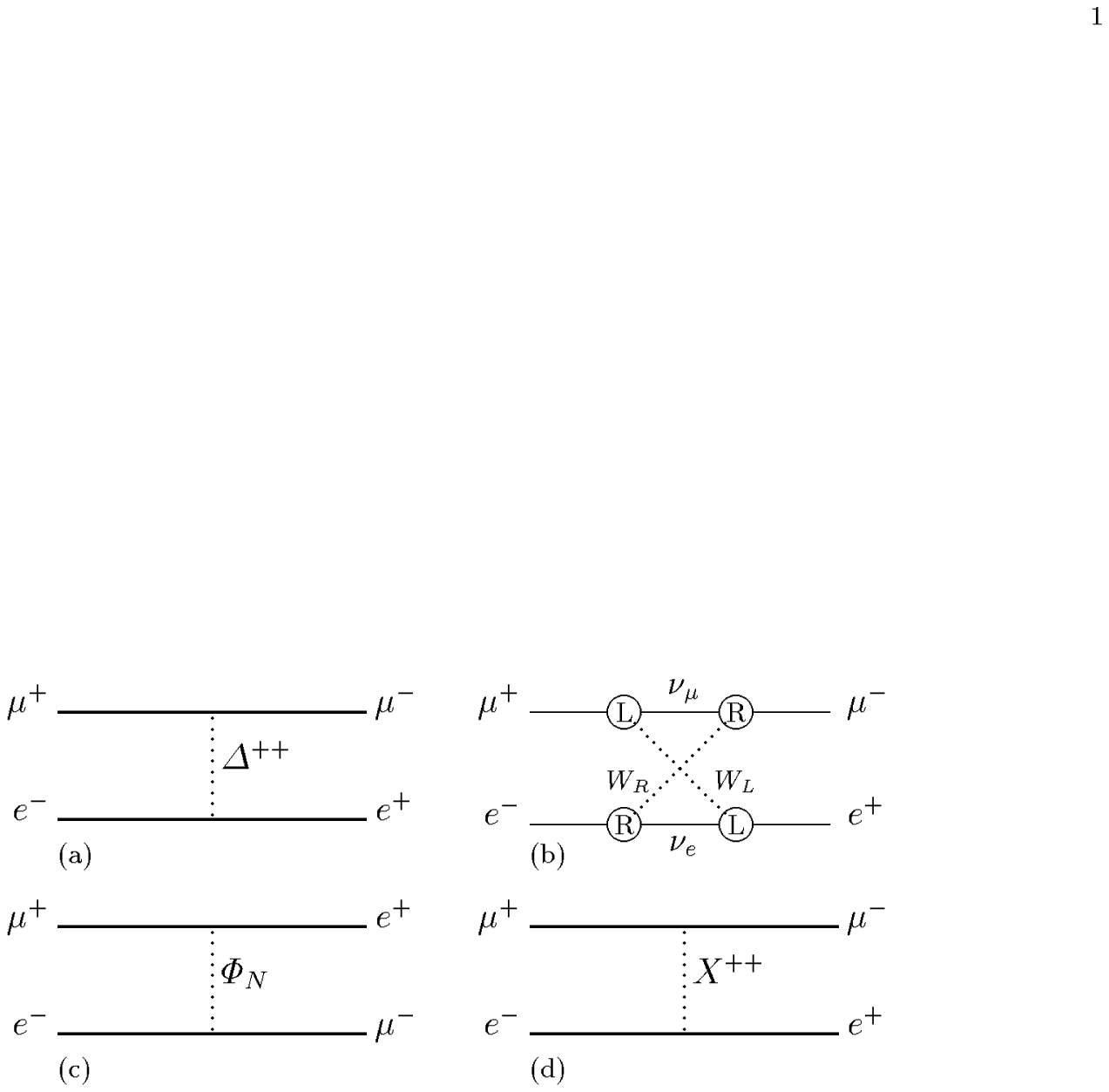,width=14cm}}
\vspace{5mm}
\caption{Examples of theoretical models to induce \mumu
conversion. They are mediated by (a) a doubly charged Higgs boson, (b)
heavy Majorana neutrinos, (c) a neutral scalar particle like a tau
sneutrino, and (d) a bilepton $X^{--}$ (after
Willmann, {\em et al.}, (1999)).}
\label{fg:l2model}
\end{figure}

In general, there could be various combinations of different types of
four fermion interactions. They are such as $(V+A)(V+A)$,
$(V-A)(V+A)$, $(S-P)(S-P)$, $(S+P)(S+P)$, $(S-P)(S+P)$, $SS$ and $PP$,
where $V$, $A$, $S$ and $P$ are vector, axial vector, scalar,
psuedoscalar effective interactions, respectively. The types of
interactions are determined by assumed theoretical models.  In
Fig.\ref{fg:l2model}, example diagrams of speculative theoretical
models responsible for the \mumu conversion are shown. They are models
of (a) the exchange of doubly charged Higgs bosons, (b) that of heavy
Majorana neutrinos, (c) that of a neutral scalar particle (including a
superparticle), and (d) that of a bilepton gauge boson. They are
briefly described in the following.

A simple example which induces the $|\Delta L_{i}| = 2$ process is
provided by a model with a doubly-charged singlet scalar boson
\cite{chan89}. In this model, only the $(V+A)(V+A)$ interaction is
generated and the \mumu conversion rate can be as large as the present
experimental limit within the constraints from the measurements of the
anomalous muon magnetic moment and the high-energy Bhabha scattering.
A more general case with the doubly charged scalar boson is also
considered \cite{swar89}.

In the left-right symmetric model with a triplet Higgs boson field,
the
\mumu conversion could be induced by a doubly charged Higgs boson,
$\triangle^{++}$ \cite{halp82}. In this model, if the mass of the muon
neutrino is greater than 35 keV/$c^2$ and less than the present direct
experimental bound of 170 keV/$c^2$, there is an upper limit of the
neutrino lifetime in order for the neutrino energy density in the
universe not to exceed its present total energy density. From this
requirement, a lower bound of $G_{\rm Mu\overline{Mu}}$ is predicted as a
function of the mass of the muon neutrino, $m_{\nu_{\mu}}$
\cite{herc92}.  For the range 35 keV/$c^2$ $\leq m_{\nu_{\mu}}
\leq$ 170 keV/$c^2$, a lower limit of
$G_{\rm Mu\overline{Mu}} \geq (1-40) \times 10^{-4} G_F$ is obtained.

Supposing that neutrinos are of Majorana nature, the \mumu conversion
could take place by an intermediate pair of neutrinos. This coupling
is related to neutrinoless double $\beta$ decays, yielding $G_{\rm
Mu\overline{Mu}}\leq 10^{-5} G_{F}$ \cite{halp82}.

There is a class of models with a neutral scalar boson, which has a
flavor-changing coupling to introduce the $|\Delta L_{i}| = 2$
processes \cite{hou96}. An important example of this kind of models is
the supersymmetric model with $R-$parity violation, where a tau
sneutrino exchange induces the \mumu conversion
\cite{moha92,halp93}. In this case, the four fermion coupling is of the
$(S-P)(S+P)$ type. The present experimental limit of the \mumu
conversion gives a constraint on the relevant coupling constant,
$|\lambda_{132}\lambda^{*}_{231}|\leq 3 \times 10^{-4}$, for a
superpartner mass of order 100 GeV/$c^2$. Also, the four fermion
coupling constant for the $\mu^{+}\rightarrow
e^{+}\nu_{\mu}\overline{\nu}_{e}$ decay is predicted to be similar in
magnitude as that for the \mumu conversion.

In some extensions of the SM gauge groups, there appear doubly charged
gauge bosons (called bileptons), $X^{--}$, which couple only to
leptons. They occur, for instance, in SU(15) GUT models \cite{fram90},
or in a gauge model with $SU(3)_C\times SU(3)_L \times U(1)_Y$ (311
model) \cite{fram92}. In these models, singly-charged and
doubly-charged bilepton gauge bosons appear from breaking of the
$SU(3)_{L}$ gauge symmetry to $SU(2)_L$ of the SM gauge groups.  The
mass bound of the bilepton gauge bosons is obtained from a precise
determination of the Michel parameters of the normal muon decay
\cite{carl92}, the muoniun hyperfine splitting, and the
$\mu^{+}\rightarrow e^{+}\nu_{\mu}\overline{\nu}_{e}$ decay
\cite{fuji94}. They give a lower bound of roughly 200 GeV/$c^2$.  In
these models, the \mumu conversion could occur by the exchange of a
doubly charged bilepton \cite{hori96,fuji93}. The effective
interaction is of the form $(V-A)(V+A)$, which is in contrast to
either the traditional $(V-A)(V-A)$ interaction by Feinberg and
Weinberg. The bilepton interaction is given by

\begin{equation}
{\lagr}=-{g_{3l}\over2\sqrt{2}}
X^{--}_{\mu}\overline{l}\gamma^{\mu}\gamma_{5}C\overline{l}^{T} +
H.c.,
\end{equation}

\noindent where $l=e,\mu,\tau$, and $C$ is the charge-conjugation
matrix. The gauge coupling constant, $g_{3l}$, is on the order of
$O(1)$, and is determined from theoretical models. The \mumu
conversion rate ($P_{\rm Mu\overline{Mu}}$) is given by

\begin{equation}
P_{\rm Mu\overline{Mu}} = 4.5 \times 10^3 \times \Bigl( {g_{3l} \over
m_{X}}\Bigr)^4,
\end{equation}

\noindent where $m_{X}$ is the mass of the bilepton gauge boson. From the
current experimental bound of the \mumu conversion, a constraint of
$m_{X}/g_{3l}\geq 2.6$ TeV/$c^2$ is obtained \cite{will99}.

To study the $|\Delta L_{i}|=2$ processes, the reaction $e^{-}e^{-}
\rightarrow \mu^{-}\mu^{-}$ has been known to be useful for a long
time \cite{glas61}. Such an experiments were carried out to place a
constraint on $G_{\rm Mu\overline{Mu}}$ \cite{barb69}. Recently,
similar scattering processes at a high energy, such as at a $ee$
linear collider or a $\mu\mu$ collider, have been discussed, based on
mostly bilepton models \cite{fram92b,hou96b,raid98b}.

\section{Normal Muon Decay}\label{sc:NMD}

Normal muon decay remains the only purely leptonic process of weak
interaction accessible to precise measurements with high
statistics. The studies are free from the complications of the strong
interaction and hadronic structure.  For this reason, precise studies
of normal muon decay would provide information unambiguously
interpreted.

The experimental progress has benefited from the high-intensity muon
beams available at the three meson factories, such as the Paul
Scherrer Institute (PSI) in Switzerland, TRIUMF in Canada, LAMPF in
U.S.A. (which was unfortunately shut down), and also lately the
Rutherford Appleton Laboratory (RAL) in England. In particular, useful
is a surface $\mu^{+}$ beam, which arises from the decay of pions
stopped at the surface of a pion production target. It has a
monochromatic kinetic energy of 4.1 MeV (29.8 MeV/$c$ in momentum) and
100\% muon polarization antiparallel to its momentum direction. This
high polarization of the surface muons is useful for various
measurements requiring muon polarization.

In the past, studies of the normal muon decay have greatly contributed
to develop the SM. It is now playing a role to probe for possible
deviations from the SM. For example, see reviews by
\citeasnoun{fets95} and \citeasnoun{herc95}. In the following, the
current status of the studies of normal muon decay, in particular
those which are being prepared or planned for the future, is
discussed. They are such as measurements of the muon lifetime, the
Michel spectrum, and the longitudinal polarization of $e^{+}$s in
polarized \muenn decay. The other important muon experiments, such as
the muon anomalous magnetic moment and the muon electric dipole
moment, are not discussed.

\subsection{Muon lifetime} \label{sc:ML}

\subsubsection{Phenomenology}

The Fermi coupling constant, $G_F$, is one of the three precisely
measured inputs of the SM, along with the fine structure constant
($\alpha$) and the $Z$ bosom mass ($m_{Z}$). Their updated values are
given in Table~\ref{tb:sm3v}. Note that the $m_{Z}$ value in the
Particle Data Group \cite{pdg98} is $m_{Z} = 91.187 \pm 0.007$ (77
ppm).  However, after a recent improvement of $m_{Z}$ \cite{lep99},
the uncertainties of $G_{F}$ and $m_{Z}$ become comparable in the
order, as can be seen in Table~\ref{tb:sm3v}. $G_F$ is determined from
the muon lifetime ($\tau_{\mu}$) from Eq.(\ref{eq:tmu}). The complete
two-loop QED corrections to an estimation of the muon lifetime was
calculated~\cite{ritb99}. The theoretical errors to derive $G_{F}$
from the muon lifetime are now reduced to be negligible, compared with
the experimental uncertainty in the measurement of the muon lifetime.
It would be necessary to improve the accuracy of $G_F$, if
experimentally possible. To test the SM, it is required to compare
$G_{F}$ from the muon lifetime with those determined from other
measurements, such as tau leptonic decays, or the $M_{W}$ and other
observables at the $M_{Z}$ pole with similar accuracies \cite{marc99}.

\begin{table}[htb!]
\caption{Three fundamental values in the Standard Model.}
\label{tb:sm3v}
\vspace{5mm}
\begin{tabular}{lll}
Parameter & Experimental value & Uncertainty \cr\hline
$1/\alpha$ & $137.0359895 \pm 0.0000061$ & 0.045 ppm \cr
$G_{F}$ & $(1.16639 \pm 0.00001) \times 10^{-5}$ GeV$^{-2}$ & 9 ppm  \cr
$m_{Z}$ & $91.1867 \pm 0.0021$ GeV/c$^2$ & 23 ppm  \cr
\end{tabular}
\end{table}

\subsubsection{Experimental status}\label{sc:MLES}

Experimentally, measurements of the muon lifetime were carried out at
TRIUMF \cite{giov84} and at Saclay \cite{bard84}. Since then, for more
than a decade, no experimental efforts to improve the situation have
been made. The present value of $\tau_{\mu}$ is $\tau_{\mu} =
2.19703(4) \times 10^{-6}$ ($\pm 18$ ppm) \cite{pdg98}.  Recently,
however, an experimental proposal at Rutherford Appleton Laboratory
(RAL) to reduce the uncertainty of $G_{F}$ by an order of magnitude
has been undertaken \cite{naka98}, and two experimental proposals to
PSI \cite{cava99,care99b} have been submitted.

An on-going experiment at the RIKEN-RAL muon facility, R77, uses a
pulsed muon beam. In previous experiments with a continuous muon beam,
only one incident muon within the time window of measurement was
allowed in order to avoid any possible confusion in the decay of other
muons occuring in the same time window. It, otherwise, would introduce
distortion of the time spectrum of muon decay. This requirement,
however, would lead to a limitation on muon beam intensity, and the
sensitivity would be statistically limited.  To overcome this problem,
R77 at RIKEN-RAL has adopted a pulsed beam, whose pulse interval of 20
msec is much longer than the measurement window. Since all muons come
at the same time, multiple muon decays within the measurement time
window are allowed. There are several sources of systematic
errors. One of them is from counting losses due to pile-up $e^{+}$
events and the dead time of detection, since the instantaneous beam
intensity is strong. To avoid this effect, segmentation of the
detectors is required. In R77, multiwire proportional chambers are
used. Another systematic error might come from a precession of muon
spin under an earth field. A special magnetic material will be adopted
for the muon stopping target to depolarize the muon spin polarization.
It aims, by accumulating $10^{11}$ muon decays, to achieve statistical
and systematic errors of about 3 ppm (a total of 4 ppm) in the
$\tau_{\mu}$ measurement. Its initial phase, planning to accumulate
$10^{10}$ muon decays, have already started in 1999.

There are two new planned experiments at PSI: R-99-06~\cite{cava99}
and R-99-07~\cite{care99b}, where R-99-07 has just been approved. The
both experiments are aiming at a factor 20 improvement over the
current world average of $\tau_{\mu}$.

The approved experiment, R-99-07, is to use a chopped surface-muon
beam at PSI. The muons are stopped in a sulfur target to reduce the
residual muon polarization, and a transverse magnetic field of 75 G is
applied to further de-phase it. The $e^{+}$ detector ($\mu Lan$
detector) consists of 180 triangular scintillating tiles distributed
within the 20 super-triangles of an icosahedral geometry centered on
the target. Each scintillating tile is viewed by a photo-multiplier,
and the signal is recorded by a waveform digitizer developed in the
$g-2$ experiment at BNL.  This 4$\pi$ geometry of the $\mu Lan$
detector, especially a sum of the point-like symmetric tile pairs with
respect to the center, would reduce further any asymmetries due to
spin rotation. The experiment has just been approved and a physics run
is expected in three years.

The experiment, R-99-06, uses a fiber-active-scintillator target
(FAST), which comprises plastic scintillating fibers of 0.5-mm
diameter in an overall active volume of $20\times 20\times 20$
cm$^{3}$. A $\pi^{+}$ beam is stopped in the FAST target, and a
$\pi^{+}\rightarrow\mu^{+}\rightarrow e^{+}$ decay chain is observed
in the target. Owing to a high granularity and fast response of the
FAST detector, many individual muon decays can be recorded in parallel
in a high event rate such as 1 MHz. Since the $\mu^{+}$s produced from
$\pi^{+}$ decays at rest have an isotropic distribution, possible
errors from the muon polarization can be eliminated.  In its first
stage, a DC $\pi^{+}$ beam is planned to use, and in the second stage,
a pulsed $\pi^{+}$ beam will be used.

\subsection{Michel decay spectrum}\label{sc:MS}

\subsubsection{Phenomenology}

The $e^{\pm}$ spectrum of the normal muon decay is given in
Eq.(\ref{eq:muenn}) with the four Michel parameters of $\rho$, $\xi$,
$\delta$, and $\eta$. It includes all possible Lorentz-invariant
interactions.  A precise determination of the Michel parameters would
serve stringent testing of the ($V-A$) structure of electroweak
interactions in the SM, and obtain a hint of new physics beyond the
SM.

\subsubsection{Experimental status}\label{sc:MSES}

In the past, each of the Michel parameters in the normal muon decay
was determined by several different experiments: for instance, the
$\rho$ parameter \cite{dere69}, the $\eta$ parameter \cite{burk85b},
the $\delta$ parameter \cite{balk88}, $P_{\mu}\xi$ \cite{belt87}, and
$P_{\mu}\xi(\delta/\rho)$ \cite{jodi86}.

A new experiment, E614 at TRIUMF, to measure the entire differential
spectrum of positrons from the decay of polarized muons is being
prepared \cite{abeg96}. By accumulating $10^{9}$ muon decays, the goal
of the E614 experiment is to measure the Michel parameters ($\rho$,
$\xi$, $\delta$ and $\eta$) to precisions of $\Delta\rho < 1 \times
10^{-4}$, $\Delta\delta < 3 \times 10^{-4}$ and $\Delta(P_{\mu}\xi) <
2 \times 10^{-4}$. The aimed precisions are 3 to 10 times better than
those previously achieved. Note that only the product $P_{\mu}\xi$ can
be experimentally determined. For example, the expected constraint on
the left-right symmetric model from E614 is presented in
Fig.\ref{fg:lrmodel}.

\begin{figure}[ht!]
\centerline{\epsfig{file=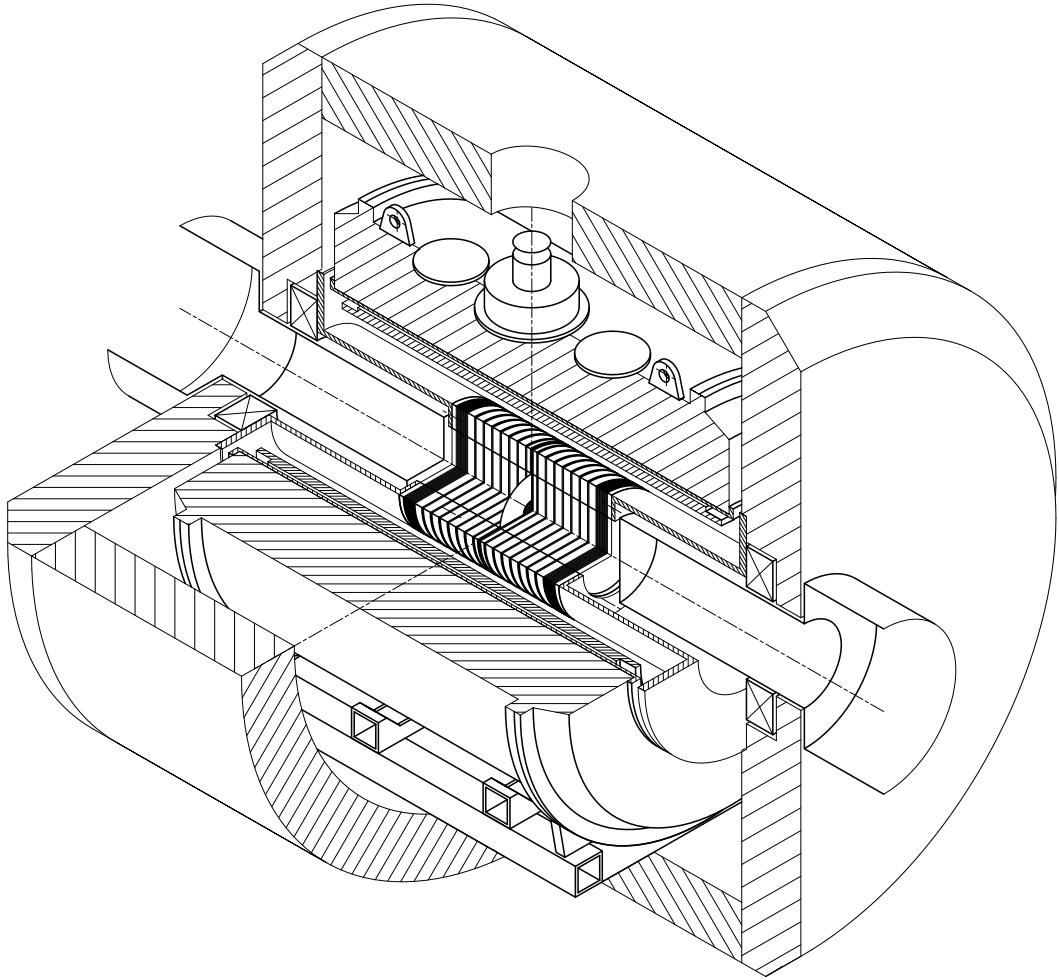,angle=270,width=14cm}}
\vspace{5mm}
\caption{Schematic view of the E614 detector at TRIUMF (provided by
D.R.~Gill).}
\label{fg:e614}
\end{figure}

A schematic view of the E614 detector is shown in
Fig.\ref{fg:e614}. The E614 spectrometer consists of a superconducting
solenoid magnet with tracking chambers. In E614, a surface muon beam
from the M13 beam channel at TRIUMF will be brought into the detector,
and stopped in a muon-stopping target made of aluminum located at the
center of the apparatus. An array of planar chambers, mounted
symmetrically upstream and downstream from the target, will track the
trajectories of $e^{+}$s from muon decays under a magnetic field of 2
T with a homogeneity of better than $10^{-4}$. The positron spectrum
will be measured over a wide range of $0.4\leq x
\leq 1.0$ for the positron energy, and $10^{\circ}\leq\theta_e \leq
70^{\circ}$ and $110^{\circ}\leq\theta_e\leq 170^{\circ}$ for the
angle between the muon spin direction and the positron momentum vector
($\theta_e$), where $x$ is defined in
Section~\ref{sc:NMMD}. 

Experimentally, it is important to keep the muon polarization fully
aligned with the magnetic field direction at the stopping target. Any
reduction of $P_{\mu}^z$, the projection of $P_{\mu}$ on the field
direction, must be minimized at a level of $10^{-4}$. Possible sources
of reduction are a misalignment between the muon-beam axis and the
magnetic field at the spectrometer, a fringing field of the
spectrometer, a contamination of non-surface muons (such as cloud
muons), multiple scattering of muons in the production target, and
possible depolarization of the muon spin in the aluminum muon-stopping
target. At the same time, as spectrometer requirements, low-mass
tracking chambers to minimize multiple scattering of positrons and
position accuracy of the detector assembly are crucial. Detector
construction has been started, and physics data-taking is expected
from 2001.

\subsection{Polarization of $e^{+}$ in \muenn decay}\label{sc:PMD}

\subsubsection{Phenomenology}

The longitudinal polarization of $e^{+}$ ($P_L$) in \muenn decay is
already given in Eq.(\ref{eq:lpol}). When the muon is not polarized
($P_{\mu}=0$) and the SM values of $\rho = \delta = 3/4$ are taken,
$P_L$ leads to

\begin{equation}
P_{L}(x,\cos\theta_{e}) = \xi^{'}.
\end{equation}

\noindent It is independent of the values of $x$ and $\cos\theta_{e}$.
Therefore, a measurement of the longitudinal polarization of $e^{+}$s
emitted by unpolarized muons would provide good a direct determination
of the parameter $\xi^{'}$.

If the muon is polarized with the SM values of the $\rho$ and $\delta$
parameters, $P_{L}$ is given by

\begin{equation}
P_{L}(x,\cos\theta_{e}) = \xi^{'} + {{\xi P_{\mu}\cos\theta_{e}(2x-1)}
\over {(3-2x)+\xi P_{\mu}\cos\theta_e (2x-1)}}\cdot{{(\xi^{''} -
\xi\xi^{'}) \over \xi}}.
\end{equation}

\noindent From this, the measurement of $P_L$ as a function of energy
($x$) and angle ($\theta_{e}$) would give other information about the
combination of parameters, $(\xi^{''}-\xi\xi^{'})/\xi$. In particular,
for $x\approx 1$ and $\cos\theta_{e}\approx -1$, it leads to

\begin{equation}
P_{L}(x=1,\cos\theta_{e}=-1)\approx \xi^{'} + {-\xi P_{\mu} \over
1-\xi P_{\mu}} \cdot {(\xi^{''} - \xi\xi^{'}) \over \xi}.
\label{eq:xidouble}
\end{equation}

\noindent The combination $(\xi^{''} - \xi\xi^{'})/\xi$ is
multiplied by an enhancement factor of $\xi P_{\mu}/(1-\xi P_{\mu})$,
which could be large when $P_{\mu}$ is close to unity.

The two transverse polarization components ($P_{T1}$ and $P_{T2}$) of
the $e^{+}$ in $\mu^{+}\rightarrow e^{+}\nu_{e}\overline{\nu}_{\mu}$
decay are given in Eqs.(\ref{eq:t1pol}) and (\ref{eq:t2pol}). If the
time-reversal invariance holds, $P_{T2}$, which is the transverse
$e^{+}$ polarization normal to the decay plane determined by $P_{\mu}$
and the $e^{+}$ momentum direction, should be vanishing. A non-zero
$P_{T2}$ would signal a violation of the time-reversal invariance. The
electromagnetic final-state interaction, which mimics a T-odd effect,
is known to be small. On the other hand, $P_{T1}$, which is not
forbidden by the fundamental symmetries, is sensitive to the Michel
parameter, $\eta$. It would be advantageous, since the determination
of $\eta$ from the isotropic part of the $e^{\pm}$ energy spectrum is
difficult owing to the small $x_{0}$ factor ($\sim 10^{-2}$)
multiplied by $\eta$.

\subsubsection{Experimental status}\label{sc:PMDES}

The latest measurement of $P_{L}$ of $e^{+}$ in $\mu^{+}\rightarrow
e^{+}\nu_{e} \overline{\nu}_{\mu}$ decay was carried out at SIN
\cite{burk85a}. A magnetized-iron foil was adopted as a polarimeter
for the $e^{+}$ polarization, and was tilted by 45$^{\circ}$ with
respect to the $e^{+}$ momentum direction. Either Bhabha scattering of
$e^{+}$ off $e^{-}$s ($e^{+}e^{-}\rightarrow e^{+}e^{-}$) or
annihilation in flight ($e^{+}e^{-}\rightarrow\gamma\gamma$) in the
magnetized iron foil was utilized, where the cross sections of those
processes have a particular dependence on the relative angle between the
$e^{\pm}$ polarization directions. Since the magnitude and direction
of $e^{-}$ polarization in the magnetized-iron foil is known, the
polarization of $e^{+}$ from
\muenn decay can be determined. Either $e^{+}e^{-}$ or $\gamma\gamma$
pairs from the polarimeter were detected by four NaI(Tl) crystal
detectors located behind. Both unpolarized and polarized muons were
studied. From the case of unpolarized muons, $P_{L} (=\xi^{'}) =
0.998\pm 0.042$ was obtained, whereas the current average value in the
Particle Data Group is $P_L = 1.00 \pm 0.04$~\cite{pdg98}, showing no
strong evidence of the right-handed current. For the case of polarized
muons, they obtained $(\xi^{''}-\xi\cdot\xi^{'})/\xi = -0.35
\pm 0.33$ which, however, did not make better constraints of the
coupling constants at this moment.

The transverse polarization components, $P_{T1}$ and $P_{T2}$, of
$e^{+}$ in $\mu^{+}\rightarrow e^{+}\nu_{e} \overline{\nu}_{\mu}$
decay were measured by the same group \cite{burk85b}. A
magnetized-iron foil was used again as a polarimeter, but was placed
perpendicular to the $e^{+}$ momentum direction. The measurement was
based on the fact that two photons from the annihilation of
transversely-polarized $e^{+}$ with $e^{-}$s in a magnetized-iron foil
are preferentially emitted in the plane determined by the bisector of
the $e^{+}$ polarization vector ($\vec{P}_{T}$) and the $e^{-}$
polarization vector in a magnetized foil. Their results were $<P_{T1}>
= 0.016 \pm 0.023$ and $<P_{T2}> = 0.007 \pm 0.023$. From the measured
value of $P_{T1}$, $\eta = -0.007 \pm 0.013$ was obtained.

A new experiment, R-94-10 at PSI \cite{barn94}, is in preparation to
measure the transverse polarization, both $P_{T_1}$ and $P_{T_2}$, of
the positrons from polarized \muenn decay, with a precision of $3
\times 10^{-3}$. The experimental principle is the same as the
previous one, mentioned before. Major improvements are expected to be
from a higher muon-stopping rate due to a higher proton current at the
PSI cyclotron, installation of two analyzing foils with additional
wire chamber in between, and replacement of four NaI(Tl) crystals by
127 hexagonal BGO crystal. An engineering run will start in 1999.

Another new experiment, R-97-06 at PSI \cite{vanh97}, is under
development to measure the longitudinal polarization, $P_{L}$, of the
positrons emitted antiparallel to the muon spin from polarized \muenn
decay. As explained in Eq.(\ref{eq:xidouble}), $P_L$ at $x\approx 1$
and $\cos\theta_{e}\approx -1$ is sensitive to the combination of
$(\xi^{''}-\xi\xi^{'})/\xi$ with the enhancement factor. R-97-06 aims
to measure this observable with an improvement of more than one order
of magnitude over the previous experiments at SIN. It will use three
solenoidal magnets to track $e^{+}$s from \muenn decay with
double-sided Si strip detectors. Two magnetized iron foils with
opposite sign of the analyzing power are used as a polarimeter,
followed by 127 BGO crystals to detect $e^{\pm}$s and also
photons. The asymmetry for two different analyzing foils are compared
for the two cases of polarized and unpolarized muons, giving a
relative measurement to reduce systematic errors. The goal is to
measure $(\xi^{''}-\xi\xi^{'})/\xi$ to about 0.5 \%. An engineer run
with the complete setup is planed for late 1999.

\section{Lepton Flavor Violating Muon Decays}\label{FMD}

The muon system is one of the best places to search for LFV, compared
with the others. In Table \ref{tb:limits}, the upper limits of various
lepton flavor violating decays are listed. The sensitivity to LFV is
superb in the muon system. It is mostly because of a large number of
muons (of about $10^{14}-10^{15}$/year) available for experimental
searches today. The theoretical frameworks for LFV were already
presented in Section~\ref{LFV}.  In this section, phenomenology and
experimental results, including the prospect for future improvements
for each of the forbidden muon LFV processes, are reviewed. They are
such as \mueg decay, \meee decay, \muec conversion in a muonic atom,
\mupc conversion, and muonium to anti-muonium conversion. In the first
three processes, lepton flavors change by one unit
($|\Delta~L_{i}|=1$).

\begin{table}[hbt!]
\caption{Limits of the lepton-flavor violating decays of muon, tau, pion,
kaon and $Z$ boson.}
\label{tb:limits}
\vspace{5mm}
\begin{tabular}{lll}
Reaction & Present limit & Reference \cr\hline
\mueg & $< 1.2 \times 10^{-11}$ & \citeasnoun{broo99} \cr
\meee & $< 1.0 \times 10^{-12}$ & \citeasnoun{bell88} \cr
$\mu^{-}Ti\rightarrow e^{-}Ti$ & $< 6.1 \times 10^{-13}$ &
\citeasnoun{wint98} \cr
$\mu^{+}e^{-} \rightarrow \mu^{-}e^{+}$ & $< 8.3 \times 10^{-11}$ &
\citeasnoun{will99} \cr
$\tau\rightarrow e\gamma$   & $< 2.7 \times 10^{-6}$ &
\citeasnoun{edwa97} \cr
$\tau\rightarrow \mu\gamma$ & $< 3.0 \times 10^{-6}$ &
\citeasnoun{edwa97} \cr
$\tau\rightarrow \mu\mu\mu$ & $< 1.9 \times 10^{-6}$ & 
\citeasnoun{blis98} \cr
$\tau\rightarrow eee$ & $< 2.9 \times 10^{-6}$ & 
\citeasnoun{blis98} \cr
$\pi^{0} \rightarrow \mu e$ & $< 8.6 \times 10^{-9}$ & 
\citeasnoun{krol94} \cr
$K^{0}_{L}\rightarrow \mu e$ & $< 4.7 \times 10^{-12}$ &
\citeasnoun{ambr98} \cr
$K^{+}\rightarrow\pi^{+}\mu^{+}e^{-}$ & $< 2.1 \times 10^{-10}$ & 
\citeasnoun{lee90} \cr
$K^{0}_{L}\rightarrow\pi^{0}\mu^{+}e^{-}$ & $< 3.1 \times 10^{-9}$ & 
\citeasnoun{aris98} \cr
$Z^{0}\rightarrow \mu e$    & $< 1.7 \times 10^{-6}$ &
\citeasnoun{aker95} \cr
$Z^{0}\rightarrow \tau e$   & $< 9.8 \times 10^{-6}$ &
\citeasnoun{aker95} \cr
$Z^{0}\rightarrow \tau\mu$  & $< 1.2 \times 10^{-5}$ &
\citeasnoun{abre97} \cr
\end{tabular}
\end{table}

\subsection{$\mu^{+} \rightarrow \lowercase{e}^{+}\gamma$ decay}\label{MEG}

\subsubsection{Phenomenology of \mueg decay}

The most popular process of lepton-flavor-violating muon decay is
$\mu^{+}\rightarrow e^{+}\gamma$.  The Lagrangian for the \mueg
amplitude is given by, as shown in Eq.(\ref{eq:muegeff}),

\begin{equation}
{\lagr}_{\mu\rightarrow e\gamma} 
     =  -{4G_F\over\sqrt{2}} \Biggl[ {m_{\mu }}{A_R}\overline{\mu_{R}}
        {{\sigma }^{\mu \nu}{e_L}{F_{\mu \nu}}}
       + {m_{\mu }}{A_L}\overline{\mu_{L}}
        {{\sigma }^{\mu \nu}{e_R}{F_{\mu \nu}}} + H.c. \Biggr].
\end{equation}

\noindent The differential angular distribution of
\mueg decay is already given, as in Eq.(\ref{eq:ang-mueg}), by

\begin{equation}  
{dB(\mu^{+}\rightarrow e^{+}\gamma)\over{d(\cos\theta_e)}} = {192\pi^2}
\Biggl[
\left| A_R \right|^2 (1-P_{\mu}\cos{\theta_e}) + 
\left| A_L \right|^2 (1+P_{\mu}\cos{\theta_e}) \Biggr],
\end{equation}

\noindent where $\theta_{e}$ is the angle between the muon polarization
and the $e^{+}$ momentum vectors. $P_{\mu}$ is the magnitude of the
muon spin polarization. The branching ratio is presented by

\begin{equation}
B(\mu^{+}\rightarrow e^{+}\gamma) 
= {\Gamma(\mu^{+}\rightarrow e^{+}\gamma) \over
\Gamma(\mu^{+}\rightarrow e^{+}\nu\overline{\nu})}
= 384\pi^{2} (|A_{R}|^2 + |A_{L}|^2).
\end{equation}

\subsubsection{Event signature and backgrounds}

The event signature of $\mu^{+} \rightarrow e^{+} \gamma$ decay at
rest is an $e^{+}$ and a photon in coincidence, moving collinearly
back-to-back with their energies equal to half of the muon mass
($m_{\mu}/2 = 52.8$ MeV). The searches in the past were carried out by
using positive muon decay at rest to fully utilize its kinematics. A
negative muon cannot be used, since it is captured by a nucleus when
it is stopped in a material.  There are two major backgrounds to a
search for $\mu^{+}\rightarrow e^{+} \gamma$. One is a physics
(prompt) background from radiative muon decay, \muenng, when $e^{+}$
and photon are emitted back-to-back with the two neutrinos carrying
off little energy. The other background is an accidental coincidence
of an $e^{+}$ in a normal muon decay, \muenn, accompanied by a high
energy photon. The sources of the latter might be either \muenng
decay, or annihilation-in-flight or external bremsstrahlung of
$e^{+}$s from normal muon decay. These backgrounds are described in
more detail in the following.
 
\subsubsection{Physics background}\label{sc:PB}

One of the major physics backgrounds to the search for
$\mu^{+}\rightarrow e^{+} \gamma$ decay is radiative muon decay,
\muenng (branching ratio = 1.4 \% for $E_{\gamma} >$ 10 MeV), when the
$e^{+}$ and photon are emitted back-to-back with two neutrinos
carrying off little energy. The differential decay width of this
radiative muon decay was calculated as a function of the $e^{+}$
energy ($E_{e}$) and the photon energy ($E_{\gamma}$) normalized to
their maximum energies, namely $x = 2E_{e}/m_{\mu}$ and $y =
2E_{\gamma}/m_{\mu}$ \cite{fron59,ecks59}.  The ranges of $x$ and $y$
are already shown in Eq.(\ref{eq:xyrange}).  As a background to
$\mu^{+}\rightarrow e^{+}\gamma$, the kinematic case when $x\approx1$
and $y\approx1$ is important. In an approximation of the limit of
$x\approx1$ and $y\approx1$ with an angle between $e^{+}$ and photon
($\theta_{e\gamma})$ of almost $180^{\circ}$, the differential decay
width of \muenng decay is given by \cite{kuno96}

\begin{eqnarray}
&&d\Gamma(\mu^{+}\rightarrow e^{+}\nu\overline{\nu}\gamma)
\cong {G_{F}^2 m_{\mu}^5 \alpha \over 3\times 2^{8}\pi^{4}} \times \cr
&&\Bigl[ (1-x)^{2}(1-P_{\mu}\cos\theta_{e}) + 
\Bigl( 4(1-x)(1-y) - {1\over2}z^{2} \Bigr)(1+P_{\mu}\cos\theta_{e})
\Bigr]dxdyzdzd(\cos\theta_{e}),
\label{eq:dif}
\end{eqnarray}

\noindent where $\theta_{e}$ is the angle between the muon spin and
the $e^{+}$ momentum direction. $G_{F}$ is the Fermi coupling
constant, $\alpha$ is the fine-structure constant, $z =
\pi-\theta_{e\gamma}$, and $\cos z$ is expanded in a polynomial of $z$,
since $z$ is small. In Eq.(\ref{eq:dif}), only the terms of up to the
second order of a combination of $(1-x)$, $(1-y)$ and $z$ are kept. At
$x\approx1$ and $y\approx 1$, the effect of the positron mass is found
to be very small, on the order of $(m_{e}/m_{\mu})^2$, and is
therefore neglected in Eq.(\ref{eq:dif}). The first term in
Eq.(\ref{eq:dif}) represents the $e^{+}$ being emitted preferentially
opposite to the muon spin direction, whereas in the second term the
$e^{+}$ is emitted along the muon-spin direction. When $x=1$ and $y=1$
exactly, this differential decay width vanishes. However, in a real
experiment, finite detector resolutions introduce background events
which would ultimately limit the sensitivity of a search for
$\mu^{+}\rightarrow e^{+}\gamma$.

Given the detector resolution, the sensitivity limitation from this
physics background can be estimated by integrating the differential
decay width over the kinematic signal box. It is given by

\begin{eqnarray}
dB(\mu^{+}\rightarrow e^{+}\nu\overline{\nu}\gamma) &=& {1 \over
\Gamma(\mu^{+}\rightarrow e^{+}\nu\overline{\nu})}
\int^{1}_{1-\delta x} dx \int^{1}_{1-\delta y} dy
\int^{\min(\delta z, 2\sqrt{(1-x)(1-y)})}_{0} dz
~~{d\Gamma(\mu^{+}\rightarrow e^{+}\nu\overline{\nu}\gamma) 
\over dxdydz}, \nonumber \\ &=&
{\alpha\over16\pi}
\Bigl[ 
J_{1}(1-P_{\mu}\cos\theta_{e}) 
+ J_{2}(1 + P_{\mu}\cos\theta_{e}) 
\Bigr]d(\cos\theta_{e}),
\label{eq:par}
\end{eqnarray}

\noindent where $\delta x$, $\delta y$ and $\delta z$ are a half width 
of the \mueg signal region for $x$, $y$ and $z$,
respectively. $\Gamma(\mu^{+}\rightarrow e^{+}\nu\overline{\nu})$ is
the total muon decay width. $J_1$ and $J_2$ are given as the sixth
power of a combination of $\delta x$ and $\delta y$. For the case of
$\delta z >2\sqrt{\delta x\delta y}$, they are presented by

\begin{equation}
J_1 = (\delta x)^4(\delta y)^2 \quad {\rm and} 
\quad J_2 = {8\over3}(\delta x)^3(\delta y)^3.
\label{eq:par1}
\end{equation}

\noindent When the angular resolution meets $\delta z \leq 2\sqrt{\delta
x \delta y}$, they are given by

\begin{eqnarray}
J_1 &=& {8\over3}(\delta x)^3(\delta y)({\delta z\over 2})^2 -
2(\delta x)^2({\delta z\over 2})^4 + {1\over 3}{1 \over (\delta
y)^2}({\delta z\over 2})^8, \\
J_2 &=& 8(\delta x)^2(\delta y)^2({\delta z\over 2})^2 - 8(\delta
x)(\delta y)({\delta z\over 2})^4 + {8\over 3}({\delta z\over 2})^6.
\label{eq:par2}
\end{eqnarray}

\noindent Experimentally, the resolution of the $e^{+}$ energy is better
than that of the photon energy, {\em i.e.} $\delta x<\delta y$.  Also,
the angular resolution, $\delta z$, has been poor in past experiments. 
Thereby, $J_2$ is much larger than $J_1$ for most cases.

\begin{figure}[ht!]
\centerline{\epsfig{file=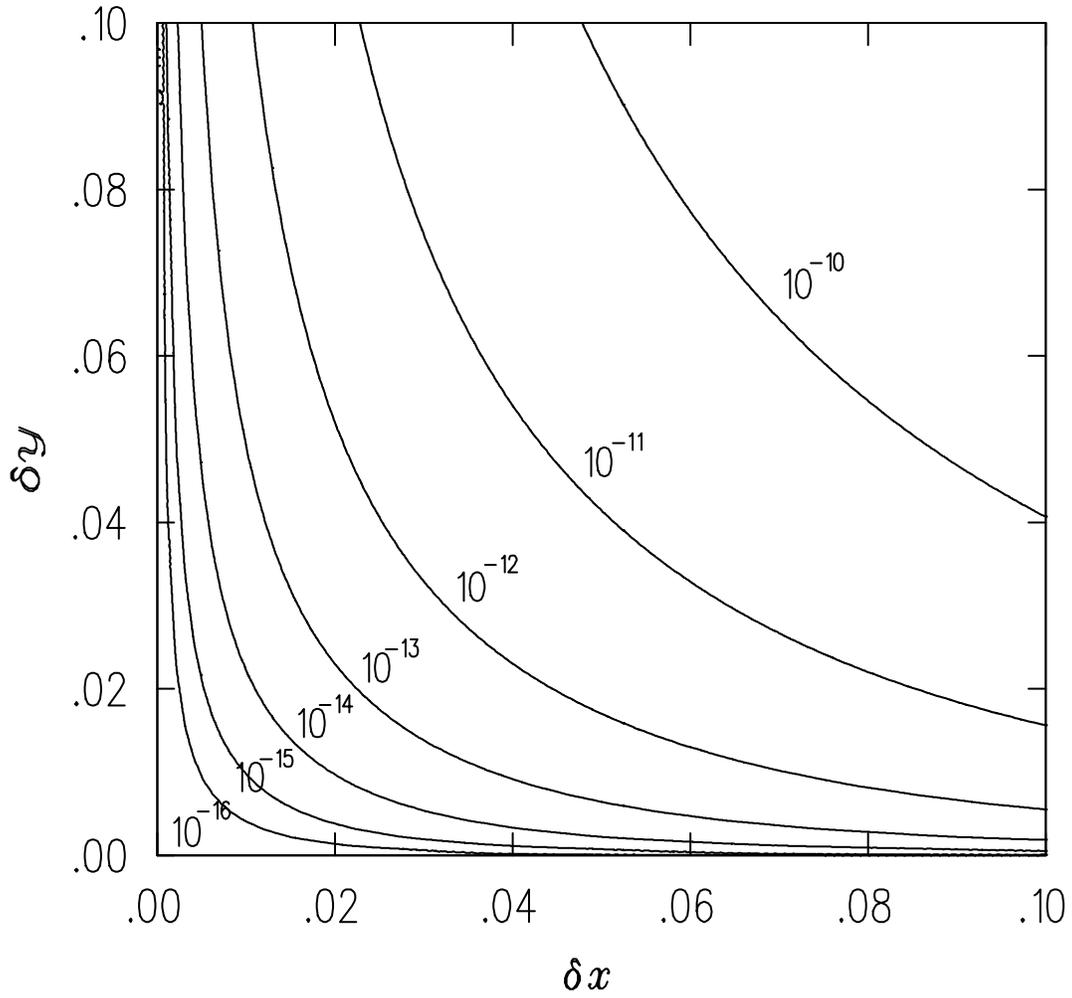,width=14cm}}
\vspace{5mm}
\caption{Effective branching ratio of the physics background from the \muenng
decay as a function of the $e^{+}$ energy resolution ($\delta x$) and
photon energy resolution ($\delta y$) (after Kuno and Okada, (1996)).}
\label{fg:rmd}
\end{figure}

Fig.\ref{fg:rmd} shows a fraction of the \muenng decay for the given
$\delta x$ and $\delta y$ values with unpolarized muons in the case of
$\delta z\geq 2\sqrt{\delta x\delta y}$. From Fig.\ref{fg:rmd}, it can
be seen that both $\delta x$ and $\delta y$ on the order of 0.01 are
needed to achieve a sensitivity limit at the level of $10^{-15}$.

Radiative corrections to the radiative muon decay for the case of the
physics background to \mueg decay have been calculated to be on the
order of several \%, depending on the detector resolution
\cite{arbu98}.

\subsubsection{Accidental background}

With a very high rate of incident muons, the accidental background
becomes more important than the physics background. This is usually
the case for the present and future experiments. The event rate of the
accidental background normalized to the total decay rate ($B_{\rm
acc}$) can be estimated by

\begin{equation}
B_{\rm acc} = R_{\mu} 
\cdot f^{0}_{e} \cdot f^{0}_{\gamma} 
\cdot (\Delta t_{e\gamma}) \cdot ({\Delta\omega_{e\gamma} \over 4\pi}),
\end{equation}

\noindent where $R_{\mu}$ is the instantaneous muon intensity.
$f^{0}_{e}$ and $f^{0}_{\gamma}$ are the integrated fractions of the
spectrum of $e^{+}$ in the normal muon decay and that of photon (such
as from \muenng~decay) within the signal region, respectively. They
include their corresponding branching ratios. $\Delta t_{e\gamma}$ and
$\Delta\omega_{e\gamma}$ are, respectively, the full widths of the
signal regions for timing coincidence and angular constraint of the
back-to-back kinematics.

Given the sizes of the signal region, $B_{acc}$ can be evaluated. Let
us take $\delta x$, $\delta y$, $\delta\theta_{e\gamma}$, and $\delta
t_{e\gamma}$ to be the half width of the signal region for $e^{+}$,
photon energies, angle $\theta_{e\gamma}$, and relative timing between
$e^{+}$ and photon, respectively. $f_{e}^{0}$ can be estimated by
integrating the Michel spectrum of the normal muon decay over
$1-\delta x \leq x \leq 1$, yielding $f_{e}^{0}~\approx 2(\delta
x)$. Given the angular resolution, $\delta\theta_{e\gamma}$, the
back-to-back resolution ($\Delta\omega_{e\gamma}/4\pi$) is presented
by ($\Delta\omega_{e\gamma}/4\pi) = {(\delta\theta_{e\gamma})^2/4}$.
As for $f_{\gamma}^{0}$, if the radiative muon decay \muenng is
considered as a source of the 52.8 MeV photon, it can be given by
integrating Eq.(\ref{eq:radmu}) over $2\pi$ for $\theta_{\gamma}$, and
then over the photon energy within the width of the signal region
($1-\delta y\leq y \leq 1$). For unpolarized muons, it is given by

\begin{equation}
f_{\gamma}^{0} = \int^{1}_{1-\delta y}dy \int 
d(cos\theta_{\gamma}) {dB(\mu^{+}\rightarrow
e^{+}\nu\overline{\nu}\gamma) \over dy d(cos\theta_{\gamma})}
\approx \Bigl({\alpha\over2\pi}\Bigr) (\delta y)^2 
\Bigl[\ln(\delta y) + 7.33 \Bigr].
\label{eq:int50}
\end{equation}

\noindent From Eq.(\ref{eq:int50}), it is shown that $f_{\gamma}^{0}$ for
\muenng~decay is roughly proportional to $(\delta y)^2$.

\begin{figure}[ht!]
\centerline{\epsfig{file=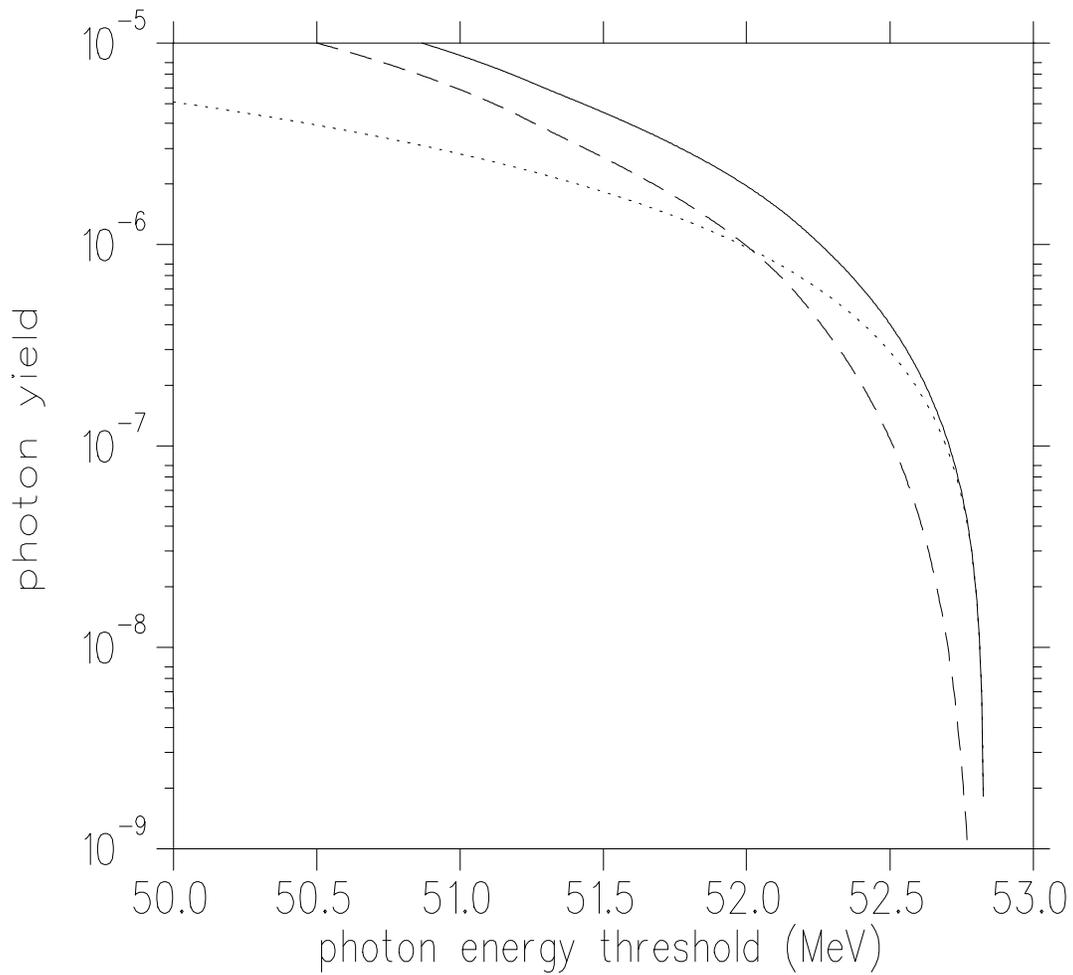,width=14cm}}
\vspace{5mm}
\caption{Integrated rates of backgrounds from annihilation-in-flight
(a dotted line) and radiative muon decay (a dashed line) as a function
of the photon energy. The sum of the two is shown by the solid line.}
\label{fg:aif}
\end{figure}

The other sources of high-energy photons are annihilation in flight of
$e^{+}$s in the normal muon decay and external bremsstrahlung. The
contribution from annihilation of $e^{+}$ in flight depends on the
materials along the $e^{+}$'s track path. Fig.\ref{fg:aif} shows, for
instance, the contribution of annihilation in flight for that case of
$e^{+}$s passing through a muon-stopping target of 50 mg in thickness.
It indicates that its contribution from the target is smaller than the
radiative muon decay, and only becomes important if the photon energy
resolution becomes extremely good. However, it is dependent on the
total amount of materials in an experimental setup.

From the above, the effective branching ratio of accidental background
is given by

\begin{equation}
B_{acc}=R_{\mu} \cdot (2\delta x) \cdot \Bigl[ {\alpha \over
2\pi}(\delta y)^2({\rm ln}(\delta y)+7.33) \Bigr]
\times \Bigl( {\delta \theta^2 \over 4} \Bigr) \cdot (2\delta t).
\end{equation}

\noindent For instance, take some reference numbers such as the $e^{+}$
energy resolution of 1\% (FWHM), the photon energy resolution of 6\%
(FWHM), $\Delta\omega_{e\gamma} = 3 \times 10^{-4}$ steradian, $\Delta
t_{e\gamma}$ = 1 nsec, and $R_{\mu} = 3 \times 10^{8}~\mu^{+}$/sec,
$B_{\rm acc}$ is $3\times 10^{-13}$. The accidental background becomes
severe. Therefore, it is critical to make significant improvements in
the detector resolution in order to reduce the accidental background.

\subsubsection{Muon polarization}\label{sc:MP}

\begin{figure}[ht!]
\centerline{\epsfig{file=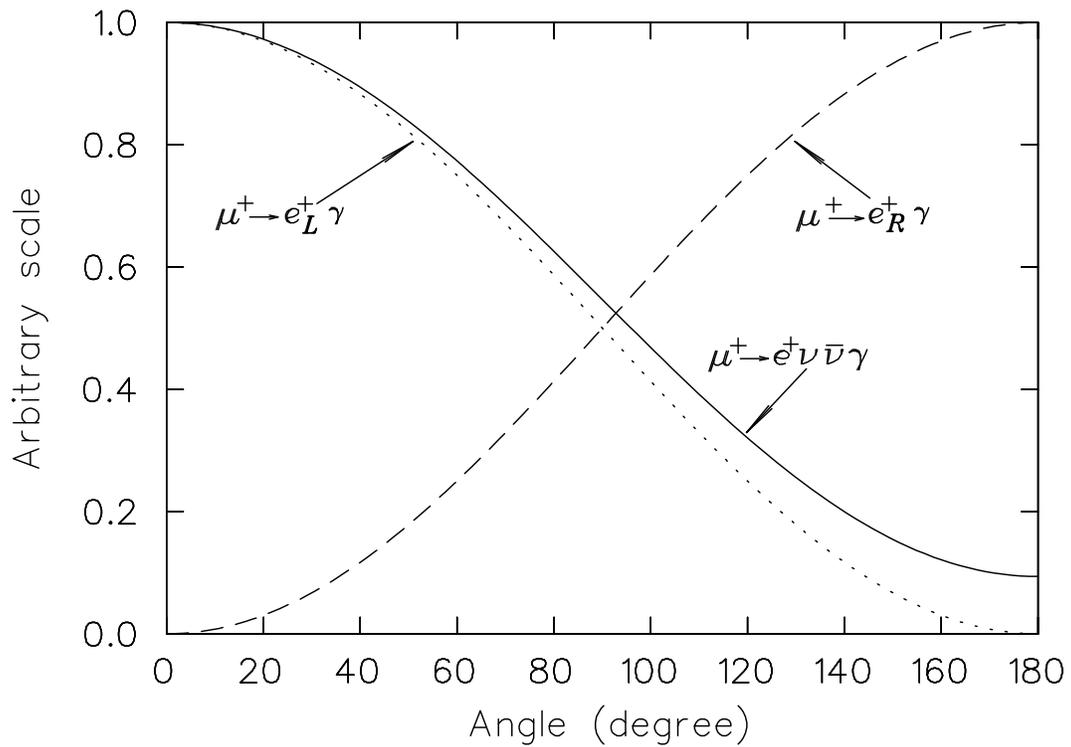,width=14cm,angle=90}}
\vspace{5mm}
\caption{Angular distribution of the physics background from the \muenng
decay from polarized muons (a solid line). $\mu^{+}\rightarrow
e^{+}_{L}\gamma$ (a dotted line) and $\mu^{+}\rightarrow
e^{+}_{R}\gamma$ (a dashed line) decays are also shown (after Kuno and
Okada, (1996)).}
\label{fg:prompt-dis}
\end{figure}

The use of polarized muons has been found to be useful to suppress
backgrounds for \mueg search \cite{kuno96,kuno97a}. For the physical
(prompt) background, as already discussed in Subsection \ref{sc:PB},
the coefficient of $J_2$ is much larger than $J_1$, since the
resolution of the photon energy is much worse than that of the $e^{+}$
energy detection. Therefore, the angular distribution of the physics
background follows approximately (1+$P_{\mu}\cos\theta$) as long as
$\delta y > \delta x$. Fig.\ref{fg:prompt-dis} shows the angular
distribution of \muenng with, for instance, $\delta y/\delta x =
4$. If we selectively measure the $e^{+}$s in $\mu^{+}\rightarrow
e^{+}\gamma$ going opposite to the muon-polarization direction, the
background from \muenng would be significantly reduced for the
$\mu^{+}\rightarrow e^{+}_{R} \gamma$ search.  Furthermore, by varying
$\delta x$ and $\delta y$, the angular distribution of the \muenng
background can change according to Eq.(\ref{eq:par}), thus providing
another means to discriminate the signal from the backgrounds.

Regarding the accidental background, the use of polarized muons has
also provided a means for its suppression \cite{kuno97a}. It happens
that the sources of accidental backgrounds have a specific angular
distribution when a muon is polarized. For instance, the $e^{+}$s in
the normal Michel $\mu^{+}$ decay are emitted preferentially along the
muon spin direction, following a ($1+P_{\mu}\cos\theta_e$)
distribution, as in Eq.(\ref{eq:simple}), whereas the inclusive
angular distribution of a high-energy photon ({\em e.g.}  $\geq 50$
MeV) from \muenng decay follows a ($1+P_{\mu}\cos\theta_{\gamma}$)
distribution, as in Eq.(\ref{eq:radmu}), where $\theta_{\gamma}$ is
the angle of the photon direction with respect to the muon spin
direction.  It should be noted that this inclusive angular
distribution was obtained after integrating the energy and direction
of the $e^{+}$s, which is in contrast to the case of the physics
background, where only the extreme kinematics of the $e^{+}$ and
photon being back-to-back in \muenng decay is relevant. It is further
noted that the other sources of high-energy photons, such as external
bremsstrahlung and annihilation in flight of $e^{+}$s from the normal
muon decay, also follow a ($1+P_{\mu}\cos\theta_{\gamma}$)
distribution.

\begin{figure}[ht!]
\centerline{\epsfig{file=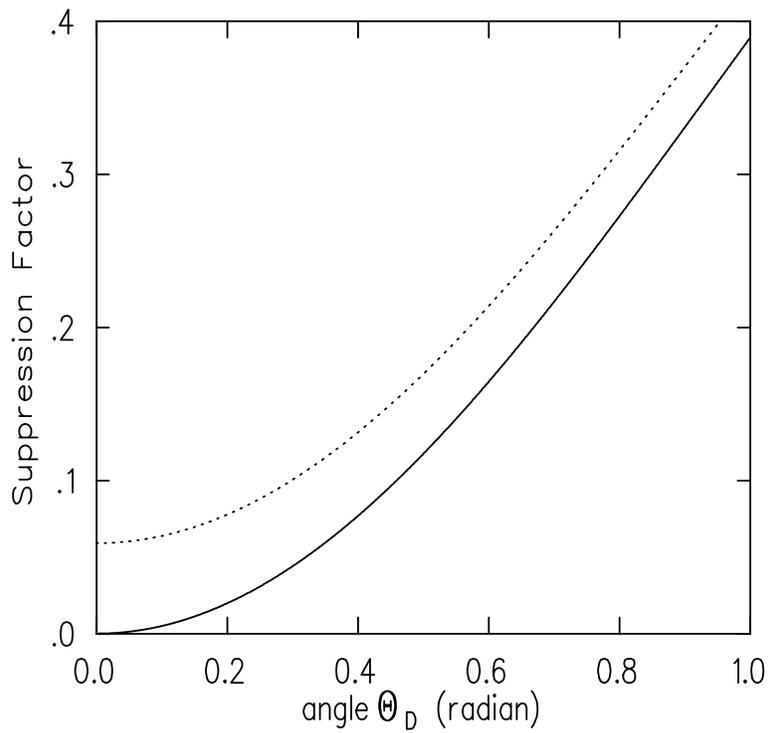,width=10truecm}}
\vspace{5mm}
\caption{Suppression factor of the accidental background in a \mueg search
as a function of (half) the detector opening angle. The solid (dotted)
line is for 100\% (97\%) muon polarization (after Kuno {\em et al.},
(1997)).}
\label{fg:accidental-suppression}
\end{figure}

This inclusive angular distribution of a high-energy photon in \muenng
implies that the accidental background could be suppressed for
$\mu^{+} \rightarrow e^{+}_{L} \gamma$, where high-energy photons must
be detected at the opposite direction to the muon polarization. A
similar suppression mechanism of accidental background can be seen for
$\mu^{+}\rightarrow e^{+}_{R} \gamma$ when high-energy positrons are
detected at the opposite direction to the muon polarization. As a
result, the selective measurements of either $e^{+}$s or photons
antiparallel to the muon spin direction would give the same accidental
background suppression for $\mu^{+} \rightarrow e^{+}_{R} \gamma$ and
$\mu^{+}\rightarrow e^{+}_{L} \gamma$ decays, respectively. This
favorable situation comes from the fact that the inclusive
distributions of both high-energy $e^{+}$s and photons, respectively,
in the normal and radiative muon decays, follow a ($1 +
P_{\mu}\cos\theta$) distribution, where $\theta$ is either
$\theta_{e}$ or $\theta_{\gamma}$.  The suppression factor, $\eta$, is
calculated for polarized muons by

\begin{eqnarray}
\eta &\equiv& \int^{1}_{\cos\theta_{D}}d(\cos\theta) (1 +
P_{\mu}\cos\theta)(1 - P_{\mu}\cos\theta) /
\int^{1}_{\cos\theta_{D}}d(cos\theta) \nonumber \\
&=& (1-P_{\mu}^2) +
{1\over3}P_{\mu}^2(1-\cos\theta_{D})(2+\cos\theta_{D}),
\label{eq:cos}
\end{eqnarray}

\noindent where $\theta_{D}$ is a half opening angle of detection with
respect to the muon polarization direction. $\eta$ is shown in
Fig.\ref{fg:accidental-suppression} as a function of $\theta_{D}$. For
instance, for $\theta_{D} = 300$ mrad, an accidental background can be
suppressed down to about $1/20$ ($1/10$) when $P_{\mu}$ is 100 (97)\%.

\subsubsection{Experimental status of \mueg decay}\label{sc:MGES}

Experimental searches for \mueg have a long history of more than 50
years. These searches have been actively promoted by intense muon
beams available at the meson factories. Experimental efforts have been
devoted to improving the detection resolutions of four variables,
namely the positron energy ($E_{e}$), the photon energy
($E_{\gamma}$), the timing between the positron and photon ($\Delta
t_{e\gamma}$), and the angle between the positron and photon ($\Delta
\theta_{e\gamma}$). Various kinds of apparatus have been tried in the
past. In Table~\ref{tb:hmueg}, several experimental results of 90\%
C.L. upper limit of \mueg decay in the past experiments are listed
along with their achieved detection resolutions.

\begin{table}[t!]
\caption{Historical Progress of search for \mueg since the era of
meson factories with 90 \% C.L. upper limits. The resolutions quoted
are given as a full width at half maximum (FWHM).}
\label{tb:hmueg}
\vspace{5mm}
\begin{tabular}{llllllll}
Place & Year & $\Delta E_{e}$ & $\Delta E_{\gamma}$ & $\Delta t_{e\gamma}$ 
& $\Delta\theta_{e\gamma}$ & Upper limit & References \cr\hline
TRIUMF & 1977 & 10\% & 8.7\% & 6.7ns & $-$ & $< 3.6 \times 10^{-9}$ 
& \citeasnoun{depo77} \cr
SIN & 1980 & 8.7\% & 9.3\% & 1.4ns & $-$ & $< 1.0 \times 10^{-9}$ 
& \citeasnoun{vand80} \cr
LANL & 1982 & 8.8\% & 8\% & 1.9ns & 37mrad & $< 1.7 \times 10^{-10}$ 
& \citeasnoun{kinn82} \cr
LANL & 1988 & 8\% & 8\% & 1.8ns & 87mrad & $< 4.9 \times 10^{-11}$ 
& \citeasnoun{bolt88} \cr
LANL & 1999 & 1.2\%$^{*}$ & 4.5\%$^{*}$ & 1.6ns & 15mrad & 
$< 1.2 \times 10^{-11}$ & \citeasnoun{broo99} \cr
\end{tabular}
\vspace{5mm}
$^{*}$ shows an average of the numbers given in \citeasnoun{broo99}.
\vspace{5mm}
\end{table}

The upper limit quoted in the Particle Data Group~\cite{pdg98} is
$B(\mu^{+}\rightarrow e^{+} \gamma) < 4.9\times 10^{-11}$, which was
obtained by an experiment with the ``Crystal Box'' detector
\cite{bolt88} at Los Alamos National Laboratory (LANL). Its apparatus
consisted of 396 NaI(Tl) crystals, cylindrical drift chambers
surrounding a muon-stopping target in a zero magnetic field.

\begin{figure}[ht!]
\centerline{\epsfig{file=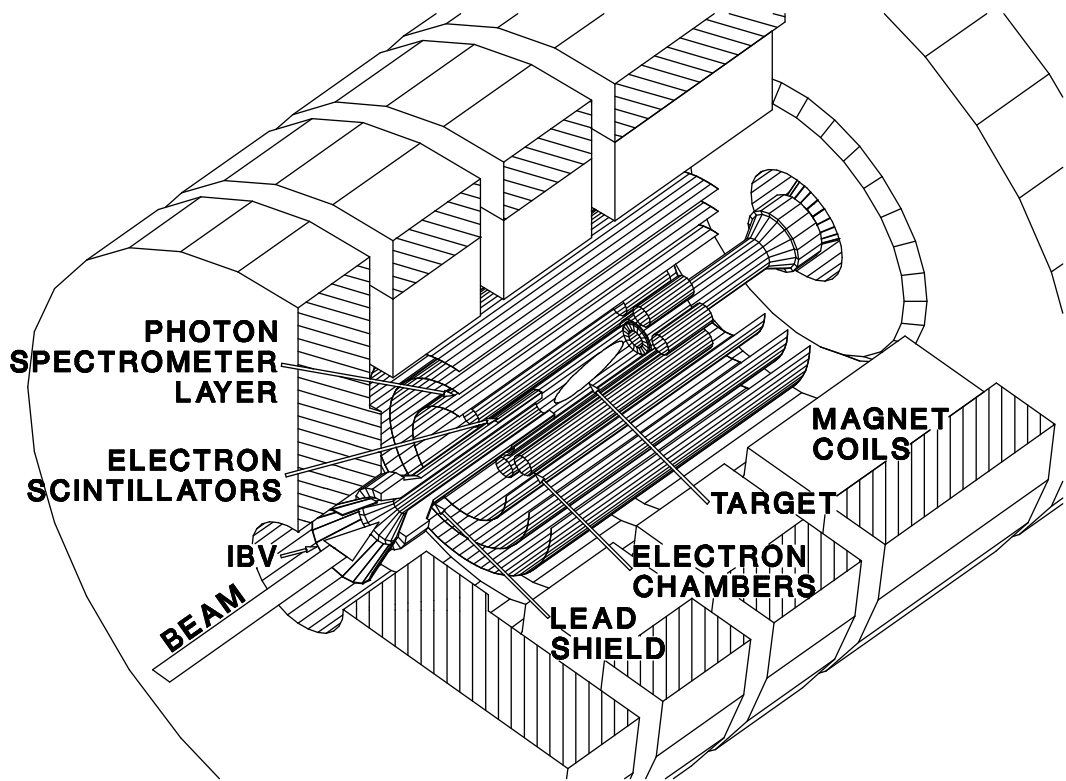,width=14cm}}
\vspace{5mm}
\caption{Schematic layout of the MEGA detector (provided by R.~Mischke).}
\label{fg:mega}
\end{figure}

Since then, a new experimental search for \mueg has been carried out
by the MEGA collaboration at LANL. A schematic view of the MEGA
spectrometer is shown in Fig.\ref{fg:mega}. The MEGA detector
consisted of a magnetic spectrometer for the positron and three
concentric pair-spectrometers for the photon. They were placed inside
a superconducting solenoid magnet of a 1.5 T field. The positron
spectrometer comprised eight cylindrical wire chambers and
scintillators for timing. The positron energy resolution (FWHM) was
from 0.5 MeV (0.95\%) to 0.85 MeV (1.6\%) for a 52.8-MeV $e^{+}$,
depending on the number of helical loops of $e^{+}$ tracks. For the
pair-spectrometer, each layer had lead converters, MWPCs, drift
chambers and scintillators. The photon energy resolutions (FWHM) were
1.7 MeV (3.3\%) and 3.0 MeV (5.7\%) for the outer and inner Pb
conversion layers, respectively.  A surface $\mu^{+}$ beam of 29.8
MeV/$c$ was introduced along the detector axis, and was stopped in the
muon-stopping target made of a thin tilted Mylar foil. All of the
charged particles from muon decays are confined within the positron
spectrometer. The intensity of the muon beam was $2.5\times
10^{8}$/sec with a macroscopic duty factor of 6\%. The total number of
muons stopped was $1.2\times 10^{14}$. By using the likelihood method,
a new limit of $1.2\times 10^{-11}$ with 90\% C.L. has been
reported~\cite{broo99}.

Recently, a new experimental proposal, R-99-05, aiming at a
sensitivity of $10^{-14}$ in the \mueg branching ratio has been
approved at PSI
\cite{bark99}. The improvement will be expected by utilizing a
continuous muon beam of 100\% duty factor at PSI. With keeping the
same instantaneous beam intensity as MEGA, the total number of muons
available can be increased by a factor of 16. Further improvement is a
novel liquid xenon scintillation detector of the ``Mini-Kamiokande''
type, which is a 0.8-m$^3$ volume of liquid xenon viewed by an array
of a total of 800 photomultipliers from all the sides. The expected
resolutions (FWHM) of the photon energy and position are about 1.4\%
and 4 mm, respectively. As the $e^{+}$ detection, a solenoidal
magnetic spectrometer with a graded magnetic field is adopted, in
which the magnetic field is arranged so that $e^{+}$ from the \mueg
decay follows a trajectory with a constant radius, independently of
its emission angle. It allows easier identification of the $e^{+}$ in
the \mueg decay. Physics data taking is expected to start in year
2003.

A search for \muegg was also undertaken simultaneously with the
\mueg search. The present 90\% C.L. upper limit of $B(\mu^{+}
\rightarrow e^{+}\gamma\gamma) < 7.2 \times 10^{-11}$ was obtained
\cite{bolt88}.

\subsection{$\mu^{+} \rightarrow \lowercase{e^{+}e^{+}e^{-}}$ decay}

\subsubsection{Phenomenology of \meee decay}\label{sc:PM3ED}

The decay width of $\mu^{+} \rightarrow e^{+}e^{-}e^{-}$ is determined
from the effective Lagrangian (at the $m_{\mu}$ scale) described in
Eqs.(\ref{eq:muegeff}) and (\ref{eq:mu3eeff}) in Section III. The
relevant interactions are

\begin{eqnarray}
{\lagr}_{\mu\rightarrow eee} &=& -{4G_{F}\over\sqrt{2}} 
\Biggl[  {m_{\mu }}{A_R}\overline{\mu_{R}}
        {{\sigma }^{\mu \nu}{e_L}{F_{\mu \nu}}}
       + {m_{\mu }}{A_L}\overline{\mu_{L}}
        {{\sigma }^{\mu \nu}{e_R}{F_{\mu \nu}}} \nonumber \\
    && + {g_1}(\overline{{{\mu }_R}}{e_L})
              (\overline{{e_R}}{e_L})
       + {g_2}(\overline{{{\mu }_L}}{e_R})
              (\overline{{e_L}}{e_R}) \nonumber \\ 
    &&   +{g_3}(\overline{{{\mu }_R}}{{\gamma }^{\mu }}{e_R})
              (\overline{{e_R}}{{\gamma }_{\mu }}{e_R})
       + {g_4}(\overline{{{\mu }_L}}{{\gamma }^{\mu }}{e_L})
              (\overline{{e_L}}{{\gamma }_{\mu }}{e_L})  \nonumber \\
    &&   +{g_5}(\overline{{{\mu }_R}}{{\gamma }^{\mu }}{e_R})
              (\overline{{e_L}}{{\gamma }_{\mu }}{e_L})
       + {g_6}(\overline{{{\mu }_L}}{{\gamma }^{\mu }}{e_L})
              (\overline{{e_R}}{{\gamma }_{\mu }}{e_R})
       +  H.c. \Biggr],
\label{eq:mu3eLag}
\end{eqnarray}

\noindent where the $f_{E0}$ and $f_{M0}$ photonic contributions in
Eq.(\ref{eq:photonic}) are included in the four fermion coupling
constants.

\begin{figure}[ht!]
\centerline{\epsfig{file=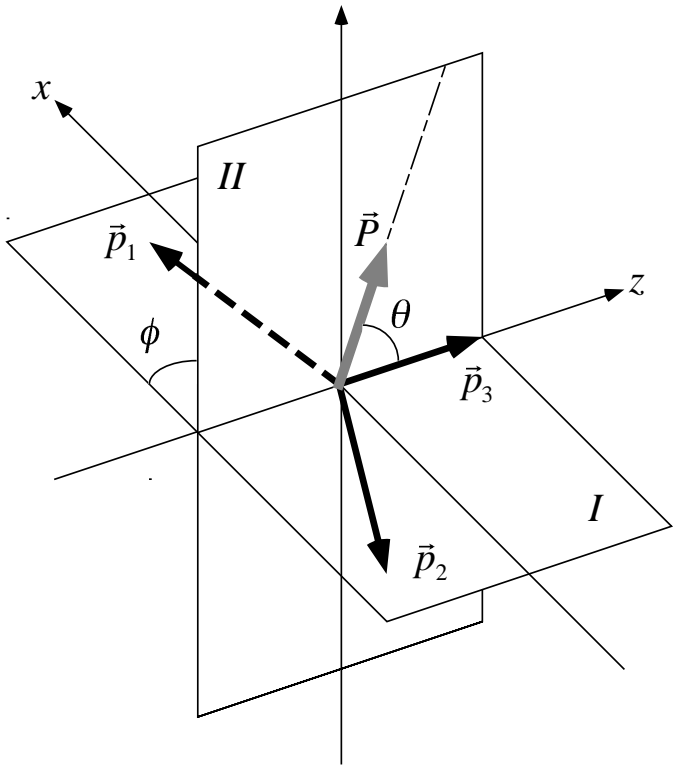,width=14cm}}
\vspace{5mm}
\caption{Kinematics of the \meee decay in the muon center-of-mass
system, in which $\vec{p}_1$, $\vec{p}_2$ are the momentum vectors of
the two $e^{+}$s and $\vec{p}_{3}$ is that of the $e^{-}$,
respectively. The plane-I is the decay plane on which $\vec{p}_1$,
$\vec{p}_2$, and $\vec{p}_3$ lie. The plane-II is the plane in which
the muon polarization vectors, $\vec{P}$ and $\vec{p}_3$, are located
(after Okada,~{\em et al.}, (1999)).}
\label{fg:coor3e}
\end{figure}

When muons are polarized, the kinematics of the \meee decay is
determined by two energy variables and two angle variables of the
decay positrons \cite{okad98,okad99}.  The energy variables are
$x_{1}=2E_{1}/m_{\mu}$ and $x_{2}=2E_{2}/m_{\mu}$, where $E_{1}$
($E_{2}$) is the higher (lower) energy of the decay positrons. The
allowed regions of $x_{1}$ and $x_{2}$ are ${1\over2}\leq x_{1}\leq 1$
and $1-x_{1}\leq x_{2} \leq x_{1}$, if $m_{e}$ is neglected compared
to $m_{\mu}$.  Let us take the coordinate as shown in
Fig.\ref{fg:coor3e}, where the $z$-axis is in the direction of the
decay electron momentum ($\vec{p_{3}}$), and the $z-x$ plane is the
decay plane. The positive direction of the $x$-axis is chosen to be in
the hemisphere of higher-energy positron.  The two angles
($\theta,\varphi$) determine the direction of the muon polarization
($\vec{P}_{\mu}$) with respect to the decay plane.

In this coordinate, the differential branching ratio of the \meee
decay is given by

\begin{eqnarray}
{dB(\mu^{+}\rightarrow e^{+}e^{+}e^{-})\over 
dx_1 dx_2 d(\cos\theta) d\varphi}&=& {3\over 2\pi}
       \Bigl[ C_{1}\alpha_{1}(x_{1},x_{2})(1 + P_{\mu}\cos \theta) +
       C_{2}\alpha_{1}(x_{1},x_{2})(1 - P_{\mu}\cos \theta)
\nonumber \\
         &+&C_{3}\{\alpha_{2}(x_{1},x_{2}) 
                   +P_{\mu}\beta_{1}(x_{1},x_{2})\cos \theta 
                   +P_{\mu}\gamma_{1}(x_{1},x_{2})\sin \theta \cos \varphi \} 
\nonumber \\
         &+&C_{4}\{\alpha_{2}(x_{1},x_{2}) 
                   -P_{\mu}\beta_{1}(x_{1},x_{2})\cos \theta 
                   -P_{\mu}\gamma_{1}(x_{1},x_{2})\sin \theta \cos \varphi \} 
\nonumber \\
         &+&C_{5}\{\alpha_{3}(x_{1},x_{2}) 
                   +P_{\mu}\beta_{2}(x_{1},x_{2})\cos \theta
                   +P_{\mu}\gamma_{2}(x_{1},x_{2})\sin \theta \cos \varphi \} 
\nonumber \\
         &+&C_{6}\{\alpha_{3}(x_{1},x_{2}) 
                   -P_{\mu}\beta_{2}(x_{1},x_{2})\cos \theta
                   -P_{\mu}\gamma_{2}(x_{1},x_{2})\sin \theta \cos \varphi \} 
\nonumber \\
         &+&C_{7}\{\alpha_{4}(x_{1},x_{2})(1-P_{\mu}\cos \theta) 
                  +P_{\mu}\gamma_{3}(x_{1},x_{2})\sin \theta \cos \varphi  \} 
\nonumber \\
         &+&C_{8}\{\alpha_{4}(x_{1},x_{2})(1+P_{\mu}\cos \theta) 
                  -P_{\mu}\gamma_{3}(x_{1},x_{2})\sin \theta \cos \varphi  \} 
\nonumber \\
         &+&C_{9}\{\alpha_{5}(x_{1},x_{2})(1+P_{\mu}\cos \theta)
                  -P_{\mu}\gamma_{4}(x_{1},x_{2})\sin \theta \cos \varphi\} 
\nonumber \\
         &+&C_{10}\{\alpha_{5}(x_{1},x_{2})(1-P_{\mu}\cos \theta)
                  +P_{\mu}\gamma_{4}(x_{1},x_{2})\sin \theta \cos \varphi\} 
\nonumber \\
         &+&C_{11}P_{\mu}\gamma_{3}(x_{1},x_{2})\sin \theta \sin \varphi
           -C_{12}P_{\mu}\gamma_{4}(x_{1},x_{2})\sin \theta \sin \varphi 
+ H.c. \Bigr], 
\label{eq:m3ediffbr}
\end{eqnarray}

\noindent where $P_{\mu}$ is the magnitude of the polarization
vector. Functions of $\alpha_{i}$, $\beta_{i}$ and $\gamma_{i}$ are
presented in the Appendix~\ref{sc:KFDBR}. The coefficients of $C_{i}$
are expressed by $g_{i}$~$(i=1-6)$, $A_{L}$ and $A_{R}$. They are
given by

\begin{eqnarray}
&& C_{1} = \frac{|g_{1}|^{2}}{16} + |g_{3}|^{2},~  
C_{2} = \frac{|g_{2}|^{2}}{16} + |g_{4}|^{2},~ \nonumber \\
&& C_{3} = |g_{5}|^{2},~ C_{4} = |g_{6}|^{2},C_{5} = |eA_{R}|^{2},~   
C_{6}   =   |eA_{L}|^{2}, \nonumber \\
&& C_{7}   =   {\rm Re}(eA_{R}g_{4}^{*}),~ 			
C_{8}   =   {\rm Re}(eA_{L}g_{3}^{*}),~
C_{9}   =   {\rm Re}(eA_{R}g_{6}^{*}),~  			
C_{10}   =   {\rm Re}(eA_{L}g_{5}^{*}),~ \nonumber \\ 			
&& C_{11}   =   {\rm Im}(eA_{R}g_{4}^{*}+eA_{L}g_{3}^{*}),~
C_{12}   =   {\rm Im}(eA_{R}g_{6}^{*}+eA_{L}g_{5}^{*}).
\end{eqnarray}

In Eq.(\ref{eq:m3ediffbr}), there are four types of contributions
which have different angular dependences with respect to the muon
polarization. They are the contributions with an isotropic angular
distribution, which are the even parity (P) and time-reversal (T),
those proportional to either $\cos\theta$ or $\sin\theta\cos\varphi$,
which are P-odd and T-even, and finally, those proportional to
$\sin\theta\sin\varphi$, which are P-even and T-odd.

The integrated branching ratio and the T-odd asymmetry for \meee decay
are given by \cite{okad99}

\begin{eqnarray}
B(\mu^{+}\rightarrow e^{+}e^{+}e^{-}) &=& 
  \int_{1\over2}^{1}dx_1
  \int_{1-x_1}^{x_1}dx_2 
  \int^{1}_{-1} d(\cos\theta) \int^{\pi}_{0} d\phi
  \frac{dB(\mu^{+}\rightarrow e^{+}e^{+}e^{-})}{dx_1 dx_2 d(\cos\theta)d\phi}, 
\nonumber\\
&=& 2(C_{1}+C_{2}) + (C_{3}+C_{4}) + 32 \lbrace {\rm ln}({m_{\mu}^2\over
m_{e}^2}) -{11\over4} \rbrace (C_{5}+C_{6}) \nonumber \\ 
& & +16(C_{7}+C_{8}) + 8(C_{9}+C_{10}), \label{eq:meee-br2} \\
A_T(\mu^{+}\rightarrow e^{+}e^{+}e^{-}) &=& 
{1\over{P_{\mu}B(\mu^{+}\rightarrow e^{+}e^{+}e^{-})}}
\Biggl[
\int_{1\over2}^{1}dx_1 \int_{1-x_1}^{x_1}dx_2 \int^1_{-1}
d(\cos\theta) \int^{\pi}_{0} d\phi 
\frac{dB(\mu^{+}\rightarrow e^{+}e^{+}e^{-})}{dx_1
dx_2 d(\cos\theta) d\phi} \nonumber \\ &-&
\int_{1\over2}^{1}dx_{1} \int_{1-x_1}^{x_1}dx_2 \int^{1}_{-1} 
d\cos\theta \int_{\pi}^{2\pi} d\phi \frac{dB(\mu^{+}
\rightarrow e^{+}e^{+}e^{-})}{dx_1 dx_2 d(\cos\theta) d\phi} \Biggr] 
\nonumber \\
&=&{64\over35}{1\over B(\mu^{+}\rightarrow e^{+}e^{+}e^{-})}\lbrace
3C_{11}-2C_{12}\rbrace. \label{eq:meee-asymmetry}
\end{eqnarray}

\noindent The T-odd asymmetry turns out to be proportional to ${\rm
Im}(eA_{R}g_{4}^{*}+eA_{L}g_{3}^{*})~(=C_{11})$ and ${\rm
Im}(eA_{R}g_{6}^{*}+eA_{L}g_{5}^{*})~(=C_{12})$. It arises from
interference between the on-shell photon-penguin terms and the
four-fermion terms.

If only the photon-penguin diagrams contribute to \meee decay (namely
in the case of $C_5\not=0$, $C_6\not=0$, and the others =0), a
model-independent relation between the two branching ratios can be
derived, as follows:

\begin{equation}
\frac{B(\mu^{+}\rightarrow e^{+}e^{+}e^{-})}{B(\mu^{+} \rightarrow
e^{+} \gamma)} \simeq
\frac{\alpha}{3\pi}(\ln(\frac{m_{\mu}^{2}}{m_{e}^{2}})-\frac{11}{4}) 
= 0.006.
\label{ratio}
\end{equation}

\subsubsection{Event signature and backgrounds}

The event signature of the \meee decay is kinematically well
constrained, since all particles in the final state are detectable.
Muon decay at rest has been used in all past experiments. In this
case, the conservation of momentum ($|\sum_{i}
\vec{p}_{i}| = 0$) and energy ($\sum_{i} E_i = m_{\mu}$) could be
effectively used together with the timing coincidence between two
$e^{+}$s and one $e^{-}$, where $\vec{p}_{i}$ and $E_{i}$ ($i=1,3$)
are the momentum and energy of each of the three $e$'s, respectively.

One of the physics background processes is an allowed muon decay,
\muennee, which becomes a serious background when $\nu_{e}$ and
$\overline{\nu}_{\mu}$ have very small energies. Its branching ratio
is $B(\mu^{+}\rightarrow e^{+} e^{+} e^{-} \nu_{e}
\overline{\nu}_{\mu}) = (3.4 \pm 0.4) \times 10^{-5}$. The other
background is an accidental coincidence of an $e^{+}$ from normal muon
decay with an uncorrelated $e^{+}e^{-}$ pair, where a $e^{+}e^{-}$
pair could be produced either from Bhabha scattering of $e^{+}$, or
from the external conversion of the photon in $\mu^{+}\rightarrow
e^{+}\nu_{e}\overline{\nu}_{\mu} \gamma$ decay. Since the $e^{+}e^{-}$
pair from photon conversion has a small invariant mass, it could be
removed by eliminating events with a small opening angle between
$e^{+}$ and $e^{-}$. This, however, causes a loss of the signal
sensitivity, in particular for theoretical models in which
\meee decay occurs mostly through photonic diagrams.

The other background, which comes especially at the trigger level,
comprises fake events with an $e^{+}$ curling back to the target,
which mimics an $e^{+}e^{-}$ pair. For this background, an
$e^{+}e^{-}$ pair forms its relative angle of 180$^{\circ}$, and can
therefore be rejected.

\subsubsection{Experimental status of \meee decay}

\begin{table}[th!]
\caption{Historical progress and summary of searches for \meee decay.}
\label{tb:meee}
\vspace{5mm}
\begin{tabular}{llll}
Place & Year & 90\%C.L. upper limit & Reference \cr\hline
JINR   & 1976 & $<1.9 \times 10^{-9}$  & \citeasnoun{kore76} \cr
LANL   & 1984 & $<1.3 \times 10^{-10}$ & \citeasnoun{bolt84} \cr
PSI    & 1984 & $<1.6 \times 10^{-10}$ & \citeasnoun{bert84} \cr
PSI    & 1985 & $<2.4 \times 10^{-12}$ & \citeasnoun{bert85} \cr
LANL   & 1988 & $<3.5 \times 10^{-11}$ & \citeasnoun{bolt88} \cr
PSI    & 1988 & $<1.0 \times 10^{-12}$ & \citeasnoun{bell88} \cr
JINR   & 1991 & $<3.6 \times 10^{-11}$ & \citeasnoun{bara91} \cr
\end{tabular}
\end{table}

After the pioneering measurement in 1976 using a cylindrical
spectrometer, which gave an upper limit of $B(\mu^{+}\rightarrow
e^{+}e^{+}e^{-}) < 1.9\times 10^{-9}$~\cite{kore76}, various
experiments to search for \meee decay have been carried out, as shown
in Table~\ref{tb:meee}. In particular, a series of experimental
measurements with the SINDRUM magnetic spectrometer at SIN
\cite{bert84,bert85,bell88} were carried out. A surface $\mu^{+}$ beam
with $5\times 10^6 \mu^{+}$/sec was used, and the muons were stopped
in a hollow double-cone target. The $e^{+}$s and $e^{-}$s were tracked
by the SINDRUM spectrometer, which consisted of five concentric
multiwire proportional chambers (MWPC) and a cylindrical array of 64
plastic scintillation counters under a solenoid magnetic field of 0.33
T. The momentum resolution was $\Delta p/p = (12.0 \pm 0.3)$ \% (FWHM)
at $p=$50 MeV/$c$. This experiment gave a 90\% C.L. upper limit of
$B(\mu^{+}\rightarrow e^{+}e^{+}e^{-}) < 1.0\times 10^{-12}$, assuming
a constant matrix element for the \meee decay
\cite{bell88}. They also observed $9070 \pm 10$ events of \muennee decay.
A detailed analysis of the differential decay rate of \muennee decay
was studied, and was found to be consistent with the $V-A$ interaction
\cite{kers88}.

Another recent experiment to search for \meee was performed at Joint
Institute for Nuclear Research (JINR), Dubna, Russia \cite{bara91}. A
magnetic 4$\pi$ spectrometer with cylindrical proportional chambers
was used.  They obtained an upper limit of 90\% CL of $B(\mu^{+}
\rightarrow e^{+}e^{+}e^{-}) < 3.6 \times 10^{-11}$, where the matrix
element of \meee was assumed to be constant.

\subsection{$\mu^{-}-\lowercase{e}^{-}$ coherent conversion in a muonic atom}

\subsubsection{Phenomenology of \muec conversion}\label{sc:PMUEC}

Another prominent process concerning lepton flavor violation is
$\mu^{-} - e^{-}$ conversion in a muonic atom. When a negative muon is
stopped in some material, it is trapped by an atom, and forms a muonic
atom. After it cascades down in energy levels in the muonic atom, a
muon is bound in its 1{\em s} ground state. The fate of the muon is then
either decay in an orbit ($\mu^{-} \rightarrow e^{-}\nu_{\mu}
\overline{\nu}_{e}$) or capture by a nucleus of mass number $A$
and atomic number $Z$, namely

\begin{equation}
\mu^{-} + (A,Z) \rightarrow \nu_{\mu} + (A,Z-1).
\end{equation}

\noindent However, in the context of physics beyond the Standard
Model, the exotic process of neutrinoless muon capture, such as

\begin{equation}
\mu^{-} + (A,Z) \rightarrow e^{-} + (A,Z),
\end{equation}

\noindent is also expected. This process is called \muec
conversion in a muonic atom. It violates the conservation of the
lepton flavor numbers, $L_{e}$ and $L_{\mu}$, by one unit, but
conserves the total lepton number, $L$.

The branching ratio of \muec conversion can be given by

\begin{equation}
B(\mu^{-}+(A,Z)\rightarrow e^{-}+(A,Z)) \equiv {\Gamma(\mu^{-} + (A,Z)
\rightarrow e^{-} + (A,Z))
\over \Gamma(\mu^{-} + (A,Z) \rightarrow capture)},
\end{equation}

\noindent where $\Gamma$ is the corresponding decay width.

The final state of the nucleus $(A,Z)$ could be either the ground
state or excited states. In general, the transition process to the
ground state, which is called coherent capture, is dominant. The rate
of the coherent capture process over non-coherent ones is enhanced by
a factor approximately equal to the number of nucleons in the nucleus,
since all of the nucleons participate in the process.

The possible contributions to \muec conversion in a muonic atom can be
grouped into two parts, which are the photonic contribution and the
non-photonic contribution. Therefore, in principle, this process is
theoretically interesting, since it does occur by mechanisms which do
not contribute to the \mueg process.  The study of the photonic
contribution was initiated by Weinberg and Feinberg \cite{wein59}. The
non-photonic contribution was studied later, for instance by
\citeasnoun{marc77b}.

Let us first discuss the photonic transition. The effective Lagrangian
for the photonic transition is written as

\begin{eqnarray}
\lagr_{\em photo}\mit = -e J^{\mu}_{photo} A_{\mu}.
\end{eqnarray}

\noindent The matrix element of the $\mu^{-}(p_{\mu}) \rightarrow e^{-}
(p_{e})\gamma^{*}(q)$ transition, where $p_{\mu}$, $p_{e}$ and
$q=p_{\mu}-p_{e}$ are the muon, electron and virtual photon
four-momenta respectively, is given by

\begin{eqnarray}
M_{photonic} &=&-e A^{*}_{\mu}(q)<e^- (p_{e})| J^{\mu}_{photo}(0)
|\mu^- (p_{\mu}) >,
\nonumber \\
&=& -e A^*_{\mu}(q) \overline{u}_{e}(p_{e})
\Bigl[ (f_{E0}(q^2) + \gamma_{5}f_{M0}(q^2))\gamma_{\nu}
(g^{\mu\nu} - {q^{\mu}q^{\nu} \over q^2}) 
\nonumber\\
&& + (f_{M1}(q^2) + \gamma_{5}f_{E1}(q^2))
{i\sigma_{\mu\nu}q^{\nu}\over m_{\mu}}\Bigr] u_{\mu}(p_{\mu}).
\label{eq:mphoto}
\end{eqnarray}

Based on Eq.(\ref{eq:mphoto}), the branching ratio of the coherent
\muec conversion through the photonic contribution is given by
\cite{wein59}

\begin{equation}
B(\mu^{-}N\rightarrow e^{-}N) = (8\alpha^{5}m_{\mu}
Z^{4}_{eff}ZF_{p}^2 \xi^2)\cdot{1\over\Gamma_{capt}},
\label{eq:wein}
\end{equation}

\noindent where $\Gamma_{cap}$ is the total muon capture
rate. $\xi^{2}$ is given by

\begin{equation}
\xi^2 = | f_{E0}(-m_{\mu}^2) + f_{M1}(-m_{\mu}^2) |^2 
+ | f_{E1}(-m_{\mu}^2) + f_{M0}(-m_{\mu}^2) |^2.
\label{eq:form_mue}
\end{equation}

\noindent It is noted that in photonic diagrams, in contrast to
\mueg, not only $f_{E1}$ and $f_{M1}$, but also $f_{E0}$ and $f_{M0}$,
can contribute to the \muec conversion.  $Z_{eff}$ is an effective
atomic charge obtained by averaging the muon wave function over the
nuclear density \cite{chia93}. This is defined as

\begin{equation}
\frac{\alpha^3 m_{\mu}^3}{\pi} \frac{Z_{eff}^4}{Z}
=\int d^3x|\Phi_{\mu}(x)|^2 \rho_{p}(x) \equiv <\Phi_{1s}>^{2},
\label{eq:zeff}
\end{equation}

\noindent where $\Phi_{\mu}(x)$ is the non-relativistic muon wave
function for the {\em 1s} state of the muonic atom and $\rho_{p}(x)$
is the proton density in the nucleus normalized as

\begin{equation}
\int d^3x\rho_p (x) =1.
\end{equation}

\noindent $F_{p}^2$ is the nuclear matrix element squared, given by

\begin{equation}
F_{p} = \int d^{3}x e^{-i p_e x} \rho_p(x)
= 4\pi \int \rho_p(r)\frac{\sin{m_{\mu} r}}{m_{\mu} r} r^2 dr.
\end{equation}

\noindent In Eq.(\ref{eq:wein}), the \muec conversion process is
roughly proportional to $(Z_{eff})^{4}Z$, whereas the normal muon
capture, $\Gamma_{capt}$, is proportional to $(Z_{eff})^{4}$. The
enhancement by a factor of $Z$ in the \muec coherent conversion is
evident.

Let us next discuss the non-photonic contribution. The general four
fermion interaction of the \muec conversion at the quark level is
given by

\begin{eqnarray}
\lagr_{\em non-photo} &=& -\frac{G_F}{\sqrt{2}}\sum_{q=u,d,s...}
\Biggl[ (g_{LS(q)} \overline{e_L}\mu_{R}+ 
    g_{RS(q)} \overline{e_R}\mu_{L})\overline{q}q
\nonumber\\
&&+  (g_{LP(q)} \overline{e_L}\mu_{R}+ 
    g_{RP(q)} \overline{e_R}\mu_{L})\overline{q}\gamma_{5}q
\nonumber\\
&&+  (g_{LV(q)} \overline{e_L}\gamma^{\mu}\mu_{L}+ 
    g_{RV(q)} \overline{e_R}\gamma^{\mu}\mu_{R})\overline{q}\gamma_{\mu}q
\nonumber\\
&&+  (g_{LA(q)} \overline{e_L}\gamma^{\mu}\mu_{L}+ 
    g_{RA(q)} \overline{e_R}\gamma^{\mu}\mu_{R})
   \overline{q}\gamma_{\mu}\gamma_{5}q
\nonumber\\
&&+\frac{1}{2}(g_{LT(q)} \overline{e_L}\sigma^{\mu \nu}\mu_{R}+ 
    g_{RT(q)} \overline{e_R}\sigma^{\mu \nu}\mu_{L})
    \overline{q}\sigma_{\mu \nu}q + H.c. \Biggr].
\label{eq:non-ph-quark}
\end{eqnarray}

At first, this effective Lagrangian at the quark level is converted
into that at the nucleon level by using the nucleon form factors
\cite{verg86,bern93,faes99}.  Since the momentum transfer in the \muec
conversion process is smaller than the size of the nucleon structure,
the momentum dependence of the nucleon form factors can be neglected.
In such a case, the matrix element of the quark currents can be
replaced by the corresponding nucleon current by using

\begin{eqnarray}
<p|\overline{q}\Gamma_K q|p>&=&G_K^{(q,p)}\overline{p}\Gamma_K p,
\nonumber\\
<n|\overline{q}\Gamma_K q|n>&=&G_K^{(q,n)}\overline{n}\Gamma_K n,
\end{eqnarray}

\noindent with
$\Gamma_K=(1,\gamma_5,\gamma_{\mu},\gamma_{\mu}\gamma_{5},\sigma_{\mu\nu}$)
for $K= (S,P,V,A,T)$. For the vector current,
$G_V^{(u,p)}=G_V^{(d,n)}=2$ and $G_V^{(d,p)}=G_V^{(u,n)}=1$.  In
general, the isospin invariance requires the relations
$G_K^{(u,p)}=G_K^{(d,n)}$, $G_K^{(u,p)}=G_K^{(d,n)}$ and
$G_K^{(s,p)}=G_K^{(s,n)}$. The effective Lagrangian at the nucleon
level is given by

\begin{eqnarray}
\lagr_{\em non-photo}\mit &=& -\frac{G_F}{\sqrt{2}}
\Biggl[ \overline{e_L}\mu_{R}
 \overline{\Psi}\{(g_{LS}^{(0)}+g_{LS}^{(1)}\tau_3)+
 (g_{LP}^{(0)}+g_{LP}^{(1)}\tau_3)\gamma_5\}\Psi
\nonumber\\
&&+\overline{e_R}\mu_{L}
 \overline{\Psi}\{(g_{RS}^{(0)}+g_{RS}^{(1)}\tau_3)+
 (g_{RP}^{(0)}+g_{RP}^{(1)}\tau_3)\gamma_5\}\Psi
\nonumber\\
&&+\overline{e_L}\gamma^{\mu}\mu_{L}
 \overline{\Psi}\gamma_{\mu}
\{(g_{LV}^{(0)}+g_{LV}^{(1)}\tau_3)+
 (g_{LA}^{(0)}+g_{LA}^{(1)}\tau_3)\gamma_5\}\Psi
\nonumber\\
&&+\overline{e_R}\gamma^{\mu}\mu_{R}
 \overline{\Psi}\gamma_{\mu}
\{(g_{RV}^{(0)}+g_{RV}^{(1)}\tau_3)+
 (g_{RA}^{(0)}+g_{RA}^{(1)}\tau_3)\gamma_5\}\Psi
\nonumber\\
&&+\frac{1}{2}\overline{e_L}\sigma^{\mu \nu}\mu_{R}
 \overline{\Psi}\sigma_{\mu \nu}
\{(g_{LT}^{(0)}+g_{LT}^{(1)}\tau_3)\}\Psi
+\frac{1}{2}\overline{e_R}\sigma^{\mu \nu}\mu_{L}
 \overline{\Psi}\sigma_{\mu \nu}
\{(g_{RT}^{(0)}+g_{RT}^{(1)}\tau_3)\}\Psi + H.c. \Biggr],
\end{eqnarray}

\noindent where $\Psi=(p,n)^T$ and the isoscalar and isovector coupling
constants $g_{XK}^{(0)}$ and $g_{XK}^{(1)}$ $( X=L, R,$ and $ K= S, P,
V,A, T)$ are respectively given by

\begin{eqnarray}
g_{XK}^{(0)}&=&\frac{1}{2}\sum_{q=u,d,s}(g_{XK(q)}G_K^{(q,p)}
+g_{XK(q)}G_K^{(q,n)}), \\
g_{XK}^{(1)}&=&\frac{1}{2}\sum_{q=u,d,s}(g_{XK(q)}G_K^{(q,p)}
-g_{XK(q)}G_K^{(q,n)}).
\end{eqnarray}

For coherent \muec conversion, only the scalar and vector coupling
constants can be kept. By using a non-relativistic approximation for
the muon wave function, the transition rate is given by

\begin{equation}
B(\mu^{-}N\rightarrow e^{-}N) = 
\frac{p_e E_e G_F^2}{8\pi}(|X_L(p_e)|^2+|X_R(p_e)|^2) 
{1\over\Gamma_{capt}},
\label{eq:nonphr}
\end{equation}

\noindent where 

\begin{eqnarray}
X_L(p_e)&=&(g_{LS}^{(0)}+g_{LS}^{(1)}+g_{LV}^{(0)}+g_{LV}^{(1)})
Z M_{p}(p_e)
+(g_{LS}^{(0)}-g_{LS}^{(1)}+g_{LV}^{(0)}-g_{LV}^{(1)})
N M_{n}(p_e), \label{eq:xl} \\ 
X_R(p_e)&=&(g_{RS}^{(0)}+g_{RS}^{(1)}+g_{RV}^{(0)}+g_{RV}^{(1)})
Z M_{p}(p_e)+(g_{RS}^{(0)}-g_{RS}^{(1)}+g_{RV}^{(0)}-g_{RV}^{(1)})
N M_{n}(p_e), \label{eq:xr}
\end{eqnarray} 

\noindent and $N\equiv A-Z$ is the number of neutrons in the nuclei.
$M_{p}(p)$ and$M_{n}(p)$ are given by

\begin{equation}
M_{p}(p) = \int d^{3}x e^{-i p x} \rho_p(x)\Phi_{\mu}(x),~~ M_{n}(p) =
\int d^{3}x e^{-i p x} \rho_n(x)\Phi_{\mu}(x),
\label{eq:mpmn}
\end{equation}

\noindent with the proton and neutron densities, $\rho_p(x)$ and
$\rho_n(x)$, normalized to unity. If it is assumed that the proton and
neutron densities are equal and that the muon wave function does not
change very much in the nucleus, by using Eq.(\ref{eq:zeff}),
Eq.(\ref{eq:nonphr}) can be deduced into

\begin{eqnarray}
B(\mu^{-}N\rightarrow e^{-}N)& =&
\frac{p_e E_e m_{\mu}^3 G_F^2 \alpha^3 Z_{eff}^4 F_{p}^2}
{8\pi^2 Z}\Biggl\lbrace |~(Z+N)(g_{LS}^{(0)}+g_{LV}^{(0)})+
(Z-N)(g_{LS}^{(1)}+g_{LV}^{(1)})~|^2 \nonumber\\
&&+|~(Z+N)(g_{RS}^{(0)}+g_{RV}^{(0)})+
(Z-N)(g_{RS}^{(1)}+g_{RV}^{(1)})~|^2\Biggr\rbrace\cdot{1\over\Gamma_{capt}}.
\label{eq:nonphr2}
\end{eqnarray}

\noindent This equation corresponds to Eq.(\ref{eq:wein}) for the
photonic case, which was derived in a similar approximation. In
Eq.(\ref{eq:nonphr2}), the coherent process is enhanced by a factor of
the number of nucleons, as can be seen in Eq.(\ref{eq:wein}).

In general, both photonic and non-photonic contributions might
exist. If the non-relativistic approximation for the muon wave
function is used and the momentum transfer of $q^2$ is replaced by
$-m_{\mu}^2$, the photonic contribution can be regarded as additional
terms to the vector-coupling constants. They are given by

\begin{eqnarray}
\Delta g^{(0)}_{LV}&=&\Delta g^{(1)}_{LV}=
\frac{2\sqrt{2}\alpha\pi}{G_F m_{\mu}^2}
(f_{E0}(-m_{\mu}^2) + f_{M1}(-m_{\mu}^2)+
f_{M0}(-m_{\mu}^2) + f_{E1}(-m_{\mu}^2)),\\
\Delta g^{(0)}_{RV}&=&\Delta g^{(1)}_{RV}=
\frac{2\sqrt{2}\alpha\pi}{G_F m_{\mu}^2}
(f_{E0}(-m_{\mu}^2) + f_{M1}(-m_{\mu}^2)-
f_{M0}(-m_{\mu}^2) - f_{E1}(-m_{\mu}^2)).
\end{eqnarray}

\noindent These contributions are added to the corresponding vector
coupling constants in Eq.(\ref{eq:nonphr2}). In such a case, the
interference terms should be taken into account as well.

So far, the non-relativistic approximation for the muon wave function
and a plane wave for the emitted electron have been used to derive the
\muec conversion rates. Possible corrections for this approximation
turn out to be important for heavy nuclei.  A relativistic treatment
based on the Dirac equation was considered and the corrections to the
Weinberg-Feinberg formulas were calculated \cite{shan79}.  Recently,
the photonic transitions due to $f_{M1}$ and $f_{E1}$ were further
examined by properly treating the electric potential in the muonic
atom \cite{czar97}.

In the case that the photonic contributions of $f_{E1}(q^2)$ and
$f_{M1}(q^2)$ dominate over the other contributions, the rate of \muec
conversion can be parameterized by \cite{czar97}

\begin{equation}
B(\mu^{-}N\rightarrow e^{-}N) =3\cdot 10^{12}
 (|f_{E1}|^2 + |f_{M1}|^2) B(A,Z),
\label{eq:baz}
\end{equation}

\noindent where $B(A,Z)$ represents the rate dependence on the mass number
($A$) and the atomic number ($Z$) of the target nucleus. This
particular case becomes important, for instance, in $SO(10)$ SUSY GUT. 
The values of $B(A,Z)$, based on different approximations, are
tabulated in Table~\ref{tb:zmuec}, where $B_{WF}(A,Z)$ is from the
Weinberg-Feinberg approximation \cite{wein59}, $B_{S}(A,Z)$ is from
\citeasnoun{shan79}, and $B_{CMK}(A,Z)$ is from
\citeasnoun{czar97}. From Eq.(\ref{eq:baz}), the ratio
of $B(\mu^{+}\rightarrow e^{+}\gamma)/B(\mu^{-}N\rightarrow e^{-}N)$
is given by

\begin{equation}
\frac{B(\mu^{+}\rightarrow e^{+}\gamma)}
{B(\mu^{-}N\rightarrow e^{-}N)} 
=\frac{96\pi^{3} \alpha}{G_F^2 m_{\mu}^4}\cdot
{1\over{3\cdot 10^{12}B(A,Z)}}\sim \frac{428}{B(A,Z)}.
\end{equation}

\noindent By using the values in Table \ref{tb:zmuec}, the ratio 
$B(\mu^{+}\rightarrow e^{+}\gamma)/B(\mu^{-}N\rightarrow e^{-}N)$ for
different target nuclei can be calculated. It varies from 389 for
$^{27}$Al, 238 for $^{48}$Ti, and 342 for $^{208}$Pb. This result
indicates that the rate of \muec conversion has a maximum around the
medium nuclei ($A \approx 60$), and flattens out or slightly decreases
for heavy nuclei. However, the calculations, which took into account
the nuclear effect, show a different $Z$ dependence
\cite{chia93,kosm96,kosm98}.

\begin{table}[htb!]
\caption{$Z$ dependence of the photonic contribution in the 
\muec conversion estimated by various theoretical models
(after Czarnecki,~{\em et al.}, (1997)).}
\label{tb:zmuec}
\vspace{5mm}
\begin{tabular}{lllll}
Models & Al & Ti & Pb & Reference \cr\hline
$B_{WF}(A,Z)$ & 1.2 & 2.0 & 1.6 & \citeasnoun{wein59} \cr
$B_{S}(A,Z)$ & 1.3 & 2.2 & 2.2 & \citeasnoun{shan79} \cr
$B_{CMK}(A,Z)$ & 1.1 & 1.8 & 1.25 & \citeasnoun{czar97} \cr
\end{tabular}
\end{table}

The \muec conversion rates to the ground state and all excited states
have been calculated by either the shell-model closure approximation
\cite{kosm90} or the quasi-particle random-phase-approximation (QRPA)
\cite{kosm94}. The fraction of the coherent transition to the ground
state is dominant. It was calculated specifically for $^{48}Ti$ to be
(95-99)\% in the QRPA approximation, which is even larger than in the
shell-model closure approximation. It is also experimentally
advantageous, since the background from excited states induced by the
reaction can be minimized. It was also found that among the
transitions to excited states, the dipole $1^{-}$ state is large both
in the photonic and non-photonic contributions
\cite{kosm94}.
   
\subsubsection{Event signature and backgrounds}

The event signature of the coherent \muec conversion in a muonic atom
is a mono-energetic single electron emitted from muon capture with
an energy of

\begin{eqnarray}
E_{\mu e} &=& m_{\mu} - B_{\mu} - E_{rec}^{0} \nonumber \\
&\approx& m_{\mu} - B_{\mu},
\label{eq:emue}
\end{eqnarray}

\noindent where $m_{\mu}$ is the muon mass, and $B_{\mu}$ and
$E_{rec}^{0}$ are the binding energy of the 1$s$ muonic atom and the
nuclear-recoil energy, respectively. The nuclear-recoil energy is
approximately $E_{rec}^{0}\approx (m_{\mu}-B_{\mu})^2/(2M_{A})$, where
$M_{A}$ is the mass of the recoiling nucleus, which is small. Since
$B_{\mu}$ is different for various nuclei, the peak energy of the
\muec conversion signal changes. For instance, it varies from $E_{\mu
e}$ = 104.3 MeV for titanium to $E_{\mu e}$ = 94.9 MeV for lead.

From an experimental point of view, \muec conversion is very
attractive. First, the $e^{-}$ energy of about 105 MeV is far above
the end-point energy of the muon decay spectrum ($\sim$ 52.8
MeV). Second, since the event signature is a mono-energetic electron,
no coincidence measurement is required. The search for this process
has a potential to improve the sensitivity by using a high muon rate
without suffering from accidental background, which would be serious
background for other processes, such as \mueg and \meee decays.

One of the major backgrounds is muon decay in orbit from a muonic atom
(also called a bound muon decay), in which the $e^{-}$ endpoint energy
is the same as the energy of the signal. It is discussed in more
detail below. The other background sources are (i) radiative pion
capture ($\pi^{-} + (A,Z) \rightarrow (A,Z-1) + \gamma$) or radiative
muon captures ($\mu^{-} + (A,Z) \rightarrow \nu_{\mu}+(A,Z-1)
+\gamma$) followed by internal and external asymmetric $e^{+}e^{-}$
conversion of the photon ($\gamma\rightarrow e^{+}e^{-}$) with $e^{+}$
undetected, (ii) electrons in the beam scattering off the target,
(iii) muon decay in flight, and (iv) cosmic rays. Note that the
maximum $e^{-}$ energy ($E^{max}_{bg}$) from the background of
radiative muon capture is given by

\begin{eqnarray}
E^{max}_{bg} &=& m_{\mu} - B_{\mu} - E_{rec} - \Delta_{Z-1}
\nonumber \\
&\approx& m_{\mu} - B_{\mu} - \Delta_{Z-1},
\label{eq:rmc}
\end{eqnarray}

\noindent where $\Delta_{Z-1}$ is the difference in the nuclear binding
energy of the initial and final nucleus involved in radiative muon
capture. Therefore, an appropriate target with a large $\Delta_{Z-1}$
can be selected so as to keep a wide background-free region for the
coherent signal. The typical values of $E^{max}_{bg}$ are 89.7 MeV and
91.4 MeV for $^{48}$Ti and $^{46}$Ti, respectively, whereas $E_{\mu
e}$ is 104.3 MeV.  In general, to eliminate these backgrounds, the
purity of the beam (with no contamination of pions and electrons) is
crucial, together with a highly efficient veto for cosmic rays.

When the muon is polarized, the angular distribution of $e^{-}$ in the
coherent \muec conversion process is given by

\begin{equation}
{dB(\mu^{-}N\rightarrow e^{-}N)\over d(\cos\theta_{e})} = 
{p_{e}E_{e}G_{F}^2 \over 16\pi} \Biggl[ 
|X_{L}(p_{e})|^2 (1-P_{\mu}\cos\theta_{e}) + 
|X_{R}(p_{e})|^2 (1+P_{\mu}\cos\theta_{e}) 
\Biggr] \cdot {1\over\Gamma_{capt}},
\end{equation}

\noindent where $\theta_e$ is the angle between the $e^{-}$ direction
and the muon spin direction. Since the nucleus does not change for the
coherent process, the conversion electron carries the original muon
spin. $X_L$ and $X_R$ are given in Eqs.(\ref{eq:xl}) and
(\ref{eq:xr}), and correspond to the emission of left-handed electrons
($e^{-}_{L}$) and right-handed electrons ($e^{-}_{R}$),
respectively. As in polarized \mueg decay, in principle, the angular
distribution would be useful to discriminate between theoretical
models.  However, even if negative muons in the beam are 100\% spin
polarized, they are depolarized during their atomic cascades down to
the {\em 1s} ground state. For a nucleus with zero nuclear spin, the
residual polarization is about 16\% \cite{evse75}. For a nucleus with
non-zero nuclear spin, it becomes much smaller. It would make a
measurement of the angular distribution difficult unless high
statistics is accumulated. If the $\mu^{-}$ polarization is restored,
however, it might provide useful information. One possible way to
repolarize a negative muon in a muonic atom is to use a polarized
nuclear target \cite{naga74,kuno86}.

\subsubsection{Muon decay in orbit}

Muon decay in orbit \cite{port51} is one of the important background
sources in the search for \muec conversion in a muonic atom, since the
end point of the electron spectrum comes close to the signal region of
\muec conversion. Only the high-energy end of the electron energy
spectrum is of interest for \muec conversion experiments. At the
high-energy end, the effect of the nuclear-recoil energy plays an
important role (on its phase space). There have been several studies
on its electron energy spectrum with nuclear-recoil energy taken into
account \cite{hang74,herz80,shan82}.  With the approximation of a
constant nuclear-recoil energy, the electron spectrum with an
expansion in powers of the electron energy ($E_{e}$) at the end-point
energy is given by \cite{shan82}

\begin{equation}
N(E_{e})dE_{e} = \Bigl( {E_{e} \over m_{\mu}} \Bigr)^2 
\Bigl( {\delta_{1} \over m_{\mu}} \Bigr)^5 
\Bigl[ D + E\cdot( {\delta_{1} \over m_{\mu}} ) 
+ F\cdot( {\delta \over m_{\mu}} ) \Bigr] dE_{e}, 
\label{eq:bdmu}
\end{equation}

\noindent where $\delta = E_{\mu e} - E_{e}$ and $\delta_{1} = (m_{\mu} 
- B_{\mu}) - E_{rec} - E_{e}$. $E_{\mu e}$ is the $e^{-}$ energy of
the \muec conversion signal defined in Eq.(\ref{eq:emue}). $E_{rec}$
is the nuclear-recoil energy given by $E_{rec}\approx
E^{2}_{e}/(2M_{A})$. It should be stressed that the spectrum falls off
sharply as the fifth power of $\delta_{1}$ towards its end point. The
coefficients $D$, $E$ and $F$ as well as the end-point energy are
given in a numerical table~\cite{shan82}. The contributions of the $E$
and $F$ terms to the total rate are about 4\% and 8\% respectively for
$Z=29$ and $E_{e}$ = 100 MeV. Eq.(\ref{eq:bdmu}) agrees with those in
\citeasnoun{hang74} and \citeasnoun{herz80}. In the evaluation of the
leading term $D$, important are (1) the use of a correct electron
wave function incorporating the finite nuclear charge distribution,
(2) the use of the Dirac muon wave function, and (3) the use of the
small component of the muon relativistic wave function. In particular,
the effect of (1) is large \cite{shan97}.

\begin{figure}[ht!]
\centerline{\epsfig{file=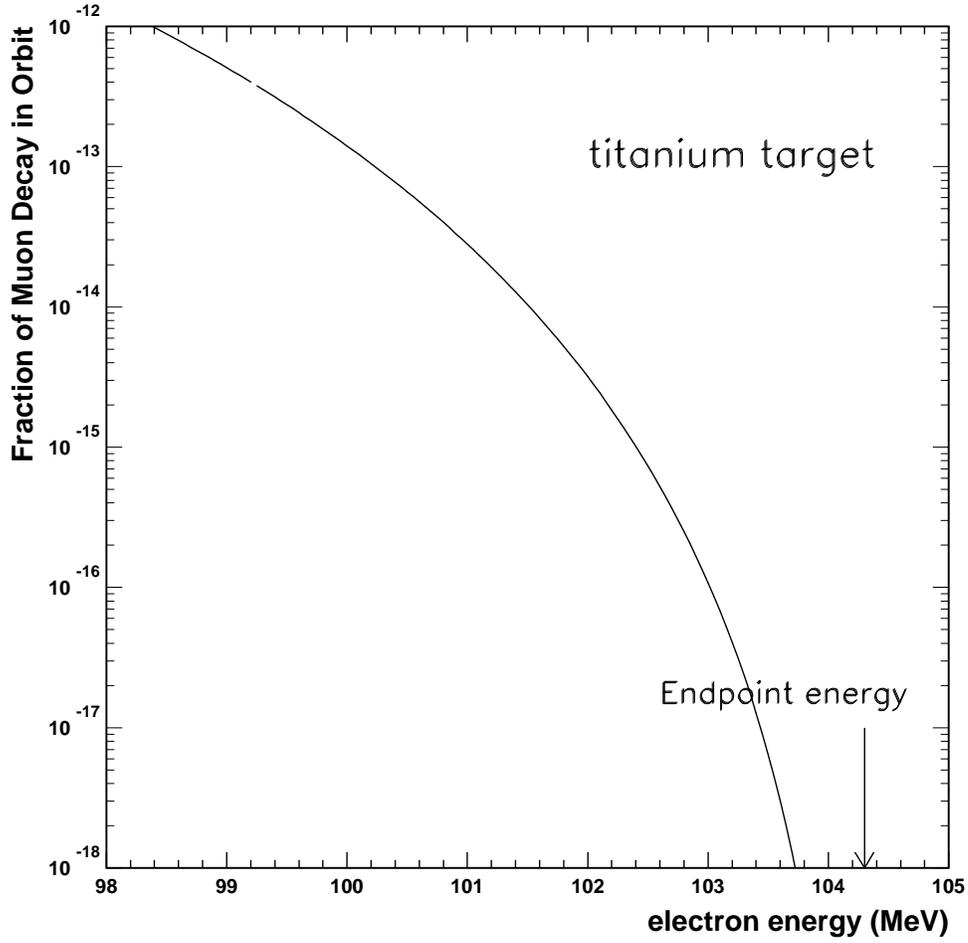,width=14cm}}
\vspace{5mm}
\caption{Fraction of muon decay in orbit normalized to the total nuclear
muon capture rate as a function of the $e^{-}$ energy for a titanium
target. It represents an effective branching ratio of muon decay in
orbit as a background to the \muec conversion. It was calculated by
Shanker's formula in Eq.(147). The energy of the \muec conversion
signal in a titanium target is $E_{\mu e}$=104.3 MeV.}
\label{fg:bmd}
\end{figure}

Experimentally, to avoid any background from muon decay in orbit, the
momentum resolution of $e^{-}$ detection must be
improved. Fig.\ref{fg:bmd} shows the effective branching ratio of the
muon decay in orbit as a function of $E_{e}$ for the case of a
titanium target, where $E_{\mu e}$ = 104.3 MeV. It was calculated
using Eq.(\ref{eq:bdmu}). For a resolution better than 2\%, the
contribution from muon decay in orbit occurs at a level below
$10^{-14}$.

What is the asymmetric angular distribution of electrons in muon decay
in orbit, if muons are polarized ? Numerical calculations can be made
by taking into account the angular distribution of electrons from
polarized muon decay in orbit \cite{wata87}.  It is given by

\begin{equation}
N(E_{e},\theta_{e})dE_{e}({d\Omega_{e}\over 4\pi}) =
N_{0}(E_{e})\Bigl( 1+\alpha(E_{e})P_{\mu}\cos\theta_{e} \Bigr)
({d\Omega_{e}\over 4\pi}),
\end{equation}

\noindent where $\alpha(E_{e})$ is the asymmetry parameter, which
becomes $\alpha(E_{e}) = -1$ at the end-point, giving a distinct ($1 -
P_{\mu}\cos\theta_{e}$) distribution; namely, electrons are likely to
be emitted opposite to the muon polarization direction. At low energy,
$\alpha(E_{e})$ becomes positive and electrons are emitted along the
muon polarization.  The Coulomb effect is significant for heavy
nuclei, like $^{208}$Pb, but very small for light nuclei, like
$^{16}$O.  The calculated results of the decay rate, emitted electron
energy spectrum, and asymmetry parameters for muon decay in orbit are
given in numerical tables for some nuclei \cite{wata93}.

\subsubsection{Experimental status of \muec conversion}\label{sc:MEES}

\begin{figure}[ht!]
\centerline{\epsfig{file=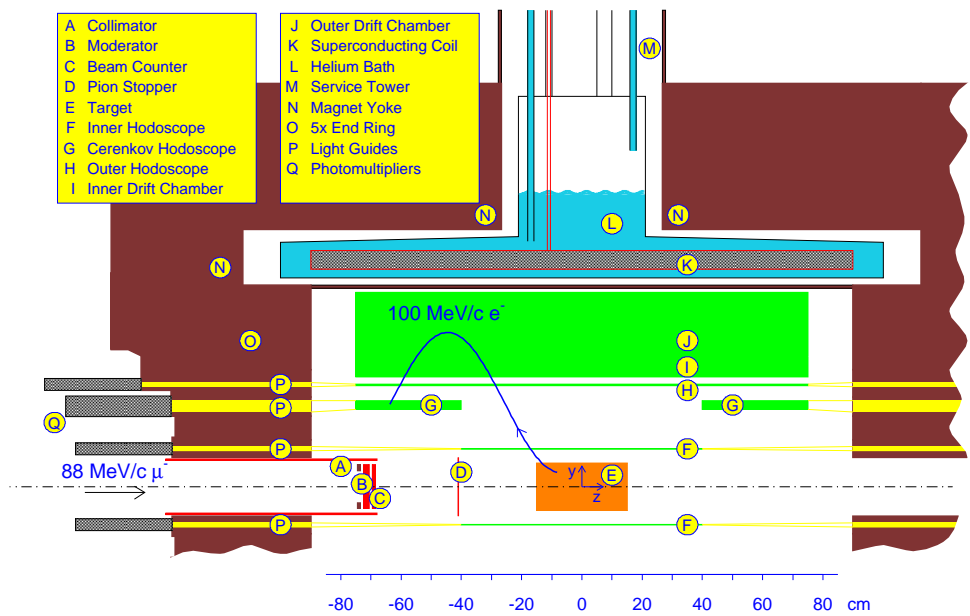,width=14cm}}
\vspace{5mm}
\caption{Schematic layout of the SINDRUM-II detector (provided by 
P.~Wintz).}
\label{fg:sindrum}
\end{figure}

The SINDRUM II collaboration at PSI is carrying out experiments to
search for \muec conversion in various nuclei. A schematic view of the
SINDRUM II spectrometer is shown in Fig.\ref{fg:sindrum}. It consists
of a set of concentric cylindrical drift chambers inside a
superconducting solenoid magnet of 1.2~T. Negative muons with a
momentum of about 90 MeV/$c$ were stopped in a target located at the
center of the apparatus, after passing a CH$_2$ moderator and a beam
counter made of plastic scintillator. Charged particles with
transverse momentum (with respect to the magnetic field direction)
above 100 MeV/$c$, originating from the target, hit two layers of
plastic scintillation arrays and then two layers of drift chambers,
and eventually hit plexiglass Cherenkov hodoscopes placed at both
ends. Charged particles having transverse momentum below about 100
MeV/$c$ were contained inside, and could not reach the tracking region
under a magnetic field of 1.2 T. A momentum resolution of about 2.8\%
(FWHM) for the energy region of conversion electrons was achieved. For
the background rejection, the $e^{-}$ energy ($E_{e}$), a time delay
between the times of charged particle tracks in the spectrometer and
the beam-counter signal ($\Delta t$), the position of the origin of
the reconstructed trajectory ($\Delta z$), the polar track angle, are
used in an off-line analysis. Events with small $\Delta t$ were
removed so as to reject prompt backgrounds, such as electron
scattering and radiative pion capture.

\begin{figure}[ht!]
\centerline{\epsfig{file=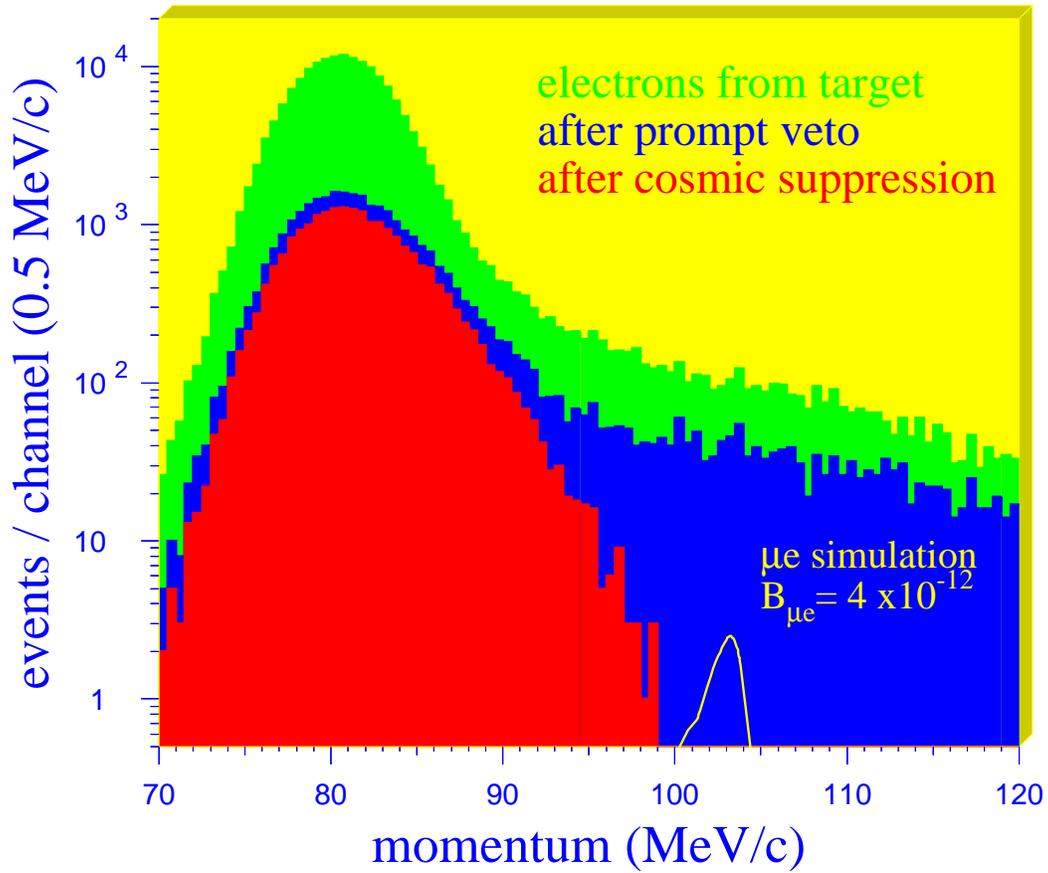,width=14cm}}
\vspace{5mm}
\caption{Electron momentum distributions for the $\mu^{-} + Ti \rightarrow
e^{-} + Ti$ reaction, measured by the SINDRUM-II detector, after the
consecutive analysis cuts. The expected signal at $B_{\mu
e}=4\times10^{-12}$ is shown (provided by P.~Wintz).}
\label{fg:sindrum-ti}
\end{figure}

In a 1993 run with a titanium target, a total of $3 \times 10^{13}$
stopped $\mu^{-}$s were accumulated at a rate of $1.2 \times
10^{7}~\mu^{-}$/sec from the $\mu E1$ beam line at PSI. The overall
efficiency was about 13 \%. The $e^{-}$ momentum spectrum for the Ti
target in the 1993 data is shown in Fig.\ref{fg:sindrum-ti}, where the
successive background rejections by prompt veto ({\em i.e.} $\Delta t$
cut) and cosmic-ray suppression are shown. Since no events were found
in the signal region, a 90\% C.L. upper limit of $6.1\times 10^{-13}$
was obtained \cite{wint98}. Also, for a lead target, it gave
$B(\mu^{-}Pb\rightarrow e^{-}Pb)<4.6\times 10^{-11}$ \cite{hone96}.

\begin{table}[bth!]
\caption{History and summary of \muec conversion in various nuclei.}
\vspace{5mm}
\begin{tabular}{lllll}
Process & 90\% C.L. upper limit & place & year & reference \cr\hline
$\mu^{-}+Cu \rightarrow e^{-}+Cu$ & $<1.6 \times 10^{-8}$ & 
SREL & 1972 & \citeasnoun{brym72} \cr
$\mu^{-}+^{32}$$S \rightarrow e^{-}+^{32}$$S$ & $<7 \times 10^{-11}$ & 
SIN & 1982 & \citeasnoun{bade82} \cr
$\mu^{-}+Ti \rightarrow e^{-}+Ti$ & $<1.6 \times 10^{-11}$ & 
TRIUMF & 1985 & \citeasnoun{brym85} \cr
$\mu^{-}+Ti \rightarrow e^{-}+Ti$ & $<4.6 \times 10^{-12}$ & 
TRIUMF & 1988 & \citeasnoun{ahma88} \cr
$\mu^{-}+Pb \rightarrow e^{-}+Pb$ & $<4.9 \times 10^{-10}$ & 
TRIUMF & 1988 & \citeasnoun{ahma88} \cr
$\mu^{-}+Ti \rightarrow e^{-}+Ti$ & $<4.3 \times 10^{-12}$ & 
PSI & 1993 & \citeasnoun{dohm93} \cr
$\mu^{-}+Pb \rightarrow e^{-}+Pb$ & $<4.6 \times 10^{-11}$ & 
PSI & 1996 & \citeasnoun{hone96} \cr
$\mu^{-}+Ti \rightarrow e^{-}+Ti$ & $<6.1 \times 10^{-13}$ & 
PSI & 1998 & \citeasnoun{wint98} \cr 
\end{tabular}
\end{table}

A next round of the SINDRUM II experiment is under preparation at the
$\pi E5$ beam line at PSI. The key element of the next stage is a
pion-muon converter (PMC) to eliminate contamination of pions and
electrons in the muon beam. It is needed since a veto of secondary
pions and electrons by a beam counter will no longer be working with a
high rate, such as $10^{8}~\mu^{-}$s/sec, at the $\pi E5$ beam
line. The PMC consists of a long-straight superconducting solenoid
magnet with length of 8.5 m and an inner diameter of 0.4 m. It is
located between the pion target and the SINDRUM II spectrometer, and
produces the same magnetic field as in the SINDRUM II spectrometer, 2
T. Low-energy negative muons (called cloud muons) from the production
target are injected into the PMC. After the 8.5-m flight length, most
of the pions in a beam would decay out, resulting in a negligible pion
contamination. Unfortunately, the original PMC magnet did not fulfill
the specification at the initial installation stage, and thus caused a
severe delay. After its new assembly, the magnet finally met the goal. 
With a lower magnetic field of the PMC magnet, data with a gold target
were taken in 1998. A new run for a Ti target is expected to start in
spring, 1999, with aimed sensitivity of $B(\mu^{-}Ti\rightarrow
e^{-}Ti)<$ a few~$\times~10^{-14}$.

\begin{figure}[ht!]
\centerline{\epsfig{file=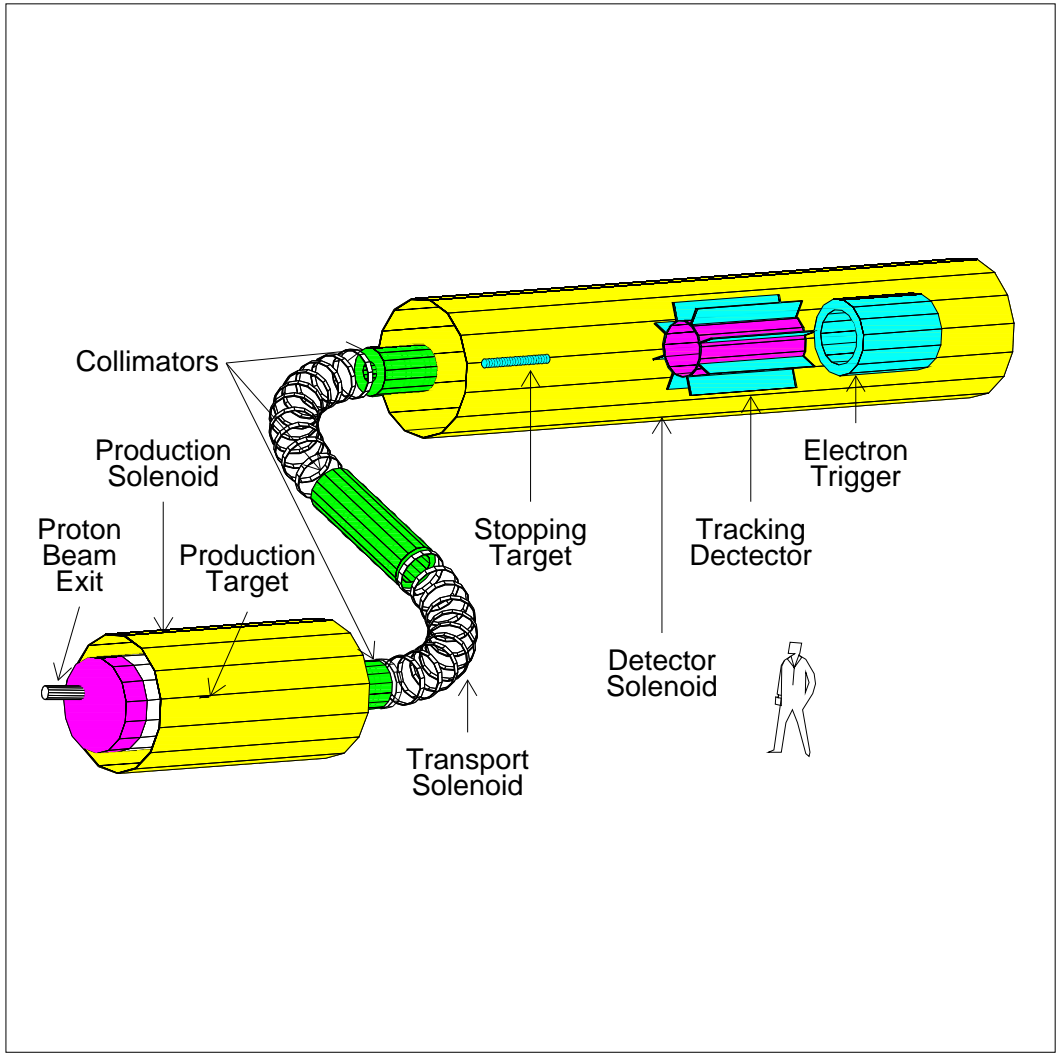,width=14cm}}
\caption{Schematic layout of the MECO detector (provided by W.R.~Molzon).}
\label{fg:meco}
\end{figure}

A new experiment, E940, at Brookhaven National Laboratory (BNL) AGS,
the MECO (Muon Electron COnversion) experiment, was prepared
\cite{bach97}.  MECO aims to search for $\mu^{-}+Al\rightarrow
e^{-}+Al$ at a sensitivity below $10^{-16}$. It will use a new
high-intensity pulsed muon beam, which could yield about
$10^{11}~\mu^{-}$s/sec stopped in a target.  A schematic layout of the
MECO detector is shown in Fig.\ref{fg:meco}. The MECO apparatus
consists of a superconducting (SC) solenoid magnet to capture pions
from the production target (production solenoid), a curved transport
SC solenoid magnet system (transport solenoid), and a SC solenoid
spectrometer, which observes only the 105-MeV signal electrons
(detector solenoid). Based on the solenoid capture scheme originally
proposed by MELC~\cite{dzhi89}, it has an axially graded magnetic
field (from 3.5 T to 2.0 T) to efficiently capture pions from a
tungsten target located on the axis of the solenoid magnet. The curved
transport solenoid will capture muons from pion decays, and select the
momentum and sign of charged particles by using collimators at three
positions. Layers of thin aluminum targets where $\mu^{-}$s are
stopped are placed in the detector solenoid with an axially graded
magnetic field. The conversion electron of 105 MeV is momentum
analyzed with a resolution of 300 keV (RMS) and an acceptance of 25\%
in a straw tracking chamber.  A pulsed proton beam of about 1 MHz
repetition with a pulse length of 30 nsec can be extracted at the AGS. 
A high extinction between the beam pulses (the ratio of a number of
protons between pulses to that in the beam pulse) of $10^{-9}$ is
needed to eliminate severe beam background at a high rate. They expect
to observe 6 signal events for $B(\mu^{-}Al\rightarrow e^{-}Al)\approx
10^{-16}$ during a one-year run, with an expected background of 0.4
events.

\subsection{$\mu^{-}-\lowercase{e}^{+}$ conversion in a muonic atom}

\subsubsection{Phenomenology of \mupc conversion}

The other neutrinoless muon-capture process is a charge-changing
reaction, such as

\begin{equation}
\mu^{-} + (A,Z) \rightarrow e^{+} + (A,Z-2)^{*},
\end{equation}

\noindent which violates the conservation of the total lepton number
as well as the lepton flavor numbers, $L_{e}$ and $L_{\mu}$. This
process is closely related to neutrinoless double $\beta-$decay
($\beta\beta_{0\nu}$), since both processes require a mechanism
involving two nucleons. The final state of the nucleus $(A,Z-2)^{*}$
could be either the ground state ($gs$) or excited states
($ex$). Since the final nucleus is not the same as the initial
nucleus, no coherent enhancement, even for the transition to the
ground state, is expected.  The branching ratio of the \mupc
conversion is defined by

\begin{equation}
B(\mu^{-}+(A,Z) \rightarrow e^{+}+(A,Z-2)^{*}) \equiv
{\Gamma(\mu^{-}+(A,Z) \rightarrow e^{+}+(A,Z-2)^{*}) \over
\Gamma(\mu^{-}+(A,Z) \rightarrow capture)}.
\end{equation}

Various theoretical models predict the rates accessible
experimentally. One is the minimum supersymmetric models (MSSM) with
R-parity breaking, which allows the predicted branching ratio of the
\mupc conversion on the level of $10^{-12}$, since the $\lambda$ and
$\lambda'$ parameters involved are not constrained \cite{babu95}. The
left-right symmetric models with a low-mass $W_{R}$ also predict the
\mupc conversion branching ratio of $10^{-14}$, estimated by the same
authors.

\subsubsection{Event signature and backgrounds}

The energy of the positron from the \mupc conversion is given by

\begin{eqnarray}
E_{\mu e^{+}} &=& m_{\mu}-B_{\mu}-E_{rec}-\Delta_{Z-2}\cr
&\approx& m_{\mu} - B_{\mu} - \Delta_{Z-2},
\end{eqnarray}

\noindent where $\Delta_{Z-2}$ is the difference in the nuclear binding
energy between the $(A,Z)$ and $(A,Z-2)$ nuclei, with the excitation
energy in the final nucleus taken into account.  In \mupc conversion,
because of the absence of coherent enhancement, the final nucleus
could be either in the ground state or in excited states. Usually, it
is assumed that a large fraction of the final nucleus could be in the
giant dipole resonance state, which has a mean energy of 20 MeV and a
width of 20 MeV. Therefore, the $e^{+}$ from
\mupc conversion will have a broad momentum distribution corresponding
to the width of giant dipole resonance excitation.

The major background is radiative muon capture (RMC) or radiative pion
capture (RPC), followed by asymmetric $e^{+}e^{-}$ conversion of the
photon. For some nuclei, the endpoint of the RMC background in
Eq.(\ref{eq:rmc}) can be selected to be much lower than the
signal. For instance, for a titanium target, the maximum endpoint of
RMC (of about 90 MeV) is about 10 MeV lower than the signal energy of
about $E_{\mu e^{+}} \approx$ 100 MeV. The background from RPC must be
reduced from the rejection of pions in the beam.

\subsubsection{Experimental status of \mupc conversion}

\begin{table}[th!]
\caption{Historical progress and summary of \mupc conversion in various nuclei.
$gs$ and $ex$ denote the transitions to the ground state and excited
states (mostly giant dipole-resonance states) respectively.}
\label{tb:mupc}
\vspace{5mm}
\begin{tabular}{lllll}
Process & 90\% C.L. upper limit & place & year & reference \cr\hline
$\mu^{-}+Cu\rightarrow e^{+}+Co$ & $2.6 \times 10^{-8}$ &
SREL & 1972 & \citeasnoun{brym72} \cr
$\mu^{-}+S \rightarrow e^{+}+Si$ & $9 \times 10^{-10}$ &
SIN & 1982 & \citeasnoun{bade82}\cr 
$\mu^{-}+Ti\rightarrow e^{+}+Ca(gs)$ & $9 \times 10^{-12}$ &
TRIUMF & 1988 & \citeasnoun{ahma88} \cr
$\mu^{-}+Ti\rightarrow e^{+}+Ca(ex)$ & $1.7 \times 10^{-10}$ &
TRIUMF & 1988 & \citeasnoun{ahma88} \cr
$\mu^{-}+Ti\rightarrow e^{+}+Ca(gs)$ & $4.3 \times 10^{-12}$ &
PSI & 1993 & \citeasnoun{dohm93} \cr
$\mu^{-}+Ti\rightarrow e^{+}+Ca(ex)$ & $8.9 \times 10^{-11}$ &
PSI & 1993 & \citeasnoun{dohm93} \cr
$\mu^{-}+Ti\rightarrow e^{+}+Ca(gs)$ & $1.7 \times 10^{-12}$ &
PSI & 1998 & \citeasnoun{kaul98} \cr
$\mu^{-}+Ti\rightarrow e^{+}+Ca(ex)$ & $3.6 \times 10^{-11}$ & 
PSI & 1998 & \citeasnoun{kaul98} \cr
\end{tabular}
\end{table}

\begin{figure}[ht!]
\centerline{\epsfig{file=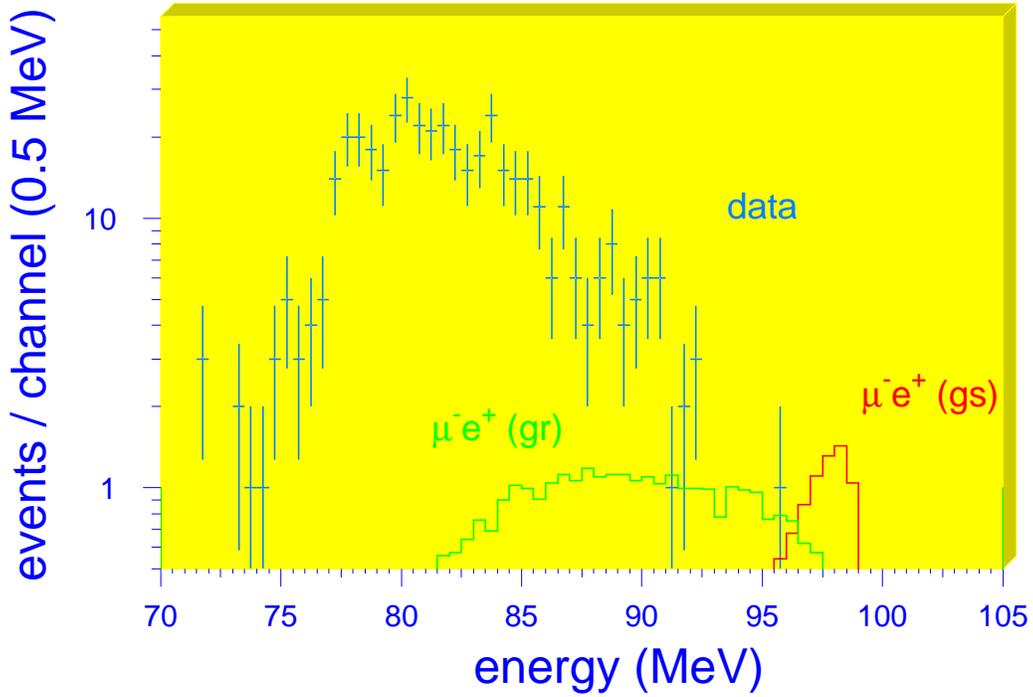,width=14cm}}
\vspace{5mm}
\caption{Positron momentum spectra of the $\mu^{-} + Ti \rightarrow e^{+}
+ Ca$ reaction. $\mu^{-}e^{+}$(gs) and $\mu^{-}e^{+}$(gr) are the
expected signals for the transitions to the ground state and to the
giant dipole resonance states, respectively (provided by P.~Wintz).}
\label{fg:mupc}
\end{figure}

The SINDRUM II Collaboration at PSI has reported on a search for the
charge-changing $\mu^{-} + Ti \rightarrow e^{+} + Ca$ in muonic atoms
\cite{kaul98}.  It was carried out simultaneously with a
measurement of $\mu^{-} + Ti \rightarrow e^{-} + Ti$. The $e^{+}$
momentum spectrum is shown in Fig.\ref{fg:mupc}. The results are given
separately for the transition to the ground state and that to the
giant dipole resonance. They are summarized in Table~\ref{tb:mupc},
together with the previous results.

\subsection{Muonium to anti-muonium conversion}

A muonium atom is a hydrogen-like bound state of $\mu^{+}$ and
$e^{-}$.  The spontaneous conversion (or oscillation) of a muonium
atom ($\mu^{+}e^{-}$ or ${\rm Mu}$) to its anti-atom, anti-muonium
atom ($\mu^{-}e^{+}$ or $\overline{\rm Mu}$) is another interesting
class of muon LFV process. In this \mumu conversion, the lepton
flavors change by two units ($\Delta L_{e/\mu}=\pm 2$) in the ordinary
law of separate additive muon and electron numbers, whereas it would
be consistent with multiplicative muon or electron number conservation
\cite{fein61}. The possibility was suggested by Pontecorvo in 1957
\cite{pont57}, even before the muonium atom was observed for the first
time at the Nevis cyclotron of Columbia University \cite{hugh60}.
 
\subsubsection{Phenomenology of \mumu conversion}

Various interactions could induce the $|\Delta L_{i}|=2$ processes,
such as \mumu conversion, as discussed in Section \ref{sc:DL2P}. To
discuss the phenomenology of the \mumu conversion, as an example, the
effective four fermion interaction of the $(V-A)(V-A)$ type
\cite{fein61} is taken. It is given by

\begin{equation}
H_{\rm Mu\overline{Mu}} = \Bigl({G_{\rm Mu\overline{Mu}} \over
\sqrt{2}} \Bigr)
\overline{\mu}\gamma_{\lambda}(1-\gamma_5){e}
\overline{\mu}\gamma^{\lambda}(1-\gamma_5){e} + H.c.
\label{eq:mu}
\end{equation}

\noindent in which $G_{\rm Mu\overline{Mu}}$ is a coupling constant
characterizing the strength of the interaction.

In the absence of an external magnetic field, the muonium and the
anti-muonium have the same ground-state energy levels. The possible
new interaction in Eq.(\ref{eq:mu}) would cause a splitting of their
energy levels of

\begin{equation}
\delta \equiv 2<\overline{M}|H_{\rm Mu\overline{Mu}}|M> = { 8 G_F \over
\sqrt{2}n^2\pi a_{0}^3} \Bigl( {G_{\rm Mu\overline{Mu}} \over G_F } \Bigr),
\end{equation}

\noindent where $n$ is the principal quantum number of the muonium atom,
and $a_{0}$ is the Bohr radius of the muonium atom. For the ground
state of the muonium atom ($n=1$),

\begin{equation}
\delta = 1.5 \times 10^{-12} \cdot \Bigl( {G_{\rm Mu\overline{Mu}} \over G_F}
\Bigr) \quad {\rm (eV)}.
\end{equation}

\begin{figure}[ht!]
\centerline{\epsfig{file=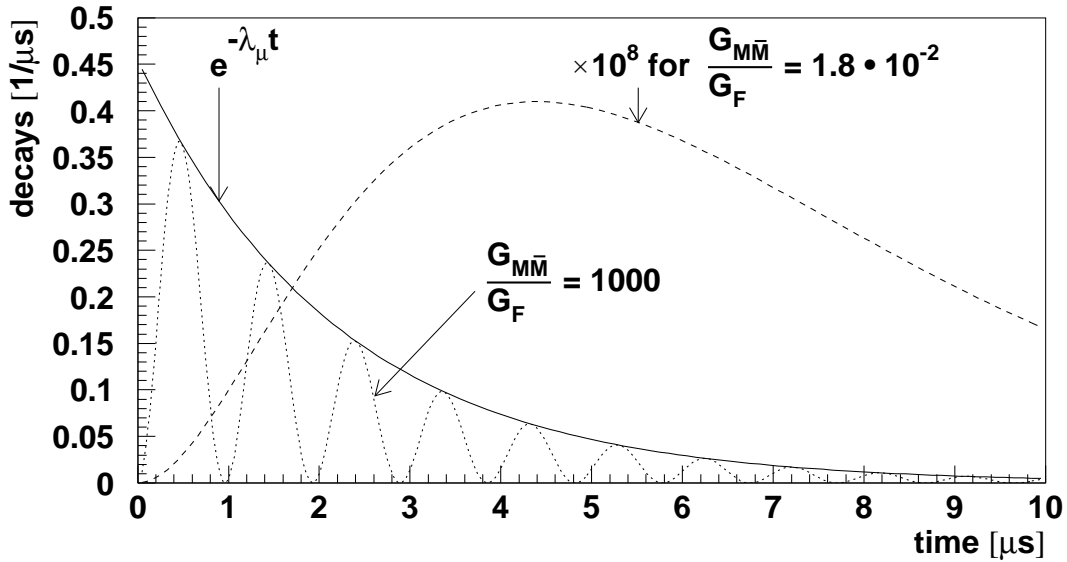,width=14cm}}
\vspace{5mm}
\caption{Time dependence of the probability of anti-muonium decay when
a pure muonium atom is created initially. The solid line represents
the exponential decay of muonium. The decay probability of
anti-muonium is given for $G_{M\overline{M}}/G_{F}$ = 1000
($G_{M\overline{M}}/G_{F} = 1.8\times 10^{-2}$) for the dotted line
(for the dashed line). In the latter case, the maximum probability
occurs at about twice the muon lifetime (after Willmann and Jungmann,
(1998)).}
\label{fg:muonium}
\end{figure}

The \mumu conversion is analogous to $K^{0}-\overline{{K}^{0}}$
mixing. If a muonium atom is formed at $t=0$ in a vacuum under no
external electromagnetic field, it could oscillate into an
anti-muonium atom with time. For a small $\delta$ value, the
probability ($\wp_{\rm Mu\overline{Mu}}$) is approximately given by
\cite{will98}

\begin{equation}
\wp_{\rm Mu\overline{Mu}}(t) = \sin^2\Bigl( {\delta t \over 2} \Bigr)
\cdot\lambda_{\mu}e^{-\lambda_{\mu}t} \approx \Bigl({\delta t \over
2}\Bigr)^2 \cdot \lambda_{\mu}e^{-\lambda_{\mu}t},
\end{equation}

\noindent where $\lambda_{\mu} = 1/\tau_{\mu} (= 2.996 \times
10^{-10}$ eV) is the muon decay width. The maximum probability of
anti-muonium decay is $t_{max} = 2\tau_{\mu}$. Fig.\ref{fg:muonium}
shows the oscillation pattern as a function of time. The total
conversion probability after integration over time ($P^{0}_{\rm
Mu\overline{Mu}}$) in a zero magnetic field is

\begin{equation}
P^{0}_{\rm Mu\overline{Mu}} = \int_{0}^{\infty} \wp_{\rm
Mu\overline{Mu}}(t) dt = { |\delta|^2\over 2( |\delta|^2 +
|\lambda_{\mu}|^2) } = 2.56 \times 10^{-5} \cdot\Bigl( {G_{\rm
Mu\overline{Mu}} \over G_F} \Bigr)^2.
\label{eq:prbmu}
\end{equation}

\noindent The experimental limit constrains the upper limit of
magnitude of $G_{\rm Mu\overline{Mu}}$. The limit of $G_{\rm
Mu\overline{Mu}}$ is improved by the square root of the conversion
probability.

\begin{figure}[ht!]
\centerline{\epsfig{file=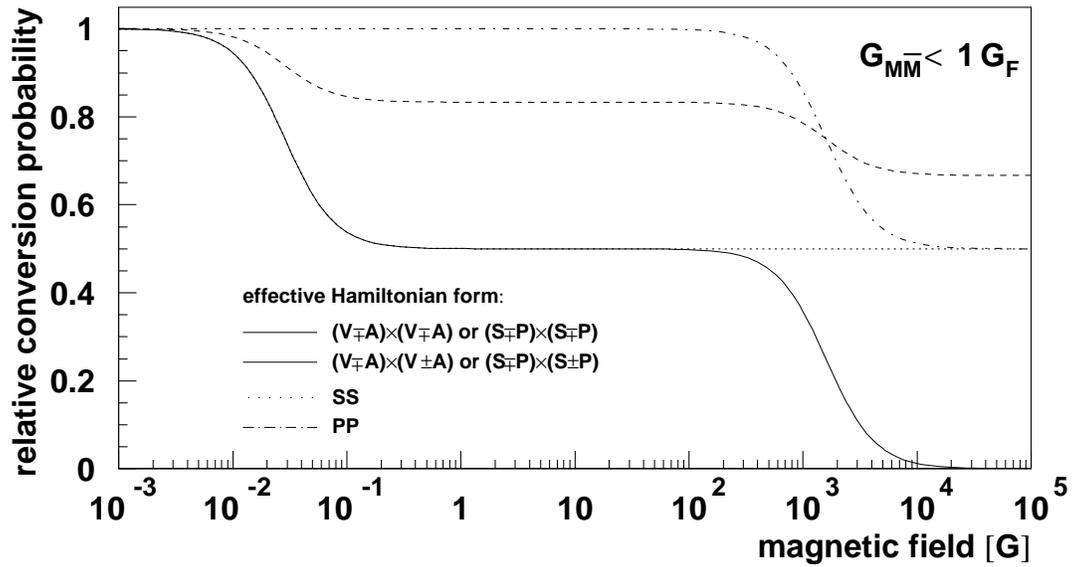,width=14cm}}
\vspace{5mm}
\caption{\mumu conversion rate for different interactions as a
function of the external magnetic field (after Willmann and Jungmann,
(1998)).}
\label{fg:mf}
\end{figure}

The presence of an external electromagnetic field would remove the
degeneracy between the muonium and the anti-muonium atoms. It would
reduce the probability of the muonium to anti-muonium conversion. The
splitting of different muonium energy levels in the presence of a
magnetic field is calculated by using the Breit-Rabi formula for the
states of their total spin, $F$, and its $z$-component, $m_{F}$. In a
magnetic field, the $(F,m_{F})=(1,\pm1)\rightarrow (1,\pm1)$
transitions become rapidly suppressed, even at a weak field, because
of the Zeeman splitting of energy levels. The transitions between
different $F$ states are also highly suppressed, even in a zero
magnetic field, owing to the muonium {\em 1s} hyperfine splitting (of
$1.846\times 10^{-5}$ eV).  By taking into account the magnetic-field
dependences of different energy levels of muonium and anti-muonium and
their transition rates, Eq.(\ref{eq:prbmu}) can be modified for
unpolarized muons by

\begin{equation}
P_{\rm Mu\overline{Mu}}(B) = {1\over4}\sum_{F,m_{F}}{ |\delta|^2 \over
2( |\delta|^2 + |\triangle|^2 + |\lambda_{\mu}|^2) } \equiv P^{0}_{\rm
Mu\overline{Mu}} \cdot S_B(B),
\end{equation}

\noindent where $\Delta \equiv E_{Mu}(F,m_{F}) -
E_{\overline{Mu}}(F,m_{F})$, and $\delta$ and $\Delta$ are functions
of the magnitude of the magnetic field ($B$). The reduction factor,
$S_B(B)$, has been calculated for possible interactions of different
types \cite{hou95,hori96}. Fig.\ref{fg:mf} shows the dependence of the
\mumu conversion probability on the external magnetic field and
different coupling types.  For example, for the traditional
$(V-A)(V-A)$ interaction, the conversion rate becomes one half at a
magnetic field of about 10 mG and is further strongly suppressed for a
magnetic field greater than $10^3$ G.

\subsubsection{Event signature and backgrounds}

In experiments of the \mumu conversion, an anti-muonium converted from
a muonium initially produced is searched.  The experimental signature
of an anti-muonium decay is the emission of an energetic $e^{-}$ from
$\mu^{-}\rightarrow e^{-}\overline{\nu}_{\mu} \nu_{e}$ decay with a
dissociated $e^{+}$ left behind with an average kinetic energy of 13.5
eV. This corresponds to the binding energy of the {\em 1s} state of a
muonium atom.

The sensitivity to \mumu conversion is known to be suppressed when the
muonium atom is in matter. This occurs since a negative muon in
anti-muonium is easily captured by surrounding atoms. Therefore,
recent experiments have been performed by using muonium atoms in a
vacuum.

There are two major sources of potential backgrounds. One is
accidental coincidences of energetic $e^{-}$ produced by Bhabha
scattering of $e^{+}$ from $\mu^{+}$ decay in a muonium and the
scattered $e^{+}$. The second is the physics (prompt) background from
the $\mu^{+}\rightarrow e^{+} \nu_{e} \overline{\nu}_{\mu} e^{+}
e^{-}$ decay (whose branching ratio is $3.4 \times 10^{-5}$), when the
$e^{-}$ becomes energetic and only one of the two $e^{+}$s is
detected.

\subsubsection{Experimental status of \mumu conversion}\label{sc:mumuex}

\begin{table}[th!]
\caption{Historical progress and summary of \mumu conversion.}
\label{tb:mumu}
\vspace{5mm}
\begin{tabular}{llll}
Place & Year & $G_{\rm Mu\overline{Mu}}/G_{F}$ & Reference \cr\hline
TRIUMF & 1982 & $<42$    & \citeasnoun{mars82} \cr
TRIUMF & 1986 & $<20$    & \citeasnoun{beer86} \cr
TRIUMF & 1990 & $<0.29$  & \citeasnoun{hube90} \cr
LANL   & 1991 & $<0.16$  & \citeasnoun{matt91} \cr
LANL   & 1993 & $<6.9$   & \citeasnoun{ni93} \cr
PSI    & 1996 & $<0.018$ & \citeasnoun{abel96} \cr
JINR   & 1997 & $<0.14$  & \citeasnoun{gord97} \cr
PSI    & 1999 & $<0.003$ & \citeasnoun{will99}
\end{tabular}
\end{table}

\begin{figure}[ht!]
\centerline{\epsfig{file=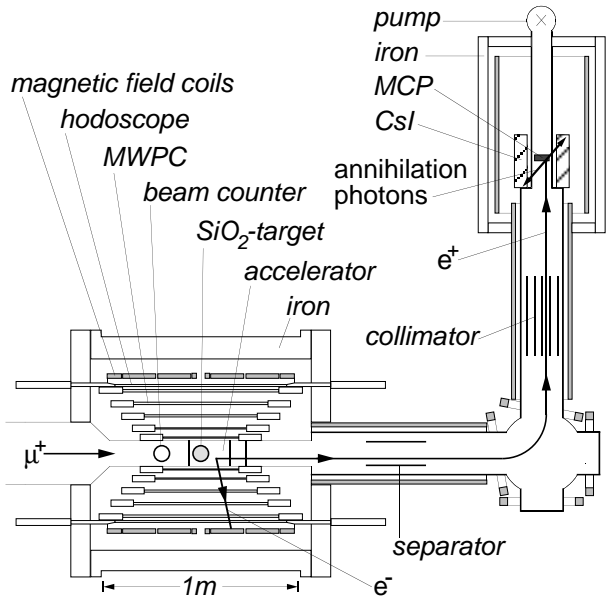,angle=90,height=8truecm}}
\vspace{5mm}
\caption{Schematic layout of the detector for muonium-antimuonium
conversion at PSI (after Willmann {\em et al.}, (1999)).}
\label{fg:mumubar}
\end{figure}

\begin{figure}[ht!]
{\epsfig{file=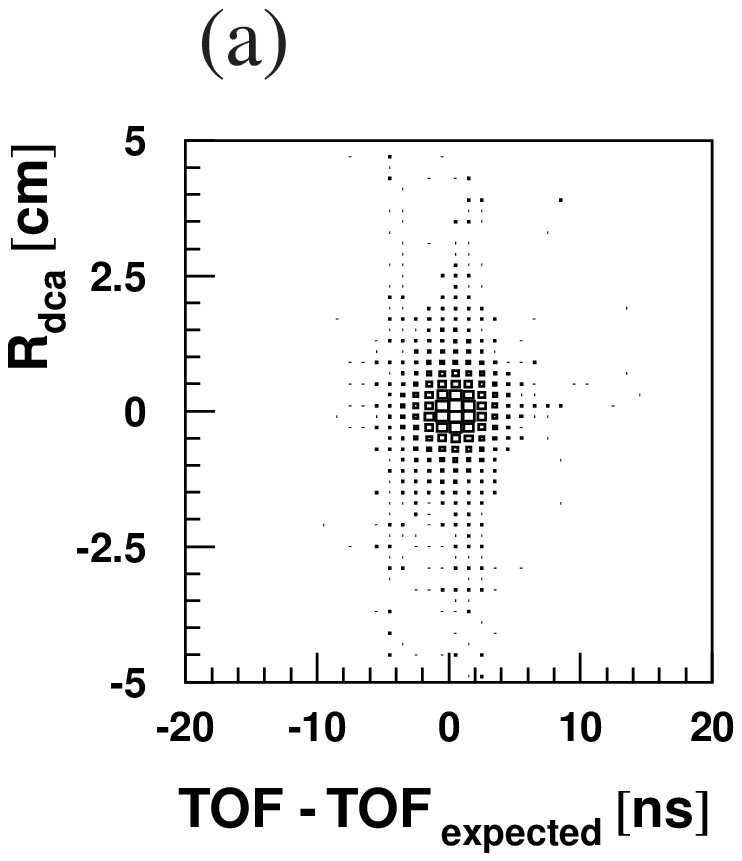,width=8cm}}
{\epsfig{file=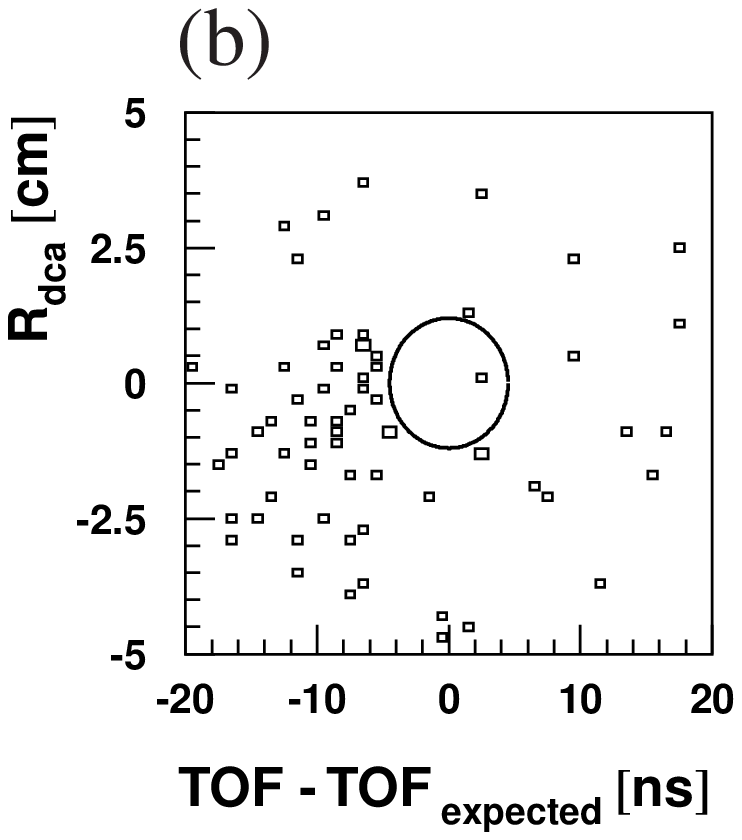,width=8cm}}
\vspace{5mm}
\caption{Distribution of the distance of closest approach between
the $e^{+}$ track and $e^{-}$ track versus their timing difference in
the experiment to search for \mumu conversion. (a) Monte Carlo data
and (b) real data (after Willmann {\em et al.}, (1999)).}
\label{fg:mumudata}
\end{figure}

The historical progress in the searches for \mumu conversion is listed
in Table~\ref{tb:mumu}.  A recent experiment was carried out at PSI
\cite{will99}. The experiment fully utilized the powerful techniques
developed at the previous experiment at LANL \cite{matt91}, which
requires the coincidence identification of both particles in the
anti-muonium decay.  Its experimental setup is shown in
Fig.\ref{fg:mumubar}.  Muonium atoms were produced by stopping surface
muons in a SiO$_2$ powder target, where some fraction diffused out
through the target surface with thermal energy in a vacuum. To detect
$e^{-}$ from $\mu^{-}$ decay, a magnetic spectrometer was used. It
consisted of five concentric multiwire proportional chambers with 64
segmented hodoscopes at a 0.1 T magnetic field. The $e^+$ with an
average kinetic energy of 13.5 eV was detected by micro-channel plate
detectors after electrostatic acceleration to 8 keV. With the
production of about $5.7\times 10^{10}$ muonium atoms, their analysis
yielded one event satisfying all of the required criteria with the
expected background events of $1.7 \pm 0.2$ due to accidental
coincidence. The Monte-Carlo data and real data are given in
Fig.\ref{fg:mumudata}. The 90\% C.L. upper limit on the conversion
probability at zero magnetic filed is

\begin{equation}
P^{0}_{\rm Mu\overline{Mu}} \leq 8.3\times 10^{-11}/S_B(B),
\end{equation}

\noindent where the factor $S_B(B)$ describes the suppression of
the \mumu conversion in an external magnetic field, $B$. It could be
translated into the upper limit on the effective coupling constant,
$G_{\rm Mu\overline{Mu}}$, which is given by

\begin{equation}
G_{\rm Mu\overline{Mu}} \leq 3.0 \times 10^{-3} G_{F}
\end{equation}

\noindent at 90\% C.L. upper limit under a 0.1 T magnetic field.

\section{Future Prospects}\label{FPS}

The field of muon decay physics is presently very productive, even
after its long history of over 60 years. Currently, there are several
new experiments which are being either prepared or planned. Some of
them, which were mentioned in this article, are R77 at RIKEN-RAL and
R-99-07 at PSI for the muon lifetime measurement (in
Section~\ref{sc:MLES}), E614 at TRIUMF to measure the Michel spectrum
and its asymmetry (in Section~\ref{sc:MSES}), R-94-10 and R-97-07 at
PSI to measure the $e^{+}$ polarization in polarized \muenn decay (in
Section~\ref{sc:PMDES}), R-99-05 at PSI for \mueg decay (in
Section~\ref{sc:MGES}), the new phase of SINDRUM-II at PSI, and E940
(MECO) at BNL for \muec conversion (in Section~\ref{sc:MEES}).  Each
of them is aiming at an improvement of about an order of magnitude or
more over the previous experiments. The potential progress expected by
each of such experiments is based not only on innovative ideas on
detection methods, but also on muon beams of high intensity and good
quality. In particular, the planned searches for muon LFV processes
strongly rely on the beam, such as the PMC in SINDRUM-II and the
superconducting solenoid capture and transport systems for the MECO
experiment. More muon fluxes with less contamination are critical for
further improvements.

Currently, two out of the three meson factories are operational. One
of the two operational machines, the PSI cyclotron, has increased its
proton current, achieving 1.5 mA, the highest proton current in the
world.  The muon beam intensities for various existing laboratories
are listed in Table~\ref{tb:mi}.  In addition, the use of higher
energy proton machines, such as the BNL AGS for negative muons, is
being considered for the MECO experiment, where a pulsed-beam
capability at the AGS and a larger cross section of negative pions at
a few 10 GeV proton energy are to be utilized. In the long-term
future, there are several new projects to construct high-intensity
proton accelerators: the JAERI/KEK Joint Project (previously JHF)
\cite{jhf99}, which consists of a 50-GeV proton synchrotron (50-GeV PS)
with a 15 $\mu$A beam intensity, and a 3-GeV proton synchrotron with a
330 $\mu$A beam intensity; the SNS (Spallation Neutron Source) at Oak
Ridge; a possible European Spallation Neutron Source (ESS).  Probably,
a proton driver for a $\mu^{+}\mu^{-}$ collider
\cite{mumu96,anke99} can be included in the long-term future. Note that
among the above, only the 50-GeV PS is planned to have a continuous
proton beam by slow beam extraction, whereas the others may have only
fast beam extraction of a low repetition rate.

\begin{table}[bht!]
\caption{Intensities of existing muon beams available.}
\label{tb:mi}
\vspace{5mm}
\begin{tabular}{lllll}
Facility & Protons & Time structure & \multicolumn{2}{c}{Muon
Intensity} \cr\hline 
PSI & 1.5 mA & continuous & $3\times 10^{8}~\mu^{+}$s/sec 
& at 28 MeV/$c$ (surface muons) \cr 
& 590 MeV & (50 MHz) & $1 \times 10^{8}~\mu^{-}$s/sec 
& at 100 MeV/$c$ \cr \hline 
TRIUMF & 150 $\mu$A & continuous & $2 \times 10^{7}~\mu^{+}$s/sec 
& at 28 MeV/$c$ (surface muons) \cr 
& 500 MeV & (50 MHz) & $3 \times 10^{6}~\mu^{-}$s/sec 
& at 100 MeV/$c$ \cr \hline
RAL & 200 $\mu$A & pulsed & $1 \times 10^{6}~\mu^{+}$s/sec~$^{1)}$ 
& at 28 MeV/$c$ (surface muons) \cr
& 800 MeV & (50 Hz) & $1 \times 10^{5}~\mu^{-}$s/sec 
& at 50 MeV/$c$ \cr\hline
MSL~$^{2)}$ & 6 $\mu$A & pulsed  &  $1 \times 10^{5}~\mu^{+}$s/sec 
& at 28 MeV/$c$ (surface muons) \cr
& 500 MeV & (20 Hz) & $1 \times 10^{4}~\mu^{-}$s/sec 
& at 55 MeV/$c$ \cr\hline
JINR~$^{3)}$ & 4 $\mu$A & continuous & $3 \times 10^{4}~\mu^{+}$s/sec 
& at 28 MeV/$c$ (surface muons) \cr
& 660 MeV &  & $1 \times 10^{3}~\mu^{-}$s/sec 
& at 100 MeV/$c$ \cr
\end{tabular}
\vspace{5mm}
$^{1)}$ The highest instantaneous intensity of 
$2 \times 10^{4}\mu^{+}$s/200 nsec.\\
$^{2)}$ Meson Science Laboratory at KEK, using the existing 500 MeV
Booster ring. \\
$^{3)}$ Phasotron, Joint Institute for Nuclear Research, Dubna, Russia.
\end{table}

Regarding LFV, besides the study of muon decays, a unique possibility
of lepton flavor changing Rutherford scattering, where the conversion
of incident electrons into muons of the same energy by scattering in
the external electric field of a massive nucleus, is also discussed
\cite{abra96}. However, the expected cross section is very small to
compete with rare muon decay processes, and technical details have not
yet been discussed.

\subsection{Towards new high-intensity muon sources}

\begin{figure}[ht!]
\centerline{\epsfig{file=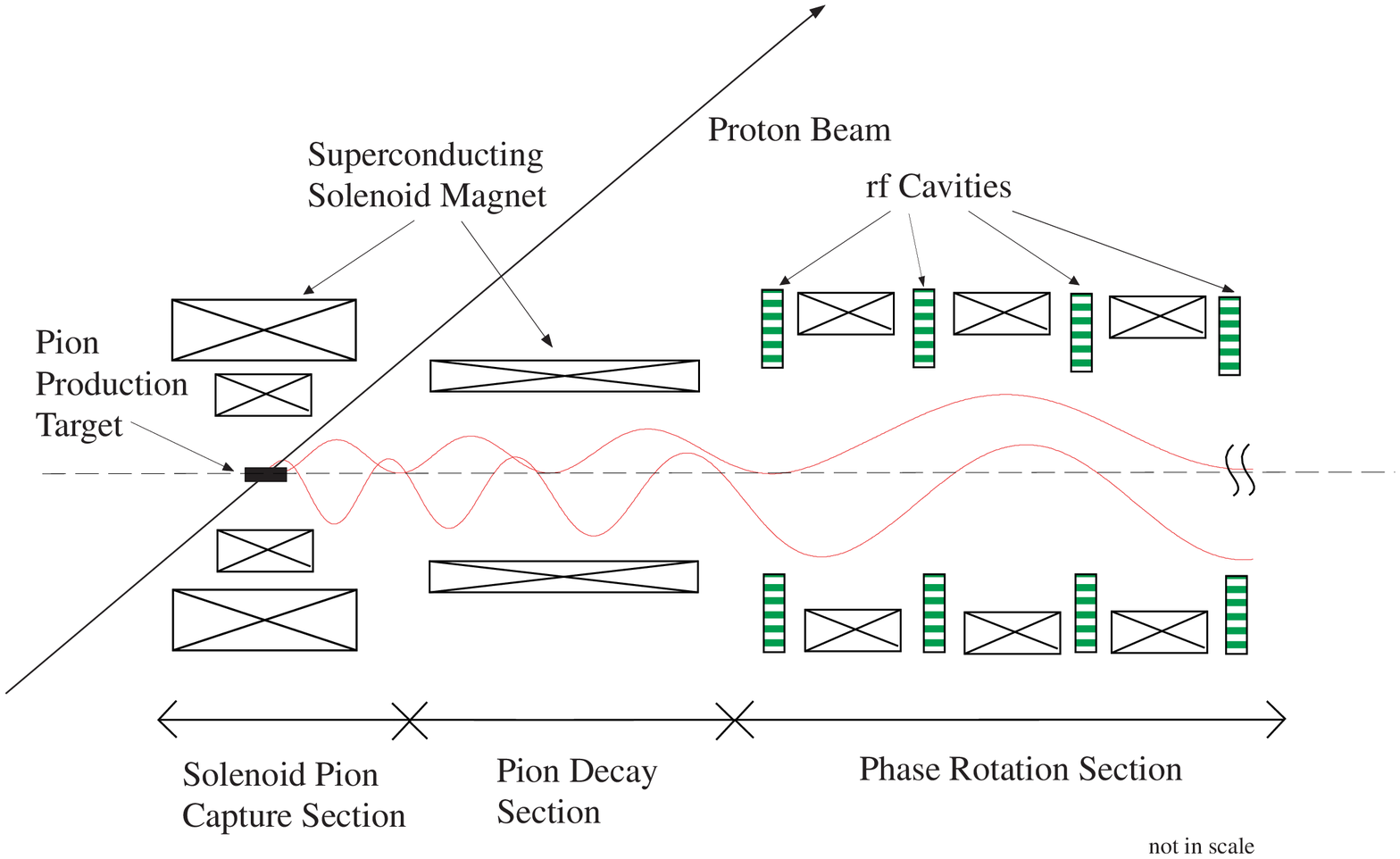,width=14cm}}
\vspace{5mm}
\caption{Schematic layout of a high-intensity muon source.}
\label{fg:prism}
\end{figure}

Significant improvements in low-energy muon physics could be expected
if a high-intensity muon source, having a beam intensity of
$10^{12}-10^{13}~\mu^{\pm}$s/sec with a narrow energy-spread and less
contamination, can be realized. The muon beam intensity envisaged
would be four or five orders of magnitude higher than that available
today. Ideas of such a high-intensity muon source are based on (i)
solenoid pion capture, (ii) phase rotation, and (iii) muon cooling. A
schematic view of the basic concept is shown in Fig.\ref{fg:prism}.

In solenoid pion capture, low-energy pions and muons are trapped in a
high solenoidal magnetic field (such as 10 T or more).  From Monte
Carlo simulations with appropriate pion production cross sections,
about 0.3 to 0.1 captured pions (of less than 0.5 GeV/$c$) per proton
are estimated for proton beam energies of 50 GeV to 10 GeV,
respectively.  For a proton intensity of the existing and planned
proton machines of about $10^{13}-10^{14}$ protons/sec, a large number
of captured pions sufficient for the aimed intensity are expected.

The phase rotation is to accelerate slow muons and to decelerate fast
muons by a strong radio-frequency (RF) electric field, yielding a
narrow longitudinal momentum spread. To identify fast and slow muons
by their time of flight from the production time, a very narrow pulsed
proton beam must be used.  An intensity enhancement of a factor of
about ten in the longitudinal energy distribution is estimated from
Monte Carlo simulations \cite{kuno97b}.

The muon cooling, which is based on ionization cooling
\cite{skri81}, is to reduce muon beam emittance. The
longitudinal cooling is based on the repetition of energy loss by
ionization and subsequent acceleration to restore the longitudinal
momentum. The ionization cooling works only for muons.

These ideas have emerged in studies of a $\mu^{+}\mu^{-}$ collider at
the high-energy frontier \cite{mumu96,anke99}. The physics potential
with low energy muons available from the front-end of the
$\mu^{+}\mu^{-}$ collider complex has been discussed.  Although there
are many common R\&D items between a low-energy muon source and a
$\mu^{+}\mu^{-}$ collider, there are discussions on whether the
front-end muon collider (FMC) could be directly used in experiments
with muons.  The FMC will run with a pulsed beam of slow repetition
(at typically 15 Hz). However, most experiments with muons require a
beam with a high duty factor, or a nearly continuous beam, because of
the reduction of the instantaneous rate \cite{molz97}. The precise
requirement on the beam time structure depends on the type of
experiments. For instance, searches for \mueg and \meee must use a
continuous beam to reduce the instantaneous rate, whereas searches for
\muec (or \mupc) conversion, \mumu conversion and a measurement of the
muon lifetime need a pulsed beam with a pulse separation of an order
of the muon lifetime ($\sim~\mu$sec). Thus, independent R\&D items, in
particular concerning phase rotation and the muon cooling system,
exist in a low-energy muon source. These technical issues must be
pursued separately.

There are several dedicated R\&D programs on low-energy muon sources
with high intensity. One of those is the PRISM project at KEK in Japan
\cite{kuno98}. The PRISM project, which is an acronym of
Phase-Rotation Intense Secondary Meson beam, would combine solenoid
capture, phase rotation and possibly modest muon cooling to produce a
cooled muon beam. The requirements on muon cooling as a secondary beam
is not as strict as that in the $\mu^{+}\mu^{-}$ collider. Its R\&D
program starts from a relatively low repetition rate ($\sim$ kHz), and
aims at a higher repetition in the future. The others are the MUONS
project at TRIUMF \cite{blac97}, and the Super-Super Muon Channel
project at the RIKEN-RAL muon facility \cite{ishi98}. For the last
project, a new scheme of the production of cooled $\mu^{+}$s by laser
ionization of thermal muonium is also proposed \cite{naga96}.

Once a highly intense muon source with a narrow energy spread and less
contamination is available, physics programs with stopped muons, in
particular searches for rare muon LFV processes, would be
significantly improved. First of all, the potential sensitivity
achievable in searches for rare processes is ultimately limited by the
number of muons available. Therefore, a high-intensity beam is
essential. Small beam contamination is necessary to further reduce any
background associated with it. A narrow energy spread of the beam will
allow a thin muon stopping target to improve the detection resolution. 
For instance, if about $10^{19}-10^{20}$ muons/year are available, a
new experiment on \muec conversion with a sensitivity of $10^{-18}$
could be possible
\cite{blac97}.

The high-intensity muon sources could be used not only for experiments
with low-energy muons, but also for experiments with energetic muons
if the muons thus produced are injected into additional accelerators
for further acceleration. Potential programs might include the
measurements of the muon anomalous magnetic moment and the muon
electric dipole moment, and also a muon accumulator ring for neutrino
sources \cite{geer98,cern99}. In addition to particle-physics
programs, a broad research field from materials science to biology
would benefit from new highly intense muon sources, which would
definitely open up a new era of muon science.

\section{Conclusions}

We have described the current theoretical and experimental status in
the field of muon decay to search for physics beyond the Standard
Model.  Among many interesting topics of physics related to muons, we
have discussed the precise measurements of normal muon decay, and the
searches for muon LFV processes. In particular, we have highly
stressed the importance of muon LFV processes.

The physics motivation for LFV is extremely strong. LFV has recently
attracted much attention from theorists and experimentalists, more
than ever.  This has happened because SUSY models predict large
branching ratios for LFV processes. Their predictions are just as
large as one or two orders of magnitude lower than the present
experimental limits. They could be accessible and tested by future
experiments. There are many scenarios of SUSY models which predict
sizeable LFV effects.  They are such as SUSY GUT, SUSY with
right-handed (heavy) Majorana neutrinos, SUSY with $R$-parity
violation and others. They would provide an opportunity to give a hint
on physics at very high energy scales, like either the GUT scale or
the mass scale of a heavy right-handed Majorana neutrino for the
see-saw mechanism. Of course, there are many other classes of
theoretical models which predict a large LFV effect besides SUSY.
Therefore, LFV searches have robust potential to uncover new physics
beyond the SM.

We have presented the phenomenology of muon LFV processes of $|\Delta
L_{i}|=1$, such as the \mueg and \meee decays, the \muec conversion in
a muonic atom, and those of $|\Delta L_{i}|=2$ such as the muonium to
anti-muonium conversion. We discussed all of the above processes of
$|\Delta L_{i}| = 1$ within the same framework of the effective
Lagrangian to illustrate how various contributions (such as photonic
and non-photonic) can be disentangled with measurements of the three
muon LFV processes. Thereby, searches for these three processes are
equally important. If the muon is polarized, additional information
could be obtained by measuring the angular distributions in the \mueg
decay and \muec conversion, and T-odd and P-odd correlations in the
\meee decay. Furthermore, for the \mueg decay, the use of polarized
muons would be useful to eliminate any background processes in the
search. Experimentally, positive muons in a surface muon beam are
known to be 100\% polarized, and thereby the use of polarized muons
will be feasible in the future. Then, we have briefly mentioned the
most recent experimental results, such as MEGA for \mueg decay,
SINDRUM-II for \muec conversion, the recent search for \mumu
conversion at PSI, and others, together with future experimental
prospects.  In addition, precision measurements of the normal muon
decay, \muenn, have attracted much interest. In the near future, new
measurements of the muon lifetime, of the Michel spectrum and its
asymmetry, and of the $e^{+}$ polarization in polarized
\muenn decay will be carried out while aiming at an order of magnitude
improvement.

A new intense muon source with $10^{12}-10^{13}~\mu^{\pm}$s/sec would
be strongly required to make substantial improvements in low-energy
muon physics. The aimed intensity is four or five orders of magnitude
higher than that available now. The ideas for such a muon source are
based on (i) solenoid pion capture, (ii) phase rotation, and (iii)
muon cooling.  These ideas came from a $\mu^{+}\mu^{-}$ collider.
However, the beam time structure must be of high duty factor for
low-energy muon physics, leading to new technical challenges which do
not exist for the $\mu^{+}\mu^{-}$ collider R\&D studies. To overcome
these issues, several R\&D programs dedicated to low-energy muons are
now being undertaken at KEK, RIKEN and TRIUMF. With increased muon
fluxes, the searches for rare muon LFV processes, as well as precision
measurements of muon decay, are expected to be significantly advanced.

Muon physics becomes important with strong physics motivations. There
are extraordinary opportunities which will allow us to explore
discovery potentials of physics beyond the SM, with low-energy muons.

\acknowledgments 

It is a great pleasure to thank the many people who helped us to write
this review article; P.~Depommier, J.~Deutsch, D.R.~Gill, J.~Hisano,
K.P.~Jungmann, J.A.~Macdonald, R.~Mischke, W.R.~Molzon, S.N.~Nakamura,
K.~Okumura, Y.~Shimizu, H.C.~Walter, and P.~Wintz. Especially, Pierre
Depommier gave many useful comments on the whole manuscript. We also
wish to thank many people for discussions, in particular,
A.~Czarnecki, T.~Goto, P.~Herczeg, R.~Kitano, A.~Maki, W.J.~Marciano,
Y.~Mori, K.~Nagamine, S.~Nagamiya, S.~Orito, S.T.~Petcov, N.~Sasao,
A.~van~der~Schaaf, J.~Vergados, and T.K.~Yokoi. One of us (Y.K.) 
acknowledges all the colleagues of the PRISM working group. The work
of Y.K. was supported in part by the Grant-in-aid of the Ministry of
Education, Science, Sports and Culture, Government of Japan
(No.~10309009 and No.~11691134). The work of Y.O. was supported in
part by the Grant-in-Aid of the Ministry of Education, Science, Sports
and Culture, Government of Japan (No.~09640381), Priority area
``Supersymmetry and Unified Theory of Elementary Particles''
(No.~707), and ``Physics of CP Violation'' (No.~09246105).

\appendix

\section{Radiative Muon Decay}\label{sc:ARMD}

The differential branching ratio of the radiative muon decay, \muenng,
is given in Eq.(\ref{eq:br_radmu}). The functions appearing in
Eq.(\ref{eq:br_radmu}), $F(x,y,d)$, $G(x,y,d)$ and $H(x,y,d)$ in the
SM, are given as follows:

\begin{eqnarray}
F&=&F^{(0)}+ r F^{(1)}+ r^2 F^{(2)}, \nonumber\\
G&=&G^{(0)}+ r G^{(1)}+ r^2 G^{(2)}, \nonumber\\
H&=&H^{(0)}+ r H^{(1)}+ r^2 H^{(2)},
\end{eqnarray}  

\noindent where  $r = (m_{e}/m_{\mu})^2$. $m_e$ and $m_{\mu}$ are
the masses of an electron and a muon, respectively. Here, $x$ and $y$
are the normalized electron and photon energies, $x = 2E_{e}/m_{\mu}$
and $y = 2E_{\gamma}/m_{\mu}$; $d$ is given by $d\equiv
1-\beta\hat{p_e}\cdot\hat{p_{\gamma}}$. $\hat{p_e}$ and
$\hat{p_{\gamma}}$ are unit momentum vectors of the electron and the
photon respectively. $\beta$ is defined as $\beta \equiv
|\vec{p_e}|/E_{e}$.

\begin{eqnarray}
F^{(0)}(x,y,d)&=&\frac{8}{d}
\{ y^2(3-2y)+6xy(1-y)+2x^2(3-4y)-4x^3\}
\nonumber \\
&&+8\{-xy(3-y-y^2)-x^2(3-y-4y^2)+2x^3(1+2y)\}
\nonumber \\
&&+2d\{x^2y(6-5y-2y^2)-2x^3y(4+3y)\}
+2d^2x^3y^2(2+y)
\\
F^{(1)}(x,y,d)&=&\frac{32}{d^2}
\{ -\frac{y(3-2y)}{x}-(3-4y)+2x\}
+\frac{8}{d}\{y(6-5y)-2x(4+y)+6x^2\}
\nonumber \\
&&+8\{x(4-3y+y^2)-3x^2(1+y)\}
+6dx^2y(2+y)
\\
F^{(2)}(x,y,d)&=&\frac{32}{d^2}
\{ \frac{(4-3y)}{x}-3\}
+\frac{48y}{d}
\\
G^{(0)}(x,y,d)&=&\frac{8}{d}
\{ xy(1-2y)+2x^2(1-3y)-4x^3\}
+4\{-x^2(2-3y-4y^2)+2x^3(2+3y)\} 
\nonumber \\
&&-4dx^3y(2+y)
\\
G^{(1)}(x,y,d)&=&\frac{32}{d^2}( -1+2y+2x)
+\frac{8}{d}(-xy+6x^2)
-12x^2(2+y)
\\
G^{(2)}(x,y,d)&=&-\frac{96}{d^2}
\\
H^{(0)}(x,y,d)&=&\frac{8}{d}
\{ y^2(1-2y)+xy(1-4y)-2x^2y\}
+4\{2xy^2(1+y)-x^2y(1-4y)+2x^3y\}
\nonumber \\
&&+2d\{x^2y^2(1-2y)-4x^3y^2\}
+2d^2x^3y^3
\\
H^{(1)}(x,y,d)&=&\frac{32}{d^2}
\{ -\frac{y(1-2y)}{x}+2y\}
+\frac{8}{d}\{y(2-5y)-xy\}
+4xy(2y-3x)
+6dx^2y^2
\\
H^{(2)}(x,y,d)&=&-\frac{96y}{d^2x}
+\frac{48y}{d}
\end{eqnarray} 

\section{MSSM Lagrangian}\label{sc:AMSSM}

The Lagrangian for the minimal supersymmetric standard model (MSSM) is
described.  In SUSY theories, elementary fields are introduced as a
pair of bosonic and fermionic fields. Such a pair is called a
supermultiplet.  There are two types of supermultiplets, a gauge
multiplet and a chiral multiplet. A gauge multiplet consists of a
gauge field ($A_{\mu}^a$) and its superpartner, a gauge fermion (or
gaugino) field ($\lambda^a$), which is a Majorana fermion field in the
adjoint representation of the gauge group. In the MSSM, we have to
introduce gaugino fields for $SU(3)_C$, $SU(2)_L$ and $U(1)_Y$ gauge
groups.  A chiral multiplet is a set of a complex scalar field ($
\phi$) and a left-handed Weyl fermion field ($\psi_{L}$). Its complex
conjugate is called an anti-chiral multiplet which consists of $
\phi^*$ and $\psi_{R}$. In the MSSM, these chiral multiplets
correspond to matter fields, namely quark, lepton and Higgs fields and
its superpartners. The fields in the same chiral multiplet have the
same quantum numbers for the gauge groups. Chiral multiplets necessary
for the MSSM are listed with their quantum numbers in Table
\ref{tb:mssm}. The right-handed squarks and sleptons are defined as follows:
$(\tilde{u}_{iL}^{c})^{*} =\tilde{u}_{iR}$, $(\tilde{d}_{iL}^{c})^{*}
=\tilde{d}_{iR}$, and $(\tilde{e}_{iL}^{c})^{*} = \tilde{e}_{iR}$.  We
sometimes use a notation $\tilde{\phi}$ and $\tilde{\psi}_L$ for
superpartners of $\phi$ and $\psi_L$ and $\Phi$ for a supermultiplet
($\phi$, $\psi_{L}$).

\begin{table}[htb!]
\caption{Chiral multiplets in MSSM.}
\label{tb:mssm}
\begin{center}
\begin{tabular}{c|ccccccc}
 & $Q_{i}$($\tilde{q}_{i L}$,$q_{i L}$) &$U_{i}^c$($\tilde{u}_{i
L}^c$,$u_{i L}^c$) &$D_{i}^c$($\tilde{d}_{i L}^c$,$d_{i L}^c$)
&$L_{i}$($\tilde{l}_{i L}$,$l_{i L}$) &$E_{i}^c$($\tilde{e}_{i
L}^c$,$e_{i L}^c$) &$H_1$($H_1$,$\tilde{H}_1$)
&$H_2$($H_2$,$\tilde{H}_2$)\cr\hline
$SU(3)_C$&\bf{3}&$\bf{\bar{3}}$&$\bf{\bar{3}}$ & \bf{1} & \bf{1}&\bf{1}&\bf{1}
\cr 
$SU(2)_L$& \bf{2}&\bf{1}&\bf{1}& \bf{2} & \bf{1}
&\bf{2}&\bf{2} \cr 
$U(1)_Y$&$\frac{1}{6}$ &$-\frac{2}{3}$ &
$\frac{1}{3}$ & $-\frac{1}{2}$ & 1 &-$\frac{1}{2}$&$\frac{1}{2}$\cr
\end{tabular}
\end{center}
\end{table}

In phenomenological applications of SUSY model, the SUSY Lagrangian
consists of two parts, namely the SUSY invariant Lagrangian and the
soft SUSY breaking terms.

\begin{equation}
\cal L \mit  = \cal L \mit _{SUSY~inv} + \cal L \mit _{SUSY~breaking} 
\end{equation}

The SUSY invariant part of the MSSM Lagrangian is given as follows.
${\lagr}_{SUSY~inv}$ can be decomposed into two parts,

\begin{equation}
\cal L \mit _{SUSY~inv}= \cal L \mit _{gauge} +\cal L \mit _{superpotential},
\end{equation} 

\noindent $\cal L \mit _{gauge}$ depends on the gauge coupling constants, and
is given by

\begin{eqnarray}
 \cal L \mit _{gauge}& =
& \sum_{gauge~multiplet} F^{(a)}_{\mu \nu} F^{(a)\mu \nu}
+\sum_{chiral~multiplet}(i {\overline {{\psi}_{i}}_{L}}
\gamma \cdot {\cal D} {\psi_{i}}_{L} + |{\cal D}_{\mu} \phi_{i}|^2)
\nonumber \\
&&+\cal L \mit _{gaugino-matter} + \cal L \mit _{D~term},
\end{eqnarray}

\noindent where for each gauge group the $\cal L \mit _{gaugino-matter} $
and $\cal L \mit _{D~term}$ terms are given by

\begin{eqnarray}
\cal L \mit _{gaugino-matter}&=-&\sum_{i} \sqrt{2} g \phi_{i}^{\dagger} 
\overline{\lambda}^a T^a \psi_{i L} + H.c. , \label{eq:lgm}\\
\cal L \mit _{D~term}&=& -\sum_{a}\frac{g^2}{2}\sum_{i}
( \phi_{i}^{\dagger} T^a \phi_{i}) ^2. \label{eq:ldt}
\end{eqnarray}

\noindent In addition to the normal gauge interactions defined in the
covariant derivatives, these two types of interactions specified by
the gauge coupling constants in Eqs.(\ref{eq:lgm}) and (\ref{eq:ldt})
are necessary to keep the SUSY invariance of the Lagrangian.  $\cal L
\mit _{superpotential}$ is determined from the superpotential
$W(\phi_i)$, which is a function of scalar fields of the chiral
multiplets,

\begin{eqnarray}
\cal L \mit _{superpotential} 
&=&  - \sum_{i} |\frac {W(\phi)}{\partial \phi_i}|^2
-{1\over2}{\partial^2  W(\phi) \over
{\partial \phi_i \partial \phi_j}} \overline {(\psi_{i L})^c}
\psi_{j L} + H.c.
\label{eq:superpotential_appendix}
\end{eqnarray}

\noindent The superpotential therefore generates a set of bosonic
and fermionic interactions.  From the gauge invariance, the
superpotential ($W(\phi_i)$) for the MSSM is given by

\begin{equation}
W_{\mit MSSM}= {(y_{e})}_{ij}H_{1} E_{i}^{c} L_{j}
               + {(y_{d})}_{ij}H_{1} D_{i}^{c} Q_{j} 
               + {(y_{u})}_{ij}H_{2} U_{i}^{c} Q_{j}
               - \mu H_{1}H_{2},
\end{equation}

\noindent where the contraction of the $SU(2)$ indices is made by
using the anti-symmetric tensor, $\varepsilon_{\alpha\beta}$. Also,
the $R$-parity conservation is required (Section~\ref{sc:smrv}). By
substituting these functions in Eq.(\ref{eq:superpotential_appendix}),
this superpotential would induce the ordinary Yukawa couplings and the
higgsino mass terms as well as the other interactions which are
necessary to ensure the SUSY invariance.

The soft SUSY breaking mass terms are defined as terms which do not
introduce quadratic divergence, and essentially serve as mass terms
for superpartners.  A general form of SUSY-breaking terms in the MSSM
is given by

\begin{eqnarray}
\cal L_{\mit SUSY~breaking}
&=&-(m^{2}_{e})_{ij}\tilde{e}_{Ri}^{*}\tilde{e}_{Rj}
   -(m^{2}_{l})_{ij}\tilde{l}_{Li}^{*}\tilde{l}_{Lj}
   -(m^{2}_{d})_{ij}\tilde{d}_{Ri}^{*}\tilde{d}_{Rj} \nonumber \\
& &-(m^{2}_{u})_{ij}\tilde{u}_{Ri}^{*}\tilde{u}_{Rj}  
   -(m^{2}_{q})_{ij}\tilde{q}_{Li}^{*}\tilde{q}_{Li}
   -m^{2}_{H_{1}}H_{1}^{*}H_{1}
   -m^{2}_{H_{2}}H_{2}^{*}H_{2}                      \nonumber \\
& &-[m_{0}(A_{e})_{ij}H_{1}\tilde{e}_{i}^{*}\tilde{e}_{j}
   + m_{0}(A_{d})_{ij}H_{1}\tilde{D}_{i}^{*}\tilde{Q}_{j} \nonumber \\
& &+m_{0}(A_{u})_{ij}H_{2}\tilde{U}_{i}^{*}\tilde{Q}_{j}
   - \mu B H_{1}H_{2} \nonumber \\ 
& &+\frac{1}{2}M_{1}\overline{\tilde{B_{R}}}\tilde{B_{L}}
   +\frac{1}{2}M_{2}\overline{\tilde{W_{R}}}\tilde{W_{L}}
   +\frac{1}{2}M_{3}\overline{\tilde{G_{R}}}\tilde{G_{L}} + h.c. ].
\end{eqnarray}

These terms are quadratic terms for scalar quarks, leptons and Higgs
fields, scalar trilinear terms ($A$ terms) and gaugino Majorana mass
terms. These terms are supposed to be generated from spontaneous
symmetry breaking of SUSY, presumably at some high energy scale in
some sector outside the MSSM dynamics, such as the supergravity or the
dynamical SUSY breaking sector. For more details, see for example
\citeasnoun{nill84} or \citeasnoun{habe85}.

\section{Differential Branching Ratio of the \meee Decay}\label{sc:KFDBR}

The differential branching ratio of the \meee decay is given in
Eq.(\ref{eq:m3ediffbr}). The kinematical functions appearing in
Eq.(\ref{eq:m3ediffbr}), $\alpha_{i}(x_1,x_2)$, $\beta_{i}(x_1,x_2)$
and $\gamma_{i}(x_1,x_2)$, are given as follows ($x_i \equiv
2E_i/m_{\mu}$ ($i=1,2$) are the normalized $e^{+}$ energies):

\begin{eqnarray}
\alpha_{1}(x_{1},x_{2}) & = & 8(2-x_{1}-x_{2})
                               (x_{1}+x_{2}-1), \nonumber \\
\alpha_{2}(x_{1},x_{2}) & = & 2\{x_{1}(1-x_{1})+x_{2}(1-x_{2})\}, \nonumber \\
\alpha_{3}(x_{1},x_{2}) & = & 8\{\frac{2x_{2}^{2}-2x_{2}+1}{1-x_{1}}
                                +\frac{2x_{1}^{2}-2x_{1}+1}{1-x_{2}}\}, 
                                \nonumber \\
\alpha_{4}(x_{1},x_{2}) & = & 32(x_{1}+x_{2}-1), \nonumber \\
\alpha_{5}(x_{1},x_{2}) & = & 8(2-x_{1}-x_{2}), \nonumber \\
\beta_{1}(x_{1},x_{2})  & = & 2\frac{ (x_{1}+x_{2})
         (x_{1}^{2}+x_{2}^{2})-3(x_{1}+x_{2})^{2}+6(x_{1}+x_{2})
                                    -4}{(2-x_{1}-x_{2})}, \nonumber \\
\beta_{2}(x_{1},x_{2})  & = & \frac{8}
                   {(1-x_{1})(1-x_{2})(2-x_{1}-x_{2})}\times \nonumber \\
                        &   &  \{
        2(x_{1}+x_{2})(x_{1}^{3}+x_{2}^{3})
     -4(x_{1}+x_{2})(2x_{1}^{2}+x_{1}x_{2}+2x_{2}^{2}) \nonumber \\
                        &   & +(19x_{1}^{2}+30x_{1}x_{2}+19x_{2}^{2})
     -12(2x_{1}+2x_{2}-1)
      \},                                \nonumber \\       
\gamma_{1}(x_{1},x_{2}) & = & 4\frac{\sqrt{(1-x_{1})(1-x_{2})(x_{1}+x_{2}-1)}
                             (x_{1}-x_{2})}{(2-x_{2}-x_{1})}, \nonumber \\
\gamma_{2}(x_{1},x_{2}) & = & 32\sqrt{\frac{(x_{1}+x_{2}-1)}
                                           {(1-x_{1})(1-x_{2})}}
                                \frac{(x_{1}+x_{2}-1)(x_{2}-x_{1})}
                                {(2-x_{1}-x_{2})}, \nonumber \\
\gamma_{3}(x_{1},x_{2}) & = & 16\sqrt{\frac{(x_{1}+x_{2}-1)}
                                           {(1-x_{1})(1-x_{2})}}
                                (x_{1}+x_{2}-1)(x_{2}-x_{1}),
                                \nonumber \\
\gamma_{4}(x_{1},x_{2}) & = & 8\sqrt{\frac{(x_{1}+x_{2}-1)}
                                           {(1-x_{1})(1-x_{2})}}
                                (2-x_{1}-x_{2})(x_{2}-x_{1}).
\end{eqnarray}

\newpage

\newpage

\newpage

\newpage

\begin{references}

\harvarditem{Abachi,~{\em et al.}}{1996}{abac96} Abachi,~S. (D0
Collaboration), 1996, Phys.~Rev.~Lett. {\bf 76}, 3271.

\harvarditem{Abdurashitov,~{\em et al.}}{1996}{abdu96}
Abdurashitov,~J.N., {\em et al.} (SAGE Collaboration), 1996,
Phys.~Rev.~Lett. {\bf 77}, 4708.

\harvarditem{Abegg,~{\em et al.}}{1996}{abeg96} Abegg,~R.,~{\em et
al.} (TRIUMF E614 Collaboration), 1996, An experimental proposal to
TRIUMF ``Precision Measurement of the Michel Parameters of $\mu^{+}$
Decay''.

\harvarditem{Abela,~{\em et al.}}{1996}{abel96} Abela, R.,~{\em et al.},
1996, Phys.~Rev.~Lett. {\bf 77}, 1950. 

\harvarditem{Abraham and Lampe}{1996}{abra96} Abraham,~K.J., and
B.~Lampe, 1996, Phys.~Lett. B {\bf 367}, 299.

\harvarditem{Abreu,~{\em et al.}}{1997}{abre97} Abreu,~P.,~{\em et al.}
(DELPHI Collaboration), 1997, Z.~Phys. C {\bf 73}, 243.

\harvarditem{Ahmad,~{\em et al.}}{1988}{ahma88} Ahmad, S.,~{\em et al.},
1988, Phy.~Rev. D {\bf 38} 2102.

\harvarditem{Akers,~{\em et al.}}{1995}{aker95} Akers,~R.,~{\em et
al.} (OPAL Collaboration), 1995, Z.~Phys. C {\bf 67}, 555.

\harvarditem{Altarelli,~{\em et al.}}{1977}{alta77} Altarelli,~G.,
L.~Baulieu, N.~Cabibbo, L.~Maiani, and R.~Petronzio, 1977,
Nucl.~Phys. B {\bf 125}, 285.

\harvarditem{Amaldi,~{\em et al.}}{1991}{amal91} Amaldi,~U.,
W.~de~Boer, and H.~Furstenau, 1991, Phys.~Lett. D {\bf 25}, 3092.

\harvarditem{Ambrose,~{\em et al.}}{1998}{ambr98} Ambrose,~D.,~{\em et
al.} (BNL E871 Collaboration), 1998, Phys.~Rev.~Lett. {\bf 81} 5734.

\harvarditem{Ankenbrandt {\em et al.}}{1999}{anke99}
Ankenbrandt,~C.M., {\em et al.} (Muon Collider Collaboration), 1999,
Phys.~Rev.~ST Accel.~Bean 2, 081001.

\harvarditem{Arbuzov,~{\em et al.}}{1998}{arbu98} Arbuzov,~A.B.,
O.~Krehl, E.A.~Kuraev, E.N.~Magar, B.G.~Shaikhatdenov, 1998,
Phys.~Lett. B {\bf 432}, 421.

\harvarditem{Arisaka,~{\em et al.}}{1998}{aris98} Arisaka,~K.,~{\em
et al.}, 1998, Phys.~Lett. B {\bf 432}, 230.

\harvarditem{Arkani-Hamed,~{\em et al.}}{1996a}{arka96a}
Arkani-Hamed,~N., H.-C. Cheng, and L.J.~Hall, 1996a, Phys.~Rev. D {\bf
53}, 413.

\harvarditem{Arkani-Hamed,~{\em et al.}}{1996b}{arka96b}
Arkani-Hamed,~N., H.-C.~Cheng, J.L.~Feng, and L.J.~Hall, 1996b,
Phys.~Rev.~Lett. {\bf 77}, 1937.

\harvarditem{Arkani-Hamed~{\em et al.}}{1997}{arka97} Arkani-Hamed,
J.L.~Feng, L.J.~Hall, and H.-C. Cheng, 1997, Nucl.~Phys. B {\bf 505}, 3.

\harvarditem{Athanassopoulos,~{\em et al.}}{1998}{atha98}
Athanassopoulos,~C., {\em et al.} (LSND Collaboration), 1998,
Phys.~ReV.~Lett. {\bf 81}, 1774.

\harvarditem{Babu and Mohapatra}{1995}{babu95}Babu, K.S., and R.N.
Mohapatra, 1995, Phys. Rev. Lett. {\bf 75}, 2276.

\harvarditem{Bachman,~{\em et al.}}{1997}{bach97}Bachman, M.,~{\em et
al.} (MECO Collaboration), 1997, An experimental proposal E940 to
Brookhaven National Laboratory AGS, ``A Search for
$\mu^{-}N\rightarrow e^{-}N$ with Sensitivity below $10^{-16}$''.

\harvarditem{Badertscher,~{\em et al.}}{1982}{bade82}Badertscher, A.,~{\em
et al.}, 1982, Nucl.~Phys. A {\bf 377}, 406.

\harvarditem{Balke,~{\em et al.}}{1988}{balk88} Balke,~B.,~{\em
et al.}, 1988, Phys.~Rev. D {\bf 37}, 587.

\harvarditem{Baranov,~{\em et al.}}{1991}{bara91}Baranov, V.A.,~{\em et
al.}, 1991, Sov. J. Nucl. Physics {\bf 53}, 802. 

\harvarditem{Barber,~{\em et al.}}{1969}{barb69}Barber,~W.C., {\em et
al.}, 1969, Phys.~Rev.~Lett. {\bf 22} 902.

\harvarditem{Barbieri and Hall}{1994}{barb94}Barbieri, R., and
L.J.~Hall, 1994, Phys. Lett. B {\bf 228} 212.

\harvarditem{Barbieri,~{\em et al.}}{1995a}{barb95a} Barbieri, R.,
L.J.~Hall, and A.~Strumia, 1995a, Nucl.~Phys. B {\bf 445}, 219.

\harvarditem{Barbieri,~{\em et al.}}{1995b}{barb95b} Barbieri, R.,
L.J.~Hall, and A.~Strumia, 1995b, Nucl.~Phys. B {\bf 449}, 437.

\harvarditem{Barbieri,~{\em et al.}}{1996}{barb96} Barbieri,~R.,
G.~Dvali, and L.J.~Hall, 1996, Phys.~Lett. B {\bf 377}, 76.

\harvarditem{Bardin,~{\em et al.}}{1984}{bard84}Bardin, G.,~{\em et
al.}, 1984, Phys.~Lett. B {\bf 137}, 135. 

\harvarditem{Barkov,~{\em et al.}}{1999}{bark99}Barkov,~L.M.,~{\em et
al.}, 1999, a research proposal to PSI ``Search for \mueg down to
$10^{-14}$ branching ratio''.

\harvarditem{Barnett,~{\em et al.}}{1994}{barn94}Barnett,~I.,~{\em et
al.}, 1994, a proposal for an experiment at PSI, ``Measurement of the
transverse polarization of positrons from the decay of polarized
muons''.

\harvarditem{Beer,~{\em et al.}}{1986}{beer86} Beer,~G.A.,
{\em et al.}, 1986, Phys.~Rev.~Lett. {\bf 57}, 671.

\harvarditem{Behrends,~{\em et al.}}{1956}{behr56} Behrends,~R.E.,
R.J.~Finkelstein, and A.~Sirlin, 1956, Phys.~Rev. {\bf 101}, 866.

\harvarditem{Bellgardt,~{\em et al.}}{1988}{bell88} Bellgardt, U.,~{\em et
al.}, 1988, Nucl. Phys. B {\bf 229}, 1. 

\harvarditem{Beltrami,~{\em et al.}}{1987}{belt87} Beltrami,~I., {\em et
al.}, 1987, Phys.~Lett. B {\bf 194}, 326.

\harvarditem{Bernabeu,~{\em et al.}}{1993}{bern93} Bernabeu,~J.,
E.~Nardi, and D.~Tommasini, 1993, Nucl.~Phys. B {\bf 409}, 69. 

\harvarditem{Bertl,~{\em et al.}}{1984}{bert84} Bertl, W.,~{\em et
al.}, 1984, Phys. Lett. B {\bf 140}, 299.

\harvarditem{Bertl,~{\em et al.}}{1985}{bert85} Bertl, W.,~{\em et
al.}, 1985, Nucl. Phys. B {\bf 260}, 1. 

\harvarditem{Bilenky,~{\em et al.}}{1977}{bile77} Bilenky,~S.M.,
S.T.~Petcov, B.~Pontecorvo, 1977, Phys.~Lett. B {\bf 67}, 309.

\harvarditem{Bilenky and Petcov}{1987}{bile87} Bilenky,~S.M.,
and S.T.~Petcov, 1987, Rev.~Mod.~Phys. {\bf 59}, 671.

\harvarditem{Bjorken,~{\em et al.}}{1977}{bjor77} Bjorken,~J.D.,
K.~Lane, and S.~Weinberg, 1977, Phys.~Rev. D {\bf 16}, 1474.

\harvarditem{Blackmore,~{\em et al.}}{1997}{blac97} Blackmore, E.,~{\em et
al.}, 1997, ``MUONS at TRIUMF'', TRIUMF internal report, unpublished.

\harvarditem{Bliss,~{\em et al.}}{1998}{blis98} Bliss,~D.W.,~{\em et
al.}, 1998, Phys.~Rev. D {\bf 57}, 5903.

\harvarditem{Bolton,~{\em et al.}}{1984}{bolt84} Bolton,~R.D.,~{\em et
al.}, 1984, Phys.~Rev.~Lett. {\bf 53}, 1415.

\harvarditem{Bolton,~{\em et al.}}{1988}{bolt88} Bolton,~R.D.,~{\em et
al.}, 1988, Phys.~Rev. D {\bf 38}, 2077.

\harvarditem{Borzumati and Masiero}{1986}{borz86} Borzumati, F., and
A. Masiero, 1986, Phys.~Rev. {\bf 57}, 961.

\harvarditem{Bouchiat and Michel}{1957}{bouc57} Bouchiat,~C., and
L.~Michel, 1957, Phys.~Rev. {\bf 106}, 170.

\harvarditem{Bowser-Chao and Keung}{1997}{bows97} Bowser-Chao,~D., 
and W.-K.~Keung, 1997, Phy.~Rev. D {\bf 56} 3924.

\harvarditem{Brooks,~{\em et al.}}{1999}{broo99} Brooks,~M.L.,~{\em et
al.} (MEGA Collaboration), 1999, hep-ex/9905013, submitted to
Phys.~Rev.~Lett.

\harvarditem{Bryman,~{\em et al.}}{1972}{brym72} Bryman,~D.A.,
M.~Blecher, K.~Gotow, and R.J.~Powers, 1972, Phys.~Rev.~Lett. {\bf
28}, 1469.

\harvarditem{Bryman,~{\em et al.}}{1985}{brym85} Bryman,~D.A., {\em et 
al.}, 1985, Phys. Rev. Lett. {\bf 55} 465.

\harvarditem{Burkard,~{\em et al.}}{1985a}{burk85a} Burkard, H.,~{\em et
al.}, 1985a, Phys.~Lett. B {\bf 150}, 242. 

\harvarditem{Burkard,~{\em et al.}}{1985b}{burk85b} Burkard, H.,~{\em et
al.}, 1985b, Phys.~Lett. B {\bf 160}, 343.

\harvarditem{Cao~{\em et al.}}{1999}{cao99} Cao,~J.-J., T.~Han,
X.~Zhang, and G.-R.~Lu, 1999, Phy.~Rev. D {\bf 59} 09501.

\harvarditem{Carena,~{\em et al.}}{1997}{carena97} Carena, M.,
G.F.~Giudice, and C.E.M. Wagner, 1997, Phys.~Lett. B {\bf 390}, 234.

\harvarditem{Carey,~{\em et al.}}{1999a}{care99a} Carey, R.M.,~{\em et
al.}, 1999a, Phys.~Rev.~Lett. {\bf 82}, 1632.

\harvarditem{Carey,~{\em et al.}}{1999b}{care99b} Carey, R.M.,~{\em et
al.}, 1999b, an experimental proposal to PSI, ``A Precision
Measurement of the Positive Muon Lifetime Using a Pulsed Muon Beam and 
the $\mu Lan$ Detector''.

\harvarditem{Carlson and Frampton}{1992}{carl92} Carlson, E.D., and
P.H. Frampton, 1992, Phys. Lett. B {\bf 283}, 123.

\harvarditem{Cavallo,~{\em et al.}}{1999}{cava99} Cavallo, F.R. {\em et
al.}, 1999, a letter of intent to PSI, ``A Preceision Measurement of
the $\mu^{+}$ Lifetime ($G_F$) with the FAST detector''.

\harvarditem{CERN Report}{1999}{cern99} CERN Report, 1999,
CERN 99-02 and ECFA 99-197, ``Prospective Study of Muon Storage Rings
at CERN'', edited by B.~Autin, A.~Blondel, and J.~Ellis.

\harvarditem{Chaichian and Huitu}{1996}{chai96} Chaichian, M., and
K. Huitu, 1996, Phys.~Lett. B {\bf 384}, 157.

\harvarditem{Chang and Keung}{1989}{chan89} Chang, D., and W.Y.~Keung,
1989, Phys.~Rev.~Lett. {\bf 62}, 2583.

\harvarditem{Chattopadhyay and Nath}{1996}{chat96} Chattopadhyay,~U.
and P.~Nath, 1996, Phys.~Rev. D {\bf 53}, 1648.

\harvarditem{Cheng and Li}{1977a}{chen77a} Cheng, T.P., and L.F.~Li,
1977a, Phys.~Rev.~Lett. {\bf 38}, 381.

\harvarditem{Cheng and Li}{1977b}{chen77b} Cheng, T.P., and L.F.~Li,
1977b, Phys.~Rev. D {\bf 16}, 1425.

\harvarditem{Cheng and Li}{1980}{chen80} Cheng, T.P., and L.F.~Li,
1980, Phys.~Rev.~Lett. {\bf 45}, 1908.

\harvarditem{Chiang,~{\em et al.}}{1993}{chia93} Chiang, H.C., E.~Oset,
T.S.~Kosmas, A.~Faessler, and J.D.~Vergados, 1993, Nucl. Phys. A {\bf
559}, 526.

\harvarditem{Choi~{\em et al.}}{1998}{choi98} Choi,~S.Y., C.S.~Kim,
Y.J.~Kwon, and S.-H.~Lee, 1998, Phy.~Rev. D {\bf 57} 7023.

\harvarditem{Ciafaloni,~{\em et al.}}{1996}{ciaf96} Ciafaloni, P., 
A.~Romanino and A.~Strumia, 1996, Nucl. Phys. {\bf B458}, 3.

\harvarditem{Cleveland,~{\em et al.}}{1998}{clev98} Cleveland, B.T.
{\em et al.}, 1998, Astrophys. J. {\bf 496}, 505.

\harvarditem{Cohen and Taylor}{1987}{cohe87} Cohen,~E.R., and
B.N.~Taylor, 1987, Rev.~Mod.~Phys. {\bf59}, 1121.

\harvarditem{Cohen,~{\em et al.}}{1996}{cohe96} Cohen,~A.G.,
D.B.~Kaplan, and A.E.~Nelson, 1996, Phys.~Lett. B {\bf 388}, 588.

\harvarditem{Colemen and Glashow}{1999}{cole99} Coleman,~S., and
S.L. Glashow, 1999, Phy.~Rev. D {\bf 59} 116008.

\harvarditem{Conversi,~{\em et al.}}{1947}{conv47}Conversi, M., E.
Pancini, and O. Piccioni, 1947, Phys.~Rev. {\bf 71}, 209.

\harvarditem{Crittenden,~{\em et al.}}{1961}{crit61} Crittenden,~R.R.,
W.D.~Walker, and J.~Ballam, 1961, Phys.~Rev. {\bf 121}, 1823.

\harvarditem{Czarnecki,~{\em et al.}}{1997}{czar97} Czarnecki, A.,
W.J. Marciano, and K. Melnikov, 1997, in {\em Proceedings of Workshop
on Physics at the First Muon Collider and at the Front End of the Muon
Collider}, edited by S.H.~Geer and R.~Raja, Fermilab, U.S.A., AIP
Conference Proceedings 435, p.409.

\harvarditem{Czarnecki and Marciano}{1998}{czar98} Czarnecki, A.,
and W.J. Marciano, 1998, ``Lepton anomalous magnetic moments $-$ a
theory update'', preprint BNL-HET-98/43, hep-ph/9810512.

\harvarditem{Danby,~{\em et al.}}{1962}{danb62} Danby, G., J.M.
Gaillard, K.~Goulianos, L.M.~Lederman, N.~Mistry, M.~Schwartz, and
J.~Steinberger, 1962, Phys.~Rev.~Lett. {\bf 9}, 36.

\harvarditem{Depommier,~{\em et al.}}{1977}{depo77} Depommier, P. 
{\em et al.}, 1977, Phys.~Rev.~Lett. {\bf 39}, 1113.

\harvarditem{Depommier}{1987}{depo87} Depommier, P., 1987, in
{\em Neutrinos} edited by H.V. Klapdor, (Springer, Berlin), p.265.

\harvarditem{Depommier and Leroy}{1995}{depo95} Depommier, P., and C.
Leroy, 1995, Reports on Progress in Phys. {\bf 58}, 61.

\harvarditem{Derenzo}{1969}{dere69} Derenzo,~S.E., 1969, Phys.~Rev. {\bf
181}, 1854.

\harvarditem{Deshpande,~{\em et al.}}{1996}{desh96} Deshpande,~N.G., 
B. Dutta and E. Keith, 1996, Phys.~Rev. D {\bf 54}, 730.

\harvarditem{Dimopoulos and Georgi}{1981}{dimo81} Dimopoulos,~S., and
H.~Georgi, 1981, Nucl.~Phys. B {\bf 193}, 150.

\harvarditem{Dimopoulos and Hall}{1995}{dimo95} Dimopoulos,~S., and
L.~Hall, 1995, Phys.~Lett. B {\bf 344}, 185.

\harvarditem{Dine and Nelson}{1993}{dine93} Dine,~M., and A.E.~Nelson,
1993, Phys.~Rev. D {\bf 48}, 1277.

\harvarditem{Dine,~{\em et al.}}{1995}{dine95} Dine,~M., A.E.~Nelson,
and Y.~Shirman, 1995, Phys.~Rev. D {\bf 51}, 1362.

\harvarditem{Dine,~{\em et al.}}{1996}{dine96} Dine,~M., A.E.~Nelson,
Y.~Nir, and Y.~Shirman, 1996,  Phys.~Rev. D {\bf 53}, 2658.

\harvarditem{Dine,~{\em et al.}}{1997}{dine97} Dine,~M., Y.~Nir, and
Y.~Shirman, 1997, Phy.~Rev. D {\bf 55} 1501.  

\harvarditem{Dohmen,~{\em et al.}}{1993}{dohm93} Dohmen, C.,~{\em et
al.} (SINDRUM II Collaboration), 1993, Phys.~Lett. B {\bf 317}, 631.

\harvarditem{Dubovsky and Gorbunov}{1998}{dubo98} Dubovsky,~S.L., and
D.S.~Gorbunov, 1998, Phys.~Lett. B {\bf 419}, 223.

\harvarditem{Duong,~{\em et al.}}{1996}{duon96} Duong,~T.V., B.~Dutta, 
and E.~Keith, 1996, Phys.~Lett. B {\bf 378}, 128.

\harvarditem{Dzhilkibaev and Lobashev}{1989}{dzhi89} Dzhilkibaev,
R.M., and V.M. Lobashev, 1989, Sov.~J.~Nucl.~Phys. {\bf 49}, 384. 

\harvarditem{Eckstein and Pratt}{1959}{ecks59} Eckstein, S.G., and
R.H. Pratt, 1959, Ann.~of~Phys.~{\bf 8}, 297 (1959).

\harvarditem{Edwards,~{\em et al.}}{1997}{edwa97} Edwards,~K.W.,~{\em et
al.} (CLEO Collaboration), 1997, Phys.~Rev. B {\bf 55}, R3919.

\harvarditem{Eichenberger,~{\em et al.}}{1984}{eich84}
Eichenberger,~W., R.~Engfer, and A.~van~der~Schaaf, 1984, Nucl.~Phys.
A {\bf 412}, 523.

\harvarditem{Eitel,~{\em et al.}}{1999}{eite99}
Eitel,~K., {\em et al.} (KARMEN Collaboration), 1999,
Nucl.~Phys.~Proc.~Suppl. {\bf 70}, 210.

\harvarditem{Ellis,~{\em et al.}}{1991}{ell91} Ellis,~J., S.~Kelly, and
D.~Nanopoulos, 1991, Phys.~Lett. B {\bf 260}, 131.

\harvarditem{Engfer and Walter}{1986}{engf86} Engfer R., and H.K.
Walter, 1986, Ann.~Rev.~Nucl.~Part.~Sci. {\bf 36}, 327. 

\harvarditem{Evseev}{1975}{evse75} Evseev,~V.S., 1975, in {\em Muon Physics
Vol.III Chemistry and Solids}, edited by V.W. Hughes and C.S.~Wu
(Academic Press), p.236.

\harvarditem{Faessler,~{\em et al.}}{1999}{faes99} Faessler,~A.,
T.S.~Kosmas, S.~Kovalemko, and J.D.~Vergados, 1999, hep-ph/9904335,
``Constraints on $R$-parity Violating Supersymmetry from \muec Nuclear 
Conversion''.

\harvarditem{Feinberg}{1958}{fein58} Feinberg, G., 1958,
Phys.~Rev. {\bf 116}, 1482.

\harvarditem{Feinberg and Weinberg}{1961}{fein61} Feinberg, G., and
S. Weinberg, 1961, Phys.~Rev.~Lett. {\bf 123}, 1439.

\harvarditem{Fetscher,~{\em et al.}}{1986}{fets86} Fetscher,~W.,
H.J.~Gerber, and K.F.~Johnson, 1986, Phys.~Lett. B {\bf 173}, 102.

\harvarditem{Fetscher and Gerber}{1995}{fets95} Fetscher,~W., and
H.J.~Gerber, 1995, in {\em Precision Tests of the Standard
Electroweak Model} edited by P.~Langacker (World Scientific), p.657.

\harvarditem{Fetscher and Gerber}{1998}{fets98} Fetscher,~W. and
H.J.~Gerber, 1998, in {\em Particle Data Group: Review of Particle
Properties}, Euro. Phys. Journal {\bf 3}, p.282.

\harvarditem{Feynman and Gell-Mann}{1958}{feyn58} Feynman, R.P., and
M. Gell-Mann, 1958, Phys.~Rev. {\bf 109}, 193.

\harvarditem{Fisher, Kayser, and McFarland}{1999}{fish99} Fisher,~P.,
B.~Kayser, and K.S.~McFarland, 1999, ``Neutrino Mass and
Oscillation'', hep-ph/9906244, submitted to Ann.~Rev.~Nucl.~Part.~Sci.

\harvarditem{Frampton and Lee}{1990}{fram90} Frampton, P.H., and
B.H. Lee, 1990, Phys.~Rev.~Lett. {\bf 64}, 619.

\harvarditem{Frampton}{1992}{fram92} Frampton, P.H., 1992,
Phys.~Rev.~Lett. {\bf 69}, 2889.

\harvarditem{Frampton}{1992b}{fram92b} Frampton, P.H., 1992b,
Phys.~Rev.~D {\bf 45}, 4240.

\harvarditem{Frankel}{1975}{fran75} Frankel, S., 1975, in {\em Muon
Physics II: Weak Interaction}, edited by V.W. Hughes and C.S. Wu
(Academic, New York), p83.

\harvarditem{Freedman, {\em et al.}}{1993}{free93} Freedman,~S.J.,
{\em et al.}, 1993, Phys.~Rev. D {\bf 47}, 811.

\harvarditem{Fronsdal and \"Uberall}{1959}{fron59} Fronsdal, C., and
H. \"Uberall, 1959, Phys.~Rev.~{\bf 118}, 654.

\harvarditem{Fujii,~{\em et al.}}{1993}{fuji93} Fujii,~H., S.~Nakamura, 
and K.~Sasaki, 1993, Phys.~Lett. B {\bf 299}, 342.

\harvarditem{Fujii,~{\em et al.}}{1994}{fuji94} Fujii,~H., Y.~Mimura,
K.~Sasaki, and T.~Sasaki, 1994,  Phys.~Rev. D {\bf 49}, 559.

\harvarditem{Fukuda,~{\em et al.}}{1996}{fuku96} Fukuda, Y.,~{\em et
al.} (Kamiokande Collaboration), 1996, Phys.~Rev.~Lett. {\bf
77}, 1683. 

\harvarditem{Fukuda,~{\em et al.}}{1998a}{fuku98a} Fukuda, Y.,~{\em et
al.} (Super-Kamiokande Collaboration), 1998a, Phys.~Rev.~Lett. {\bf
81}, 1158, and Erratum: 1998, {\em ibid.} {\bf 81} 4279.

\harvarditem{Fukuda,~{\em et al.}}{1998b}{fuku98b} Fukuda, Y.,~{\em et
al.} (Super-Kamiokande Collaboration), 1998b, Phys.~Rev.~Lett. {\bf
81}, 1562. 

\harvarditem{Fukugita and Yanagida}{1994}{fuku94} Fukugita,~M. and
T.~Yanagida, 1994, in {\em Physics and Astrophysics of Neutrinos},
edited by M.~Fukugita and A.~Suzuki (Springer-Verlag), p.1.

\harvarditem{Gabrielli and Sarid}{1997}{gabr97} Gabrielli,~E., and
U.~Sarid, 1997, Phys.~Rev.~Lett. {\bf 79}, 4752.

\harvarditem{Geer}{1998}{geer98} Geer, S., 1998, Phys.~Rev. D {\bf 57},
6989, and Erratum: 1999, {\em ibid.} D {\bf 59}, 039903.

\harvarditem{Gell-Mann,~{\em et al.}}{1979}{gell79} Gell-Mann, M.,
P. Ramond, and R. Slansky, 1979, in {\em Supergravity}, edited by
D. Freedman and P. van Nieuwenhuizen, (North Holland, Amsterdam),
p.315.

\harvarditem{Giovanetti,~{\em et al.}}{1984}{giov84}Giovanetti,
K.L.,~{\em et al.}, 1984, Phys.~Rev. D {\bf 29}, 343.

\harvarditem{Giudice and Rattazzi}{1998}{giud98} Giudice,~G.F., and
R.~Rattazzi, 1998, ``Theories with Gauge Mediated Supersymmetry
Breaking'', hep-ph/9801271, submitted to Phys.~Rep.

\harvarditem{Glashow}{1961}{glas61} Glashow,~S.L., 1961,
Phys.~Rev.~Lett. {\bf 6}, 196.

\harvarditem{Glashow and Krauss}{1987}{glas87} Glashow,~S.L., and
L.M.~Krauss, 1987, Phys.~Lett. B {\bf 190}, 199.

\harvarditem{Glashow, {\em et al}}{1999}{glas99} Glashow,~S.L.,
P.J.~Kerman, and L.M.~Krauss, 1999, Phys.~Lett. B {\bf 445}, 412.

\harvarditem{G\'{o}mez and Goldberg}{1996}{gome96} Gomez,~M.E. and 
H.~Goldberg, 1996, Phys.~Rev. D {\bf 53}, 5244.

\harvarditem{Gordeev, {\em et al.}}{1997}{gord97} Gordeev,~V.A., {\em
et al.}, 1997, Phys. Atom. Nucl. {\bf 60}, 1164: 1997, Yad. Fiz. {\bf
60}, 1291.

\harvarditem{Haber and Kane}{1985}{habe85} Haber,~H.E., and
G.L.~Kane, 1985, Phys.~Rep. {\bf 117}, 75.

\harvarditem{Hall,~{\em et al.}}{1986}{hall86}Hall, L.J.,
V.A. Kostelecky and S. Raby, 1986, Nucl.~Phys. B {\bf 267}, 415.

\harvarditem{Halprin}{1982}{halp82} Halprin, A.,
1982, Phys.~Rev.~Lett. {\bf 48}, 1313.

\harvarditem{Halprin and Masiero}{1993}{halp93} Halprin, A. and 
A. Masiero, 1993, Phys.~Rev. D {\bf 48}, 2987.

\harvarditem{Hampel,~{\em et al.}}{1999}{hamp99} Hampel, W.,~{\em et
al.} (Gallex Collaboration), 1999, Phys.~Lett. B {\bf 447}, 127.

\harvarditem{H\"anggi,~{\em et al.}}{1974}{hang74}H\"anggi, P., R.D.
Viollier, U. Raff, and K. Adler, 1974, Phys. Lett. B {\bf 51}, 119.

\harvarditem{Herczeg}{1986}{herc86}Herczeg, P., 1986, Phys.~Rev.~D
{\bf 34}, 3449.

\harvarditem{Herczeg and Mohapatra}{1992}{herc92} Herczeg, P., and
R.N. Mohapatra, 1992, Phys.~Rev.~Lett. {\bf 69}, 2475.

\harvarditem{Herczeg}{1995}{herc95} Herczeg, P., 1995, in {\em
Precision Tests of the Standard Electroweak Model} edited by
P.~Langacker (World Scientific), p.786.

\harvarditem{Herzog,~{\em et al.}}{1980}{herz80}Herzog, F., and K.
Adler, 1980, Helv. Phys. Acta {\bf 53}, 53. 

\harvarditem{Hincks and Pontecorvo}{1947}{pont47} Hincks, E.P., and B. 
Pontecorvo, 1947, Phys.~Rev.~Lett. {\bf 73}, 246.

\harvarditem{Hirouchi and Tanaka}{1998}{hiro98} Hirouchi,~M., and
M.~Tanaka, 1998, Phy.~Rev. D {\bf 58} 032004.

\harvarditem{Hisano,~{\em et al.}}{1995}{hisa95} Hisano, J., T.~Moroi,
K.~Tobe, M.~Yamaguchi, and T.~Yanagida, 1995, Phys.~Lett. B {\bf 357},
576.

\harvarditem{Hisano,~{\em et al.}}{1996}{hisa96} Hisano, J., T.~Moroi,
K.~Tobe, and M.~Yamaguchi, 1996, Phys.~Rev. D {\bf 53}, 2442.

\harvarditem{Hisano,~{\em et al.}}{1997}{hisa97} Hisano, J., T.~Moroi,
K.~Tobe, and M.~Yamaguchi, 1997, Phys.~Lett. B {\bf 391}, 341, and
Errata: 1997, {\em ibid.} B {\bf 397}, 357.

\harvarditem{Hisano,~{\em et al.}}{1998a}{hisa98a} Hisano, J.,
D.~Nomura, and T.~Yanagida, 1998a, Phys.~Lett. B {\bf 437}, 351. 

\harvarditem{Hisano,~{\em et al.}}{1998b}{hisa98b} Hisano, J.,
D.~Nomura, Y.~Okada, Y.~Shimizu, and M.~Tanaka, 1998b, Phys.~Rev. D
{\bf 58}, 116010.

\harvarditem{Hisano and Nomura}{1999a}{hisa99a} Hisano, J., and
D.~Nomura, 1999a, Phys.~Rev. D {\bf 59}, 116005.

\harvarditem{Hisano~{\em et al.}}{1999b}{hisa99b} Hisano,~J.,
M.M.~Nojiri, Y.~Shimizu, and M~.Tanaka, 1999b, Phy.~Rev. D {\bf 60}
055008.

\harvarditem{Honecker,~{\em et al.}}{1996}{hone96} Honecker, W.,~{\em
et al.} (SINDRUM II Collaboration), 1996, Phys. Rev. Lett. {\bf 76},
200.

\harvarditem{Horikawa and Sasaki}{1996}{hori96} Horikawa, K., and
K. Sasaki, 1996, Phys.~Rev. D {\bf 53}, 560.

\harvarditem{Hou and Wong}{1995}{hou95} Hou,~W.S., and G.G.~Wong,
1995, Phys.~Lett. B {\bf 357}, 145.

\harvarditem{Hou and Wong}{1996}{hou96} Hou,~W.S., and G.G.~Wong,
1996, Phys.~Rev. D {\bf 53}, 1537.

\harvarditem{Hou}{1996b}{hou96b} Hou,~W.S., 1996b, Nucl. Phys. B {\bf
51A}, 40. 

\harvarditem{Huber,~{\em et al.}}{1990}{hube90} Huber,~T.M., {\em et
al.}, 1990, Phys.~Rev. D {\bf 41}, 2709.

\harvarditem{Hughes,~{\em et al.}}{1960}{hugh60} Hughes, V.W.,
D.W. McColm, K. Ziock and R. Prepost, 1960, Phys.~Rev.~Lett. {\bf 5},
63.

\harvarditem{Hughes and Kinoshita}{1999}{hugh99} Hughes, V.W.,
and T.~Kinoshita, 1999, Rev.~Mod.~Phys. {\bf 71}, S133.

\harvarditem{Huitu,~{\em et al.}}{1998}{huit98} Huitu,~K., J.~Maalampi,
M.~Raidal, and A.~Santamaria, 1998, Phys.~Lett. B {\bf 430}, 355.

\harvarditem{Ishida and Nagamine}{1998}{ishi98} Ishida,~K., and
K.~Nagamine, 1998, in {\em Proceedings of International Workshop on
JHF Science}, edited by J.~Chiba, M.~Furusaka, H.~Miyatake and
S.~Sawada, Vol.II, p.12.

\harvarditem{JAERI/KEK Joint Project}{1999}{jhf99} JAERI/KEK Joint
Project Proposal, 1999, unpublished.

\harvarditem{Jodidio,~{\em et al.}}{1986}{jodi86}Jodidio, A.,~{\em et
al.}, 1986, Phys.~Rev. D {\bf 34}, 1967, and Erratum: 1988, {\em
ibid.} {\bf 37}, 237. 

\harvarditem{Kaulard,~{\em et al.}}{1998}{kaul98} Kaulard,~J.,~{\em et al.},
1998, Phys. Lett. B {\bf 422}, 334.

\harvarditem{Kersch,~{\em et al.}}{1988}{kers88} Kersch,~A. N.~Kraus,
and R.~Engfer, 1988, Nucl.~Phys. A {\bf 485}, 606.

\harvarditem{Kim,~{\em et al.}}{1982}{kim97} Kim,~J.E., P.~Ko, and
D.G.~Lee, 1997, Phys.~Rev. D {\bf 56}, 100.

\harvarditem{King and Oliveira}{1999}{king98} King,~S.F., and
M.~Oliveira, 1999, Phys.~Rev. D {\bf 60}, 035003.

\harvarditem{Kinnison,~{\em et al.}}{1982}{kinn82} Kinnison, 
W.W.,~{\em et al.}, 1982, Phys. Rev. D {\bf 25}, 2846.

\harvarditem{Kinoshita and Sirlin}{1957a}{kino57a} Kinoshita,~T., and
A.~Sirlin, 1957a, Phys.~Rev. {\bf 107}, 593.

\harvarditem{Kinoshita and Sirlin}{1957b}{kino57b} Kinoshita,~T., and
A.~Sirlin, 1957b, Phys.~Rev. {\bf 108}, 844.

\harvarditem{Kinoshita and Sirlin}{1959}{kino59} Kinoshita,~T., and
A.~Sirlin, 1959, Phys.~Rev. {\bf 113}, 1652.

\harvarditem{Kitano and Yamomoto}{1999}{kita99} Kitano,~R., and
K.~Yamamoto, 1999, KEK preprint KEK-TH-62, hep-ph/995459, ``Lepton
flavor violation in a supersymmetic $E_6$ type model''.

\harvarditem{Kobayashi and Maskawa}{1973}{koba73} Kobayashi, M., and
T. Maskawa, 1973, Prog. Theor. Phys. {\bf 49}, 652.

\harvarditem{Korenchenko,~{\em et al.}}{1976}{kore76} Korenchenko,
S.M.,~{\em et al.}, 1976, Sov. Phys. JETP {\bf 43}, 1.

\harvarditem{Kosmas,~{\em et al.}}{1990}{kosm90} Kosmas, T.S., and
J.D.~Vergados, 1990, Nucl.~Phys. A {\bf 510}, 641.

\harvarditem{Kosmas,~{\em et al.}}{1994}{kosm94} Kosmas, T.S.,
J.D.~Vergados, O.~Civitarese, and A. Faessler, 1994, Nucl.~Phys. A
{\bf 570}, 637.

\harvarditem{Kosmas and Vergados}{1996}{kosm96} Kosmas, T.S., and
J.D.~Vergados, 1996, Phys.~Rep. {\bf 264}, 251.

\harvarditem{Kosmas,~{\em et al.}}{1998}{kosm98} Kosmas, T.S.,
J.D.~Vergados, and A. Faessler, 1998, Phys. Atom. Nucl. {\bf 61},
1161: 1998, Yad. Fiz. {\bf 61}, 1261. 

\harvarditem{Krasnikov}{1994}{kras94} Krasnikov,~N.V., 1994,
Mod.~Phys.~Lett. A {\bf 9} 791.

\harvarditem{Krasnikov}{1996}{kras96} Krasnikov,~N.V., 1996,
Phys.~Lett. B {\bf 388}, 783.

\harvarditem{Krolak,~{\em et al.}}{1994}{krol94} Krolak,~P., {\em et
al.}, 1994, Phys.~Lett. B {\bf 320}, 407.

\harvarditem{Kuno, {\em et al.}}{1986}{kuno86} Kuno,~Y., K.~Nagamine,
and T.~Yamazaki, 1986, Nucl.~Phys. A {\bf 475}, 615.

\harvarditem{Kuno and Okada}{1996}{kuno96} Kuno,~Y. and Y.~Okada,
1996, Phys.~Rev.~Lett. {\bf 77}, 434.

\harvarditem{Kuno,~{\em et al.}}{1997a}{kuno97a} Kuno, Y., A.~Maki, and
Y.~Okada, 1997a, Phys.~Rev. D {\bf 55}, 2517.

\harvarditem{Kuno}{1997b}{kuno97b} Kuno, Y., 1997b, in {\em
Proceedings of Workshop on Physics at the First Muon Collider and at
the Front End of the Muon Collider}, edited by S.H.~Geer and R.~Raja
(AIP Conference Proceedings 435), p.261.

\harvarditem{Kuno}{1998}{kuno98} Kuno, Y.,~{\em et al.},
1998, in {\em Proceedings of Workshop on High Intensity Secondary Beam
with Phase Rotation}, edited by Y.~Kuno and N.~Sasao, p.71.

\harvarditem{Lagarrigue and Peyrou}{1952}{laga52} Lagarrigue, A., and
C. Peyrou, 1952, Acad. Sci. Paris, {\bf 234}, 873.

\harvarditem{Langacker and Sankar}{1989}{lang89} Langacker,~P., and
S.U.~Sankar, 1989, Phys.~Rev. D {\bf 40}, 1569.

\harvarditem{Langacker and Luo}{1991}{lang91} Langacker,~P., and
M.~Luo, 1991, Phys.~Rev. D {\bf 44}, 817.

\harvarditem{Lee,~{\em et al.}}{1977a}{leep77} Lee,~B.W., S.~Pakvasa,
R.E.~Shrock, and H.~Sugawara, 1977a, Phys.~Rev.~Lett. {\bf 38}, 937.

\harvarditem{Lee and Shrock}{1977b}{lee77} Lee,~B.W., and
R.E.~Shrock, 1977b, Phys.~Rev. D {\bf 16}, 1444.

\harvarditem{Lee,~{\em et al.}}{1990}{lee90} Lee,~A.M., {\em et al.},
1990, Phys.~Rev.~Lett. {\bf 64}, 165.

\harvarditem{Lenard}{1953}{lena53} Lenard,~A., 1953, Phys.~Rev. {\bf
90}, 968.

\harvarditem{LEP and SLD Electroweak Working Group}{1999}{lep99} LEP
Electroweak Working Group and SLD Heavy Flavor and Electroweak Goups,
1999, CERN-EP/99-15 preprint ``A Combination of Preliminary
Electroweak Measurements and Constraints on the Standard Model''.

\harvarditem{Lobashev}{1998}{loba98} Lobashev,~V.M., 1998, in {\em
Proceedings of 18th Workshop on Physics in Collisions''}, edited by
S.~Bianco, A.~Calcaterra, P.~de Simone, F.L.~Fabbri (Frascati Physics
series, 11), p.179.

\harvarditem{Lokonathan and Steinberger}{1955}{loko55} Lokonathan, S. and
J.~Steinberger, 1955, Phys.~Rev. {\bf 98}, 240.

\harvarditem{Lopez,~{\em et al.}}{1994}{lope94} Lopez,~J.L.,
D.V.~Nanopoulos, and X.~Wang, 1994, Phys.~Rev. D {\bf 49}, 366.

\harvarditem{Maki,~{\em et al.}}{1962}{maki62} Maki, Z., M.~Nakagawa,
and S.~Sakata, 1962, Prog. Theor. Phys. {\bf 28}, 870.

\harvarditem{Marciano and Sanda}{1977a}{marc77a} Marciano, W.J., and
A.I.~Sanda, 1977a, Phys.~Lett. B {\bf 67}, 303.

\harvarditem{Marciano and Sanda}{1977b}{marc77b} Marciano, W.J., and
A.I.~Sanda, 1977b, Phys.~Rev.~Lett. {\bf 38}, 1512.

\harvarditem{Marciano and Sirlin}{1988}{marc88} Marciano, W.J., and
A.~Sirlin, 1988, Phys.~Rev.~Lett. {\bf 61}, 1815.

\harvarditem{Marciano}{1999}{marc99} Marciano, W.J., 1999, preprint
hep-ph/9903451, ``Fermi Constants and New Physics''.

\harvarditem{Marshall,~{\em et al.}}{1982}{mars82} Marshall,~G.M.,
{\em et al.}, 1982, Phys.~Rev. D {\bf 25}, 1174.

\harvarditem{Matthias,~{\em et al.}}{1991}{matt91} Matthias, B.,~{\em et al.},
1991, Phys.~ReV.~Lett. {\bf 66}, 2716.

\harvarditem{Michel}{1950}{mich50} Michel,~L., 1950, Proc.~Phys.~Soc.
A {\bf 63}, 514.

\harvarditem{Mikheyev and Smirnov}{1985}{mikh85} Mikheyev, S.P., and
A.Yu. Smirnov, 1985, Yad. Fiz. {\bf 42}, 1441 [Sov. J. Nucl. Phys.
{\bf 42}, 913.]

\harvarditem{Mohapatra}{1992}{moha92} Mohapatra,~R.N., 1992,
Z.~Phys. C {\bf 56}, 117.

\harvarditem{Mohapatra and Pal}{1998}{moha98} Mohapatra,~R.N., and
P.B.~Pal, 1998, {\em Massive Neutrinos in Physics and Astrophysics
(Second Edition)}, (World Scientific).

\harvarditem{Molzon}{1997}{molz97} Molzon,~W.R., 1997, in {\em
Proceedings of Workshop on Physics at the First Muon Collider and at
the Front End of the Muon Collider}, edited by S.H.~Geer and R.~Raja
(AIP Conference Proceedings 435), p.152.

\harvarditem{Moroi}{1996}{moro96} Moroi, T., 1996, Phys.~Rev. D {\bf 53},
6565.

\harvarditem{Muon Collider Collaboration}{1996}{mumu96} Muon Collider
Collaboration, 1996, ``$\mu^{+}\mu^{-}$ Collider $-$ A Feasibility
Study'', preprint BNL-52503, Fermilab-Conf.-96/092, LBNL-38946. 

\harvarditem{Nagamine and Yamazaki}{1974}{naga74} Nagamine,~K., and
T.~Yamazaki, 1974, Nucl.~Phys. A {\bf 219}, 104.

\harvarditem{Nagamine}{1996}{naga96} Nagamine,~K., 1996, Nucl.~Phys. B
(Proc. Suppl.) {\bf 51A}, 115.

\harvarditem{Nakamura,~{\em et al.}}{1998}{naka98} Nakamura, S.,~{\em et
al.}, 1998, A research proposal to RIKEN-RAL (R77), ``Precise
Measurement of the $\mu^{+}$ Lifetime and Test of the Exponential
Law''.

\harvarditem{Neddermeyer and Anderson}{1937}{nedd37}Neddermeyer,~S.H,
and C.D.Anderson, 1937, Phys.~Rev. {\bf 51}, 884.

\harvarditem{Ni,~{\em et al.}}{1993}{ni93} Ni,~B., {\em et al.},
1993, Phys.~Rev. D {\bf 48}, 1976.

\harvarditem{Nilles}{1984}{nill84} Nilles, H.P., 1984,
Phys.~Rep. {\bf 110}, 1.

\harvarditem{Nir and Seiberg}{1993}{nir93} Nir,~Y., and N.~Seiberg,
1993, Phys.~Lett. B {\bf 309}, 337.

\harvarditem{Nishijima}{1957}{nish57} Nishijima, K., 1957, Phys.~Rev.
{\bf 108}, 907. 

\harvarditem{Okada,~{\em et al.}}{1998}{okad98} Okada,~Y., K.~Okamura,
and Y.~Shimizu, 1998, Phys.~Rev. D {\bf 58}, 051901.

\harvarditem{Okada,~{\em et al.}}{1999}{okad99} Okada,~Y., K.~Okamura,
and Y.~Shimizu, 1999, KEK Preprint KEK-TH-623, ``\mueg and \meee
Processes with Polarized Muons and Supersymmetric Grand Unified
Theories'', hep-ph/9906446.

\harvarditem{Otten and Weinheimer}{1998}{otte98} Otten, ~E.W. and
Ch. Weinheimer, 1998, in {\em Proceedings of the First International
Symposium on Lepton and Baryon Number Violation}, edited by
H.V.~Klapdor-Kleingrothaus and I.V.~Krivosheina (Institute of Physics
Publishing, Bristol and Philadelphia), p.309.

\harvarditem{Particle Data Group}{1998}{pdg98} Particle Data Group,
1998, Euro. Phys. Journal, {\bf 3}, 1.

\harvarditem{Petcov}{1977}{petc77} Petcov,~S.T., 1977, Yad. Fiz. {\bf
25}, 641 [1977, Sov. J. Nucl. Phys. {\bf 25}, 340].

\harvarditem{Pich and Silva}{1995}{pich95} Pich,~A., and J.P.~Silva,
1995, Phys.~Rev. D {\bf 52}, 4006.

\harvarditem{Pich}{1997}{pich97} Pich,~A., 1997, in {\em Proceedings
of NATO Advanced Study Institute on Masses of Fundamental Particles},
edited by M.~Levy, J.~Iliopoulos, R.~Gastmans, J.-M.~Gerard (New York,
Plenum, NATO ASI series B, Physics 363), p.173.

\harvarditem{Pontecorvo}{1957}{pont57} Pontecorvo, B., 1957,
Zh. Eksp. Teor. Fiz. {\bf 33} 549 [1958, Sov. Phys. JETP, {\bf 6}, 429].

\harvarditem{Porter and Primakoff}{1951}{port51} Porter, C.E., and
H. Primakoff, 1951, Phys.~Rev. {\bf 83}, 849.

\harvarditem{Pratt}{1958}{prat58} Pratt,~R., 1958, Phys.~Rev. {\bf
111}, 646.

\harvarditem{Raidal and Santamaria}{1998}{raid98} Raidal, M., and
A. Santamaria, 1998, Phys. Lett. {\bf B421}, 250.

\harvarditem{Raidal}{1998b}{raid98b} Raidal, M., 1998b, Phys.~Rev.~D
{\bf 57}, 2013.

\harvarditem{Rattazzi and Sarid}{1996}{ratt96} Rattazzi, R., and
U.~Salid, 1996, Nucl.~Phys. {\bf B475}, 27.

\harvarditem{Sakai}{1981}{saka81} Sakai,~N., 1981, Z.~Phys. {\bf C11},
153.

\harvarditem{Scheck}{1978}{sche78} Scheck, F., 1978, Phys. Rep. {\bf
44}, 187. 

\harvarditem{Schwinger}{1957}{schw57} Schwinger, J., 1957, Ann.~Phys.
{\bf 2}, 407. 

\harvarditem{Shanker}{1979}{shan79} Shanker, O., 1979, Phys. Rev. {\bf
D20}, 1608.

\harvarditem{Shanker}{1982}{shan82} Shanker, O., 1982, Phys. Rev. {\bf
D25}, 1847.

\harvarditem{Shanker and Roy}{1997}{shan97} Shanker, O., and R. Roy,
1997, Phys. Rev. {\bf D55}, 7307.

\harvarditem{Skrinskii and Parkhomuchuk}{1981}{skri81}
Skrinskii,~A.N., and V.V.~Parkhomchuk, 1981, Fiz.~Elem.~Chastits At.
Yadra, {\bf 12}, 557, [1981, Sov. J. Part. Nucl. {\bf 12} 223].

\harvarditem{Steinberger}{1948}{stei48}Steinberger, J., 1948,
Phys.~Rev. {\bf 74}, 500. 

\harvarditem{Steinberger and Wolfe}{1955}{stei55}Steinberger, J., and
H.B. Wolfe, 1955, Phys.~Rev. {\bf 100}, 1490.

\harvarditem{Swartz}{1989}{swar89} Swartz, M.L, 1989, 
Phys.~Rev. D {\bf 40}, 1521.

\harvarditem{Treiman,~{\em et al.}}{1977}{trei77} Treiman,~S.B.,
F.~Wilczek, and A.~Zee, 1977, Phys.~Rev. D {\bf 16}, 152.

\harvarditem{Van der Schaaf,~{\em et al.}}{1980}{vand80}Van der Schaaf,
A.,~{\em et al.}, 1980, Nucl. Phys. A {\bf 340}, 249.

\harvarditem{Van der Schaaf}{1993}{vand93}Van der Schaaf,
A., 1993, Progress in Particle and Nuclear Physics {\bf 31}, 1.

\harvarditem{Van Hove,~{\em et al.}}{1997}{vanh97} Van Hove,~P., {\em
et al.}, 1997, An experimental proposal R-97-06 to PSI, ``A Precision
Measurement of the Michel Parameter $\xi^{''}$ in Polarized Muon
Decay''.

\harvarditem{Van Ritbergen and Stuart}{1999}{ritb99} Van Ritbergen,
T., and R.G. Stuart, 1999, Phys.~Rev.~Lett. {\bf 82}, 488.

\harvarditem{Vergados}{1986}{verg86}Vergados, J.D., 1986, Phys.~Rep.
{\bf 133} 1. 

\harvarditem{Yanagida}{1979}{yana79}Yanagida, T., 1979, in {\em
Proceedings of Workshop on Unified Theory and Baryon Number in the
Universe}, edited by O. Sawada and A. Sugamoto.

\harvarditem{Yukawa}{1935}{yuka35}Yukawa, H., 1935, Prog. Phys. Math.
Soc. Japan, {\bf 17}, 48.

\harvarditem{Watanabe,~{\em et al.}}{1987}{wata87} Watanabe,~R.,
M.~Fukui, H.~Ohtsubo, and M.~Morita, 1987, Prog. Theor. Phys. {\bf
78}, 114.

\harvarditem{Watanabe,~{\em et al.}}{1993}{wata93} Watanabe,~R.,
K.~Muto, T.~Oda, T.~Niwa, H.~Ohtsubo, R.~Morita, and M.~Morita, 1993,
Atomic Data and Nucl. Data Table, {\bf 54}, 165.

\harvarditem{Weinberg and Feinberg}{1959}{wein59} Weinberg, S., and
G. Feinberg, 1959, Phys.~Rev.~Lett. {\bf 3}, 111, and Erratum: 1959,
{\em ibid.}, 244.

\harvarditem{Wilczek and Zee}{1977}{wilc77} Wilczek,~F., and A.~Zee,
1977, Phys.~Rev.~Lett. {\bf 38}, 531.

\harvarditem{Willmann and Jungmann}{1998}{will98} Willmann,~L., and
K.P.~Jungmann, 1998, Physics {\bf 499}, 43.

\harvarditem{Willmann,~{\em et al.}}{1999}{will99} Willmann, L.,~{\em et
al.}, 1999, Phys.~Rev.~Lett. {\bf 82}, 49.

\harvarditem{Wintz}{1998}{wint98} Wintz,~P., 1998, in {\em Proceedings of
the First International Symposium on Lepton and Baryon Number
Violation}, edited by H.V.~Klapdor-Kleingrothaus and I.V.~Krivosheina
(Institute of Physics Publishing, Bristol and Philadelphia), p.534.

\harvarditem{Wolfenstein}{1978}{wolf78} Wolfenstein, L., 1978,
Phys.~Rev. D {\bf 17}, 2369.

\harvarditem{Zee}{1985}{zee85} Zee,~A., 1985, Phys.~Rev.~Lett. {\bf
55}, 2382.

\end{references}
\end{document}